\documentclass[11pt]{article}

\title{8D $\mathcal{N}=2$ SUGRA}

\global\arraycolsep=1pt
\usepackage[colorlinks=true,backref=true,linkcolor=black,anchorcolor=black,citecolor=black,filecolor=black,menucolor=black,pagecolor=black,urlcolor=black]{hyperref}

\usepackage{amsmath}
\usepackage{amsthm}
\usepackage{amssymb}
\usepackage{mathrsfs}
\usepackage[Symbol]{upgreek}
\usepackage{dsfont}
\usepackage[vcentermath]{youngtab}
\usepackage{textcomp}

\usepackage{tikz}
\usetikzlibrary{arrows}
\usepackage{pgfplots}

\usepackage{latexsym}
\usepackage{graphicx}

\newcommand{\eprint}[1]{{\href{http://arxiv.org/abs/#1}{\texttt{[#1}]}}}
\newcommand{\eprintN}[1]{{\href{http://arxiv.org/abs/#1}{\texttt{#1 [hep-th]}}}}

\def\DJo{$\;$\kern-.4em \hbox{D\kern-.8em\raise.15ex\hbox{--}\kern.35em okovi\'c}}

\def\DEVIII#1#2#3#4#5#6#7#8{{\tiny $ { \left[ \begin{array}{ccccccc}  & & \mathfrak{#2} \hspace{-1.2mm}&&&& \vspace{ -1.2mm} \\ \mathfrak{#1}\hspace{-1.2mm} &  \mathfrak{#3} \hspace{-1.2mm}& \mathfrak{#4} \hspace{-1.2mm} & \mathfrak{#5}\hspace{-1.2mm}&\mathfrak{#6}\hspace{-1.2mm}&\mathfrak{#7}\hspace{-1.2mm}&\mathfrak{#8} \end{array}\right] }$}}
\def\DSOXVI#1#2#3#4#5#6#7#8{{\tiny $ {  \vspace{-2mm} \left[ \begin{array}{ccccccccc}  && \mathfrak{#8} \hspace{-1.2mm}&&&&&& \vspace{ -1.3mm} \\ \cdot \hspace{-1.0mm}& \mathfrak{#7}\hspace{-1.2mm} &\mathfrak{#6}\hspace{-1.2mm} &  \mathfrak{#5} \hspace{-1.2mm}& \mathfrak{#4} \hspace{-1.2mm} & \mathfrak{#3}\hspace{-1.2mm}&\mathfrak{#2}\hspace{-1.2mm}&\mathfrak{#1}  \hspace{-1.2mm}\end{array}\right] }$}}

\def\DEVII#1#2#3#4#5#6#7{{\tiny $ { \left[ \begin{array}{cccccc}  & & \mathfrak{#2} \hspace{-1.2mm}&&& \vspace{ -1.0mm} \\ \mathfrak{#1}\hspace{-1.2mm} &  \mathfrak{#3} \hspace{-1.2mm}& \mathfrak{#4} \hspace{-1.2mm} & \mathfrak{#5}\hspace{-1.2mm}&\mathfrak{#6}\hspace{-1.2mm}&\mathfrak{#7} \end{array}\right] }$}}
\def\DSOXII#1#2#3#4#5#6#7{{\tiny $ {   \left[ \begin{array}{ccccccc}  && & \mathfrak{#6} \hspace{-1.2mm}&&& \vspace{ -1.5mm} \\  \mathfrak{#1}\hspace{-1.2mm} &\mathfrak{#2}\hspace{-1.2mm} &  \mathfrak{#3} \hspace{-1.2mm}& \mathfrak{#4} \hspace{-1.2mm} & \mathfrak{#5}\hspace{-1.2mm}&\cdot \hspace{-0.5mm}& \mathfrak{#7} \end{array}\right] }$}}

\def\DSOX#1#2#3#4#5{{\tiny $ {   \biggl[ \begin{array}{ccc}  &&\mathfrak{#3}  \vspace{ -1.5mm} \\  \mathfrak{#1}\hspace{0.2mm}\mathfrak{#2}\hspace{-0.6mm} &\mathfrak{#4} \hspace{-0.9mm}&\vspace{-1.5mm}\\ && \mathfrak{#5}  \end{array}\biggr] }$}}

\def\WSOXVI#1#2#3#4#5#6#7#8{{\tiny $ {   \biggl[ \begin{array}{ccc}  &&\mathfrak{#3}  \vspace{ -1.5mm} \\  \mathfrak{#6}\hspace{0.2mm}  \mathfrak{#2}\hspace{0.2mm}  \mathfrak{#3}\hspace{0.2mm}  \mathfrak{#4}\hspace{0.2mm}\mathfrak{#5}\hspace{-0.6mm} &\mathfrak{#7} \hspace{-0.9mm}&\vspace{-1.5mm}\\ && \mathfrak{#8}  \end{array}\biggr] }$}}
\def\WSOXVIn#1#2#3#4#5#6#7#8{{\tiny $ {   \biggl[ \begin{array}{ccc}  &&\mathfrak{#3}  \vspace{ -1.5mm} \\  \mathfrak{#6}\hspace{0.2mm}  \mathfrak{#2}\hspace{0.2mm}  \mathfrak{#3}\hspace{0.2mm}  \mathfrak{#4}\hspace{0.2mm}\mathfrak{#5}\hspace{-0.6mm} &\mathfrak{#7} \hspace{-0.9mm}&\vspace{-1.5mm}\\ && {#8}  \end{array}\biggr] }$}}
\def\WSOXVInk#1#2#3#4#5#6#7#8{{\tiny $ {   \biggl[ \begin{array}{ccc}  &&\mathfrak{#3}  \vspace{ -1.5mm} \\  \mathfrak{#6}\hspace{0.2mm}  \mathfrak{#2}\hspace{0.2mm}  \mathfrak{#3}\hspace{0.2mm}  {#4}\hspace{0.2mm}\mathfrak{#5}\hspace{-0.6mm} &\mathfrak{#7} \hspace{-0.9mm}&\vspace{-1.5mm}\\ && {#8}  \end{array}\biggr] }$}}
\def\DSON#1#2#3#4#5#6#7#8{{\tiny $ {   \left[ \begin{array}{cccccccc}  &&&&& & \mathfrak{#7} \hspace{-1.2mm}& \vspace{ -1.5mm} \\  \mathfrak{#1}\hspace{-1.2mm} &\mathfrak{#2}\hspace{-1.2mm} &  \mathfrak{#3} \hspace{-1.2mm}& \mathfrak{#4} \hspace{-1.2mm} & \cdot \hspace{-0.8mm}\cdot\hspace{-0.8mm}\cdot \hspace{-1.2mm} & \mathfrak{#5}\hspace{-1.2mm}& \mathfrak{#6}\hspace{-1.2mm} & \mathfrak{#8} \end{array}\right] }$}}

\def\SU{SU_{\hspace{-0.5mm}\scriptscriptstyle \rm c}(8)}
\def\Sp{Sp_{\scriptscriptstyle \rm c}(4)}

\def\Evector#1{E_{\scriptscriptstyle [#1\mathfrak{0}\mathfrak{0}\mathfrak{0}]}}
\def\Etensor#1{E_{\scriptscriptstyle [\mathfrak{0}\mathfrak{0}#1\mathfrak{0}]}}
\def\EA#1{E_{\scriptscriptstyle [#1\mathfrak{0}]}}

\def\EiEVII#1{{E_{\fontsize{6.35pt}{6pt}\selectfont   \left[ \begin{array}{cccccc}  & & \mathfrak{0} \hspace{-0.6mm}&&& \vspace{ -1.0mm} \\ #1 \hspace{-0.6mm} &  \mathfrak{0} \hspace{-0.6mm}& \mathfrak{0} \hspace{-0.6mm} & \mathfrak{0}\hspace{-0.6mm}&\mathfrak{0}\hspace{-0.6mm}&\mathfrak{0} \end{array}\right] \fontsize{12.35pt}{12pt}\selectfont }}}

\def\EiEVI#1{{E_{\fontsize{6.35pt}{6pt}\selectfont   \left[ \begin{array}{cccccc}  & & \mathfrak{0} \hspace{-0.6mm}&& \vspace{ -1.0mm} \\ #1 \hspace{-0.6mm} &  \mathfrak{0} \hspace{-0.6mm}& \mathfrak{0} \hspace{-0.6mm} & \mathfrak{0}\hspace{-0.6mm}&\mathfrak{0} \end{array}\right] \fontsize{12.35pt}{12pt}\selectfont }}}

\def\EiEVIItxt#1{{E{\fontsize{6.35pt}{6pt}\selectfont   \left[ \begin{array}{cccccc}  & & \mathfrak{0} \hspace{-0.6mm}&&& \vspace{ -1.0mm} \\ #1 \hspace{-0.6mm} &  \mathfrak{0} \hspace{-0.6mm}& \mathfrak{0} \hspace{-0.6mm} & \mathfrak{0}\hspace{-0.6mm}&\mathfrak{0}\hspace{-0.6mm}&\mathfrak{0} \end{array}\right] \fontsize{12.35pt}{12pt}\selectfont }}}

\newfont{\bbbold}{msbm10 scaled \magstep1}

\def\cA{{\cal A}}

\def\cD{{\cal D}}
\def\cE{{\cal E}}
\def\cF{{\cal F}}
\def\cG{{\cal G}}

\def\cJ{{\cal J}}

\def\cL{{\cal L}}

\def\cN{{\cal N}}
\def\cO{{\cal O}}

\def\cT{{\cal T}}

\def\cV{{\cal V}}

\newfont{\goth}{eufm10 scaled \magstep1}

\def\gl{\mbox{\goth l}}

\def\adt{{\dot \alpha}}
\def\bdt{{\dot \beta}}
\def\cdt{\dot\gamma}
\def\ddt{\dot\delta}
\def\e{ \varepsilon}\def\ve{\varepsilon}

\def\be{\begin{equation}}\def\ee{\end{equation}}
\def\bea{\begin{eqnarray}}\def\eea{\end{eqnarray}}
\def\barr{\begin{array}}\def\earr{\end{array}}

\let\la=\label

\def\nn{\nonumber}
\def\bd{\begin{document}}
\def\ed{\end{document}}
\def\ba{\begin{array}}
\def\ea{\end{array}}
\def\bea{\begin{eqnarray}}
\def\eea{\end{eqnarray}}
\def\ft#1#2{{\frac{\scriptstyle #1}{\scriptstyle #2}}}
\def\fft#1#2{\frac{#1}{#2}}
\def\sst#1{{\scriptscriptstyle #1}}
\def\oneone{\rlap 1\mkern4mu{\rm l}}
\def\DEVIII#1#2#3#4#5#6#7#8{{\tiny $ { \left[ \begin{array}{ccccccc}  & & \mathfrak{#2} \hspace{-0.7mm}&&&& \vspace{ -1.5mm} \\ \mathfrak{#1}\hspace{-0.7mm} &  \mathfrak{#3} \hspace{-0.7mm}& \mathfrak{#4} \hspace{-0.7mm} & \mathfrak{#5}\hspace{-0.7mm}&\mathfrak{#6}\hspace{-0.7mm}&\mathfrak{#7}\hspace{-0.7mm}&\mathfrak{#8} \end{array}\right] }$}}
\def\DSOXVI#1#2#3#4#5#6#7#8{{\tiny $ {  \vspace{-2mm} \left[ \begin{array}{ccccccccc}  && \mathfrak{#8} \hspace{-0.7mm}&&&&&& \vspace{ -1.5mm} \\ \cdot \hspace{-0.5mm}& \mathfrak{#7}\hspace{-0.7mm} &\mathfrak{#6}\hspace{-0.7mm} &  \mathfrak{#5} \hspace{-0.7mm}& \mathfrak{#4} \hspace{-0.7mm} & \mathfrak{#3}\hspace{-0.7mm}&\mathfrak{#2}\hspace{-0.7mm}&\mathfrak{#1} \end{array}\right] }$}}

\def\DEVII#1#2#3#4#5#6#7{{\tiny $ { \left[ \begin{array}{cccccc}  & & \mathfrak{#2} \hspace{-0.7mm}&&& \vspace{ -1.5mm} \\ \mathfrak{#1}\hspace{-0.7mm} &  \mathfrak{#3} \hspace{-0.7mm}& \mathfrak{#4} \hspace{-0.7mm} & \mathfrak{#5}\hspace{-0.7mm}&\mathfrak{#6}\hspace{-0.7mm}&\mathfrak{#7} \end{array}\right] }$}}
\def\DSOXII#1#2#3#4#5#6#7{{\tiny $ {   \left[ \begin{array}{ccccccc}  && & \mathfrak{#6} \hspace{-0.7mm}&&& \vspace{ -1.5mm} \\  \mathfrak{#1}\hspace{-0.7mm} &\mathfrak{#2}\hspace{-0.7mm} &  \mathfrak{#3} \hspace{-0.7mm}& \mathfrak{#4} \hspace{-0.7mm} & \mathfrak{#5}\hspace{-0.7mm}&\cdot \hspace{-0.5mm}& \mathfrak{#7} \end{array}\right] }$}}

\def\DSON#1#2#3#4#5#6#7#8{{\tiny $ {   \left[ \begin{array}{cccccccc}  &&&&& & \mathfrak{#7} \hspace{-0.7mm}& \vspace{ -1.5mm} \\  \mathfrak{#1}\hspace{-0.7mm} &\mathfrak{#2}\hspace{-0.7mm} &  \mathfrak{#3} \hspace{-0.7mm}& \mathfrak{#4} \hspace{-0.7mm} & \cdot \hspace{-0.8mm}\cdot\hspace{-0.8mm}\cdot \hspace{-0.7mm} & \mathfrak{#5}\hspace{-0.7mm}& \mathfrak{#6}\hspace{-0.7mm} & \mathfrak{#8} \end{array}\right] }$}}

\def\DlacedLeft{{\fontsize{0.004pt}{0.0005pt}\selectfont  \scriptscriptstyle \mbox{$= \hspace{-2.2mm}  \langle$} \fontsize{12pt}{14.5pt}\selectfont }}

\def\DSpIV#1#2#3#4{{\tiny $ { \left[   \mathfrak{#1}\,  \mbox{-} \mathfrak{#2} \, \mbox{-} \mathfrak{#3} \hspace{-0.2mm}\DlacedLeft \hspace{0.2mm} \mathfrak{#4} \right] }$}}

\def\DSOVIII#1#2#3#4{{\tiny $ {   \biggl[ \begin{array}{ccc}  &&\mathfrak{#3}  \vspace{ -1.5mm} \\  \mathfrak{#1}\hspace{-0.6mm} &\mathfrak{#2} \hspace{-0.9mm}&\vspace{-1.5mm}\\ && \mathfrak{#4}  \end{array}\biggr] }$}}

\newcommand{\eq}[1]{(\ref{#1})}
\newcommand{\w}[1]{\\[0.#1cm]}
\def\eqs#1#2{(\ref{#1}-\ref{#2})}
\def\det{{\rm det\,}}
\def\tr{{\rm tr}}

\def\gl{\mathfrak{gl}}
\def\sl{\mathfrak{sl}}
\def\so{\mathfrak{so}}
\def\su{\mathfrak{su}}
\def\sp{\mathfrak{sp}}
\def\e{\mathfrak{e}}

\newcommand{\hoch}[1]{$\, ^{#1}$}
\newcommand{\imperial}{\it\small Theoretical Physics Group, Imperial College London\\ Prince Consort Road, London SW7 2AZ, UK}
\newcommand{\kings}
{\it\small Department of Mathematics, King's College, University of London\\ Strand, London WC2R 2LS, UK}
\newcommand{\uu}
{\it\small Department of Theoretical Physics, Uppsala, Sweden}
\newcommand{\hip}
{\it\small HIP-Helsinki Institute of Physics, P.O. Box 64 FIN-00014
University of Helsinki, Suomi-Finland}
\newcommand{\stock}
{\it\small Department of Theoretical Physics, Stockholm, Sweden}
\newcommand{\cpht}
{\it\small Centre de Physique Th{\'e}orique, Ecole Polytechnique, CNRS\\ 91128 Palaiseau Cedex, France}
\makeatletter
\renewcommand\theequation{\thesection.\arabic{equation}}
\@addtoreset{equation}{section} \makeatother

\newcommand{\sa}{/ \hspace{-1.2ex}}
\newcommand{\saa}{/ \hspace{-1.4ex}}
\newcommand{\saaa}{\, / \hspace{-1.6ex}}
\newcommand{\Scal}[1]{\Bigl ({#1} \Bigr )}
\newcommand{\scal}[1]{\bigl ({#1} \bigr )}

\newcommand{\CR}{\nonumber \\*}

\newcommand{\trace}{\hbox {tr}~}
\newcommand{\traceS}{\hbox {tr}_{\scriptscriptstyle \mathfrak{S}}~}

\DeclareMathAlphabet{\mathpzc}{OT1}{pzc}{m}{it}
\def\BRST{\,\mathpzc{s}\,}
\def\aBRST{{\scriptstyle (\mathpzc{s})}}
\def\q{{{\scriptscriptstyle (Q)}}}
\def\qs{{\scriptscriptstyle (Q\mathpzc{s})}}
\def\Qsla{{\mathcal{S}_{\q}}}
\def\Slav{{\mathcal{S}_\aBRST}}
\def \varepsilonb{{\overline{ \varepsilon}}}
\def\bulletup{{\scriptstyle \bullet}}

\newcommand{\grad}[3]{{\scriptscriptstyle (#1 , #2, #3 )}}
\newcommand{\gra}[2]{{\scriptscriptstyle (#1 , #2 )}}
\newcommand{\ord}[1]{{\scriptscriptstyle (#1)}}

\def\cL{{\cal L}}
\def\cN{\mathcal{N}}
\def\cO{\mathcal{O}}

\def\ie{{\it i.e.}\ }
\def\eg{{\it e.g.}\ }

\newcommand{\sfrac}[2]{{\scriptstyle \frac{#1}{#2}}}
\newcommand{\stfrac}[2]{{\scriptscriptstyle \frac{#1}{#2}}}

 \def\balpha{{\overline{\alpha}}}
 \def\bbeta{{\overline{\beta}}}
 \def\bgamma{{\overline{\gamma}}}
 \def\bdelta{{\overline{\delta}}}
 \def\bepsilon{{\overline{ \varepsilon}}}
 \def\bvarepsilon{{\overline{\varepsilon}}}
 \def\bzeta{{\overline{\zeta}}}
 \def\bareta{{\overline{\eta}}}
 \def\btheta{{\overline{\theta}}}
 \def\bvartheta{{\overline{\vartheta}}}
 \def\biota{{\overline{\iota}}}
 \def\bkappa{{\overline{\kappa}}}
 \def\blambda{{\overline{\lambda}}}
 \def\bmu{{\overline{\mu}}}
 \def\bnu{{\overline{\nu}}}
 \def\bxi{{\overline{\xi}}}
 \def\bpi{{\overline{\pi}}}
 \def\brho{{\overline{\rho}}}
 \def\bvarrho{{\overline{\varrho}}}
 \def\bsigma{{\overline{\sigma}}}
 \def\bvarsigma{{\overline{\varsigma}}}
 \def\btau{{\overline{\tau}}}
 \def\bphi{{\overline{\phi}}}
 \def\bvarphi{{\overline{\varphi}}}
 \def\bchi{{\overline{\chi}}}
 \def\bpsi{{\overline{\psi}}}
 \def\bomega{{\overline{\omega}}}

\def\thalf{{\textrm{\tiny\textonehalf}}}
\def\tquarter{{\textrm{\tiny\textonequarter}}}
\def\Ko{{\scriptscriptstyle K}}
\def\tKo{\scriptscriptstyle k }
\def\N{{\mathcal{N}}}
\def\csN{{\fontsize{9.35pt}{9pt}\selectfont \mbox{$\cN$} \fontsize{12.35pt}{12pt}\selectfont }}
\def\cssN{{\fontsize{6.35pt}{6pt}\selectfont \mbox{$\cN$} \fontsize{12.35pt}{12pt}\selectfont }}
\def\csssN{{\fontsize{4.35pt}{4pt}\selectfont \mbox{$\cN$} \fontsize{12.35pt}{12pt}\selectfont }}

\def\ai{{\hat{\imath}}}
\def\aj{{\hat{\jmath}}}
\def\ak{{\hat{k}}}

\def\inv{{ \scriptscriptstyle \rm{\mbox{\tiny-1}}}}
\def\un{{\mathpzc{1}}}
\def\deux{{\mathpzc{2}}}
\def\trois{{\mathpzc{3}}}
\def\quatre{{\mathpzc{4}}}
\def\cinq{{\mathpzc{5}}}

\newcommand{\red}[1]{ {\color{red} #1 }} 
\newcommand{\blue}[1]{{\color{blue} #1 }}
\newcommand{\green}[1]{{\color{green} #1 }}
\newcommand{\bleu}[1]{ {\color{cyan} #1 }} 

\renewcommand{\thefootnote}{\arabic{footnote}}

\usepackage{color}
\usepackage{colordvi}
\definecolor{mygreen}{rgb}{0,0.75,0}

\def\guillaume#1#2{{\color{blue} G: #1 \color{black}}{\scriptsize #2}}
\def\valentin#1#2{{\color{mygreen} V: #1 \color{black}}{\scriptsize #2}}

\begin{document}

\renewcommand{\thefootnote}{\arabic{footnote}}
\setcounter{footnote}{0}

\allowdisplaybreaks[1]
\renewcommand{\thefootnote}{\fnsymbol{footnote}}
\def\corr{$\spadesuit $}
\def\trefle{ $\clubsuit$}
\begin{titlepage}
\begin{flushright}
CPHT-RR038.0614\\
\end{flushright}

\bigskip
\bigskip
\centerline{\Large \bf Minimal unitary representations from supersymmetry}
\centerline{\Large \bf }
\bigskip
\bigskip
\centerline{{\bf Guillaume Bossard and Valentin Verschinin}}
\bigskip
\centerline{Centre de Physique Th\'eorique, Ecole Polytechnique, CNRS}
\centerline{91128 Palaiseau cedex, France \footnote{email: bossard@cpht.polytechnique.fr,  valentin.verschinin@cpht.polytechnique.fr}}
\bigskip
\bigskip

\begin{abstract}
We compute the supersymmetry constraints on the $R^4$ type corrections in maximal supergravity in dimension 8, 6, 4 and 3, and determine the tensorial differential equations satisfied by the function of the scalar fields multiplying the $R^4$ term in the corresponding invariants. The second-order derivative of this function restricted to the Joseph ideal vanishes in dimension lower than six. These results are extended to the $\nabla^4 R^4$ and the $\nabla^6 R^4$ corrections, based on the harmonic superspace construction of these invariants in the linearised approximation. We discuss the solutions of these differential equations and analysis the consequences on the non-perturbative type II  low energy string theory effective action.

\end{abstract}

\end{titlepage}
\renewcommand{\thefootnote}{\arabic{footnote}}
\setcounter{footnote}{0}

\tableofcontents 

\setcounter{page}{1}

\section{Introduction}
Type II string theory on $\mathds{R}^{1,9-d} \times T^{d}$ is extremely constrained by supersymmetry and duality symmetries. The various formulations of the theory are conjectured to be related by U-duality, an arithmetic $E_{d(d)}(\mathds{Z})$ subgroup of the split real form of the Lie group of type $E_{d}$ \cite{Hull:1994ys}. In particular, the exact low energy expansion of the effective action is expected to exhibit this symmetry \cite{Green:1997tv,Green:1997as,Kiritsis:1997em}. However there is no non-perturbative formulation of superstring theory that would permit to  derive directly the low energy expansion of the amplitudes, and one must use perturbative string theory \cite{D'Hoker:2005jc,Green:2008uj,Gomez:2013sla,Green:2014yxa,D'Hoker:2014gfa} and eleven-dimensional supergravity \cite{Green:1997as,Green:1999pu} together with U-duality to derive their  non-perturbative completion. One can deduce the superstring effective action from the amplitude by inverse Legendre transform  (up to field redefinition ambiguities), which can then be expressed in the low energy limit as the supergravity 1PI generating functional computed with the complete (appropriately renormalised) string theory Wilsonian effective action. The supersymmetric Wilsonian effective action admits the following expansion in the reduced Newton constant $\kappa^2$ in $10-d$ dimensions 
\be S = \frac{1}{\kappa^2} S^\ord{0} + \kappa^{2 \frac{d-2}{8-d}} S^\ord{3}[ \cE_\gra{0}{0}] + \kappa^{2 \frac{d+2}{8-d}} S^\ord{5}[ \cE_\gra{1}{0}]  + \kappa^{2 \frac{d+4}{8-d}} S^\ord{6}[ \cE_\gra{0}{1}] + \sum_{n=7}^\infty \kappa^{2 \frac{d-8+2n}{8-d}} S^\ord{n} \ ,\ee 
where $S^\ord{0}$ is the supergravity classical action, and $S^\ord{n+3}[ \cE_\gra{p}{q}]$ with $2p+3q=n$ is a $\partial^{2n} R^4$ type supersymmetric correction to the effective action depending on a function $ \cE_\gra{p}{q}$ of the scalar fields parametrizing the symmetric space  $E_{d(d)}/ K_d$ \cite{Green:2010wi};\footnote{Here the functions $ \cE_\gra{p}{q}$ are defined as in  \cite{Green:2010wi}, up to the subtlety that they are not necessarily U-duality invariant in our conventions when there is a non-trivial mixing with the 1PI generating functional.} although starting from $n\ge 5$ one has independent corrections in $\partial^{2n-2} R^5$ and etcetera at higher orders \cite{Green:2013bza}. 

It was shown in \cite{Green:1998by} that supersymmetry implies that the function $ \cE_\gra{0}{0}$ characterising $S^\ord{3}[ \cE_\gra{0}{0}]$ in type IIB supergravity in ten dimensions is an eigenfunction of the Laplace operator with eigenvalue $-\frac{3}{4}$, consistently with the analysis carried out in \cite{Green:1997tv}. As a consequence, supersymmetry and duality invariance entirely determine the function $\cE_\gra{0}{0}$ in ten dimensions. The constraints from supersymmetry have been computed for higher order invariants \cite{Sinha:2002zr} and the same conclusion holds for the $\nabla^4 R^4$ type corrections \cite{Green:1999pu}. The realisation of these functions as Eisenstein functions \cite{Green:1997tv,Kiritsis:1997em} has been generalised in lower dimensions \cite{Obers:1999um}, and to higher order $\nabla^6 R^4$ type corections \cite{Green:2005ba}, leading to more developments in lower dimensions \cite{Basu:2007ru,Basu:2007ck}. 

We start by considering  $R^4$ invariants in lower dimensions. We carry out this program within the formalism of superforms in superspace developed in \cite{Voronov,Gates:1997kr,Gates:1997ag}. We concentrate in a first section on $R^4$ type invariants in $\cN=2$ supergravity in eight dimensions. Computing the complete invariant is out of reach, and we concentrate on the components of the superform that carry the maximal R-symmetry weight representations, similarly as in \cite{Green:1998by,Basu:2011he}. We find in this way that the function of the scalar fields must satisfy a tensorial second-order differential equation consistent with the explicit Eisenstein function computed in \cite{Kiritsis:1997em}. 

We extend these results in dimension $6,\, 4$ and $3$ and show that the function defining the $R^4$ type invariant satisfies a unique tensorial  second-order differential equation associated to the minimal unitary representations of $SO(5,5)$,\, $E_{7(7)}$ and $E_{8(8)}$, respectively. The function multiplying $R^4$ must satisfy the constraint that its second-order derivative vanishes when restricted to the Joseph ideal \cite{Joseph}
\be \cJ(\cD,\cD) \, \cE_\gra{0}{0} = 0 \ .\label{Joseph}  \ee
The relation between the minimal unitary representations and the $R^4$ type threshold function has been argued from several perspectives \cite{Pioline:2001jn,Kazhdan:2001nx,Gunaydin:2001bt,Pioline:2004xq,Pioline:2010kb} and it is in particular conjectured that the function can be defined as the exceptional theta series associated to the minimal unitary representation of $E_{d(d)}$ \cite{Pioline:2010kb}. Our results strongly support this conjecture by showing that supersymmetry implies  indeed \eqref{Joseph}, whose solutions with appropriate boundary conditions define the minimal unitary representation of the corresponding exceptional group. Using the harmonic superspace construction of the higher order invariants in the linearised approximation \cite{Galperin:1984av,Berkovits:1997pj,Drummond:2003ex,Bossard:2009sy}, we extend these results to the $\nabla^4 R^4$ type invariants. In four dimensions we also determine the equation satisfied by the function defining the $\nabla^6 R^4$ type invariant, relying on properties derived in \cite{Green:2010kv} to fix the free coefficients. We find that the threshold functions satisfy higher order differential equations attached to certain nilpotent coadjoint orbits exhibiting their relation to next to minimal unitary representations as proposed in \cite{Pioline:2010kb}.

We study the corresponding differential equations in some detail in six and four dimensions, and find perfect agreement with the definition of the threshold functions as Eisenstein series \cite{Green:2010wi,Green:2010kv,Green:2011vz,Fleig:2013psa}. We discuss in particular the two Eisenstein functions defining the $\nabla^4 R^4$ type correction in six dimensions \cite{Green:2010kv}, and show that these two functions are associated to two independent invariants, and solve independent differential equations associated to the two next to minimal nilpotent orbits of $D_5$ (that both only include the closure of the minimal nilpotent orbit in their topological closure). Working out the general solutions to these differential equations, we extend the results of  \cite{Green:2011vz} on the structure of the Fourier modes of these functions.

Because the $R^4$ type corrections to the effective action are defined in the linearised approximation as superspace integrals over half of the Grassmann coordinates \cite{Berkovits:1997pj}, the property that they only receive corrections from non-perturbative effects associated to 1/2 BPS instantons has been conjectured to be a consequence of supersymmetry \cite{Green:1997tv}. The differential equation that we find to be a consequence of supersymmetry implies indeed strong restrictions on the possible perturbative corrections that the effective action can receive in string theory, and moreover implies through the dependence on the scalar fields that the non-perturbative corrections associated to instantons must also be 1/2 BPS by supersymmetry. The generalisation of these results for $\nabla^4 R^4$ to only receive corrections from (at least) 1/4 BPS instantons go through as well, in agreement with the analysis carried out in \cite{Green:2011vz}, and the differential equation we propose for the $\nabla^6 R^4$ type invariant in four dimensions implies that it can only receive corrections from (at least) 1/8 BPS instantons, as expected from its harmonic superspace construction in the linearised approximation.

In this paper we distinguish the Wilsonian effective action that preserves local supersymmetry from the 1PI generating functional satisfying to the quadratic BRST master equation. In particular we show that the logarithmic contributions to the threshold functions responsible for the constant right-hand-side in the Poisson equation satisfied by these functions \cite{Green:2010sp}, do not appear in the Wilsonian effective action, but are consequences of duality anomalies. We discuss this property in particular in eight dimensions, where the $R^4$ threshold gets one contribution associated to the chiral 1-loop $U(1)$ anomaly similarly as in four dimensions \cite{Bossard:2013rza}, whereas the second is associated to an incompatibility between supersymmetry and $SL(3,\mathds{R})$ duality invariance. We also exhibit that the $\nabla^4 R^4$ threshold function in six dimensions satisfies a Poisson equation with a right-hand-side proportional to the $R^4$ threshold function, which is attributed to the duality transformation of the $R^4$ superform insertion (\ie form factor) in the supergravity 1PI generating functional. The anomalies associated to the incompatibility between duality and  supersymmetry Ward identities bypass the analysis carried out in \cite{Bossard:2010dq} (although their possible existence was not overlooked), but they can only arise by construction when the threshold function is constrained to satisfy to the Laplace equation (\ie with zero eigenvalue) from supersymmetry Ward identities. Therefore such anomalies can only arise when the supergravity amplitude exhibits a logarithm divergence  \cite{Green:2010sp}, such that they do not affect the non-renormalisation theorems established in \cite{Bossard:2010bd,Beisert:2010jx} regarding the absence of logarithm divergence in $\cN=8$ supergravity before seven-loop order based on the absence of $E_{7(7)}$ anomalies, consistently with the factorisation of eight additional external momenta in the explicit 4-loop four-graviton supergravity amplitude \cite{Bern:2009kd}. Our work does not give new insights on the ultra-violet behaviour of maximal supergravity amplitudes, but it does give predictions on the logarithmic divergences of supersymmetric densities form factors. The integrated invariants are observables of the theory, and therefore the zero momentum limit of the associated form factors are BRST invariant observables. Generalising the argument of \cite{Green:2010sp} to these cases we find that the supersymmetric $R^4$ form factor should diverge at one loop in $\nabla^4R^4$ in six dimensions, and similarly that the $\nabla^4 R^4$ form factor should diverge at one loop in $\nabla^6 R^4$ in four dimensions, whereas the $R^4$ form factor must be finite until 4-loop order by supersymmetry.

The paper is organised in four sections devoted to the analysis of maximal supergravity in eight, six, four and three dimensions, respectively. It is in eight dimensions that we work out the supersymmetry constraints on the $R^4$ type invariants in most detail. For this purpose we start by deriving the superspace geometry, including cubic terms in the fermions that are relevant to our analysis. The latter can be found in Appendix \ref{appendix3/2Dim}. From six dimensions and below, the algebraic constraints on the consistent second-order differential equations on $E_{d(d)}/ K_d$ are so strong that it is enough to work out the supersymmetry constraints on the maximal R-symmetry weight terms of order sixteen in the  fermion fields to determine them. This is due to the property that \eqref{Joseph} determines uniquely the eigenvalue of the Laplace operator.

More generally we find that the differential equations satisfied by the scalar pre-factors of the $ R^4$, $\nabla^4 R^4$ and $\nabla^6 R^4$ type invariants can be deduced from their harmonic superspace construction in the linearised approximation, up to a potential free parameter that is fixed for $R^4$ and $\nabla^4 R^4$ in dimension lower than six. The harmonic variables parametrise a homogeneous space $K_d/(U(1)\times H_{d})$ where the $U(1)$ factor determines the G-analytic superfield $W$ as the component of the scalar field  of highest $U(1)$ weight. The harmonic superspace integrands are therefore in one to one correspondence with the symmetric order $n$ monomials in the G-analytic superfield, that are associated to a set of irreducible representations $R_{d,n,k}$ of $K_d$. The algebraic restriction on the symmetric monomials of the G-analytic superfield define a subspace of the vector space of monomials of a generic coset element. Assuming that the non-linear invariants are in one to one correspondence with the harmonic superspace integral invariants, the same restriction must hold on the jet space of $n^{\rm th}$ order derivative acting on a generic function $ \cE$ defining these invariants, \ie 
\be \cD^n \cE_\gra{p}{q} \in \sum_{k} R_{d,n,k} \ . \ee
This assumption is justified in four dimensions by the complete classification of $SU(2,2|8)$ chiral primary operators \cite{Drummond:2003ex,Drummond:2010fp}, which proves that all supersymmetry invariants are realised as harmonic superspace integrals. Although there is no theorem, is seems that all supersymmetry invariants can indeed be defined as harmonic superspace integrals in the linearised approximation in dimension lower than six.\footnote{From seven dimensions and above there are counter examples, and one must at least consider Lorentzian harmonics to cover all possible invariants \cite{Bossard:2009sy}.} This $U(1)$ factor lies inside a $GL(1,\mathds{C})$ subgroup of the complexication of $K_d$ that determines a graded decomposition of the complex Lie algebra $\mathfrak{k}_d(\mathds{C})$ as well as $\mathfrak{e}_d$. The highest weight component of $\mathfrak{e}_d\ominus \mathfrak{k}_d(\mathds{C})$ determines a nilpotent element, that characterises a unique nilpotent orbit of the real Lie group $E_{d(d)}$ according to the Kostant--Sekiguchi correspondence \cite{Sekiguchi}. It follows that a nilpotent element ${\bf Q}$ satisfies an algebraic constraint that is such that 
\be {\bf Q}^{\otimes n} \in \sum_{k} R_{d,n,k}(\mathds{C}) \ . \ee
We conclude that the same algebraic constraint satisfied by the nilpotent element ${\bf Q}$ is satisfied by the symmetrised product of derivatives acting on $\cE_\gra{p}{q}$. For the $R^4$ type invariant, the relevant nilpotent orbit is always the minimal nilpotent orbit of $E_{d(d)}$, and the quadratic algebraic constraint is the Joseph ideal \cite{Joseph}. In general the solutions to the corresponding differential equation with the appropriate boundary conditions define the unitary representation associated to the corresponding nilpotent orbits. Because the nilpotent orbits are classified by the $K_d(\mathds{C})$ weighted Dynkin diagram characterising the subgroup $GL(1,\mathds{C})$, it is straightforward to read of the nilpotent orbit associated to a given harmonic superspace in the classification \cite{Levi}. For $E_{6(6)},\, E_{7(7)},\, E_{8(8)}$ the 1/2 BPS and 1/4 BPS couplings correspond to the minimal and next to minimal nilpotent orbits, which $K_d$ weighted Dynkin diagram carry zeros on the maximal semi-simple $H_d$  subgroup Dynkin diagram and 1 on the other nodes. The 1/8 BPS couplings correspond to the nilpotent orbits which $K_d$ weighted Dynkin diagram carry zeros on the maximal semi-simple $H_d$  subgroup Dynkin diagram and 2 on the other nodes. 

\section{$\cN=2$ supergravity in eight dimensions}
In this section we shall discuss the $R^4$ type invariants in $\cN=2$ supergravity in eight dimensions, and prove that the $R^4$ threshold function must satisfy differential equations consistent with the explicit $SL(2,\mathds{Z})\times SL(3,\mathds{Z})$ threshold computed in \cite{Kiritsis:1997em}. We will consider the problem in the superspace formulation of the theory, and we shall therefore compute the geometrical tensors of $\cN=2$ supergravity in superspace in a first subsection. Our strategy is inspired from the idea proposed in \cite{Green:1998by} to concentrate on the fermion monomials of maximal weight, as was used in  \cite{Basu:2011he} in eight dimensions. However we will go beyond this results, and show that the function satisfies a stronger equation than the Laplace equation already exhibited in \cite{Kiritsis:1997em}.

\subsection{Supergravity in superspace}
In order to determine supersymmetry invariants we shall use the superspace formalism. In this section we will derive the structure of the supergeometry in eight dimensions, following the same construction as in \cite{Brink:1979nt,Howe:1981gz}. The R-symmetry group is $U(2)$, and is represented such that the covariant derivatives $D_\alpha^i,\, \bar D_{\dot{\alpha} i}$ have respectively weights $1$ and $-1$ with respect to the axial $U(1)$, and the indices $i$ correspond to the fundamental of $SU(2)$, whereas $\alpha$ and $\adt$ are respectively in the chiral and the anti-chiral Weyl representation of $Spin(1,7)$, which are complex conjugate.  The complete set of fields is depicted in figure \ref{fig:8DFieldCont}.
 \def\xmin{1}
 \def\ymin{0}
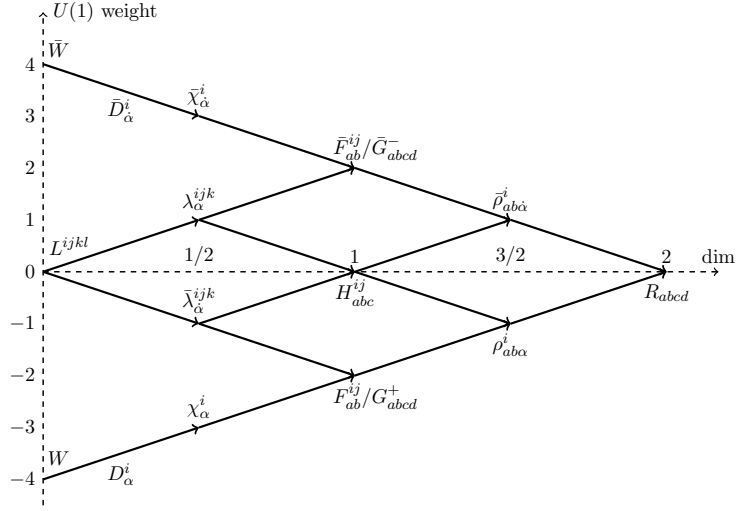
\begin{figure}[htbp]
\begin{center}
\resizebox{10 cm}{!}{
 \begin{tikzpicture}
\draw[->,draw=black,very thick] (\xmin,\ymin) -- (\xmin + 3,\ymin - 1);
\draw[->,draw=black,very thick] (\xmin + 3,\ymin - 1) -- (\xmin + 6,\ymin - 2); 
\draw[->,draw=black,very thick] (\xmin + 6,\ymin - 2) -- (\xmin + 9,\ymin - 3); 
 
\draw[->,draw=black,very thick] (\xmin + 9,\ymin - 3) -- (\xmin + 12,\ymin - 4); 
\draw[->,draw=black,very thick] (\xmin + 3,\ymin - 3) -- (\xmin + 6,\ymin - 2); 

\draw[->,draw=black,very thick] (\xmin + 3,\ymin - 3) -- (\xmin + 6,\ymin - 4); 
\draw[->,draw=black,very thick] (\xmin + 6,\ymin - 4) -- (\xmin + 9,\ymin - 5); 

\draw[->,draw=black,very thick] (\xmin,\ymin - 4) -- (\xmin + 3,\ymin - 3); 
\draw[->,draw=black,very thick] (\xmin,\ymin - 4) -- (\xmin + 3,\ymin - 5); 
\draw[->,draw=black,very thick] (\xmin + 3,\ymin - 5) -- (\xmin + 6,\ymin - 6); 

\draw[->,draw=black,very thick] (\xmin + 3,\ymin - 5) --  (\xmin + 6,\ymin - 4); 
\draw[->,draw=black,very thick] (\xmin + 6,\ymin - 4) --  (\xmin + 9,\ymin - 3) ; 

\draw[->,draw=black,very thick] (\xmin,\ymin - 8) -- (\xmin + 3,\ymin - 7);
\draw[->,draw=black,very thick] (\xmin + 3,\ymin - 7) -- (\xmin + 6,\ymin - 6);
\draw[->,draw=black,very thick] (\xmin + 6,\ymin - 6) -- (\xmin + 9,\ymin - 5); 
\draw[->,draw=black,very thick] (\xmin + 9,\ymin - 5) -- (\xmin + 12,\ymin - 4);  
 
\draw[->,draw=black,dashed,thick] (\xmin,\ymin - 4) -- (\xmin + 13,\ymin - 4);   
\draw[->,draw=black,dashed,thick] (\xmin,\ymin - 8.5) -- (\xmin,\ymin + 1);

 \draw (\xmin + 1.2,\ymin + 1) node{$U(1)$ weight};
\draw (\xmin - 0.25,\ymin) node{$4$};
\draw (\xmin - 0.25,\ymin - 1) node{$3$};
\draw (\xmin - 0.25,\ymin - 2) node{$2$};
\draw (\xmin - 0.25,\ymin - 3) node{$1$};
\draw (\xmin - 0.25,\ymin - 4) node{$0$};
\draw (\xmin - 0.4,\ymin - 5) node{$-1$};
\draw (\xmin - 0.4,\ymin - 6) node{$-2$};
\draw (\xmin - 0.4,\ymin - 7) node{$-3$};
\draw (\xmin - 0.4,\ymin - 8) node{$-4$};

\draw (\xmin + 3,\ymin - 4 + 0.3) node{$1/2$};
\draw (\xmin + 6,\ymin - 4 + 0.3) node{$1$};
\draw (\xmin + 9,\ymin - 4 + 0.3) node{$3/2$};
\draw (\xmin + 12,\ymin - 4 + 0.3) node{$2$};
\draw (\xmin + 13,\ymin - 4 + 0.3) node{dim};

\draw (\xmin + 3,\ymin - 0.6) node{$\bar{\chi}^{i}_{\dot \alpha}$};
\draw (\xmin + 3,\ymin - 0.6 - 2) node{$\lambda^{i j k}_{\alpha}$};
\draw (\xmin + 3,\ymin - 0.55 - 4) node{$\bar \lambda^{i j k}_{\dot \alpha}$};
\draw (\xmin + 3,\ymin - 0.6 - 6) node{$\chi^{i}_{\alpha}$};

\draw (\xmin + 6 + 0.4,\ymin - 0.6 - 1) node{$\bar{F}^{ij}_{a b}$/$\bar G^{-}_{a b c d}$};
\draw (\xmin + 6,\ymin  - 4.4) node{${H}^{ij}_{a b c}$};
\draw (\xmin + 6 + 0.4,\ymin  - 6.4) node{$F^{ij}_{a b}$/$G^{+}_{a b c d}$};

\draw (\xmin + 9,\ymin  - 5.4) node{$\rho^{i}_{a b \alpha}$};
\draw (\xmin + 9,\ymin  + 0.4 - 3) node{$\bar \rho^{i}_{a b \dot \alpha}$};

\draw (\xmin + 12,\ymin  - 4.4) node{$R_{a b cd}$};

\draw (\xmin + 0.3,\ymin+0.3) node{$\bar{W}$};
\draw (\xmin + 1.5,\ymin- 0.9) node{$\bar{D}^{i}_{\dot \alpha}$};
\draw (\xmin + 1.5,\ymin- 0.9 - 7) node{$D^{i}_{\alpha}$};
\draw (\xmin + 0.5,\ymin+0.5 - 4) node{$L^{i j k l}$};
\draw (\xmin + 0.3,\ymin+0.4 - 8) node{$W$};

\end{tikzpicture}
}
\end{center}
\caption{\small Structure of the supergravity supermultiplet in the linearised approximation. It includes a chiral superfield $W$ and a tensor superfield $L^{ijkl}$ related through their second derivative. The symmetry with respect to the horizontal axe defines complex conjugation. 
}
\label{fig:8DFieldCont}
\end{figure}

The superspace coordinates $z^M$ include 8 bosonic spacetime coordinates and 32 Grassmann coordinates, and the associated vielbein $E_M{}^A$ decompose as $E_M{}^{a}, \, E_{M}{}_i^{\alpha},  \, E_M{}^{\dot \alpha i}$, where $a$ is the vector index of $SO(1,7)$. The graded commutator of two covariant derivatives on a tensor $\Phi$ gives by definition 
\be \scal{ D_A \, D_B - (-1)^{AB} D_B \, D_A } \Phi = - T_{AB}{}^C D_C \Phi - R_{ABC}{}^D \, t^{(\Phi)}_D{}^C \, \Phi \ ,  \ee
where $T_{AB}{}^C$ is the torsion, and the Riemman curvature $R_{ABC}{}^D\, t^{(\Phi)}_D{}^C$ is valued in the Lie algebra $\so(1,7)\oplus \mathfrak{u}(2)$, with  appropriate generators $ t^{(\Phi)}_D{}^C$ in the representation of the field $\Phi$. The consistency of the commutation relations implies the Bianchi identities 
\begin{equation}
d_\omega T^A = E^B \wedge R_B{}^A 
\ , \qquad  d_\omega  R_B{}^A = 0 \ ,  \end{equation}
where $d_\omega$ is the covariant exterior derivative in superspace, with $\omega_{M\, B}{}^A$ itself valued in $\so(1,7)\oplus \mathfrak{u}(2)$. The Bianchi identities read in components
\begin{equation}  \label{T-bianchi} 
D_{A} T_{B C}{}^{D} + T_{A B}{}^{F} T_{F C}{}^{D} + \circlearrowleft \ = R_{A B C}{}^{D} + \circlearrowleft \qquad   D_{A} R_{B C D}{}^{E} + T_{A B}{}^{F} R_{F C D}{}^{E} + \circlearrowleft \ = 0 \qquad 
\end{equation}
where $\circlearrowleft$ denotes the sum over cyclic permutations of $A,B,C$. Moreover the internal connexion in $ \mathfrak{u}(2)$ is determined from the Maurer--Cartan superform of scalar superfields parametrizing the symmetric space $SL(2,\mathds{R})/ SO(2) \times SL(3,\mathds{R})/SO(3)$, one complex superfield $T$ and five real superfields $\phi^\upmu$. We represent $SL(2,\mathds{R})$ in terms of the $SU(1,1)$ matrices 
\be
\cV=\left(\barr{cc} U & UT \cr \bar U\bar T & \bar U\earr\right) \ , \label{su11}
\ee
satisfying to 
\be U \bar U ( 1 - T \bar T ) = 1 \ .\ee
The Maurer--Cartan form
\be
d\cV \cV^{-1}=\left(\barr{cc} -2\omega^{\mathfrak{u}(1)}  & P\cr \bar P & 2\omega^{\mathfrak{u}(1)}\earr\right)\ ,
\la{2.4}
\ee
defines the $\mathfrak{u}(1)$ connexion and scalar momenta. Similarly one defines the $SL(3,\mathds{R})$ matrices 
\be\cV^{*\, ij I} = \varepsilon^{ik} \varepsilon^{jl} \cV_{kl}{}^I\ ,  \ee
with $i=1,2$ of the gauge group $SU(2)$ and $I=1,2,3$ of the rigid $SL(3,\mathds{R})$. We will not provide an explicit parametrization of this matrix in terms of the five scalars $\phi^\upmu$, because this will not be required in our analysis. One decomposes the Maurer--Cartan form as
\be d\cV_{ij}{}^I \, \cV^{-1}{}_{I}{}^{kl} = P_{ij}{}^{kl} - 2 \delta_{(i}^{(k} \omega_{j)}{}^{l)}\ . \label{Maurer-Cartan} \ee
The momentum $P$  and the $\su(2)$ connexion $\omega_i{}^j$ are defined in this way as
\be P_{ijkl} = d\cV_{(ij}{}^I \, \cV^{-1}{}_{\hspace{-3mm} I\, kl)}  \ , \qquad  \omega_i{}^j =- \frac{1}{2} d\cV_{ik}{}^I \, \cV^{-1}{}_{I}{}^{jk} \ , \label{CosetDecom} \ee
where $SU(2)$ indices are raised and lowered with the $\varepsilon_{ij}$ tensor. It follows from the Maurer--Cartan equations that 
\begin{equation}
d_\omega P = 0\ ,  \qquad d_\omega \bar{P} = 0 \ ,  \qquad d_\omega P^{ijkl} = 0 \ , 
\end{equation}
and that the $\mathfrak{u}(2)$ components of the Riemann tensor are determined as 
\be  R^{\mathfrak{u}(1)}  = P \wedge \bar{P} \ , \qquad \qquad  R^{i}_{\;j} = P^{iklm} \wedge P_{jklm}  \ . \ee
In components, these identities read 
\be\begin{split}
D_{A} P_{B}  - (-1)^{AB} D_B P_A+ T_{A B}{}^C P_{C} &= 0 \ ,  \\
D_{A} P^{ijkl}_{B} - (-1)^{AB} D_{B} P^{ijkl}_{A}+ T_{A B}{}^{C} P^{ijkl}_{C}  &= 0 \  
\end{split}\qquad\begin{split}
R_{A B}^{\mathfrak{u}(1)} &= P_{A} \bar{P}_{B}  - \bar P_A P_B \ , \\
R_{A B}{}^{i}_{\;j} &= 2 P_{A}^{iklm} P_{Bjklm} - \delta^{i}_{j} P_{A}^{klmn} P_{Bklmn}\ . \end{split}
\ee
To complete the definition of superspace, we enforce the existence of superform field strengths transforming in linear representations of $SL(2,\mathds{R})\times SL(3,\mathds{R})$. They are 6 1-form potentials $A^1_{I},\, A^2_I$ in the ${\bf 2}\otimes {\bf 3}$ that define the complex 2-forms $F^{ij}$, 3 2-forms potentials $B^I$ in the $\overline{\bf 3}$  that define the three form field strengths $ H^{ij}$ and one 3-form potential $C$ that defines a complex 4-form $G$ and its complex conjugate, transforming together in the ${\bf 2}$ of $SL(2,\mathds{R})$  \cite{Salam:1984ft}. They satisfy to the Bianchi identities 
\bea 
d_\omega \bar{F}^{i j} &=& P^{i j}_{\;\;\;\; p q} \wedge \bar{F}^{p q} + \bar P \wedge F^{i j} \ , \CR
d_\omega H^{i j} &=& - P^{i j k l} \wedge H_{k l} + F^{k (i} \wedge \bar{F}^{j)}{}_k\ , \CR
d_\omega \bar{G} &=& \bar{P} \wedge G + H_{i j} \wedge \bar{F}^{i j} \  .
\eea
Here we allow ourselves to fix the Chern--Simons couplings $H_{i j} \wedge \bar{F}^{i j}$ and $F^{k (i} \wedge \bar{F}^{j)}{}_k$, which determine the respective normalisation of the fields with non-canonical kinetic terms. One obtains in components 
\bea   \label{HGF-bianchi} 
 D_{A} \bar{F}_{B C}^{i j} + T_{A B}^{\;\;\;\;\;\;E} \bar{F}_{E C}^{i j} \, + \circlearrowleft   \ &=& P^{i j}_{A \; p q} \bar{F}_{B C}^{p q} + \bar P_{A} F_{B C}^{i j} \, + \circlearrowleft \ ,  \CR
2 D_{A} H^{i j}_{B C D} +3 T_{A B}^{\;\;\;\;\;\;E} H^{i j}_{E C D}\,  + \circlearrowleft\  &=& - 2 P^{i j k l}_{A} H_{BCD\;kl} + 3 F^{k (i}_{A B} \bar{F}_{C D}{}^{j)}{}_{k} \, + \circlearrowleft\  , \CR
 D_{A} \bar{G}_{B C D E} + 2 T_{A B}^{\;\;\;\;\;\;F} \bar{G}_{F C D E} \, + \circlearrowleft  \ &=& \bar{P}_{A} G_{BCDE} + 2 H_{ABC\, i j } \bar{F}^{i j}_{D E} \, + \circlearrowleft \ , 
\eea
where $\circlearrowleft $ states for the sum over alternated permutations of all tangent indices $ABC\dots$, such that the result is a graded antisymmetric tensor. 
 
 The solution to these superspace identities determines the covariant superfields of the theory, which first components  at $\theta=0$ (\ie the pull back to the bosonic space embedded in superspace) correspond to the supercovariant fields of the theory in components. By construction, these fields satisfy to the equations of motion. In this paper we shall consider the classical superspace solution solving the classical  (two derivatives action) equations of motion. Restricting ourselves to the classical superspace, one can use dimensional analysis to determine the various components of the superfields. Moreover, the dimension-zero components must necessarily be invariant tensors. It follows for example that the only dimension-zero components of the torsion are  
 \be \label{T0-term}  T_{\alpha \dot \beta j}^{i\;\;\;\;c} = - i (\gamma^c)_{\alpha \dot \beta} \delta^i_j \ , \ee
and its complex conjugate. One can use the same argument to restrict the decomposition of the superforms, such that no more than two of the tangent indices $AB\dots$ can be fermionic. Moreover $\bar F^{ij}$ and $\bar G$ have an overall $U(1)$ weight $u=2$, whereas $H^{ij}$ is neutral. Using that the dimension-zero component must be $U(1)$ invariant, one gets the decompositions 
\bea\label{F-form}
 \bar{F}^{i j} &=& \frac{1}{2} E^b \wedge E^a \bar{F}_{a b}{}^{i j} + E^b \wedge E^{\alpha}_{l} \bar{F}^{l}_{\alpha b}{}^{ij} + E^b \wedge E^{\dot{\alpha}l} \bar{F}_{\dot{\alpha}l b}{}^{ij} + \frac{1}{2} E^{\dot \beta k} \wedge E^{\dot \alpha l} \bar{F}_{\dot \alpha l \dot \beta k}{}^{ij} \\
H^{i j} &=& \frac{1}{6} E^c \wedge E^b \wedge E^a H_{a b c}{}^{i j} + \frac{1}{2} E^c \wedge E^b \wedge E^{\alpha}_{l} H^{l}_{\alpha b c}{}^{ij} + \frac{1}{2} E^c \wedge E^b \wedge E^{\dot{\alpha} l} H_{\dot{\alpha} l b c}{}^{ij}  \CR
&& \hspace{90mm} + E^c \wedge E^{\beta}_{k} \wedge E^{\dot{\alpha} l} H_{\dot{\alpha}l}{}_{\beta c}^k{}^{ij}  \label{H-form} \\
\bar{G} &=& \frac{1}{24} E^d \wedge E^c \wedge E^b \wedge E^a    \bar{G}_{a b c d} + \frac{1}{6} E^d \wedge E^c \wedge E^b \wedge E^{\alpha}_{i} \bar{G}^{i}_{\alpha b c d}  + \frac{1}{4} E^d \wedge E^c \wedge E^{\dot \beta j} \wedge E^{\dot \alpha i} \bar{G}_{\dot \alpha i \dot \beta j c d}\CR
&&\label{G-form}
 \eea
where we moreover used the property that $\bar{G}_{\adt i b c d}  =0$. This last condition is true because the only dimension 1/2 field of $U(1)$ weight $1$ is the fermion field with three symmetric $SU(2)$ indices $\lambda_\alpha^{ijk}$. In principle this property can be proved in general following \cite{Brink:1979nt,Howe:1981gz}, here we already assume the knowledge of the field content of $\cN=2$ supergravity \cite{Salam:1984ft}. One computes that the dimension-zero components of the form fields are 
 \begin{equation}
\bar{F}_{\dot \alpha i \dot \beta j}{}^{kl} = - 2 C_{\dot \alpha \dot \beta} \delta^{k}_{(i} \delta^{l}_{j)}\ ,  \qquad H_{\alpha \dot \beta j c}^{i}{}^{kl} = - i \left(\gamma_c\right)_{\dot \beta \alpha} \ve^{i (k} \delta_j^{l)}
\ , \qquad \bar G_{\dot \alpha i \dot \beta j a b} = \ve_{ij} \left(\gamma_{ab}\right)_{\dot \alpha \dot \beta}\ .
\end{equation}
 Indeed one straightforwardly checks that they are the only invariant tensors satisfying to the appropriate symmetry properties, and the specific coefficients are determined modulo an overall rescaling by the Bianchi identities \eqref{HGF-bianchi}, \ie
  \bea \label{dim0_1} &&T_{\dot \gamma k}{}_{\alpha }^{i\ e} H_e{}_{\beta \dot \delta l}^{j}{}^{mn} +  T_{\dot \gamma k}{}_{\beta }^{j\ e} H_{e}{}_{\alpha \dot \delta l}^{i}{}^{mn}+T_{\ddt l}{}_{\alpha }^{i\ e} H_e{}_{\beta \cdt k}^{j}{}^{mn} +  T_{ \ddt l}{}_{\beta }^{j\ e} H_{e}{}_{\alpha \cdt k}^{i}{}^{mn}  = F_{\alpha \beta}^{i j}{}^{p(m} \bar F_{\dot \gamma k \dot \delta l}{}^{n)}{}_p\ ,  \hspace{10mm} \CR
 \label{dim0_2} &&T^i_{\alpha \bdt j}{}^{b} \bar G_{b \cdt k \ddt l  a} \, + \circlearrowleft_{\bdt\cdt\ddt}^{jkl}  \  =  H_{a}{}_{\alpha \bdt j}^{i}{}^{pq} \bar F_{\cdt k \ddt l\;pq}\, + \circlearrowleft_{\bdt\cdt\ddt}^{jkl}  \ ,\eea
where the symbol $ \circlearrowleft_{\bdt\cdt\ddt}^{jkl}$ indicates the sum over cyclic permutations of the three pairs of indices. At dimension 1/2 one gets that there is no fermionic field of $U(1)$ weight $5$, such that $T$ in \eqref{su11} must be a chiral superfield, \ie $\bar D_{\adt i} T = 0$. 
%
%
 Therefore, the scalar momenta decompose into
 \be\label{P-form}
P^{i j k l} = E^a P_{a}^{i j k l} + E^{\alpha}_{m} P_{\alpha}^{m \; i j k l} + E^{\dot{\alpha} m} P_{\dot{\alpha} m}^{i j k l}
\ , \qquad \bar{P} = E^a \bar{P}_{a} + E^{\dot{\alpha} i} \bar{P}_{\dot \alpha i}  \ , 
\ee
with $P_{\alpha}^{m \; i j k l}$ and $P_{\dot{\alpha} m}^{i j k l}$ having dimension $1/2$ and $U(1)$ weight $1, -1$, whereas $\bar{P}_{\dot \alpha i}$ has dimension $1/2$ and $U(1)$ weight $3$. One computes that all components of $U(1)$ weight $3$ are determined in terms of one single field $\bar \chi_\adt^i$, as
 \bea \label{eq:D1/2U3Results}
T_{\alpha \beta}^{i\;j \; \dot{\gamma} k} &=& 2 C_{\alpha \beta} \varepsilon^{k (i} \bar{\chi}^{j) \dot{\gamma}} + \frac{1}{4} \varepsilon^{i j} (\gamma_{a b})_{\alpha \beta} (\gamma^{a b})^{\dot{\gamma} \dot{\beta}} \bar{\chi}^{k}_{\dot{\beta}} \ , \\
\bar{F}_{\alpha b}^i{}^{k l} &=& 2 i (\gamma_b)_{\alpha}^{\;\;\dot \beta} \varepsilon^{i(k} \bar \chi_{\dot \beta}^{l)}\ ,   \qquad \bar G_{\alpha bcd}^i = i \left(\gamma_{bcd} \right)_{\alpha}^{\;\;\dot \beta} \bar \chi_{\dot \beta}^i \ , \qquad \bar P_{\dot \alpha i} = 2 \bar \chi_{\dot \alpha i} \ ,
\eea
using the Bianchi identities
\bea \label{eqT}  T_{\alpha \beta}^{i j}{}^{\ddt l} T_{\ddt l \gamma}^{\;\;\;k\;d}  \, + \circlearrowleft_{\alpha\beta\gamma}^{ijk}  \  &=& 0\ , \\
 T_{\dot \alpha i}{}_{\beta}^{j\; a} \bar F_a{}_{\gamma}^{k\;mn} + T_{\dot \alpha i}{}_{\gamma}^{k\; a} \bar F_a{}_{\beta}^{j\;mn} +  T_{\beta \gamma}^{jk}{}^{\dot \delta l} \bar F_{\dot \delta l \dot \alpha i}{}^{mn} &=& \bar P_{\dot \alpha i} F_{\beta \gamma}^{jk}{}^{mn}  \ , \CR
  T_{\alpha \beta}^{ij}{}^{\dot \delta l} H_{\dot \delta l}{}_{\gamma a}^{k}{}^{mn}  \, + \circlearrowleft_{\alpha\beta\gamma}^{ijk}  \ &=& F_{\alpha \beta}^{ij\;p(m} \bar F^k_{\gamma a}{}^{n)}{}_p \, + \circlearrowleft_{\alpha\beta\gamma}^{ijk}  \ , \ \CR
  T_{\alpha \beta}^{ij\;\dot \delta l} \bar G_{\dot \delta l \dot \gamma k a b} +  T_{\alpha \dot \gamma k}^{i}{}^{e} \bar G_{e}{}^j_{\beta ab} +  T_{\beta \dot \gamma k}^{j}{}^{e} \bar G_{e}{}^i_{\alpha ab}  &=& \bar P_{\dot \gamma k} G_{\alpha \beta a b}^{ij} + 2 H_{\dot \gamma k}{}_\alpha^i{}_{[a}{}^{pq} \bar F_{b]}{}_{\beta}^j{}_{pq} + 2 H_{\dot \gamma k}{}_\beta^j{}_{[a}{}^{pq} \bar F_{b]}{}_{\alpha}^i{}_{pq}  \ .\nn
 \eea
In the same way one use the Bianchi identities to show that all the dimension 1/2 component of $U(1)$ weight $1$ are determined in terms of a single field $\lambda_\alpha^{ijk}$ as
\bea
 T_{\alpha \dot{\beta} j}^{i \;\;\;\;\; \dot{\gamma} k} &=& - \frac{3}{4} \delta_{\dot \beta}^{\dot \alpha} \lambda_{\alpha\,j}^{i k} + \frac{1}{2} \left(\gamma^a \right)_{\dot \beta \alpha} \left(\gamma_a \right)^{\beta \dot \gamma} \lambda_{\beta\,j}^{ik} \ , \qquad T_{\alpha \beta \; k}^{i\;j \; \gamma} = C_{\alpha \beta} \lambda^{\gamma i j}{}_{k} - \frac{1}{2} \delta^{\gamma}_{(\beta} \lambda^{i j}_{\alpha) k} \ , \CR
 \bar{F}_{\dot \alpha i b}{}^{k l} &=&  - i \left(\gamma_b\right)_{\dot \alpha}^{\;\;\beta} \lambda_{\beta\;i}^{\;kl} \ , \qquad H_{\alpha bc}^{i}{}^{kl} = \frac{1}{2}  \left(\gamma_{bc}\right)_{\alpha}^{\;\;\beta} \lambda^{ikl}_{\beta} \ , \qquad P_{\alpha}^{i\;jklm} = -  \varepsilon^{i(j} \lambda_{\alpha}^{klm)}\ . 
\eea
The computation goes on then at dimension $1$, with new independent fields associated to the scalar momenta $P_a,\, P_a^{ijkl}$ and the field strengths $\bar F_{ab}^{ij},\, H_{abc}^{ij}$ and $\bar G_{abcd_-}$, although it turns out that the sefldual component of the 4-form $\bar G$ is determined in terms of the fermions as \footnote{Note that in Minkowski signature $\gamma_{abcd}{}^{\alpha\beta} = -\frac{i}{24} \varepsilon_{abcd}{}^{e\hspace{-0.1mm}f\hspace{-0.3mm}gh} \gamma_{e\hspace{-0.1mm}f\hspace{-0.3mm}gh}{}^{\alpha\beta}$ whereas $\bar G^{-}_{abcd} = \frac{i}{24} \varepsilon_{abcd}{}^{e\hspace{-0.1mm}f\hspace{-0.3mm}gh} \bar G^{-}_{e\hspace{-0.1mm}f\hspace{-0.3mm}gh}$.}
\be
\bar G_{abcd} = \bar G_{abcd}^{-} - \frac{1}{8} \scal{ \lambda^{ijk}\gamma_{abcd}\lambda_{ijk } }\ . 
\ee 
This is consistent with the property that there is only one 3-form potential in eight-dimensions, and its complex selfdual and anti-sefldual components transform in the fundamental of $SL(2,\mathds{R})$.  From dimension 1 and beyond the solution to the constraints is rather complicated, and we only display the dimension 1 and 3/2 components in Appendix  \ref{appendix1Dim} and \ref{appendix3/2Dim}, respectively. 

Now we need to discuss the definition of supersymmetry invariants in superspace. In this section we will only consider the first corrections to the Wilsonian effective action, therefore it is enough to consider  corrections to the action that are invariant with respect to supersymmetry subject to the classical equations of motion. In the superspace framework, such a correction to the action is determined by a cohomology class in superspace, \ie a $d$-closed superform in classical superspace, defined modulo the addition of a $d$-exact superform {\cite{Gates:1997kr,Gates:1997ag}}. A superform decomposes in tangent frame as
\bea
  \cL & =&  \frac{1}{8!} E^H \wedge E^G  \wedge E^F  \wedge E^E  \wedge E^D  \wedge E^C  \wedge E^B  \wedge E^A \ \cL_{A B C D E F G H} \\
  & =& \hspace{-2mm}\sum_{\substack{m,n,p = 0 \\\scriptscriptstyle  m + n + p = 8}}^{8} \frac{1}{m! n! p!} E^{\dot \beta_p j_p} \wedge ... \wedge E^{\dot \beta_1 j_1} \wedge E^{\alpha_n}_{i_n} \wedge ... \wedge E^{\alpha_1}_{i_1} \wedge E^{a_m} \wedge ... \wedge E^{a_1}  \cL_{a_1\dots a_m}{}_{\alpha_1}^{i_1}{}_{\dots}^{\dots}{}_{\alpha_n}^{i_n}{}_{\bdt_1 j_1\dots \bdt_p j_p}\nn
\eea
where each component  will be referred to as $\cL_\grad{m}{n}{p}$, and for an order  $\kappa^{2(\ell - 1)}$  correction
\begin{equation}
  {\rm{dim}}\left[\cL_\grad{8 - p - q}{p}{q}\right] = 2 + 6 \ell -\tfrac{1}{2}  p -\tfrac{1}{2} q \qquad u\left[ \cL_\grad{8 - p - q}{p}{q}\right] = p - q  \ , 
\end{equation}
with $u$ the $U(1)$ weight. One understands that all bosonic indices are antisymmetrised whereas fermionic indices are symmetrised in pairs $\alpha_k i_k$ (respectively $\adt_k i_k$). The condition $d \cL=0$ ensures that the pull-back of this closed form to the bosonic subspace 
\be \iota^* \cL = \hspace{-2mm}\sum_{\substack{m,n,p = 0 \\\scriptscriptstyle  m + n + p = 8}}^{8} \frac{1}{m! n! p!} \psi^{\dot \beta_p j_p} \wedge ... \wedge \psi^{\dot \beta_1 j_1} \wedge \psi^{\alpha_n}_{i_n} \wedge ... \wedge \psi^{\alpha_1}_{i_1} \wedge e^{a_m} \wedge ... \wedge e^{a_1}  \cL_{a_1\dots a_m}{}_{\alpha_1}^{i_1}{}_{\dots}^{\dots}{}_{\alpha_n}^{i_n}{}_{\bdt_1 j_1\dots \bdt_p j_p}\big|_{\theta=0}
\ee
is invariant with respect to supersymmetry, modulo a total derivative and the classical equations of motion {\cite{Gates:1997kr,Gates:1997ag}}. In this form the components $\cL_\grad{m}{n}{p}|_{\theta=0}$ only depend on the supercovariant field strengths and their supercovariant derivatives. $d \cL=0$ decomposes in tangent frame in
\begin{multline}
\scal{ d \cL}_\grad{m}{n}{p} 
=  T_\grad{2}{0}{0}{}^\grad{0}{0}{1}  \cL_\grad{m\mbox{\tiny-}2}{n}{p+1} +T_\grad{2}{0}{0}{}^\grad{0}{1}{0}  \cL_\grad{m\mbox{\tiny-}2}{n+1}{p}\\+ T_\grad{1}{1}{0}{}^\grad{0}{0}{1}   \cL_\grad{m\mbox{\tiny-}1}{n\mbox{\tiny-}1}{p+1}+  \scal{  D_\grad{1}{0}{0} +T_\grad{1}{1}{0}{}^\grad{0}{1}{0} + T_\grad{1}{0}{1}{}^\grad{0}{0}{1}  } \cL_\grad{m\mbox{\tiny-}1}{n}{p} + T_\grad{1}{0}{1}{}^\grad{0}{1}{0}   \cL_\grad{m\mbox{\tiny-}1}{n+1}{p\mbox{\tiny-}1}\\
+T_\grad{0}{2}{0}{}^\grad{0}{0}{1} \cL_\grad{m}{n\mbox{\tiny-}2}{p+1} + \scal{ D_\grad{0}{1}{0} + T_\grad{0}{2}{0}{}^\grad{0}{1}{0}+T_\grad{0}{1}{1}{}^\grad{0}{0}{1}}  \cL_\grad{m}{n\mbox{\tiny-}1}{p}\\ 
+ \scal{ D_\grad{0}{0}{1} + T_\grad{0}{1}{1}{}^\grad{0}{1}{0}+T_\grad{0}{0}{2}{}^\grad{0}{0}{1}}  \cL_\grad{m}{n}{p\mbox{\tiny-}1} + T_\grad{0}{0}{2}{}^\grad{0}{1}{0} \cL_\grad{m}{n+1}{p\mbox{\tiny-}2}\\
+ T_\grad{0}{1}{1}{}^\grad{1}{0}{0} \cL_\grad{m+1}{n\mbox{\tiny-}1}{p\mbox{\tiny-}1}  \end{multline}
where we defined 
\be D_\grad{1}{0}{0} \sim D_a \ , \qquad D_\grad{0}{1}{0} \sim D_\alpha^i \ , \qquad D_\grad{0}{0}{1} \sim \bar D_{\adt i} \  ,\ee
and
\begin{gather} T_\grad{0}{1}{1}{}^\grad{1}{0}{0} \sim T_\alpha^i{}_{\bdt j}{}^c \ , \CR
T_\grad{0}{2}{0}{}^\grad{0}{0}{1} \sim T_{\alpha\beta}^{ij}{}^{\cdt k} \ , \qquad T_\grad{0}{2}{0}{}^\grad{0}{1}{0} \sim T_{\alpha\beta}^{ij}{}^{\gamma}_k\ , \qquad 
T_\grad{0}{1}{1}{}^\grad{0}{0}{1} \sim T_\alpha^i{}_{\bdt j}{}^{\cdt k} \ , \CR
T_\grad{1}{1}{0}{}^\grad{0}{0}{1} \sim T_a{}_\beta^j{}^{\cdt k}\ ,  \qquad T_\grad{1}{1}{0}{}^\grad{0}{1}{0} \sim T_a{}_\beta^j{}^{\gamma}_k \ , \CR
T_\grad{2}{0}{0}{}^\grad{0}{0}{1} \sim T_{ab}{}^{\cdt k} \ , 
 \end{gather}
together with their complex conjugate, and such that the indices of uppercase grades are understood to be contracted with indices of lowercase grades. Note that the components $T_a{}_\beta^j{}^c ,\, T_a{}_{\bdt j}{}^c$ and $T_{ab}{}^c$ vanish. In this paper we will only consider the component 
\begin{multline}
\scal{ d \cL}_\grad{8}{1}{0} 
=  D_\grad{0}{1}{0}  \cL_\grad{8}{0}{0} + T_\grad{1}{1}{0}{}^\grad{0}{0}{1}   \cL_\grad{7}{0}{1}+  \scal{  D_\grad{1}{0}{0} +T_\grad{1}{1}{0}{}^\grad{0}{1}{0}  } \cL_\grad{7}{1}{0} \\
 + T_\grad{2}{0}{0}{}^\grad{0}{0}{1}  \cL_\grad{6}{1}{1}+T_\grad{2}{0}{0}{}^\grad{0}{1}{0}  \cL_\grad{6}{2}{0} \label{811dclosed} \end{multline}
and its complex conjugate. We will indeed find out that these equations alone permit to determine the differential constraints on the function of the scalar fields characterising the $d$-closed superform. 
\subsection{The chiral $R^4$ type invariant}
\label{FR48D} 
As explained in \cite{Berkovits:1997pj}, one can define an invariant from an arbitrary holomorphic functions of the chiral superfield $T\sim W$ in the linearised approximation 
\be
\bar D^{16} \bar W^{4 + n} \sim \bar  W^{n} \biggl( \Scal{ t_8 + \frac{i}{48} \varepsilon }^2 R^4 + ... \biggr)  + \bar  W^{n-1} \left( ... \right) + ... + c_{n} \bar  W^{n-12} \bar \chi^{16} \label{LinearisedU1} 
\ee
where $t_8$ is the standard tensor defined such that 
\be t_8 F^4 = \tr F^4 - \frac{1}{4}\scal{  \tr F^2 }^2 \ , \ee
and the terms in $W^{n-k}$  vanish if $k>n$. However the torsion component \eqref{eq:D1/2U3Results} implies that the chiral vectors $E^\alpha_i{}^M \partial_M$ do not close among themselves, and there is no chiral measure in eight dimensions (as in type IIB supergravity \cite{deHaro:2002vk}). Therefore one cannot directly rely on the chiral superspace integral to define the non-linear invariant, but one can still extract information from it as we are going to discuss. 

Supposing for simplicity that the invariant is $SL(3,\mathds{R})$ symmetric, such that it only depends on the scalar fields $\phi^\upmu$ through the covariant derivative $P_a{}^{ijkl}$ and the definition of the field strengths, each component $\cL_\grad{m}{n}{p}$ decomposes into several sub-components of various $U(1)$ weight multiplying $\bar U$ to the appropriate power
\be \cL_\grad{m}{n}{p} = \sum_{q} \bar U^{-2q}  \cL_\grad{m}{n}{p}^\ord{q} \  . \ee
If one considers an invariant that reduces to \eqref{LinearisedU1} in the linearised approximation,
\be   \cL_\grad{8}{0}{0}[\bar T^n] \big|_{k\text{-point}} = 0 \ | \ k<4+n \ , \qquad    \cL_\grad{8}{0}{0}[\bar T^n] \big|_{(4+n)\text{-point}} \propto \bar D^{16} \bar W^{4+n} \ . \ee
one will have by construction 
\be   \cL_\grad{m}{n}{p}^\ord{q}[\bar T^n] \big|_{n\text{-point}} \propto  \bar T^{n-q}   \cL_\grad{m}{n}{p}^\ord{q}[\bar T^q] \big|_{q\text{-point}} \ .  \ee 
The covariance of the superspace constraints with respect to $SL(2,\mathds{R})$, implies that the derivatives of a function must necessarily be K\"{a}hler covariant derivatives
\be \bar \cD^n \cF(\bar T,T) = \prod_{k=0}^{n-1} \Scal{ \frac{\partial \ }{\partial \bar T} - 2 k \frac{T}{1-T \bar T} } \cF(\bar T,T)  = \Scal{ \frac{\partial \ }{\partial \bar T} - 2 (n-1) \frac{T}{1-T \bar T} } \cdots \Scal{ \frac{\partial \ }{\partial \bar T} } \cF(\bar T,T) \ .  \ee
Expanding in the number of fields, one can consider the term in $\cD^m \bar \cD^n \cF(\bar T,T) $, as counting for $-m-n$ fields, such that the linearised invariant corresponds to the 4-point approximation. With this convention, one gets that the superform should take the form 
\be \cL[\cF]  = \sum_{m,n\ge 0} \cD^m \bar \cD^n \cF(\bar T,T) \cL^\gra{m}{n} \ , \label{LinearSuggest} \ee
where the $\cL^\gra{m}{n}$ are $SL(2,\mathds{R})$ invariant. In the four-point approximation, one would therefore get 
\be \cL[\cF]  \big|_{4\text{-point}} = \sum_{n= 0}^{12}  \bar \cD^n \cF(\bar T,T) \cL^\gra{0}{n} \big|_{(4+n)\text{-point}}  \ , \ee
where the $\cL^\gra{0}{n} \big|_{(4+n)\text{-point}} $ are the $SL(2,\mathds{R})$ invariant components of the linearised invariant.

Let us consider this invariant more explicitly, without yet assuming the form \eqref{LinearSuggest}. The component $\cL_\grad{8}{0}{0}$ is a Lorentz scalar that can be written as
\begin{equation}
  \cL_{abcdefgh} = \ve_{a b c d e f g h} \sum_{n,\mathpzc{a}} \bar{U}^{-2 n} \mathcal{F}^{\mathpzc{a}}_{n}(T,\bar{T}) \,  \mathcal{I}^\mathpzc{a}_{4n} 
\end{equation}
where $\mathcal{I}^\mathpzc{a}_{4n}$ are $SL(2)\times SL(3)$ invariant monomials in the covariant superfields of $U(1)$ weight $4n$ and dimension $8$, and $\mathcal{F}^{\mathpzc{a}}_{n}(T,\bar{T}) $ are functions (or more precisely $(0,n)$-tensors on $SU(1,1)/U(1)$) of the scalar $T,\, \bar T$ that multiply them in the invariant. The independent such monomials are labeled by the index $\mathpzc{a}$. In this section we shall consider the monomials of maximal $U(1)$ weight in order to simplify the computation. To check the possible terms, it is convenient to consider the ratio of the $U(1)$ weight by the dimension. The largest ratio is for $\bar \chi_\alpha^i$, that has $u=3$ and dimension 1/2, and therefore the maximal $U(1)$ weight term is the unique $\bar \chi^{16}$ monomial as in \eqref{LinearisedU1}. We define its normalisation such that 
\begin{equation}\label{chiBar16}
  \mathcal{I}^\mathpzc{1}_{48} = \bar \chi^{16} \equiv \bar \chi^1_{1} \bar \chi^1_{2} \bar \chi^1_{3} \bar \chi^1_{4} \bar \chi^1_{5} \bar \chi^1_{6} \bar \chi^1_{7} \bar \chi^1_{8} \bar \chi^2_{1} \bar \chi^2_{2} \bar \chi^2_{3} \bar \chi^2_{4} \bar \chi^2_{5} \bar \chi^2_{6} \bar \chi^2_{7} \bar \chi^2_{8}  \ , 
\end{equation}
The next field is the dimension 1 field $\bar P_a$ that has $u=4$, however, note that a term of the form $\bar P_a \cD \cF_n^\mathfrak{a}$ can always be eliminated by adding a trivial cocycle to the superform without modifying the invariant, and one can therefore disregard such terms.  The next important fields are therefore the dimension 1 field strength $\bar F_{ab}^{ij},\, \bar G^{-}_{abcd}$ of $U(1)$ weight $2$ and the dimension 1/2 field $\lambda_\alpha^{ijk}$ of $U(1)$ weight 1. There is a unique monomial in $\bar\chi^{15}$ and three inequivalent monomials in $\bar\chi^{14}$, two isovectors in the irreducible $SO(1,7)$ representations  \DSOVIII0020 and \DSOVIII0000 and one $SU(2)$ singlet in the \DSOVIII0100. It is convenient to define their normalisation from the Grassmann derivative of \eqref{chiBar16} as a function of ordinary Grassmann variables (rather than fields)
\be\label{DefChiBar15}
\begin{split}
  \left(\bar \chi^{15}\right)^{j}_{\dot \alpha} &\equiv - \ve^{j k}\frac{\partial}{\partial \bar \chi^{{\dot \alpha}k}}\left(\bar \chi^{16}\right) \\
   \left(\bar \chi^{14}\right)^{ij}_{ab} &\equiv \ve^{ik} \ve^{jl} \left(\gamma_{ab}\right)_{\dot \alpha \dot \beta} \frac{\partial}{\partial \bar \chi_{\dot \alpha}^{k}} \frac{\partial}{\partial \bar \chi_{\dot \beta}^{l}}\left(\bar \chi^{16}\right) \end{split}\qquad \begin{split}
  \left(\bar \chi^{14}\right)_{abcd}&\equiv \left(\gamma_{abcd}\right)_{\dot \alpha \dot \beta} \frac{\partial}{\partial \bar \chi_{\dot \alpha}^j} \frac{\partial}{\partial \bar \chi_{\dot \beta j}}\left(\bar \chi^{16}\right)  \CR
 \left(\bar \chi^{14}\right) &\equiv \frac{\partial}{\partial \bar \chi_{\dot \alpha}^j} \frac{\partial}{\partial \bar \chi^{\dot \alpha}_{j}}\left(\bar \chi^{16}\right)  \end{split}\ee
With these definitions, we write a general ansatz for the $\mathcal{I}^\mathpzc{a}_{44}$, as
\begin{equation}
\begin{split}
   \mathcal{I}^\mathpzc{1}_{44} &\equiv \bar{G}_{abcd}^{-} \left(\bar \chi^{14}\right)^{abcd} \  , \\
 \mathcal{I}^\mathpzc{2}_{44} &\equiv \bar{F}^{ij}_{ab} \left(\bar \chi^{14}\right)^{ab}_{ij} \ , 
 \end{split}\qquad 
 \begin{split}
 \mathcal{I}^\mathpzc{3}_{44} &\equiv (\gamma_{ab})^{\alpha\beta} \lambda^{ikl}_\alpha  \lambda_\beta^{j}{}_{kl}  \left(\bar \chi^{14}\right)^{ab}_{ij}=(\lambda  \lambda)^{ij}_{ab} \left(\bar \chi^{14}\right)^{ab}_{ij}\ ,  \\
\mathcal{I}^\mathpzc{4}_{44} &\equiv \lambda^{ijk}_\alpha \lambda^\alpha_{ijk} \, \left(\bar \chi^{14}\right)= (\lambda \lambda)\, \left(\bar \chi^{14}\right) \ . 
\end{split}
\end{equation}
Note that we could also consider a term in $(\bar\chi^{13})^{ijk}_{a\alpha} \lambda^\alpha_{ijk} \bar P^a$, but one can always remove such a term by adding to the superform $\cL$ a $d$-exact form $d \Psi$ with $\Psi_\grad{7}{0}{0}$ equal to 
\be \Psi_{abcdefg}  = \ve_{abcdefg}{}^{h} \bar U^{-20} \cG_{10}(T,\bar T)  (\bar\chi^{13})^{ijk}_{h\alpha} \lambda^\alpha_{ijk}\ , \ee
while affecting only therms in $\bar U^{-20}$. Therefore we will not consider such a term that would not lead to any constraints by construction, since $ \cG_{10}(T,\bar T)$ is clearly arbitrary in  $\Psi_\grad{7}{0}{0}$. One could also guess the appearance of a term in $\bar \lambda^{ijk}  \left(\bar \chi^{15}\right)^\alpha_k$, but there is no $SU(2)$ singlet such a monomial.  Our ansatz for $\cL_\grad{8}{0}{0}$ will therefore be
\begin{multline}\label{I800}
 \cL_{abcdefgh} = \ve_{a b c d e f g h} \biggl( \sum_{n = 0,\mathpzc{a}}^{11} \bar{U}^{-2 n} \mathcal{F}^{\mathpzc{a}}_{n}(T,\bar{T})  \mathcal{I}^\mathpzc{a}_{4n}\\
 + \bar{U}^{-24} \mathcal{F}^{\mathpzc{1}}_{12}(T,\bar{T}) (\bar \chi^{16}) + \bar{U}^{-22} \mathcal{F}^{\mathpzc{1}}_{11}(T,\bar{T})  \bar{G}_{abcd}^{-} \left(\bar \chi^{14}\right)^{abcd}  + \bar{U}^{-22} \mathcal{F}^{\mathpzc{2}}_{11}(T,\bar{T}) \bar{F}^{ij}_{ab} \left(\bar \chi^{14}\right)^{ab}_{ij} \\
+ \bar{U}^{-22} \mathcal{F}^{\mathpzc{3}}_{11}(T,\bar{T}) \left(\lambda \lambda \right)^{ij}_{ab} \left(\bar \chi^{14}\right)^{ab}_{ij}  + \bar{U}^{-22} \mathcal{F}^{\mathpzc{4}}_{11}(T,\bar{T}) \left(\lambda \lambda\right) \left(\bar \chi^{14}\right)  \biggr)
\end{multline}
Writing down \eqref{811dclosed}, one sees, however, that the equation $d\cL=0$ also includes mixing of $\cL_\grad{8}{0}{0}$ with $\cL_\grad{7}{1}{0},\, \cL_\grad{7}{0}{1},\, \cL_\grad{6}{2}{0},\, \cL_\grad{6}{1}{1}, \, \cL_\grad{6}{0}{2}$, so we must also consider an ansatz for these components. In the formalism in components (as opposed to superspace), this amounts to distinguish the terms that are written in terms of supercovariant field strengths, from the ones that carry naked gravitnino fields.  Let us consider first $\cL_\grad{7}{1}{0}$, which is a spinor valued 7-form in the fundamental of $SU(2)$ with $U(1)$ weight $u=1$. It can include two irreducible representations of $Spin(1,7)$, the \DSOVIII1001 and the \DSOVIII0010. The maximal $U(1)$ weight component one can get is $u= 45$, with the term $\left(\bar \chi^{15}\right)^{i}_{\dot \alpha}$. We shall only check terms up to order $\bar U^{-22}$ in $d \cL=0$, and therefore this is the only term that will be relevant in our computation, so we consider the ansatz 
\begin{equation}
\cL_{abcdefg}{}_{\alpha}^{i} = \ve_{abcdefg}{}^{h} \biggl(  (\gamma_{h})_{\alpha}^{\;\;\dot \beta} \bar{U}^{-22} \mathcal{F}^{\mathpzc{5}}_{11}(T,\bar{T})  \left(\bar \chi^{15}\right)^{i}_{\dot \beta} + \sum_{n = 0,\mathpzc{a}}^{10} \bar{U}^{-2 n} \mathcal{F}^{\mathpzc{a}}_{n}(T,
\bar T)\  \mathcal{I}^{\mathpzc{a}}_{4n+1\,}{}_{h\alpha}^i \biggr) 
\end{equation}
with again other functions $\mathcal{F}_n^{\mathpzc{a}}$ depending on $T$ and $\bar T$. $\cL_\grad{7}{0}{1}$ has $U(1)$ weight $-1$, and decomposes into the irreducible representations  \DSOVIII1010 and the \DSOVIII0001 of $Spin(1,7)$, therefore it cannot include terms in $\bar\chi^{15}$ and the maximal $U(1)$ weight terms one can have are in $\bar{U}^{-22} \bar \chi^{14} \lambda$ and $\bar{U}^{-22} \bar \chi^{13} \bar P$. Moreover most of the latter can be reabsorbed in a trivial cocycle and lower $U(1)$ weight terms such that one obtains the ansatz 
\begin{multline}
 \cL_{abcdefg \dot \alpha i } = \ve_{abcdefg}^{\;\;\;\;\;\;\;\;\;\;\;\;\;h} \biggl( (\gamma^r)_{\adt\beta} \bar{U}^{-22} \cF^{\mathpzc{6}}_{11}(T,\bar T) \,  (\bar \chi^{14})_{hr}^{kl}  \lambda^\beta_{ikl} +(\gamma_h{}^{rs})_{\adt\beta} \bar{U}^{-22} \cF^{\mathpzc{7}}_{11}(T,\bar T) \,  (\bar \chi^{14})_{rs}^{kl}  \lambda^\beta_{ikl}   \\  +\bar{U}^{-22} \cF^{\mathpzc{8}}_{11}(T,\bar T) \,  (\bar \chi^{13})_{\adt i} \bar P_h + \sum_{n = 0,\mathpzc{a}}^{10} \bar{U}^{-2 n} \mathcal{F}^{\mathpzc{a}}_{n}(T,\bar{T})\,  \mathcal{I}^\mathpzc{a}_{4n-1 h\dot \alpha i} \biggr) \label{L701}  \end{multline}
The same idea holds for
$\cL_\grad{6}{2}{0},\, \cL_\grad{6}{1}{1}$ and $\cL_\grad{6}{0}{2}$ of dimension $7$, and of $U(1)$ weight $2,\, 0$ and $-2$, respectively. One checks that $\cL_\grad{6}{2}{0}$ and $ \cL_\grad{6}{1}{1}$ carry at most terms in $\bar U^{-20}$, whereas $\cL_\grad{6}{0}{2}$ carries terms in $\bar U^{-22} \bar \chi^{14}$, \ie
\begin{multline}
 \cL_{abcdef\dot \alpha i \bdt j} = \ve_{abcdefgh} \biggl( C_{\adt\bdt}  \bar{U}^{-22} \cF^{\mathpzc{9}}_{11}(T,\bar T)  (\bar \chi^{14})_{ij}^{gh}  +(\gamma^{gh}{}_{rs})_{\adt\bdt}  \bar{U}^{-22} \cF^{\mathpzc{10}}_{11}(T,\bar T)  (\bar \chi^{14})_{ij}^{rs} \\ \qquad +\varepsilon_{ij} (\gamma^{gh})_{\adt\bdt}    \bar{U}^{-22} \cF^{\mathpzc{11}}_{11}(T,\bar T)  (\bar \chi^{14})+\varepsilon_{ij} (\gamma_{rs})_{\adt\bdt}    \bar{U}^{-22} \cF^{\mathpzc{12}}_{11}(T,\bar T)  (\bar \chi^{14})^{ghrs}\\  + \sum_{n = 0,\mathpzc{a}}^{10} \bar{U}^{-2 n} \mathcal{F}^{\mathpzc{a}}_{n}(T,\bar{T}) \mathcal{I}^\mathpzc{a}_{4n-2}{}_{\dot \alpha\bdt ij}^{gh} \biggr) \end{multline}
Considering the terms of maximal $U(1)$ weight, $( d \cL)_\grad{8}{0}{1}=0$ simplifies to  
\be
  D_\grad{0}{0}{1}  \cL_\grad{8}{0}{0} +     D_\grad{1}{0}{0}  \cL_\grad{7}{0}{1} = \mathcal{O}(\bar U^{-22})  \ . \ee
 The terms in $\bar U^{-24}$ in $D_\grad{1}{0}{0}  \cL_\grad{7}{0}{1} $ are computed using 
 \begin{equation}
  D_a  \left(\bar{U}^{-2n}  \mathcal{F}^{\mathpzc{a}}_{n}\right) = \bar{U}^{-2(n+1)} \left(\bar{\cD}\mathcal{F}^{\mathpzc{a}}_{n}\right) \bar P_{a} + \bar{U}^{-2(n-1)} \left(1 - T\bar{T}  \right)^2 \left(\cD\mathcal{F}^{\mathpzc{a}}_{n}\right) P_{a}\ , 
\end{equation}
as \bea&&  8 D_{[a}   \cL_{bcdefgh] \dot \alpha i }  + \mathcal{O}(\bar U^{-22}) \\
 & =& - \varepsilon_{abcdefgh} \bar U^{-24}  \bar P^r \biggl( (\gamma^s)_{\adt\beta} \bar \cD \cF^{\mathpzc{6}}_{11}\,  (\bar \chi^{14})_{rs}^{kl}  \lambda^\beta_{ikl} +(\gamma_r{}^{st})_{\adt\beta} \bar\cD \cF^{\mathpzc{7}}_{11} \,  (\bar \chi^{14})_{st}^{kl}  \lambda^\beta_{ikl}  +\bar \cD  \cF^{\mathpzc{8}}_{11}\,  (\bar \chi^{13})_{\adt i} \bar P_r\biggr)\nn \eea
 whereas $  D_\grad{0}{0}{1}  \cL_\grad{8}{0}{0}$ does not depend on $\bar P_a$ at this order, and we conclude that they must cancel by themselves. However they do not, and the functions  $\cF^{\mathpzc{a}}_{11}$ must be holomorphic forms for $\mathpzc{a}=\mathpzc{6},\mathpzc{7},\mathpzc{8}$. Going further in the analysis one would in fact conclude that they vanish.
 
 Therefore we can consider the equation $  D_\grad{0}{0}{1}  \cL_\grad{8}{0}{0} =0$ at this order in $\bar U$. The order $\bar U^{-26}$ term vanishes trivially 
 \be D_{\adt i} \cL_{abcdefgh} = 2 \ve_{a b c d e f g h} \bar{U}^{-26} \bar{\cD} \mathcal{F}^{1}_{12} \, \bar \chi_{\dot \alpha i} (\bar \chi^{16})   + \mathcal{O}(\bar U^{-24} ) = \mathcal{O}(\bar U^{-24} ) \ee
whereas the order  $\bar U^{-24}$ terms give the equation  
 \begin{multline}
\mathcal{F}^{\mathpzc{1}}_{12} \bar D_{\dot \alpha i}(\bar \chi^{16})  + 2 \left(\bar{\cD}\mathcal{F}^{\mathpzc{1}}_{11}\right) \bar{G}_{abcd}^{-} \bar{\chi}_{\dot \alpha i} \left(\bar \chi^{14}\right)^{abcd}  + 2 \left(\bar{\cD}\mathcal{F}^{\mathpzc{2}}_{11}\right) \bar{F}^{kl}_{ab}  \bar{\chi}_{\dot \alpha i}  \left(\bar \chi^{14}\right)^{ab}_{k l}   \\
  +2 \left(\bar{\cD}\mathcal{F}^{\mathpzc{3}}_{11} \right) \left(\lambda \lambda \right)^{kl}_{ab}  \bar{\chi}_{\dot \alpha i} \left(\bar \chi^{14}\right)^{ab}_{k l} + 2 \left(\bar{\cD}\mathcal{F}^{\mathpzc{4}}_{11}\right) \left(\lambda \lambda\right)  \bar{\chi}_{\dot \alpha i} \left(\bar \chi^{14}\right) =0 \ . \label{DL801} \end{multline}
Solving this equation requires to consider the explicit derivative of the field $\bar \chi_\alpha^i$ computed in Appendix \ref{appendix1Dim} 
\begin{multline}
 \bar D_{\dot \alpha i} \bar \chi_{\dot \beta}^j =  -\frac{1}{8} (\gamma^{ab} )_{\dot \alpha \dot \beta} \Scal{ \bar F^{\;\;\;\;\;\;j}_{ab\;i}- \frac{1}{4} \scal{\lambda_{ikl}  \gamma_{ab} \lambda^{jkl} }} 
  +\frac{1}{192}  (\gamma^{abcd})_{\dot \alpha \dot \beta}  \delta_i^j \bar G^{-}_{a b c d}- \frac{1}{4} \bar \lambda_{\adt ki}{}^j \bar \chi_\bdt^k \\ -C_{\dot \alpha \dot \beta}\Scal{ \frac{3}{32}  \delta_i^j ( \lambda \lambda) +\frac{1}{2}   \scal{\bar \chi^k \bar \lambda_{ki}{}^j }}\ . 
\end{multline}
Using Fierz identities related to the uniqueness of $(\bar \chi^{15})_\adt^i$ and the property that the terms in $(\bar \chi^{16}) \bar \lambda_\adt^{ijk}$ cancel by themselves because $ D_\grad{0}{0}{1}  \cL_\grad{8}{0}{0}$ is in the fundamental of $SU(2)$, one computes that \eqref{DL801} is satisfied if and only if 
\begin{equation}
  \bar{\cD}\mathcal{F}^{\mathpzc{1}}_{11} = \frac{1}{768} \mathcal{F}^{\mathpzc{1}}_{12} \ , \qquad \bar{\cD}\mathcal{F}^{\mathpzc{2}}_{11} = \frac{1}{32} \mathcal{F}^{\mathpzc{1}}_{12}\ , \qquad \bar{\cD}\mathcal{F}^{\mathpzc{3}}_{11} = - \frac{1}{128}  \mathcal{F}^{\mathpzc{1}}_{12}\ ,  \qquad \bar{\cD}\mathcal{F}^{\mathpzc{4}}_{11} = - \frac{3}{128} \mathcal{F}^{\mathpzc{1}}_{12} \ . 
\end{equation}
Therefore $\mathcal{F}^{\mathpzc{a}}_{11}$ are determined up to holomorphic forms 
\be c_{\mathpzc{a}}(T,\bar T) = ( 1- T \bar T)^{-22} \tilde{c}_{\mathpzc{a}}(T) \ , \ee
in terms of a single function $ \mathcal{F}_{11}$ as
\begin{equation}
 \mathcal{F}_{11}^{\mathpzc{1}} = \frac{1}{768} \mathcal{F}_{11}\ , \quad  \mathcal{F}^{\mathpzc{2}}_{11} = \frac{1}{32} \left(\mathcal{F}_{11}  + c_\deux\right)\  ,  \quad \mathcal{F}^{\mathpzc{3}}_{11} = - \frac{1}{128} \left(\mathcal{F}_{11}  + c_\trois\right)\ ,  \quad \mathcal{F}^{\mathpzc{4}}_{11} = - \frac{3}{128} \left(\mathcal{F}_{11}  + c_\quatre \right)\ , 
\end{equation}
 where we set $c_\un=0$, such that
\begin{equation}
  \mathcal{F}^{\mathpzc{1}}_{12} = \bar{\cD} \mathcal{F}_{11} \ . \end{equation}
Similarly, restricting ourselves to the terms of maximal $U(1)$ weight, $( d \cL)_\grad{8}{1}{0}=0$ simplifies to  
\be
D_\grad{0}{1}{0}  \cL_\grad{8}{0}{0}+\scal{  D_\grad{1}{0}{0} +T_\grad{1}{1}{0}{}^\grad{0}{1}{0}  } \cL_\grad{7}{1}{0} = \mathcal{O}(\bar U^{-20}) \ , \label{USplit} \ee
where we used moreover that the terms of order $\bar U^{-22}$ of $\cL_\grad{7}{0}{1}$ in \eqref{L701} vanish. We start with the terms of order $\bar U^{-24}$ that further reduce to
\bea && D_{\alpha}^i \cL_{abcdefgh} + 8 D_{[a} \cL_{bcdefgh] \alpha}^{\;\;\;\;\;\;\;\;\;\;\;\;\;\;i}  + \mathcal{O}(\bar U^{-22}) \CR
&=&  \ve_{a b c d e f g h} \bar{U}^{-24} (\bar{\cD}\mathcal{F}_{11}) D_{\alpha}^{i} (\bar \chi^{16})  +8 \ve_{[bcdefgh}^{\;\;\;\;\;\;\;\;\;\;\;\;\;r} \left(\gamma_{r}\right)_{\alpha}^{\;\;\dot \beta} D_{a]}\left( \bar{U}^{-22} \mathcal{F}^{\mathpzc{5}}_{11}\right) \left(\chi^{15}\right)^{i}_{\dot \beta}+ \mathcal{O}(\bar U^{-22}) \CR
&=& \ve_{a b c d e f g h} \bar{U}^{-24} \Scal{ \bar{\cD}\mathcal{F}_{11} \, D_{\alpha}^{i} (\bar \chi^{16})  -  \bar{\cD} \mathcal{F}^{\mathpzc{5}}_{11}\,  (\gamma^r)_\alpha{}^\bdt  \bar{P}_{r}  \left(\chi^{15}\right)^{i}_{\dot \beta} }+ \mathcal{O}(\bar U^{-22})\CR
&=& \mathcal{O}(\bar U^{-22})\ . 
\eea
The covariant derivative $D \bar \chi$ is determined from  \eqref{eq:DChiBar} as
\begin{equation}
 D_{\alpha}^i \bar \chi_{\dot \beta}^j =  \frac{1}{2} (\gamma^a)_{\alpha \dot \beta} \Scal{  -i  \ve^{i j}  \bar P_a  +  \scal{\bar\chi_k \gamma_a \lambda^{ijk}}} + \frac{3}{4} \lambda_\alpha^{ijk} \bar \chi_{\bdt k}\ ,  
\end{equation}
so once again the terms in  $(\bar \chi^{16})  \lambda_\alpha^{ijk}$ cancel by themselves and we get the constraint 
\begin{equation}
  \bar{\cD}\mathcal{F}^{\mathpzc{5}}_{11} = -\frac{i}{2} \bar{\cD}\mathcal{F}_{11} \qquad \Rightarrow \qquad \mathcal{F}^{\mathpzc{5}}_{11} = -\frac{i}{2} \left(\mathcal{F}_{11} + c_{\mathpzc{5}} \right) \ . 
\end{equation}
Now we must consider the order $\bar U^{-22}$ components of \eqref{USplit}, however, the computation involves many terms and we shall simplify the problem by neglecting all the terms that depend explicitly on $\lambda_\alpha^{ijk}$ and $\bar P_a$. This permits in particular to neglect terms of order $\bar U^{-20}$ in $\cL_\grad{7}{1}{0}$ that we have not computed. Using this simplification, one obtains 
\bea
 && D_{\alpha}^i \cL_{abcdefgh}  + 8 D_{[a} \cL_{bcdefgh] \alpha}^{\;\;\;\;\;\;\;\;\;\;\;\;\;\;i} + 8 T_{[a|}{}_{\alpha}^i{}_j{}^\beta \cL_{bcdefgh]\beta}^{\;\;\;\;\;\;\;\;\;\;\;\;\;\;j}  + \mathcal{O}(\bar U^{-20}) \CR
 &=& \ve_{a b c d e f g h}  \bar U^{-22} \biggl( 2 (1-T\bar T)^2 \cD  \bar{\mathcal D} \mathcal{F}_{11} \,  \chi_\alpha^i  (\bar \chi^{16})+ \frac{1}{768} \mathcal{F}_{11} (D_{\alpha}^{i}  \bar{G}_{abcd}^{-}) \left(\bar \chi^{14}\right)^{abcd}  \CR&& + \frac{1}{32}\left(\mathcal{F}_{11} + c_\deux\right) (D_{\alpha}^{i}\bar{F}^{kl}_{ab}) \left(\bar \chi^{14}\right)^{ab}_{k l}  + \frac{i}{2} \left(\mathcal{F}_{11} + c_{\mathpzc{5}} \right) \Scal{  \left(\gamma^{r}\right)_{\alpha}^{\;\;\dot \beta}D_{r}\left(\bar \chi^{15}\right)^{i}_{\dot \beta} + T_{r}{}_\alpha^i{}_j^\beta \left(\gamma^r \right)_{\beta}^{\;\;\dot \beta} \left(\bar \chi^{15}\right)^{j}_{\dot \beta} } \biggr) \CR &=&  \mathcal{O}(\bar U^{-20})  \ . 
\eea
To carry out this computation we need the covariant derivative $D_\alpha^i$ of both $\bar G^{-}_{abcd}$ and $\bar F^{ij}_{ab}$  given in Appendix  \ref{appendix3/2Dim} in \eqref{eq:DGBar} and \eqref{eq:DFBar}, as well as the dimension 1 torsion $T_{a}{}_\beta^j{}_k^\gamma$ given in  \eqref{eqn:D1Torsion}, for which we neglect all terms in $\lambda_\alpha^{ijk}$ and $\bar P_a$. Moreover, the equation can only be satisfied modulo the classical equations of motion, and we must distinguish in $D_a \bar \chi_\adt^i$, its gamma trace that is equal to a polynomial in the other fields through the Dirac equation \eqref{eq:PartialChiBar}. We will write $(D_{a} \bar \chi_\adt^i)'$ its component projected to the irreducible representation \DSOVIII1010 of $Spin(1,7)$ (\ie such that $(\gamma^{a})^{\alpha\bdt} (D_{a} \bar \chi_\bdt^i)' = 0 $). Combining all these terms one obtains finally 
\bea
 && D_{\alpha}^i \cL_{abcdefgh}  + 8 D_{[a} \cL_{bcdefgh] \alpha}^{\;\;\;\;\;\;\;\;\;\;\;\;\;\;i} + 8 T_{[a|}{}_{\alpha}^i{}_j{}^\beta \cL_{bcdefgh]\beta}^{\;\;\;\;\;\;\;\;\;\;\;\;\;\;j}  + \mathcal{O}(\bar U^{-20}) \CR
 &=& \ve_{a b c d e f g h}  \bar U^{-22} \biggl( \Scal{    (1-T\bar T)^2 \cD  \bar{\mathcal D} \mathcal{F}_{11}  + 132   \mathcal{F}_{11} +  \frac{315}{4} c_\deux - 8 c_{\mathpzc{5}} } 2 \chi_\alpha^i (\bar \chi^{16}) \CR
 &&  \hspace{40mm} +  \Scal{ \frac{13 i}{192} c_\deux - \frac{7 i}{288} c_\cinq } ( \gamma^{abc})_{\alpha\bdt} H_{abc}^{ij} \left( \bar \chi^{15} \right)_{j}^\bdt   \CR
 && \hspace{20mm}+ \frac{i}{192}  c_\cinq  (\gamma^{abc})_{\alpha}^{\;\;\bdt} (D^{d} \bar \chi^i_{\bdt})' (\bar \chi^{14})_{abcd}+ \frac{i}{8}  (c_\deux - c_\cinq) (\gamma^{a})_{\alpha\bdt} (D^{b} \bar \chi_j^{\bdt})' \left(\bar \chi^{14}\right)_{ab}^{ij}  \biggr)\CR
 &=&  \mathcal{O}(\bar U^{-20}) \ . 
 \eea
We conclude therefore that the harmonic forms $c_\deux$ and $c_\cinq$ vanish as expected, and the form $\cF_{11}$ satisfies the differential equation 
\be (1-T\bar T)^2 \cD  \bar{\mathcal D} \mathcal{F}_{11}(T,\bar T)   =  - 132 \,   \mathcal{F}_{11}(T,\bar T) \ .  \label{Equa11} \ee
It is rather clear that if we had computed the terms in $\lambda_\alpha^{ijk}$ one would have obtained similarly that $c_\trois=c_\quatre=0$, and we conclude therefore that 
\begin{multline}\label{I800Solve}
 \cL_{abcdefgh} = \ve_{a b c d e f g h} \biggl( \bar{U}^{-24} \bar \cD \mathcal{F}_{11}(T,\bar{T}) (\bar \chi^{16}) \\+ \frac{1}{128} \bar{U}^{-22} \mathcal{F}_{11}(T,\bar{T})  \Scal{ \frac{1}{6}  \bar{G}_{abcd}^{-} \left(\bar \chi^{14}\right)^{abcd}  + 4 \bar{F}^{ij}_{ab} \left(\bar \chi^{14}\right)^{ab}_{ij} -  \left(\lambda \lambda \right)^{ij}_{ab} \left(\bar \chi^{14}\right)^{ab}_{ij}  -3  \left(\lambda \lambda\right) \left(\bar \chi^{14}\right)  } \\+ \sum_{n = 0,\mathpzc{a}}^{11} \bar{U}^{-2 n} \mathcal{F}^{\mathpzc{a}}_{n}(T,\bar{T})  \mathcal{I}^\mathpzc{a}_{4n} \biggr) \  . 
\end{multline}
It is important to note that this superform indeed reproduces the structure explained in the beginning of this section, \ie each covariant combination of fields  multiplying $\bar U^{-2n} \bar \cD^{n} \cF $ is approximated by the linearised invariant as
\be \bar D^{16} \bar W^{16} \Big|_{\bar W=0} \propto (\bar \chi^{16}) \ , \qquad  \bar D^{15} \bar W^{15} \Big|_{\bar W=0} \propto \frac{1}{24}  \bar{G}_{abcd}^{-} \left(\bar \chi^{14}\right)^{abcd}  +  \bar{F}^{ij}_{ab} \left(\bar \chi^{14}\right)^{ab}_{ij} \ . \ee
The relation to the linearised invariant implies indeed that each covariant combination of fields multiplying $\bar U^{-2n} \bar \cD^{n} \cF $ must be of the form $ \bar D^{n+4} \bar W^{n+4} \big|_{\bar W=0} $ such that (similarly as in \cite{Green:1998by})
\be \mathcal{F}_{11}(T,\bar{T}) = \bar \cD^{11} \cF(T,\bar T) \ , \ee
with $\cF(T,\bar T) $ the function multiplying the $SL(2,\mathds{R})$ invariant of type $R^4$. Using \eqref{Equa11}, it follows that $\cF(T,\bar T)$ satisfies itself to the equation
\be  (1-T\bar T)^2 \cD  \bar{\mathcal D}^{12} \cF(T,\bar T)   =  - 132  \, \bar \cD^{11} \mathcal{F} (T,\bar T) \  . \label{Equation132}  \ee
Using the commutation relations between $\cD$ and $\bar \cD$, one computes in general that 
\bea
(1-T\bar T)^2 \cD \bar{\mathcal D}^{n+1} \mathcal{F}  &=& - n(n-1) \bar{\mathcal D}^n \mathcal{F} + (1-T\bar T)^2\bar \cD^{n+1} \cD \cF \  \CR
&=& - n(n-1) \bar{\mathcal D}^n \mathcal{F} + \bar \cD^{n} \, \Delta \cF  \ , 
\eea
and therefore in particular that 
\be   (1-T\bar T)^2 \bar{\mathcal D}^{12} \cD   \cF(T,\bar T)  = 0 \ . \ee
At each order in $\bar U^{-2n} \bar \cD^{n} \cF(T,\bar T)$ one will get equations generalising the linearised equations of the form 
\be  (1-T\bar T)^2 \cD  \bar{\mathcal D}^{n+1} \cF(T,\bar T)   =  - n(n-1)  \, \bar \cD^{n} \mathcal{F} (T,\bar T) \  ,\ee
where the coefficient is determined to be the unique one consistent with \eqref{Equation132}, therefore we conclude that supersymmetry must imply eventually that the function  $\cF(\bar T)$ is anti-holomorphic.

There are two comments we would like to make on this computation, to be compared with the computations carried out in components in \cite{Green:1998by,Basu:2011he}. Here we implicitly used the Dirac equation satisfied by $\bar \chi_\adt^i$ in several places, by removing the gamma trace appearing in $D_a \chi_\adt^i$ when this term appeared explicitly, and when it appeared in the derivative of the field strengths $\bar F_{ab}^{ij}$ and $\bar G_{abcd}$. Indeed, in components one would consider instead the supersymmetry variation of their potentials. One concludes that considering $\int \iota^*\cL$ as a correction to the effective action, the accordingly corrected covariant derivative $D_\alpha^i \chi_\beta^j$ would be modified by terms of the form
\be D_\alpha^i \chi_\beta^j = \dots + \kappa^2\,  \bar U^{-22} \bar D^{11} \cF(\bar T) \Scal{ a_\un (\gamma^{ab})_{\alpha\beta} (\bar \chi^{14})^{ij}_{ab} + a_\deux C_{\alpha\beta} \varepsilon^{ij} (\bar \chi^{14}) } + \dots \ , \ee 
although we did not compute the coefficients explicitly. In components the correction to the Lagrange density takes the form
\begin{multline}\label{I800Solve}
 \iota^*\cL = e \biggl( \bar{U}^{-24} \bar \cD^{12} \mathcal{F}(\bar{T}) (\bar \chi^{16})  + \frac{i}{2} \bar{U}^{-22} \bar \cD^{11}  \mathcal{F}(\bar{T})   \psi_a{}^{\alpha}_i (\gamma^a)_{\alpha\bdt} (\bar \chi^{15})^{\bdt i} \\+ \frac{1}{128} \bar{U}^{-22}\bar \cD^{11}  \mathcal{F}(\bar{T})  \Scal{ \frac{1}{6}  \bar{G}_{abcd}^{-} \left(\bar \chi^{14}\right)^{abcd}  + 4 \bar{F}^{ij}_{ab} \left(\bar \chi^{14}\right)^{ab}_{ij} -  \left(\lambda \lambda \right)^{ij}_{ab} \left(\bar \chi^{14}\right)^{ab}_{ij}  -3  \left(\lambda \lambda\right) \left(\bar \chi^{14}\right)  } \\+\dots \biggr) \  . 
\end{multline}
where $\bar F^{ij}$ and $\bar G$ are supercovariant field strengths, that include respectively terms in $- 2 i e^a \wedge ( \psi^{(i} \gamma_a \bar \chi^{j)}) $ and $i e^a\wedge e^b 
\wedge e^c \wedge ( \psi_i \gamma_{abc} \bar \chi^i ) $. There is therefore three different contributions to the term in $\bar{U}^{-22} \bar \cD^{11}  \mathcal{F}(\bar{T})   (\psi_{a i} \gamma^a  (\bar \chi^{15})^{i})$, and they must all be there with their respective coefficients. 

%

\subsection{The parity symmetric $R^4$ type invariant}
In the linearised approximation, the scalar fields $\phi^\upmu$ parametrizing $SL(3,\mathds{R})/SO(3)$ are conveniently represented by an isospin $2$ field $L^{ijkl}$, such that the covariant derivative 
\be D_\alpha^p \cV^{ij I} = - \varepsilon^{p(i} \lambda_\alpha^{jkl)} \cV_{kl}{}^{I}  \ , \ee
simplifies to
\be D_\alpha^p L^{ijkl} = - \varepsilon^{p(i} \lambda_\alpha^{jkl)} \ , \ee
and similarly for the complex conjugate. As explained in \cite{Berkovits:1997pj}, one can define an invariant in the linearised approximation from an arbitrary holomorphic functions of the G-analytic superfield 
\be L^{1111} = u^1{}_i u^1{}_ju^1{}_k u^1{}_l L^{ijkl} \ , \ee
 as the harmonic integral of
\be
 (D^2)^8 (\bar D_1)^8 (L^{1111})^{4 + n} \sim  (L^{1111})^{ n}  \biggl( \Scal{ t_8 t_8 +  \frac{1}{48^2} \varepsilon  \varepsilon} R^4 + ... \biggr)  + ... + c_{n} (L^{1111})^{ n-12} (\lambda^{111})^8 (\bar \lambda^{111})^8  \label{LinearisedSU2}  \ . 
\ee
In this section we will repeat the computations of the last section to determine the dependence of this invariant in the scalar fields $\phi^\upmu$ at the non-linear level. One can already infer from the linearised analysis that the function of $\phi^\mu$ must satisfy to the Laplace equation \cite{Pioline:1998mn}. However, because the harmonic  measure does not extend to the non-linear theory this construction had no reason to give the correct answer. To start with we need to discuss some properties of the differential operators on the symmetric space $SL(3,\mathds{R})/SO(3)$ that are perhaps  less standard than for the special K\"{a}hler space $SU(1,1)/U(1)$. 
\subsubsection*{Differential operators on $SL(3,\mathds{R})/SO(3)$}
The superfield momentum $P^{ijkl}$ defined in  (\ref{Maurer-Cartan},\ref{CosetDecom}) determines the vielbein $P_\upmu{}^{ijkl}$ on $SL(3,\mathds{R})/SO(3)$ in function of $\phi^\upmu$ as
\be P^{ijkl} = d\phi^\upmu P_\upmu{}^{ijkl} \ . \ee
Considering $\phi^\upmu$ as coordinates rather than fields in this discussion, the Maurer--Cartan equation 
\bea dP_{ijkl} + 4 \omega_{(i}{}^{p} \wedge P_{jkl)p} &=& 0\ ,  \CR
d \omega_i{}^j +\omega_i{}^k \wedge \omega_k{}^j &=& \frac{1}{2} P_{ikpq}\wedge P^{jkpq} \ , \label{SymmMC} \eea
indeed gives the torsion free condition, and the definition of the constant Riemann tensor on $SL(3,\mathds{R})/SO(3)$ in tangent frame.  One defines accordingly the metric
\be G_{\upmu\upnu}(\phi) = 2 P_{\upmu\, ijkl} P_\upnu{}^{ijkl} \ , \ee
and its inverse $G^{\upmu\upnu}$ such that the inverse vielbein read 
\be E_{ijkl}{}^{\upmu} = P_{\upnu\, ijkl} G^{\upmu\upnu} \ . \ee
In these conventions one has 
\be  P_{\upmu}{}^{pqrs} E_{ijkl}{}^{\upmu}   = \frac{1}{2}  \delta_{ijkl}^{pqrs} \  ,\qquad P_{\upmu}{}^{ijkl} E_{ijkl}{}^{\upnu}   = \frac{1}{2} \delta_\upmu^\upnu \ ,  \ee
where we use the symmetrised Kronecker symbol 
\be \delta^{i_1i_2\dots i_n}_{j_1j_2\dots j_n} \equiv \delta^{(i_1}_{(j_1} \delta_{j_2}^{i_2} \dots \delta^{i_n)}_{j_n)} = \frac{1}{n!} \Scal{ \delta^{i_1}_{j_1} \delta_{j_2}^{i_2} \dots \delta^{i_n}_{j_n}+ \circlearrowleft  } \ . \ee 
One defines the covariant derivative of a function and its subsequent covariant derivatives as 
\bea \cD_{ijkl} \cE &=& E_{ijkl}{}^{\upmu} \partial_{\upmu} \cE \CR
\cD_{ijkl} \cD_{pqrs} \cE &=& E_{ijkl}{}^{\upmu} \Scal{ \partial_\upmu  \scal{  E_{pqrs}{}^{\upnu} \partial_{\upnu} \cE }+ 4 \omega_{\upmu\, (p}{}^t  E_{qrs)t}{}^{\upnu} \partial_{\upnu} \cE} \eea
and etcetera. For a generic symmetric tensor, the covariant derivative is defined accordingly as
\be \cD_{ijkl} \cE_{i_1i_2\dots i_n} = E_{ijkl}{}^{\upmu} \Scal{ \partial_\upmu \cE_{i_1i_2\dots i_n}  + n\,  \omega_{\upmu\, (i_1}{}^p \cE_{i_2\dots i_n)p}} \ , \ee
and one computes using \eqref{SymmMC} that 
\be [ \cD_{ijkl} , \cD^{pqrs} ] \cE_{i_1i_2\dots i_n}  =  \frac{n}{4} \delta_{ijk)(i_1}^{pqrs} \cE_{i_2\dots i_n)(l}  - \frac{n}{8} \delta_{ijkl}^{pqrs} \cE_{i_1i_2\dots i_n}    \ . \label{CommuteSL3} \ee 
In particular 
\be  [ \cD_{ijkl} , \cD^{pqrs} ] \cD_{tuvw} \cE =  \delta_{ijk)(t}^{pqrs} \cD_{uvw)(l} \cE - \frac{1}{2} \delta_{ijkl}^{pqrs} \cD_{tuvw} \cE \ , \ee
where the notation means that $ijkl$ and $tuvw$ are symmetrised in the first term of the right-hand-side, and similarly in \eqref{CommuteSL3}. 

The covariant derivative $\cD_{ijkl} \cD_{pqrs} \cE$ of a function $\cE$ decomposes into irreducible representations of $SU(2)$, as a singlet, an isospin 2 component and an isospin 4 component. We want to consider as a differential equation the property that the isospin 2 component is related to the first order derivative, \ie
\be \cD_{(ij}{}^{pq} \cD_{kl)pq}\,  \cE_s = -  \frac{4 s-3}{12}  \cD_{ijkl} \cE_s  \ . \ee
This equation can be rewritten 
\be  \cD_{ij}{}^{pq} \cD_{klpq} \,  \cE_s =  - \frac{4 s-3}{12}  \cD_{ijkl} \cE_s + \frac{1}{12} ( \varepsilon_{ik} \varepsilon_{jl} +  \varepsilon_{il} \varepsilon_{jk} ) \cG_s \ , \label{GeneQuad} \ee
for some function  $\cG$ to be determined. This equation implies that 
\be \Delta \,  \cE_s \equiv  2 \cD_{ijkl} \cD^{ijkl} \,  \cE_s = \cG_s \ . \ee
Because there is a unique scalar fourth order differential operator, one has the constraint 
\be 2 \cD_{ijpq} \cD^{pqrs} \cD_{rskl} \cD^{klij} \cE = \frac{1}{4} \Delta \Scal{ \Delta + \frac{1}{4} } \cE \ , \ee
for any function $\cE$, and one can therefore deduce from  \eqref{GeneQuad} that 
\be  \Delta \cG_s = \frac{2s(2s-3)}{3} \cG_s \ . \label{Gcons} \ee 
For $s\ne 0$ or $\frac{3}{2}$, one obtains immediately that the function $\cE_s$ satisfies to
\be \cD_{ij}{}^{pq} \cD_{klpq} \cE_{s} = - \frac{4 s-3}{12} \cD_{ijkl} \cE_{s}  + \frac{s ( 2 s-3)}{18}  ( \varepsilon_{ik} \varepsilon_{jl} +  \varepsilon_{il} \varepsilon_{jk} )  \cE_{s} \ , \label{EisenQuadra} \ee
and in particular 
\be \Delta \cE_s =  \frac{2s(2s-3)}{3} \, \cE_s \ . \ee
The reader might recognise at this point that this Poisson equation is satisfied by the Eisenstein series 
\be E_{[s0]} \equiv \sum_{n_I \in \mathds{Z}^3_*}  \Scal{ \cV_{ij}{}^I n_I \cV^{ij\, J} n_J }^{-s} \, \ , \ee
in the domain of absolute convergence of the series (\ie for $s>\frac{3}{2}$). One straightforwardly computes that the function $( \cV_{ij}{}^I n_I \cV^{ij\, J} n_J )^{-s}$ indeed satisfies the quadratic equation \eqref{EisenQuadra} for any vector $n_I \in \mathds{R}^3_*$, and one concludes that for $s>\frac{3}{2}$ 
\be \cD_{ij}{}^{pq} \cD_{klpq} E_{[s0]} = - \frac{4 s-3}{12} \cD_{ijkl} E_{[s0]}  + \frac{s ( 2 s-3)}{18}  ( \varepsilon_{ik} \varepsilon_{jl} +  \varepsilon_{il} \varepsilon_{jk} )  E_{[s0]} \ .\ee
We are going to prove in this section that supersymmetry requires this equation to be satisfied for the function $\cE$ multiplying the $R^4$ type term in the invariant for the value $s=\frac{3}{2}$, consistently with the string theory computation \cite{Kiritsis:1997em}. However, the series actually diverges for this value, and one must consider the regularised Eisenstein series   \cite{Green:2010wi}
\be \hat{E}_{[{\scriptscriptstyle \frac{\scriptscriptstyle 3}{\scriptscriptstyle2}} 0]} = \lim_{ \epsilon \rightarrow 0 } \Scal{  E_{[{\scriptscriptstyle \frac{\scriptscriptstyle 3}{\scriptscriptstyle2} +  \epsilon} \, 0]} - \frac{2\pi}{ \epsilon}  + 4\pi ( 1 - \gamma)} \ . \ee  
By continuity, and because the constant term drops out when acted on by the covariant derivative, one obtains that the regularised series satisfies the inhomogeneous equation 
\be \cD_{ij}{}^{pq} \cD_{klpq} \hat{E}_{[{\scriptscriptstyle \frac{\scriptscriptstyle 3}{\scriptscriptstyle2}} 0]} =-  \frac{1}{4} \cD_{ijkl} \hat{E}_{[{\scriptscriptstyle \frac{\scriptscriptstyle 3}{\scriptscriptstyle2}} 0]}  + \frac{\pi}{3}  ( \varepsilon_{ik} \varepsilon_{jl} +  \varepsilon_{il} \varepsilon_{jk} )   \ ,\label{InHomo} \ee
consistently with \cite{Green:2010wi}. Note that the constant term is indeed consistent with \eqref{Gcons}, because for $s=\frac{3}{2}$ the inhomogeneous term can in principe be any function satisfying to the Laplace equation $\Delta \cG_\stfrac{3}{2} = 0$. However the constraint from supersymmetry is by construction a homogeneous linear equation, and is in fact 
\be \cD_{ij}{}^{pq} \cD_{klpq} \cE_{\stfrac{3}{2}} = - \frac{1}{4} \cD_{ijkl}  \cE_{\stfrac{3}{2}}  \ .\label{SusySL3} \ee
The inhomogeneous term in \eqref{InHomo} is due to the logarithm $\log( \cV_{ij}{}^I n_I \cV^{ij\, J} n_J)$ that satisfies 
\be \cD_{ij}{}^{pq} \cD_{klpq}\,  \log\scal{ \cV_{rs}{}^I n_I \cV^{rs\, J} n_J}= - \frac{1}{4} \cD_{ijkl} \, \log \scal{ \cV_{rs}{}^I n_I \cV^{rs\, J} n_J} - \frac{1}{6}  ( \varepsilon_{ik} \varepsilon_{jl} +  \varepsilon_{il} \varepsilon_{jk} )   \ , \ee
and which appears explicitly in the expansion of $\hat{E}_{[{\scriptscriptstyle \frac{\scriptscriptstyle 3}{\scriptscriptstyle2}} 0]} $  at large  $\cV_{ij}{}^I n_I \cV^{ij\, J} n_J$ (for any chosen vector $n_I$),
\be \hat{E}_{[{\scriptscriptstyle \frac{\scriptscriptstyle 3}{\scriptscriptstyle2}} 0]}  \sim  -2\pi \log \scal{ \cV_{ij}{}^I n_I \cV^{ij\, J} n_J}  + \dots \label{LogE32}  \ee
We shall explain that this logarithm term is associated to an anomaly, and does not appear in the supersymmetric Wilsonian effective action. 

To prove that \eqref{SusySL3} is indeed required by supersymmetry, we shall consider the terms of maximal isospin. Because these terms will carry a large number of $SU(2)$ indices, we will use the short-hand notation 
\be  \cD^{n}_{[4n]} \cE \equiv \cD_{(i_1i_2i_3i_4}\cD_{i_5i_6i_7i_8} \cdots \cD_{i_{4n-3} i_{4n-2} i_{4n-1} i_{4n})} \cE \ , \ee
and repeated representations will be understood to correspond to contracted indices, as for example in
\be \cD^{12}_{[48]} \cE \ ( \lambda^8)^{[24]} ( \bar \lambda^8)^{[24]} \equiv  \cD_{(i_1i_2i_3i_4} \cdots \cD_{i_{45} i_{46} i_{47} i_{48})} \cE\, \lambda_{1}^{(i_1 i_2 i_3} \cdots \lambda_{8}^{i_{22} i_{23} i_{24})} \,  \bar \lambda_{1}^{(i_{25} i_{26} i_{27} } \cdots \bar \lambda_{8}^{i_{46} i_{47} i_{48})} \ . \ee
Using the commutation relations \eqref{CommuteSL3}, one computes that in general 
\begin{multline} \cD_{ijkl}  \cD^{n}_{[4n]} \cE = \cD^{n+1}_{ijkl[4n]}  \cE + \frac{12 n}{4n+3}  \varepsilon_{(i[1]} \varepsilon_{j[1]} \cD^{n-1}_{[4n-4]} \cD_{kl)}{}^{pq} \cD_{[2]pq} \cE  \\
- \frac{16 n(n-1)}{(2n+1)(4n+3)} \varepsilon_{(i[1]} \varepsilon_{j[1]}\varepsilon_{k[1]} \cD^{n-2}_{[4n-8]} \cD_{l)[2]r} \cD_{[1]}{}^{rpq} \cD_{[2]pq}\cE \\ -   \frac{n(8n+5)(4n-1)}{(2n+1)(4n+3)(4n+1)} \varepsilon_{i[1]} \varepsilon_{j[1]}  \varepsilon_{k[1]} \varepsilon_{l[1]}  \cD^{n-1}_{[4n-4]}  \cD_{pqrs} \cD^{pqrs} \cE  \\- \frac{8n(n-1)(n-2)}{(2n+1)(4n+3)(4n+1)}  \varepsilon_{i[1]} \varepsilon_{j[1]}  \varepsilon_{k[1]} \varepsilon_{l[1]}  \cD^{n-3}_{[4n-12]} \cD_{[2]}{}^{pq} \cD_{[2]pq} \cD_{[2]}{}^{rs} \cD_{[2]rs} \cE\\ -  \frac{ n(n-1) (4 n^2 + 3 n + 2)}{(4n+1)(4n+2)} \varepsilon_{i[1]} \varepsilon_{j[1]}  \varepsilon_{k[1]} \varepsilon_{l[1]}  \cD^{n-1}_{[4n-4]}  \cE\ ,  \label{DD_reduction} \end{multline} 
where $\cD^{n-k}_{[4n-4k]}$ and $ \cD^{n+1}_{ijkl[4n]}$ are respectively in the isospin $2(n-k)$ and $2n+2$ irreducible representations. Using this equation, one obtains that for a function $\cE_s$ satisfying to equation \eqref{EisenQuadra}, one has moreover  
\begin{multline} \label{DDE_reduction} \cD_{ijkl}  \cD^{n}_{[4n]} \cE_{s} = \cD^{n+1}_{ijkl[4n]}  \cE_{s} - \frac{n ( 4 s-3)}{4 n + 3} \varepsilon_{(i[1]} \varepsilon_{j[1]} \cD^{n}_{kl)[4n-2]} \cE_{s}\\  -\frac{n(2 n- 1)(2n + 1 - 2 s)(n - 1 + s)}{(4n-1)(4n+1)}   \varepsilon_{i[1]} \varepsilon_{j[1]}  \varepsilon_{k[1]} \varepsilon_{l[1]} \cD^{n-1}_{[4n-4]} \cE_{s} \ ,  \end{multline}
where $\cD^{n+1},\, \cD^n$ and $\cD^{n-1}$ are in the irreducible representations of maximal isospin $2n+2,\, 2n$ and $2n-2$, respectively. 

%
%
%
%
%

\subsubsection*{Constraining the superform}
Similarly as for the chiral superform $\cL[\cF]$ discussed in the last section, the linearised analysis suggests that the super-form $\cL[\cE]$ admits the following expansion 
\be \cL[\cE] = \sum_{n=0}^{12} \cD^n_{[4n]} \cE \, \cL^{[4n]} \ , \ee
where $\cL^{[4n]}$ are $SL(2,\mathds{R})\times SL(3,\mathds{R})$ invariant isospin $2n$ tensors superforms, that coincide with the linearised invariant at $4+n$ order in the fields 
\be \cL_\grad{8}{0}{0}^{[4n]} \propto (D^8)_{[8]} (\bar D^8)_{[8]} L^{4+n\, [16+4n]}\Big|_{L=0} + \mathcal{O}((5+n)\mbox{-points}) \  . \ee
Using this general structure, one is led to a general ansatz for $ \cL_\grad800$ 
\begin{multline} \label{L800} -\frac{1}{8!} \varepsilon^{abcdefgh} \cL_{abcdefgh}[\cE] \\
= \cD^{12}_{[48]} \cE  \, \lambda^{8[24]} \bar\lambda^{8[24]} + a_2 \cD^{11}_{[44]} \cE  \,  \bar F_{ab}^{[2]} \lambda^{6 ab[18]} \bar\lambda^{8[24]} + a_3  \cD^{11}_{[44]} \cE  \,   H_{abc}^{[2]} (\lambda^{7[21]} \gamma^{abc} \bar\lambda^{7[21]})  \\
+  a_2 \cD^{11}_{[44]} \cE  \,   F_{ab}^{[2]} \bar\lambda^{6 ab[18]} \lambda^{8[24]} + a_4 \cD^{11}_{[44]} \cE  \,  \varepsilon_{ij} P_a^{[3i]} (\lambda^{7[21]} \gamma^{a} \bar\lambda^{7[20j]})\big|_{[42]} \\ + b_1 \cD^{11}_{[44]} \cE  \,  \varepsilon_{ij} \varepsilon_{kl}   \lambda^{8[ik22]} \bar\lambda^{8[jl22]}   + b_2  \cD^{11}_{[44]} \cE  \,    \lambda^{8[22]}_{ab} \bar\lambda^{8ab[22]} \\+ b_3 \cD^{11}_{[44]} \cE  \,  \lambda^{6 ab [18]} ( \bar\lambda^{9[25]} \gamma_{ab} \bar\chi^{[1]}) + b_4 \cD^{11}_{[44]} \cE  \,  ( \lambda^{7[21]} \gamma^a  \bar \lambda^{7[21]})(\bar\chi^{[1]} \gamma_a \chi^{[1]}) \\ + b_5  \cD^{11}_{[44]} \cE  \, ( \lambda^{7[21]} \gamma^{abc } \bar \lambda^{7[21]})(\bar\chi^{[1]} \gamma_{abc} \chi^{[1]}) + b_6   \cD^{11}_{[44]} \cE  \,  \bar\lambda^{6 ab [18]} ( \lambda^{9[25]} \gamma_{ab} \chi^{[1]}) 
 +  \cD^{10}_{[40]} \cE  \cdots 
\end{multline}
and similarly for $ \cL_\grad710$
\begin{multline}  \label{torsion-terms} \frac{1}{7!} \varepsilon^{abcdefgh} \cL_{bcdefgh}{}_\alpha^i[\cE] = c_1 \cD^{11}_{[44]} \cE  \lambda^{8[i23]}  \gamma^a_{\alpha\bdt} \bar\lambda^{7 \bdt [21]} 
 + \varepsilon^{ij} \cD^{11}_{[j43]} \cE  \Bigl(  c_2 \lambda^{8[24]} \ \bar\lambda^{7 a [19]}_\alpha\\  + c_3 \lambda^{8abcd[22]} \gamma_{bcd\alpha\bdt} \bar\lambda^{7 \bdt [21]}  + c_4 \lambda^{8ab[22]} \gamma_{b\alpha\bdt} \bar\lambda^{7 \bdt [21]}+ c_5 \lambda^{8[22]}_{bc} \gamma^{abc}_{\alpha\bdt} \bar\lambda^{7 \bdt [21]} \Bigr)  +  \cD^{10}_{[40]} \cE  \cdots   \end{multline}
 and its complex conjugate. Note that this ansatz is completely general provided one replaces each derivative term $\cD^n_{[4n]} \cE$ by a generic isospin $2n$ tensor $\cE^{\mathpzc{a}}_{[4n]}$, and the computation we shall carry out does not require such an assumption. It particular, there is no candidate monomial in the fields of odd isospin at this order, and we did not avoid such terms in the ansatz.  It will turn out to be enough to look at terms of isospin $24$ in $d\cL[\cE]=0$ to determine the properties of the function $\cE$, and because $\cL_\grad710$ only contributes at this order through a space-time derivative, one can neglect the contribution from $\cL_\grad710$ if one disregards terms including the momentum $P_a^{ijkl}$. At this order $d\cL[\cE]=0$ simplifies drastically to
 \be D_\alpha^i \Scal{ -\frac{1}{8!} \varepsilon^{abcdefgh} \cL_{abcdefgh}[\cE] } = \mathcal{O}(\cD^{11} \cE)\ , \quad \bar D_{\adt i}  \Scal{ -\frac{1}{8!} \varepsilon^{abcdefgh} \cL_{abcdefgh}[\cE] } = \mathcal{O}(\cD^{11} \cE) \ . \ee
Moreover, the superform being real, these two equations are equivalent. Restricting ourselves to the components of $D_\alpha^i \cL_\grad800$ of isospin $24$, the components of isospin $22$ of $ \cL_\grad800$ only contribute through the derivative of their tensor $\cE_{[44]}^{\mathpzc{a}}$, and therefore only mix with the isospin component $\cE_{[48]}  \lambda^{8[24]} \bar\lambda^{8[24]} $ through the covariant derivative acting on the fermions, but for the terms that are themselves in $\lambda^8 \bar \lambda^8$.  It follows that most of these contributions simply constrain these tensors to satisfy to 
\be \cD_{ijkl} \cE_{[44]}^{\mathpzc{a}} \propto \cE_{ijkl[44]} + \dots \ee 
in agreement with the ansatz \eqref{L800}. Computing these terms one would determine the coefficients $a_k$ and $b_k$ for $k\ge 3$ in  \eqref{L800}, but one would not get any constraint on the function $\cE$. The only terms constraining the function itself are the ones in $\lambda^9 \bar \lambda^8$, and we will therefore focus on the restricted ansatz 
\be-\frac{1}{8!} \varepsilon^{abcdefgh} \cL_{abcdefgh}[\cE] 
= \cD^{12}_{[48]} \cE   \, \lambda^{8[24]} \bar\lambda^{8[24]} +  \cE^\un_{[44]}   \,  \varepsilon_{ij} \varepsilon_{kl}   \lambda^{8[ik22]} \bar\lambda^{8[jl22]}   + \cE^\deux_{[44]}  \,    \lambda^{8[22]}_{ab} \bar\lambda^{8ab[22]} + \cdots 
\ee
where we do not assume that the two other $SO(3)$ tensors are also derivatives of the same function. At this point we need to precise the normalisation of the fermionic monomials
\bea
 \left(\lambda^{8}\right)^{(i_1 i_2 \dots i_{24})} &\equiv& \lambda_{1}^{(i_1 i_2i_3} \lambda_{2}^{i_4 i_5 i_6} \dots  \lambda_{8}^{i_{22} i_{23} i_{24})} \ , \hspace{8mm}  \left(\bar \lambda^{8}\right)^{(i_1 i_2 \dots i_{24})} \equiv \Scal{  \left(\lambda^{8}\right)^{(i_1 i_2 \dots i_{24})}}^*  \ , \CR
 \left(\lambda^{8}\right)^{(i_1 i_2 \dots  i_{22})}_{ab} &\equiv& \lambda_{\gamma}^{j (i_1 i_2} \lambda^{\gamma\;i_3 i_4}_{j} \left(\lambda^{6}\right)^{i_5 \dots i_{22})}_{ab} \ , \hspace{11mm}  \left(\bar \lambda^{8}\right)^{(i_1 i_2 \dots  i_{22})}_{ab} \equiv \Scal{  \left(\lambda^{8}\right)^{(i_1 i_2 \dots  i_{22})}_{ab} }^*  \ , \CR
 \left(\lambda^{6}\right)^{(i_1 i_2 \dots  i_{18})}_{ab} &\equiv& \frac{1}{4} (\gamma_{ab})^{\alpha\beta}  \frac{1}{6!}  \varepsilon_{\alpha \beta}^{\;\;\;\;\gamma \dots  \zeta}\lambda_{\gamma}^{( i_1 i_2i_3}  \dots  \lambda_{\zeta}^{i_{16} i_{17} i_{18})} \ . 
\eea
The first contribution comes from  
\be  \left(D_{\alpha}^i \cD^{12}_{[48]}\cE \right) \lambda^{8[24]} \bar\lambda^{8[24]} \ee
Using \eqref{DD_reduction} one obtains 
\bea D_\alpha^p  \cD^{12}_{[48]} \cE &=& - 2 \varepsilon^{pi} \lambda_\alpha^{jkl}  \cD_{ijkl}  \cD^{12}_{[48]} \cE \CR
&=&  - 2 \varepsilon^{pi} \lambda_\alpha^{jkl}  \biggl( \cD^{13}_{ijkl[48]}   + \frac{48}{17}  \varepsilon_{(i[1]} \varepsilon_{j[1]} \cD^{11}_{[44]} \cD_{kl)}{}^{pq} \cD_{[2]pq} \CR&&- \frac{704}{425} \varepsilon_{(i[1]} \varepsilon_{j[1]}\varepsilon_{k[1]} \cD^{10}_{[40]} \cD_{l)[2]r} \cD_{[1]}{}^{rpq} \cD_{[2]pq} +\varepsilon_{i[1]} \varepsilon_{j[1]}  \varepsilon_{k[1]} \varepsilon_{l[1]}   ( \cdots)   \biggr)  \cE\ , \qquad  \eea
and using the property that the maximal isospin monomial in $\lambda^9$ is of isospin $\frac{25}{2}$, one gets that the isospin $24$ contribution in $\cD^{13}_{[52]} \cE$ cancels out such that 
\be \label{D12E} \left(D_{\alpha}^i \cD^{12}_{[48]}\cE \right) \lambda^{8[24]} \bar\lambda^{8[24]}  = - \frac{96}{17}  \varepsilon_{j[1]} \varepsilon_{k[1]} \cD^{11}_{[44]} \cD_{lr}{}^{pq} \cD_{[2]pq} \cE \, \varepsilon^{i(j} \lambda_\alpha^{klr)} \lambda^{8[24]} \bar\lambda^{8[24]} + \dots \ee
where we neglect the terms of lower isospin. Using the covariant derivatives computed in Appendix \ref{appendix1Dim}
\bea D_\alpha^i \bar \lambda_\bdt^{jkl} &=& ( \gamma^a)_{\alpha\bdt} \Scal{ - i P_a^{ijkl} + \frac{1}{2} ( \lambda^{p(ij} \gamma_a \bar \lambda^{kl)}{}_p) - \varepsilon^{i(j} ( \chi^k \gamma_a \bar \chi^{l)})}  + \frac{i}{12} (\gamma^{abc})_{\alpha\bdt} \varepsilon^{i(j} H_{abc}^{kl)} - \frac{3}{4} \lambda^{pi(j}_\alpha \bar \lambda_\bdt^{kl)}{}_p\CR
D_\alpha^i \lambda_\beta^{jkl} &=& - \frac{1}{4} (\gamma^{ab})_{\alpha\beta} \varepsilon^{i(j} \Scal{ \bar F_{ab}^{kl)} + ( \bar \chi_p \gamma_{ab}\bar  \lambda^{kl)p} )}  + \frac{1}{4} \lambda_\alpha^{pi(j} \lambda_\beta^{kl)}{}_p - \frac{1}{2} C_{\alpha\beta} ( \lambda^{p(ij} \lambda^{kl)}{}_p) \CR
&& \hspace{90mm} + (\gamma^a)_{\alpha\bdt} \bar \chi^{\adt i} \bar \lambda^{\bdt\, jkl} (\gamma_a)_{\adt\beta} 
 \eea
 and concentrating on the terms in $\lambda^9 \bar \lambda^8$, one obtains after using Fierz identities 
 \bea &&  D_\alpha^i \Scal{ \cD^{12}_{[48]} \cE   \, \lambda^{8[24]} \bar\lambda^{8[24]} +  \cE^\un_{[44]}   \,  \varepsilon_{ij} \varepsilon_{kl}   \lambda^{8[ik22]} \bar\lambda^{8[jl22]}   + \cE^\deux_{[44]}  \,    \lambda^{8[22]}_{ab} \bar\lambda^{8ab[22]} }\CR
 &=& -\frac{12}{17} \varepsilon_{j[1]} \varepsilon_{k[1]} \cD^{11}_{[44]} \scal{ \cD_{lr[2]} \cE +4 \cD_{lr}{}^{pq} \cD_{[2]pq} \cE} \, \varepsilon^{i(j} \lambda_\alpha^{klr)} \lambda^{8[24]} \bar\lambda^{8[24]}\CR
 &&\quad  -  2 \Scal{  \frac{672}{47}\cD^{12}_{[jklr44]} \cE + \cD_{jklr} \cE^\un_{[44]}}  \varepsilon^{i(j} \varepsilon_{mp} \varepsilon_{nq} \lambda_{\alpha}^{klr)}\, \lambda^{8[mn22]} \bar\lambda^{8[pq22]} \CR
 &&\qquad  - 2 \Scal{ \frac{19}{184} \cD^{12}_{[jklr44]} \cE   + \cD_{jklr} \cE^\deux_{[44]}}  \varepsilon^{i(j} \lambda_{\alpha}^{klr)} \lambda^{8[22]}_{ab} \bar\lambda^{8ab[22]}  + \dots \eea
 These three combinations being linearly independent, one concludes that 
 \be \cE^\un_{[44]} = -  \frac{672}{47}\cD^{11}_{[44]} \cE \ , \qquad \cE^\deux_{[44]} = - \frac{19}{184} \cD^{12}_{[44]} \cE \ , \ee
 assuming that there is no inhomogeneous term satisfying to 
 \be \cD_{(i_1i_2i_3i_4} \cG_{i_5i_6i_7\dots i_{4n})} = 0 \ . \ee
 One can indeed convince oneself that there is no solution to this differential equation, which defines $4n+1$ independent first order equations for only $4n-3$ variables, \ie $4$ more equations at each order, equivalently as
\be \cD_{ijkl} \cG = 0 \, ,\ee
which only solution is a constant. Because there is no higher rank symmetric tensor, there is no solution for $n>1$. The most important equation is the constraint 
 \be \cD^{11}_{(i_1i_2i_3i_4\dots i_{44}} \Scal{ \cD_{i_{45} i_{46} i_{47} i_{48})} \cE +4 \cD_{i_{45} i_{46}}{}^{kl} \cD_{i_{47}i_{48})kl} \cE } = 0 \ . \ee 
 It follows from the structure of the linearised invariants that the terms of lower isospin will be all related, such that they will satisfy to similar equations of the form 
  \be \cD^{n}_{(i_1i_2i_3i_4\dots i_{4n}} \Scal{ \cD_{i_{4n+1} i_{4n+2} i_{4n+3} i_{4n+4})} \cE + 4 \cD_{i_{4n+1} i_{4n+2}}{}^{kl} \cD_{i_{4n+3}i_{4n+4})kl} \cE } = 0 \ . \ee 
such that one gets eventually 
\be \cD_{(ij}{}^{pq} \cD_{kl)pq} \cE = -\frac{1}{4}  \cD_{ijkl} \cE  \ , \label{SusySL3Tr} \ee
as in \eqref{SusySL3}. Let us prove now that \eqref{SusySL3} must indeed be strictly satisfied. Because of equation \eqref{SusySL3Tr}, the complete superform admits an expansion in derivatives of $\cE$ as
\be \cL[\cE ] = \cE \cL + \cD_{ijkl} \cE \, \cL^{ijkl} + \cD_{(ijkl} \cD_{pqrs)} \cE\, \cL^{ijklpqrs} + \dots \ee 
Expanding $d\cL[\cE]=0$ in the same way, one gets 
\be \cE d \cL + \frac{1}{5} \Delta \cE \, P_{ijkl} \wedge \cL^{ijkl}  = 0 \ee
 but because $\Delta \cE$ is necessarily a solution to the Laplace equation, \ie $\Delta^2 \cE=0$, the two terms must vanish independently. One deduces from the linearised analysis that $\cL^{ijkl}$ carries terms of the form 
 \be \cL^{ijkl} \sim t_8 t_8 R^3 \scal{ \lambda^{(ijk} \rho^{l)} + H^{(ij} H^{kl)} + \bar F^{(ij} F^{kl)}} + \dots \ee
  and  $P_{ijkl} \wedge \cL^{ijkl} $ does not vanish, so we conclude that supersymmetry indeed requires 
 \be \Delta \cE = 0 \ , \ee
 and therefore \eqref{SusySL3} is satisfied. Using this constraint, the tensor superforms $\cL^{[4n]}$ satisfy to the differential equation 
 \be d_\omega \cL^{[4n]} - \frac{6n}{4n+3} P^{[2]}{}_{ij} \wedge \cL^{[4n-2]ij}  + 2 P^{[4]} \wedge \cL^{[4n-4]} - \frac{2n(n+1)(2n+3)(2n+1)}{(4n+5)(4n+3)} P_{ijkl} \wedge \cL^{[4n]ijkl} = 0 \ee
 and the equation we have checked explicitly in this section is the $\lambda^9 \bar \lambda^8$ component of 
 \be d_\omega \cL^{[48]} - \frac{24}{17} P^{[2]}{}_{ij} \wedge \cL^{[46]ij}  + 2 P^{[4]} \wedge \cL^{[44]} = 0 \ . \ee 
 Note moreover that this equation must satisfy the consistency condition 
 \be  d_\omega^{\; 2}  \cL^{[4n]}  = - 2 n P^{[1]ijk} \wedge P_{ijkl} \wedge  \cL^{[4n-1]l} \label{Sl3Cs} \ . \ee 
One finds that the general solution to
 \be d_\omega \cL^{[4n]}  + 2 P^{[4]} \wedge \cL^{[4n-4]}  = a_n P^{[2]}{}_{ij} \wedge \cL^{[4n-2]ij} +  b_n  P_{ijkl} \wedge \cL^{[4n]ijkl} \ee
 satisfying to \eqref{Sl3Cs} is determined up to an integration constant $s$, as
\bea && d_\omega \cL^{[4n]}  + 2 P^{[4]} \wedge \cL^{[4n-4]} \CR
&=&   \frac{2n(4s-3)}{4n+3} P^{[2]}{}_{ij} \wedge \cL^{[4n-2]ij} +  \frac{(n+1)(2n+1)(2n+3-s)(2n+2s)}{(4n+5)(4n+3)} P_{ijkl} \wedge \cL^{[4n]ijkl}  \label{RepresentationUSL3}  \ . \hspace{10mm} \eea
One recognises that the coefficients are the same as in \eqref{DDE_reduction}, and therefore they are the equations satisfied by a closed superform $\cL[\cE_s]$ associated to a function $\cE_s$ satisfying to \eqref{EisenQuadra} in general. Equation \eqref{RepresentationUSL3} defines by construction a representation of $\mathfrak{sl}_3$ through the definition of the coset generators on the infinite sum $\oplus_{n=0}^\infty ({\bf 4n+1})$, which corresponds to the unitary representation of $SL(3,\mathds{R})$ on the set of functions satisfying to  \eqref{EisenQuadra}, with appropriate boundary conditions.  
 
\subsection{Anomalies}
We have proved in this section that the function multiplying $R^4$ in the supersymmetry invariant is the sum of a harmonic function of the complex scalar $T$ and a function of the $SL(3,\mathds{R})/SO(3)$ scalars solution to the quadratic equation \eqref{SusySL3}. However, the string theory threshold function appearing in the four-graviton amplitude \cite{Kiritsis:1997em} does not solve these equations strictly, and solve  inhomogeneous equations \eqref{InHomo} \cite{Green:2010wi}. The contributions responsible for these inhomogeneous terms come from the non-analytic component of the amplitudes, and are only captured by the supergravity 1-loop 1PI generating functional $\Gamma_{\mbox{\tiny 1-loop}} $. Therefore these terms do not appear in the string theory Wilsonian effective action 
\be S = \frac{1}{\kappa^2} S^\ord{0} + S^\ord{3} + \kappa^{\frac{4}{3}} S^\ord{5} + \kappa^2 S^\ord{6} + \mathcal{O}(\kappa^{\frac{10}{3}})  \ee
invariant with respect to local supersymmetry, but only in the 1PI effective action 
\be \Gamma =  \frac{1}{\kappa^2} S^\ord{0} + \scal{ S^\ord{3} +\Gamma_{\mbox{\tiny 1-loop}} }+ \kappa^{\frac{4}{3}} S^\ord{5}  + \kappa^2 \scal{ S^\ord{6} + \bigl[ S^\ord{3} \cdot\Gamma_{\mbox{\tiny 1-loop}}  ] + \Gamma_{\mbox{\tiny 2-loop}} } +  \mathcal{O}(\kappa^{\frac{10}{3}}) \ee
satisfying to the BRST master equation.

The discussion of the inhomogeneous term in the Laplace equation on $SL(2,\mathds{R})/SO(2)$ is very similar to the one of $\cN=4$ supergravity in four dimensions \cite{Bossard:2013rza}. The complex superform $\cL[\cF(T)]$ discussed in section \ref{FR48D} admits by construction the $R^4$ type terms
\begin{multline} \cL[\cF(T)] = \cF(T)\biggl(   e \,  t_8 t_8 R^4  - \frac{1}{48^2} \varepsilon_{abcdefgh} R^{ab} \wedge R^{cd} \wedge R^{ef} \wedge R^{gh} \\ - \frac{i}{24} \Scal{ R_a{}^b \wedge R_b{}^c \wedge R_c{}^d \wedge R_d{}^a - \tfrac{1}{4} R_{ab} \wedge R^{ab} \wedge R_{cd} \wedge R^{cd} } \biggr)  +\dots \ . \label{SuperFform} \end{multline} 
In this discussion it will be convenient to consider the upper complex half plan coordinate 
\be \uptau = i \frac{1-T}{1+T} \ , \ee
that transforms with respect to $SL(2,\mathds{R})$ as (with $ad-bc =1$)
\be \uptau \rightarrow \frac{ a \uptau + b }{c \uptau + d } \ .\ee
For the specific choice $\cF(T) = \uptau$, the imaginary part of the superform  \eqref{SuperFform} coincides with the dimensional reduction of the $R^4$ type invariant in eleven dimensions on $T^3$, where the imaginary part of $\uptau$ defines the $T^3$ volume modulus and its real part the pull-back of the 3-form potential on $T^3$. This exhibits by consistency with gauge invariance in eleven dimensions that one must have 
\be \mbox{Re}\bigl[ \cL[i] \bigr] = \frac{1}{24} \Scal{ R_a{}^b \wedge R_b{}^c \wedge R_c{}^d \wedge R_d{}^a - \tfrac{1}{4} R_{ab} \wedge R^{ab} \wedge R_{cd} \wedge R^{cd} } \ , \label{TopoInv} \ee
where $R_{ab}$ is the Riemann tensor superform. One can prove this property directly in eight dimensions  by studying the structure of the superform similarly as in \cite{Bossard:2013rza} in $\cN=4$ supergravity in four dimensions, although we will only report on this analysis in a forthcoming paper. 

It follows from \cite{Kiritsis:1997em} that the complete string theory Wilsonian action includes non-perturbative corrections in M-theory corresponding to Euclidean M2 branes wrapping $T^3$ such that the associated contribution to the Wilsonian effective action is 
\be S^\ord{1} = -\frac{3}{(2\pi)^4} \int \iota^* \mbox{Re}\bigl[ \cL[ \log \eta(\uptau) ]  \bigr] \ . \ee
The logarithm of the Dedekind eta function admits the expansion 
 \be - i \log \eta(\tau)   = \frac{\pi}{12} \tau - \sum_{n=1}^\infty \Scal{ \sum_{r|n} \frac{1}{r} } e^{2\pi i n \tau} \ , \ee
 in which the first term appears in the dimensional reduction of the eleven-dimensional $R^4$ type invariant on $T^3$ whereas the contributions in $e^{2\pi i n \tau}$ are associated to  $M2$ branes wrapping altogether $n$ times $T^3$. This function is not $SL(2,\mathds{Z})$ invariant, \ie 
 \be  \log\eta\Scal{ \frac{a \uptau + b}{c \uptau + d}} =  \log\eta(\uptau) + \frac{1}{2} \log( c \uptau + d) + i \pi \frac{\tilde b}{12} \ , \ee
 where $\tilde{b}$ is an integer, and therefore the $S^\ord{3}$ correction  to the Wilsonian action is not duality invariant. However, the supergravity theory admits a $U(1)$ anomaly in eight dimensions such that the supergravity 1-loop effective action is not $SL(2,\mathds{R})$ invariant, and neither does it preserve $SL(2,\mathds{Z})$. Using the family index theorem \cite{Singer} for the chiral fields $\chi_\alpha^i,\, \lambda_\alpha^{ijk},\, G_{abcd}^+$ and $\rho_{ab}{}_\alpha^i$, one computes the anomaly to the axial $U(1)$ current conservation  as in \cite{Marcus:1985yy} 
  \bea \partial_\mu J^\mu_9 &=& - \scal{ 2 \times (-3) + 4 \times (1) } \frac{7p_1^{\; 2} - 4 p_2}{5760} + (-2) \frac{ - p_1^{\; 2}+ 7 p_2}{90} - (-2) \frac{ 289\,  p_1^{\; 2} - 988}{5760} \CR
  &=& \frac{1}{8} \scal{ p_1^{\; 2} -4  p_2}  \CR
&=& \frac{1}{8(2\pi)^4} \Scal{ \tr R^4 - \tfrac{1}{4} ( \tr R^2 )^2 } \ . \label{U1Anomaly} \eea
Strictly speaking, the fermions contribute to the anomaly for the gauge axial $U(1)$, but one can compensate for it \cite{deWit} by introducing a correction to the effective action defined in term of the holomorphic function 
 \be \log\scal{ U(1+T) } \rightarrow \log\scal{ U(1+T) }  - 2 i \alpha + \log\scal{ c \, \uptau + d} \ , \ee
 such that the supergravity 1-loop 1PI generating functional transforms with respect to $SL(2,\mathds{R})$ as
 \be \Gamma_{\mbox{\tiny 1-loop}} \rightarrow \Gamma_{\mbox{\tiny 1-loop}}  + \frac{3}{2(2\pi)^4} \int \iota^* \mbox{Re}\bigl[ \cL[ \log(c \uptau + d)]\bigr] \ .  \ee 
 It follows that the sum of the 1PI supergravity effective action and the string theory Wilsonian effective action $\Gamma$ transforms with respect to $SL(2,\mathds{Z})$ as
 \be \Gamma \rightarrow \Gamma - 2\pi \tilde{b} \, \frac{1}{12(4\pi)^4} \int \Scal{  \tr R^4 - \tfrac{1}{4} ( \tr R^2 )^2 } \ . \ee
 Therefore the complete effective action is indeed duality invariant in the eight-dimensional Minkowski background. It is a non-trivial consistency check that the same Pontryagin classes combination defining the $U(1)$ anomaly \eqref{U1Anomaly} also supports the M5 brane gravitational anomaly \cite{Duff:1995wd}, and it follows that on a general Riemmanian spin manifold
 \be \frac{1}{12(4\pi)^4} \int \Scal{  \tr R^4 - \tfrac{1}{4} ( \tr R^2 )^2 } = - 2\hat{A} + \frac{\sigma}{4} \ , \ee
 where $\hat{A}$ is the integral roof genus and $\sigma$ is the signature. If one were to consider gravitational instanton corrections, $SL(2,\mathds{Z})$ invariance would require the effective action $\Gamma$ to be invariant modulo $2\pi$, and therefore the corresponding geometry to admit a signature multiple of four. This potential $Z_4$ obstruction is identical to the tadpole cancelation requirement studied on Calabi--Yau 4-folds in \cite{Sethi:1996es}. 
 
Note that the real part of the anomalous variation is the variation of a local functional because 
\be \log\mbox{Im}\Bigl[ \frac{ a \uptau + b}{c \uptau + d} \Bigr] = \log \mbox{Im}[\uptau]  - \frac{1}{2}\log\scal{ c \, \uptau + d} - \frac{1}{2} \log\scal{ c \, \bar \uptau + d}  \ , \ee
and the $t_8 t_8 R^4$  threshold depends on the duality invariant function \cite{Kiritsis:1997em} 
\be  \hat{E}_{[1]}(\uptau)  = - \pi \log\scal{ \mbox{Im}(\uptau)\, | \eta(\uptau)|^4} \ . \ee
The log of the dilaton is responsible for the inhomogeneous term in the Laplace equation 
\be \Delta  \hat{E}_{[1]}(\uptau)  = \pi \ . \ee

Similarly, the regularised $SL(3,\mathds{R})$ Eisenstein function $\hat{E}_{[{\scriptscriptstyle \frac{\scriptscriptstyle 3}{\scriptscriptstyle2}} 0]} $  includes a logarithm term \eqref{LogE32} that cannot be part of the Wilsonian effective action by supersymmetry. To understand this, let us define the BRST-like nilpotent operator defining the $\sl_3$ action 
\be \delta^{\sl_3} \, \cV_{ij}{}^I = \cV_{ij}{}^J C_J{}^I \ , \qquad \delta^{\sl_3}\, C_J{}^I  = -C_J{}^K C_K{}^I \ , \ee
where $C_J{}^I$ is a constant anticommuting traceless matrix. The non-trivial consistent anomaly for the $\sl_3$ Ward identities are in one to one correspondence with the $\su(2)$ anomalies in the bosonic theory \cite{Bossard:2010dq}. Therefore there is no anomaly for the rigid $SL(3,\mathds{R})$ in the theory independently of supersymmetry. However, one must take care that a potential naively trivial anomaly can be removed by a local counter-term without violating supersymmetry Ward identities themselves. Consider for example the variation of the logarithm function 
\be \delta^{\sl_3} \log \scal{ \cV_{ij}{}^I n_I \cV^{ij\, J} n_J}   = C_J{}^I   \frac{ 2 \cV_{ij}{}^J n_I \cV^{ij\, K}  n_K}{\cV_{kl}{}^L n_L \cV^{kl\, P} n_P} \ . \ee
By construction it satisfies equation \eqref{SusySL3}, and therefore one can define the supersymmetry invariant 
\be \cA_I{}^J \equiv \int \iota^*\cL\biggl[  \frac{ 2 \cV_{ij}{}^J n_I \cV^{ij\, K}  n_K}{\cV_{kl}{}^L n_L \cV^{kl\, P} n_P}\biggr] \ , \ee
which satisfies by construction to the Wess--Zumino consistency condition
\be \delta^{\sl_3} \scal{ C_J{}^I \cA_I{}^J }=0 \ . \ee
However it cannot be eliminated by adding a supersymmetric counter-term because the logarithm function itself does not satisfy to \eqref{SusySL3}. In this case one cannot compute the coefficient of the anomaly using the family index theorem because it is not related to a chirality anomaly, and one would need in fact to compute the soft limit of the 1-loop six point amplitude to compute the explicit coefficient. Nonetheless it is a consistent correction, and the string theory computation \cite{Kiritsis:1997em} indicates that it  indeed appears.

The appearance of these two anomalies is directly related to the appearance of a logarithm singularity in the four-point scattering amplitudes at 1-loop \cite{Fradkin}. The relation between the logarithm of the dilaton and the logarithmic divergence is explained in string theory \cite{Green:2010sp}. Rather naively, one can understand this property in field theory by noting that supersymmetry determines the power of the dilaton multiplying the $R^4$ type invariant counter-term in function of the dimension. Assuming the existence of some kind of supersymmetric regularisation valid at 1-loop order, one would naturally get an invariant counter-term in 
\be \frac{1}{\epsilon} e^{-\epsilon \phi} t_8 t_8 R^4  \ee
such that the finite term in $\epsilon$ would define the anomaly \cite{Bossard:2013rza}.

\section{$\cN=(2,2)$ supergravity in six dimensions}
In six dimensions, the Lorentz group is $SU^*(4)$ and the internal symmetry of maximal supergravity is $Sp(2) \times Sp(2)$. The scalar fields parametrise a symmetric space $SO(5,5)/( SO(5) \times SO(5))$ through $SO(5,5)$ matrices $ \cV_{ij}{}^I  , \  \cV_{\hat{\imath}\hat{\jmath}}{}^I $ satisfying to
\bea\eta_{IJ} \cV_{ij}{}^{I} \cV_{kl}{}^J &=& \frac{1}{2} \Omega_{ik} \Omega_{jl} - \frac{1}{2} \Omega_{il} \Omega_{jk} - \frac{1}{4} \Omega_{ij} \Omega_{kl}\ , \qquad  \cV_{ij}{}^I \cV^{ijJ} -  \cV_{\hat{\imath}\hat{\jmath}}{}^I\cV^{\hat{\imath}\hat{\jmath}J} = \eta^{IJ}   \ ,  \CR
\eta_{IJ} \cV_{\hat{\imath}\hat{\jmath}}{}^{I} \cV_{\hat{k}\hat{l}}{}^J &=&- \frac{1}{2} \Omega_{\hat{\imath}\hat{k}} \Omega_{\hat{\jmath}\hat{l}} + \frac{1}{2} \Omega_{\hat{\imath}\hat{l}} \Omega_{\hat{\jmath}\hat{k}} + \frac{1}{4} \Omega_{\hat{\imath}\hat{\jmath}} \Omega_{\hat{k}\hat{l}}\ , \qquad \eta_{IJ}   \cV_{\hat{\imath}\hat{\jmath}}{}^{I} \cV_{\hat{\imath}\hat{\jmath}}{}^{J}= 0 \ ,   \eea
that are antisymmetric symplectic traceless in the pairs of $Sp(2)$ indices $ij$ and $\hat{\imath}\hat{\jmath}$, and $I=1,10$ is in the vector representation of $SO(5,5)$, such that $\eta_{IJ}$ is the $SO(5,5)$ metric and $\Omega_{ik} \Omega^{jl} = \delta_i^j$ is the $Sp(2)$ symplectic matrix, and respectively is $\Omega_{\hat{\imath}\hat{\jmath}} $ for the second $Sp(2)$. Recall that the gamma matrices in five dimensions are such that both  the  conjugation charge matrix $\Omega_{ij}$ and the gamma matrices  are antisymmetric.  They define the momenta and the $\mathfrak{sp}(2)\oplus \mathfrak{sp}(2)$ connexion through the coset decomposition of the Maurer--Cartan form
\bea d\phi^\upmu P_\upmu{}^{ij\hat{\imath}\hat{\jmath}} &=&  d\cV^{ijI}  \, \cV^{\inv}_{I}{}^{\hat{\imath}\hat{\jmath}} = - \eta_{IJ} d\cV^{ijI}  \, \cV^{\hat{\imath}\hat{\jmath}J}   \ ,\CR
   d\phi^\upmu \omega_\upmu{}^i{}_j &=& - d\cV^{ik I}\, \cV^\inv_{I jk} = - \eta_{IJ}d\cV^{ik I}\, \cV_{jk}{}^J \ , \qquad d\phi^\upmu \omega_\upmu{}^{\hat{\imath}}{}_{\hat{\jmath}} = - d\cV^{\hat{\imath}\hat{k} I}\, \cV^\inv_{I \hat{\jmath}\hat{k}} = \eta_{IJ}   d\cV^{\hat{\imath}\hat{k} I}\, \cV_{\hat{\jmath}\hat{k}}{}^J \ . \qquad \eea
The covariant derivative $\cD_{ij\hat{\imath}\hat{\jmath}}$ is defined in the $[0,1]\times [0,1]$ of $Sp(2)\times Sp(2)$, \ie antisymmetric symplectic traceless in both pairs of indices, such that 
\be d_\omega  \cT(\phi) = 2  d\phi^\upmu P_\upmu{}^{ij\hat{\imath}\hat{\jmath}} \, \cD_{ij\hat{\imath}\hat{\jmath}} \, \cT(\phi) \ee
for any $Sp(2)\times Sp(2)$ tensor function of $\phi^\upmu$. The Dirac fermion fields are $\chi_\alpha^{i\hat{\jmath}\hat{k}}$ and $\bar\chi^{\alpha\, ij\hat{k}}$ that are also symplectic traceless in the $[1,0]\times[0,1]$ and $[0,1]\times[1,0]$ respectively, and
\be P^{ij\hat{\imath}\hat{\jmath}} = E^a P_a{}^{ij\hat{\imath}\hat{\jmath}} + 2 E^{\alpha [ i} \chi_\alpha^{j]\hat{\imath}\hat{\jmath}}- \tfrac{1}{2} \Omega^{ij} \Omega_{kl} E^{\alpha k} \chi_\alpha^{l\hat{\imath}\hat{\jmath}}+2 E_\alpha^{[\hat{\imath}}\bar  \chi^{\alpha \, ij\hat{\jmath}]}- \tfrac{1}{2} \Omega^{\hat{\imath}\hat{\jmath}} \Omega_{\hat{k}\hat{l}} E_\alpha^{\hat{k}} \bar \chi^{\alpha \, ij\hat{l}} \ . 
\ee
Here we write $\chi$ and $\bar\chi$ for convenience, but recall that they are both symplectic Majorana--Weyl and not complex conjugate. The only non-vanishing dimension-zero torsion components are
 \be T_\alpha^i{}_\beta^j{}^{\, a} = - i \Omega^{ij} \sigma^{a}{}_{\alpha\beta}\  ,  \qquad T^{\alpha \ai\, \beta\aj\, a} = - i \Omega^{\ai\aj} \sigma^{a \, \alpha\beta}\ ,  \ee
 where $\alpha=1$ to $4$ is in the fundamental of $SU^*(4)$ and $\sigma^{a\, \alpha\beta} = \frac{1}{2} \varepsilon^{\alpha\beta\gamma\delta} \sigma^a{}_{\gamma\delta}$.  One computes that the non-zero dimension 1/2 components of the torsion are
 \be\begin{split}  T_\alpha^i{}_\beta^j{}_\gamma^\ak &= \varepsilon_{\alpha\beta\gamma\delta} \bar \chi^{\delta\, ij\ak} \ ,  \\
 T_\alpha^i{}^{\beta\aj\, \gamma k} &= \delta_\alpha^\beta \bar \chi^{\gamma\, ik\aj} - \frac{1}{2} \delta_\alpha^\gamma \bar \chi^{\beta\, ik\aj} \ , 
 \end{split}\qquad\begin{split} 
T^{\alpha\ai\, \beta\aj\, \gamma k}  &= \varepsilon^{\alpha\beta\gamma\delta} \chi_\delta^{k \ai\aj} \ ,  \\
 T_\alpha^i{}^{\beta\aj}{}_\gamma^\ak  &= \delta_\alpha^\beta \chi_\gamma^{i\aj\ak}  - \frac{1}{2} \delta_\gamma^\beta \chi_\alpha^{i\aj\ak} \ . 
 \end{split} \label{Torsion6D} \ee
We refer to \cite{Tanii:1984zk,Cowdall:1998rs} for the complete set of fields of the theory. 
\subsection{The $R^4$ type  invariant}
Let us recall in a first step the structure of the linearised $R^4$ type invariants. The relevant harmonic variables parametrise $Sp(2)/U(2)$ with the split ${\bf 4} \cong {\bf 2}^\ord{-1} \oplus {\bf 2}^\ord{1}$. We define  $u^r{}_i , u_{r\, i}  $ such that 
\be \Omega^{ij} u^r{}_i  u_{s\, j}   = 2 \delta^r_s \ , \qquad u^r{}_i u_{r\, j} = \Omega_{ij} \ . \ee 
The linearised superfield $L^{ij\hat{\imath}\hat{\jmath}}$ satisfies 
\bea D^{\alpha k} L^{ij\hat{\imath}\hat{\jmath}} &=& 2 \Omega^{k[i} \bar\chi^{\alpha\, j]\hat{\imath}\hat{\jmath}} + \frac{1}{2} \Omega^{ij}  \bar\chi^{\alpha\, k\hat{\imath}\hat{\jmath}} \CR
D_\alpha^{\hat{k}} L^{ij\hat{\imath}\hat{\jmath}} &=& 2 \Omega^{\hat{k}[\hat{\imath}} \chi_\alpha^{\hat{\jmath}]ij} + \frac{1}{2} \Omega^{\hat{\imath}\hat{\jmath}}  \chi_\alpha^{\hat{k}ij} \eea
The superfield 
\be W = u^1{}_i u^2{}_j u^{\hat{1}}{}_{\hat{\imath}} u^{\hat{2}}{}_{\hat{\jmath}} L^{ij\hat{\imath}\hat{\jmath}} \  , \ee
is then G-analytic, \ie
\be u^{\hat{r}}{}_{\hat{\imath}} D_\alpha^{\hat{\imath}} W = 0 \ , \qquad u^{{r}}{}_{i} \bar D^{\alpha i} W = 0 \ .\label{11GA}  \ee
One can then define linearised invariants of the form 
\be \int d^8\theta d^8\bar\theta  du F_u^{[0,n]} F_{\hat{u}}^{[0,n]}  W^{4+n} \ee
where $F^{[0,n]}_u$ is the $2n$ order monomial in the harmonic variables in the corresponding  $[0,n]$ representation of $Sp(2)$, \ie 
\be F^{i_1j_1,i_2j_2,\dots i_nj_n}_u = \varepsilon^{r_1 s_1} u_{r_1}{}^{i_1}u_{s_1}{}^{j_1}   \varepsilon^{r_2 s_2} u_{r_2}{}^{i_2}u_{s_2}{}^{j_2} \cdots    \varepsilon^{r_n s_n} u_{r_n}{}^{i_n}u_{s_n}{}^{j_n} \ , \ee  
and respectively is $ F_{\hat{u}}^{[0,n]}$ for the second $Sp(2)$ factor.  Equivalently, one can think of this invariant in the superaction formalism \cite{Howe:1981xy} as being obtained from
\bea && \int d^8\theta_{[0,4]} d^8\bar\theta_{[0,4]}  L^{4+n\, [0,4+n],[0,4+n]} \CR
& \sim& L^{n\, [0,n],[0,n]} t_8 t_8 R^4 + \dots +  L^{n-12\, [0,n-12],[0,n-12]}  \chi^{8\, [0,4],[0,8]} \bar \chi^{8\, [0,8],[0,4]}\ . \eea
However the corresponding measure does not exist at the non-linear level, and the G-analicity condition \eqref{11GA} admits obstructions, \eg  
  \be u^r{}_i u^s{}_j T_\alpha^i{}_\beta^j{}_\gamma^\ak = \varepsilon_{\alpha\beta\gamma\delta} \bar \chi^{\delta\, rs\ak} \ ,  \qquad u^r{}_i u^{\hat{s}}{}_\aj T_\alpha^i{}^{\beta\aj\, \gamma k} = \delta_\alpha^\beta \bar \chi^{\gamma\, rk\hat{s}} - \frac{1}{2} \delta_\alpha^\gamma \bar \chi^{\beta\, rk\hat{s}} \ . \ee
The structure of the linearised invariant nonetheless suggests that the non-linear invariant admits an expansion in the derivatives of a function $\cE$ of the scalar fields in the $[0,n]\times [0,n]$.
  
%
%
The only term in a $\cE R^4$ type invariant involving the twelfth derivative of the function $\cE$ in the maximal highest weight representation is 
\be \cD^{12}_{[0,12],[0,12]}\cE\,  \chi^{8\, [0,4],[0,8]} \bar \chi^{8\, [0,8],[0,4]} \ , \ee
which means that each of the two sets of $Sp(2)$ indices are symmetrised according to the Young tableau  ${ \Yboxdim4pt  {\yng(12,12)}}$, with all symplectic traces projected out. 
The covariant derivative of this term gives two contributions that cannot be compensated by other terms
\bea&& D_\alpha^{[1,0],[0,0]} \scal{  \cD^{12}_{[0,12],[0,12]}\cE\,  \chi^{8\, [0,4],[0,8]} \bar \chi^{8\, [0,8],[0,4]}} \CR
&\sim & \cD^{13}_{[0,13],[0,11]}\cE\,  \chi^{9\, [1,3],[0,9]}_\alpha \bar \chi^{8\, [0,8],[0,4]} + \cD^{13}_{[2,11],[2,11]}\cE\,  \chi^{9\, [1,4],[2,7]}_\alpha \bar \chi^{8\,  [0,8],[0,4]} + \dots \eea
Counting the number of independent equations as in the last section for $SL(3,\mathds{R})$, one can convince oneself that the equations
\be \cD^{11}_{[0,11],[0,11]}\,  \cD^2_{[0,2],[0,0]}\, \cE \Big|_{[0,13],[0,11]} = 0 \ , \qquad \cD^{11}_{[0,11],[0,11]}\,  \cD^2_{[2,0],[2,0]}\, \cE \Big|_{[2,11],[2,11]} = 0 \ , \ee
imply respectively that 
\be \cD^2_{[0,2],[0,0]}\cE = 0 \ , \qquad  \cD^2_{[2,0],[2,0]}\cE = 0 \ . \label{RepreSO55}  \ee
It will be more convenient in the following to write the derivative $\cD_{ij\hat{\imath}\hat{\jmath}}$ in terms of vector indices of $SO(5)\times SO(5)$, \ie
\be \cD_{a\hat{b}} = \frac{1}{4} (\gamma_a)^{ij} (\gamma_{\hat{b}})^{\hat{\imath}\hat{\jmath}} D_{ij\hat{\imath}\hat{\jmath}} \ , \qquad P_\upmu{}^{a\hat{b}} = \frac{1}{4} (\gamma^a)_{ij} (\gamma^{\hat{b}})_{\hat{\imath}\hat{\jmath}}  P_\upmu{}^{ij\hat{\imath}\hat{\jmath}} \ , \ee
such that 
\be \cD_{a\hat{b}}{}^\upmu P_{\upmu}{}^{c\hat{d}} = \frac{1}{2} \delta_a^c \delta_{\hat{b}}^{\hat{d}} \ , \qquad P_{\upmu}{}^{a\hat{b}}  \cD_{a\hat{b}}{}^\upnu   = \frac{1}{2} \delta_\upmu^\upnu \ . \ee
Take care that we use the same letter $a$ for the internal $SO(5)$ vector representation, as for the Lorentz vector representation. There should be no confusion however, because we shall now on only use $a$ as an $SO(5)$ vector index. More explicitly, \eqref{RepreSO55} read
\be \cD_a{}^{\hat{c}} \cD_{b\hat{c}} \, \cE = \tfrac{1}{5} \delta_{ab} \cD_{c\hat{d}} \cD^{c\hat{d}} \, \cE \ , \qquad \cD_{[a}{}^{[\hat{c}} \cD_{b]}{}^{\hat{d}]} \, \cE = 0 \ . \label{R46Dc} \ee
Altogether with the similar equation obtained using $D^{\alpha \hat{\imath}} \cL_\grad{6}{0}{0}$ instead, \ie 
\be \cD^c{}_{\hat{a}} \cD_{c\hat{b}} \, \cE = \tfrac{1}{5} \delta_{\hat{a}\hat{b}} \cD{}_{c\hat{d}} \cD^{c\hat{d}} \,  \cE \ . \label{R46Dc2}\ee
The first equation implies that $ {\bf D}_{10}^{\; 2} \cE = \tfrac{1}{10}  \mathds{1}_{10} \Delta \cE$ in the vector representation, with the normalisation $\Delta = 2   \cD_{a\hat{b}}  \cD^{a\hat{b}}$. Using the spinor representation 
\be \frac{1}{2} \cD_{a\hat{b}} \gamma^a \gamma^{\hat{b}}\, \frac{1}{2}  \cD_{c\hat{d}} \gamma^c \gamma^{\hat{d}}  \, \cE = \frac{1}{4}  \cD_{a\hat{b}}  \cD^{a\hat{b}}\, \cE  + \frac{1}{4} \gamma^{ab} \gamma_{\hat{c}\hat{d}}\cD_{[a}{}^{[\hat{c}} \cD_{b]}{}^{\hat{d}]} \, \cE\ ,   \ee
and the second equation is equivalent to $ {\bf D}_{16}^{\; 2} \cE = \tfrac{1}{8}  \mathds{1}_{16} \Delta \, \cE$ in the Majorana--Weyl representation of $\so(5,5)$. Using the relations between the Casimir operators 
\be \tr   {\bf D}_{16}^{\; 2} =  2 \tr   {\bf D}_{10}^{\; 2} \ , \qquad  \tr   {\bf D}_{16}^{\; 4} =  - \tr   {\bf D}_{10}^{\; 4} + \frac{3}{4}   \scal{ \tr   {\bf D}_{10}^{\; 2} }^2+ 3  \tr   {\bf D}_{10}^{\; 2} \ , \label{D5Casimirs} \ee
one proves that 
\be \Delta \Scal{ \Delta + \frac{15}{2} } \cE = 0 \ . \ee
We can moreover fix this ambiguity by considering the general structure of the $d$-closed superform $ \cL[\cE]$. Similarly as in the preceding section, (\ref{R46Dc},\ref{R46Dc2}) imply that the symmetric traceless tensors $\cD_{(a_1}{}^{(\hat{a}_1} \dots \cD_{a_n)^\prime}{}^{\hat{a}_n)\prime} \cE$ define a complete base of the independent tensors one can obtain from the function $\cE$ and its covariant derivatives, such that the superform $ \cL[\cE] $ expands as 
\be \cL[\cE]  =  \cE \cL + \cD_a{}^{\hat{a}}  \cE\, \cL^{a}{}_{\hat{a}}  + \sum_{n=2}^{12} \cD_{a_1}{}^{\hat{a}_1} \dots \cD_{a_n}{}^{\hat{a}_n} \cE \, \cL^{a_1\dots a_n}{}_{\hat{a}_1\dots \hat{a}_n} \ , \ee
where each $\cL^{a_1\dots a_n}{}_{\hat{a}_1\dots \hat{a}_n}$ is symmetric traceless in the indices $a_1\dots a_n$ and $\hat{a}_1\dots \hat{a}_n$. Decomposing $d \cL[\cE]=0$ in the base of  $\cD_{(a_1}{}^{(\hat{a}_1} \dots \cD_{a_n)^\prime}{}^{\hat{a}_n)\prime} \cE$, one obtains equations of the form
\be d_\omega \cL^{a_1\dots a_n}{}_{\hat{a}_1\dots \hat{a}_n} = - 2 P^{(a_1}{}_{(\hat{a}_1} \wedge \cL^{a_2\dots a_n)^\prime}{}_{\hat{a}_2\dots \hat{a}_n)^\prime} + A_n P_b{}^{\hat{b}} \wedge  \cL^{a_1\dots a_n b}{}_{\hat{a}_1\dots \hat{a}_n \hat{b}} \ , \ee
where $A_n$ are constants  that remain to be determined and the first term is understood to be symmetric traceless in both sets of indices, \ie 
\begin{multline} P^{(a_1}{}_{(\hat{a}_1} \wedge \cL^{a_2\dots a_n)^\prime}{}_{\hat{a}_2\dots \hat{a}_n)^\prime}  \equiv P^{(a_1}{}_{(\hat{a}_1} \wedge \cL^{a_2\dots a_n)}{}_{\hat{a}_2\dots \hat{a}_n)} - \frac{n-1}{2n+1} \delta_{(\hat{a}_1 \hat{a}_2} P^{(a_1|\hat{b}} \wedge \cL^{a_2\dots a_n)}{}_{\hat{a}_3\dots \hat{a}_n)\hat{b}} \\ -  \frac{n-1}{2n+1} \delta^{({a}_1{a}_2} P_{b(\hat{a}_1} \wedge   \cL^{a_3\dots a_n)b}{}_{\hat{a}_2\dots \hat{a}_n)}  + \Scal{\frac{n-1}{2n+1}}^2 \delta^{({a}_1{a}_2} \delta_{(\hat{a}_1 \hat{a}_2}  P_{b}{}^{\hat{b}} \wedge   \cL^{a_3\dots a_n)b}{}_{\hat{a}_3\dots \hat{a}_n)\hat{b}}\ .  \end{multline} 
Using the Maurer--Cartan equation 
\be d\omega^a{}_b + \omega^a{}_c \wedge \omega^c{}_b = P^{a\hat{c}} \wedge P_{b\hat{c}} \ , \qquad d\omega_{\hat{a}}{}^{\hat{b}} + \omega_{\hat{a}}{}^{\hat{c}} \wedge \omega_{\hat{c}}{}^{\hat{b}} = P_{c\hat{a}} \wedge P^{c\hat{b}}  \ , \ee
one obtains the integrability condition 
\be d_\omega^{\; 2} \cL^{a_1\dots a_n}{}_{\hat{a}_1\dots \hat{a}_n} =-n P^{(a_1|\hat{c}} \wedge P_{b\hat{c}} \wedge \cL^{a_2\dots a_n)b}{}_{\hat{a}_1\dots \hat{a}_n} -n P_{b(\hat{a}_1} \wedge P^{b\hat{c}}  \wedge \cL^{a_1\dots a_n}{}_{\hat{a}_2\dots \hat{a}_n)\hat{c}} \ , \ee 
that determines the $A_n$ uniquely such that 
\be d_\omega \cL^{a_1\dots a_n}{}_{\hat{a}_1\dots \hat{a}_n} = - 2 P^{(a_1}{}_{(\hat{a}_1} \wedge \cL^{a_2\dots a_n)^\prime}{}_{\hat{a}_2\dots \hat{a}_n)^\prime} +\frac{(n+1)^2(2n+3)}{2(2n+5)} P_b{}^{\hat{b}} \wedge  \cL^{a_1\dots a_n b}{}_{\hat{a}_1\dots \hat{a}_n \hat{b}} \ . \label{dL6D} \ee
Using  (\ref{R46Dc},\ref{R46Dc2}) altogether with this equation, and in particular
\be d \cL= \frac{3}{10} P_a{}^{\hat{a}} \wedge  \cL^{a}{}_{\hat{a}} \ , \ee
one obtains that $d \cL[\cE]=0$ if and only if 
\be \Delta \cE = - \frac{15}{2} \cE \ , \ee
consistently with \cite{Green:2010kv}. For completeness we give the equations satisfied by $\cE$ in $Sp(2)\times Sp(2)$ representations 
\bea\Omega^{\hat{p}\hat{r}} \Omega^{\hat{q}\hat{s}}  D_{ij\hat{p}\hat{q}} D_{kl\hat{r}\hat{s}} \cE &=& - \frac{3}{10} \Scal{ \Omega_{ik}\Omega_{jl} - \Omega_{il}\Omega_{jk} - \tfrac{1}{2} \Omega_{ij} \Omega_{kl} } \cE \ , \CR
\Omega^{pr} \Omega^{\hat{q}\hat{s}}  D_{ip\hat{\jmath}\hat{q}} D_{kr\hat{l}\hat{s}} \cE &=& - \frac{15}{16} \Omega_{ik}\Omega_{\hat{\jmath}\hat{l}} \ \cE \ , \CR
\Omega^{pr} \Omega^{qs}  D_{pq\hat{\imath}\hat{\jmath}} D_{rs\hat{k}\hat{l}} \cE &=& - \frac{3}{10} \Scal{ \Omega_{\hat{\imath}\hat{k}}\Omega_{\hat{\jmath}\hat{l}} -\Omega_{\hat{\imath}\hat{l}}\Omega_{\hat{\jmath}\hat{k}}- \tfrac{1}{2} \Omega_{\hat{\imath}\hat{\jmath}} \Omega_{\hat{k}\hat{l}} } \cE \ .
 \eea
but it will be more convenient in the following to write them as
\be {\bf D}_{10}^{\; 2} \cE = -\frac{3}{4} \mathds{1}_{10} \cE\ , \qquad  {\bf D}_{16}^{\; 2} \cE = -\frac{15}{16}  \mathds{1}_{16} \cE \ . \label{12BPS6D} \ee
By construction \eqref{dL6D} defines a representation of $\so(5,5)$, which corresponds to the unitary representation of $SO(5,5)$ on the set of functions satisfying to \eqref{12BPS6D} with appropriate boundary conditions. This turns out to be the minimal unitary representation of  $SO(5,5)$  as we are going to exhibit in the next section. 
\subsection{Minimal unitary representation}
Let us solve these differential equations in the parabolic gauge associated to the decompactification limit. In this case one considers the decomposition
\be \mathfrak{so}(5,5) \cong \overline{\bf 10}^\ord{-2} \oplus \scal{ \gl_1 \oplus \sl_5}^\ord{0} \oplus {\bf 10}^\ord{2} \ . \ee
The representative in the vector representation can be written \be \cV_{10} = \left(\begin{array}{cc}\ e^{2\phi} v^{\inv}_{\, J}{}^a \ & \ e^{2\phi} v^{\inv}_{\, K}{}^a \, a^{KJ}\ \\
\ 0 \ & \ e^{-2\phi} v_a{}^J \ \end{array}\right) \ . \ee
Here both $a$ and $I$ run from 1 to 5, and correspond respectively to $SO(5)$ and $SL(5)$ indices. We shall not consider a specific gauge for the $SL(5)/SO(5)$ representative $v_a{}^I$. The associated momentum is 
\be P_{10} = \left(\begin{array}{cc}\ 2 d\phi \delta^a_b - P^a{}_b \ & \ \frac{1}{2} e^{4\phi} v^{\inv}_{\, I}{}^a v^{\inv}_{\, J}{}^b \, da^{IJ}\ \\
 \ -\frac{1}{2} e^{4\phi} v^{\inv}_{\, Ia} v^{\inv}_{\, Jb} \, da^{IJ}\ &\ - 2 d\phi \delta^b_a + P_a{}^b \ \end{array}\right) \ . \ee
The metric on the symmetric space is 
\be \tr P^2 = 40 d\phi^2 + 2 P^{ab} P_{ab} + \frac{1}{2} e^{8\phi} M^\inv_{IK} M^{\inv}_{JL} da^{IJ} da^{KL} \ , \ee
where $M^{IJ} = v_a{}^I v^{a\, J} $ and the coordinates on the symmetric space $SL(5)/SO(5)$ are defined such that 
\be P_\upmu^{ab} \cD_{ab}{}^\upnu = \frac{1}{2} \delta_{\upmu}^{\upnu} \ , \qquad P_\upmu^{ab} \cD_{cd}{}^\upmu = \frac{1}{4} \Scal{ \delta^a_c \delta^b_d + \delta^a_d \delta^b_c - \tfrac{2}{5} \delta^{ab}\delta_{cd}} \ . \ee
The corresponding differential operator is 
\be {\bf D}_{10} =  \left(\begin{array}{cc}\ \frac{1}{20} \partial_\phi \delta^a_b - \cD^a{}_b \ & \ e^{-4\phi} v^{aI} v^{bJ} \partial_{IJ} \ \\
 \ -  e^{-4\phi} v_a{}^{I} v_b{}^{J} \partial_{IJ}\ &-  \frac{1}{20} \partial_\phi \delta^a_b + \cD^a{}_b  \ \end{array}\right) \ . \ee
The repeated action of the covariant derivative on a function, which we write formally as a square even if the left derivative includes a connexion component, reads
\be {\bf D}_{10}^{\; 2} =  \left(\begin{array}{cc}\ \scal{ \frac{1}{20^2} \partial^{\; 2}_\phi + \frac{1}{10} \partial_\phi} \delta^a_b + \cD^a{}_c \cD^c{}_b - \scal{ \tfrac{1}{10} \partial_\phi + \tfrac{3}{4}}  \cD^a{}_b + e^{-8\phi} v^{aI} v^{cJ} v_b{}^K v_c{}^L \partial_{IJ} \partial_{KL}  \ & \ \dots \ \\
 \ 2 e^{-4\phi} v_{(a}{}^I v^{c\, J} \cD_{b)c} \partial_{IJ} \ &\ \dots  \ \end{array}\right) \ . \ee

We shall also consider the derivative operator in the spinor representation. The coset representative is then
\be \cV_{16} = \left(\begin{array}{ccc}\ e^{5\phi}  \ & \ \frac{1}{\sqrt{2} }  e^{5\phi}  a^{KL}\ & \ e^{5\phi} \frac{1}{8} \varepsilon_{KPQRS} a^{PQ} a^{RS} \  \\
\ 0 \ & \ e^{\phi} v_{[a}{}^K v_{b]}{}^L   \ & \ e^{\phi} v_{[a}{}^R v_{b]}{}^S   \frac{1}{2\sqrt{2}} \varepsilon_{RSKPQ} a^{PQ}  \ \\
\ 0 \ & \ 0 \ & \ e^{-3\phi} v^\inv_{\; K}{}^a
 \end{array}\right) \ . \ee
The associated momentum is
\be P_{16} = \left(\begin{array}{ccc}\ 5 d\phi   \ & \  \frac{1}{2\sqrt{2}} e^{4\phi} v^\inv_{\; I}{}^c v^\inv_{\; J}{}^d d a^{IJ}\ & \ 0 \  \\
\  \frac{1}{2\sqrt{2}} e^{4\phi} v^\inv_{\; I\, a} v^\inv_{\; J\, b} d a^{IJ} \ & \ d\phi \delta_{ab}^{cd} + 2 \delta_{[a}^{[c} P_{b]}{}^{d]}  \ & \ \frac{1}{4\sqrt{2}} \varepsilon_{abcef} e^{4\phi} v^\inv_{\; I}{}^e v^\inv_{\; J}{}^f d a^{IJ}\ \\
\ 0 \ & \ \frac{1}{4\sqrt{2}} \varepsilon^{acdef} e^{4\phi} v^\inv_{\; I\, e} v^\inv_{\; J\, f}d a^{IJ}\ \ & \ - 3 d\phi \delta^a_c - P^a{}_c 
 \end{array}\right) \ . \ee
The derivative operator reads
\be {\bf D}_{16} = \left(\begin{array}{ccc}\ \frac{1}{8} \partial_\phi  \ & \  \frac{1}{\sqrt{2}} e^{-4\phi}v^{cI} v^{dJ} \partial_{IJ} \ & \ 0 \  \\
\ \frac{1}{\sqrt{2}} e^{-4\phi}v_a{}^{I} v_b{}^{J} \partial_{IJ} \ & \ \frac{1}{40} \delta_{ab}^{cd} \partial_\phi   + 2 \delta_{[a}^{[c} \cD_{b]}{}^{d]}  \ & \ \frac{1}{2\sqrt{2}} \varepsilon_{abcef} e^{-4\phi} v^{eI} v^{fJ} \partial_{IJ}\ \\
\ 0 \ & \ \frac{1}{2\sqrt{2}} \varepsilon^{acdef} e^{-4\phi} v_e{}^{I} v_f{}^{J} \partial_{IJ}\ \ & \ - \frac{3}{40} \delta^a_c \partial_\phi - \cD^a{}_c
 \end{array}\right) \ , \ee
and acting twice on a function gives 
\bea && {\bf D}_{16}^{\; 2}  =  \left(\begin{array}{cc}\ \frac{1}{64} \partial^{\; 2}_\phi + \frac{1}{4} \partial_\phi + \frac{1}{2} e^{-8\phi} M^{IK} M^{JL} \partial_{IJ} \partial_{KL}   \ \\
\sqrt{2} e^{-4\phi}\Scal{  \frac{3}{40}   v_a{}^{I} v_b{}^{J} \partial_\phi + v_{[a}{}^{I} v^{cJ} \cD_{b]c} + \frac{3}{4}  v_a{}^{I} v_b{}^{J}  } \partial_{IJ}\  \\
\ \frac{1}{4} e^{-8\phi} v^\inv_{\; P}{}^a \varepsilon^{PIJKL} \partial_{IJ} \partial_{KL}  \ 
 \end{array}\right .  \\ 
 && \hspace{25mm}  \left . \begin{array}{cc}\ 
 \sqrt{2}   e^{-4\phi}\Scal{  \frac{3}{40}   v^{cI} v^{dJ} \partial_\phi + v^{[cI} v^{eJ} \cD_e{}^{d]} + \frac{3}{4}   v^{cI} v^{dJ}  } \partial_{IJ} \  &  \   \frac{1}{4} e^{-8\phi} v^\inv_{\; Pc}\varepsilon^{PIJKL} \partial_{IJ} \partial_{KL}  
  \  \\
\ A_{ab}^{cd} \ & \ C_{c,ab}  \ \\
 \ C^{a,cd} \ & \ B_c^a \ 
 \end{array}\right)\nn \ , \eea
 where 
 \bea A_{ab}^{cd} &=& \delta_{ab}^{cd} \Scal{ \frac{1}{40^2} \partial_\phi^{\;2 }+ \frac{1}{10} \partial_\phi} + 2 \delta_{[a}^{[c} \cD_{b]}{}^e \cD_e{}^{d]}  + 2 \cD_{[a}{}^{[c} \cD_{b]}{}^{d]} + \frac{1}{10}   \delta_{[a}^{[c} \cD_{b]}{}^{d]} \partial_\phi + \frac{1}{2} \delta_{[a}^{[c} \cD_{b]}{}^{d]}\CR
 && \hspace{10mm} +  e^{-8\phi} \Scal{ \frac{1}{2} \delta_{ab}^{cd} M^{IK} M^{JL}+  v_a{}^I v_b{}^J v^{cK} v^{dL}  -2 \delta_{[a}^{[c} v_{b]}{}^I v_e{}^{J}  v^{d]K} v^{eL}  } \partial_{IJ} \partial_{KL}  \CR
 C^{a,cd} &=& - \sqrt{2}  e^{-8\phi} v_e{}^I v_f{}^J  \Scal{ \frac{1}{80} \varepsilon^{acdef} \partial_\phi + \frac{1}{2} \varepsilon^{aefg[c} \cD_g{}^{d]} + \frac{1}{4} \varepsilon^{cdefg} \cD_g{}^a + \frac{1}{8} \varepsilon^{acdef}} \partial_{IJ} \CR
 B^a_b &=&\delta^a_b \Scal{ \frac{9}{40^2} \partial_\phi^{\; 2} + \frac{3}{20} \partial_\phi } + \cD^a{}_c \cD^c{}_b + \frac{3}{4} \cD^a{}_b \scal{ 1 + \tfrac{1}{5} \partial_\phi} \CR
 && \hspace{20mm}  + e^{-8\phi} \Scal{ \frac{1}{2} \delta^a_b M^{IK} M^{JL} -  v_b{}^I v_c{}^J v^{aK} v^{cL} } \partial_{IJ} \partial_{KL}  \qquad  .  \eea
 We can now solve equations \eqref{12BPS6D}. Let us consider in a first place solutions that do not depend on $a^{IJ}$. To solve these equations, we shall use the existence of functions $\Evector{s}$ on $SL(5)/SO(5)$ satisfying to  
 \bea \cD_a{}^c \cD_c{}^b \, \Evector{s} &=& \frac{3(4s-5)}{20} \cD_a{}^b \, \Evector{s} +  \frac{2s(2s-5)}{25} \delta_a^b \, \Evector{s} \CR
\Scal{ 2 \delta_{[a}^{[c} \cD_{b]}{}^e \cD_e{}^{d]}  + 2 \cD_{[a}{}^{[c} \cD_{b]}{}^{d]} } \Evector{s} &=& \frac{4s-5}{20} 2 \delta_{[a}^{[c} \cD_{b]}{}^{d]}  \, \Evector{s}  +  \frac{3s(2s-5)}{25} \delta_{ab}^{cd} \, \Evector{s} \eea
Here the notation we use is to exhibit that the corresponding Eisenstein function of $SL(5)$
\be  \Evector{s} = \sum_{n^I\in \mathds{Z}^5_*}  \scal{ n^I v^{\inv}_I{}^a v^\inv_{Ja} n^J}^{-s} \ , \label{SeriesE5} \ee
do satisfy to these differential equations (whenever the series \eqref{SeriesE5} converges), as can  straightforwardly be checked on their generating character $( n^I v^{\inv}_I{}^a v^\inv_{Ja} n^J)^{-s}$.

 Solving the spinorial equation ${\bf D}_{16}^{\; 2} \cE = - \frac{15}{16} \mathds{1} \cE$ one finds the solution 
 \be \cE = c_0 e^{-10\phi} +  e^{-6\phi} \Evector{\frac{3}{2}} +  e^{-10\phi} \Evector{\frac{5}{2}}  \label{SolDS2} \ee
 Solving then the vector equation ${\bf D}_{10}^{\; 2} \cE = - \frac{3}{4} \mathds{1} \cE$ one gets that the last function is not solution. For the Fourier modes $\propto e^{iq_{IJ} a^{IJ}} $ one gets directly from the spinor equation the 1/2 BPS constraint 
 \be \varepsilon^{IJKLP} q_{IJ} q_{KL} = 0 \ee
 and defining 
 \be Z_2 = 2 M^{IK} M^{JL} q_{IJ} q_{KL} \ee
 one obtains the two solutions
 \be \cE^\pm_q = \frac{e^{-6\phi}}{\sqrt{Z_2}} e^{\mp e^{-4\phi} \sqrt{Z_2} +  i q_{IJ} a^{IJ}} \ . \ee
Requiring a convergent behaviour in the large radius limit $e^{-2\phi} \rightarrow \infty$, the generic solution takes the form 
  \be \cE =  \int_{\frac{SL(5)}{SL(2)\times SL(3)\ltimes \mathds{R}^{2\times3}}}  \hspace{-10mm}d\mu(q)\hspace{2mm} F(q)  \ \frac{e^{-6\phi}}{\sqrt{Z_2}} e^{-e^{-4\phi} \sqrt{Z_2} +  i q_{IJ} a^{IJ}}  +  e^{-6\phi} \Evector{\frac{3}{2}}[G(p)]  \ , \ee
such that it is determined by a function $F(q)$ of seven variables, the general solution $ \Evector{\frac{3}{2}}$ being itself determined by a function $G(p)$ of four variables. These functions are not square integrable on $SO(5,5)/(SO(5)\times SO(5))$ because the Fourier mode of momentum $q_{IJ}$ does not depend on the flat directions of $q_{IJ}$ in $SL(5,\mathds{R})$ and the integral diverges as the infinite volume of $SL(2)/SO(2) \times SL(3)/SO(3) \ltimes \mathds{R}^{2\times 3}$. Nonetheless these solutions match precisely the solution obtained from the spherical vector of the minimal unitary representation of $SO(5,5)$ in \cite{Kazhdan:2001nx}. One should be able to factor out the infinite volume such that these functions are square integrable with respect to an appropriate measure, to show that the minimal representation of $SO(5,5)$ is indeed unitary. 
  
 We see that supersymmetry constrains each component of the Eisenstein function defining the $R^4$ coupling, in perfect agreement with the  explicit form of this function \cite{Kazhdan:2001nx}
\be E_{\mbox{\DSOX{\hspace{-0.5mm}\frac{3}{2}}0000}} = \frac{2\pi^2}{3} e^{-10\phi} +  e^{-6\phi} \Evector{\frac{3}{2}} + 4\pi \sum_{\ \vspace{-5mm}\begin{array}{c}\vspace{-3mm}  \scriptstyle q\in \mathds{Z}^{10} \\  \scriptstyle q\times q=0\end{array}} \Scal{ \sum_{n|q_{IJ}}n}  \frac{e^{-6\phi}}{ \sqrt{Z_2}} e^{-2\pi e^{-4\phi}  \sqrt{Z_2} + 2\pi i q_{IJ}   a^{IJ}} \ . \ee

\subsection{Relation to BPS instantons}
The differential equations \eqref{12BPS6D} implies a non-renormalisation theorem such that the instantons that contribute to the $R^4$ type correction in the effective action are 1/2 BPS. To see this, let us consider a supergravity instanton determined by the scalar fields only. In this case we consider the Euclidean theory for which the $SO(5,5)$ symmetry requires to consider a non-compact complex real form of the divisor group, \ie $SO(5,5)/SO(5,\mathds{C})$. This real form is suggested in six Euclidean dimensions because there is no self-dual 3-form in Euclidean signature, and the five 3-form field strengths must decompose into complex selfdual and complex-antiselfdual in the complex five dimensional representation of $SO(5,\mathds{C})$ and is complex conjugate. In this case the instanton can decouple from gravity and the metric is chosen to be flat. The scalar fields then lie in a nilpotent subgroup, which is characterised by the number of preserved supersymmetries. For a 1/2 BPS solution, one splits $Sp(4,\mathds{C})$ into
 \be \sp(4,\mathds{C}) \cong ( {\bf 3}_\mathds{C})^\ord{-1} \oplus \scal{ \mathfrak{gl}_1(\mathds{C}) \oplus \sl_2(\mathds{C})}^\ord{0} \oplus ( {\bf 3}_\mathds{C})^\ord{1} \ . \ee
 The fundamental representation in which lies the supersymmetry spinor parameters then decomposes as
 \be {\bf 4}_\mathds{C} \cong ( {\bf 2}_\mathds{C})^\ord{-\frac{1}{2}} \oplus ( {\bf 2}_\mathds{C})^\ord{\frac{1}{2}} \ee
 such that the grad 1/2 components carries the preserved half of supersymmetries. The coset component of $SO(5,5)$ decomposes accordingly such that 
 \be  ({\bf 5}\times {\overline{\bf 5}})_\mathds{R} \cong {\bf 1}^\ord{-2} \oplus ( {\bf 3} \oplus \overline{\bf 3})_\mathds{R}^\ord{-1} \oplus \scal{ \mathds{C} \oplus ({\bf 3}\otimes \overline{\bf 3})_{\mathds{R}}}^\ord{0} \oplus ( {\bf 3} \oplus \overline{\bf 3})_\mathds{R}^\ord{1}\oplus {\bf 1}^\ord{2} \ee
 The grad 2 component contains a single Lie algebra element that squares to zero in both the vector and the spinor representation. Defining the scalar fields with such a generator, the solution automatically preserves one half of supersymmetry because the Dirac spinors $\chi, \, \bar \chi$ do not carry a grad 5/2 component within this decomposition. The associated function is then simply a harmonic function on $\mathds{R}^6$. More explicitly, the 1/2 BPS instanton with a charge $q_{IJ}$ satisfying to the condition $\varepsilon^{IJKLP} q_{IJ} q_{KL}=0$ defines a rank 2 antisymmetric tensor 
 \be Z_{ab} =  v_{\circ a}{}^I  v_{\circ b}{}^J q_{IJ} \ , \ee  
 where the zero subscript indicates that this is the asymptotic value of the scalar at infinity. One can normalise it such that 
 \be J_{ab} = \frac{Z_{ab}}{\sqrt{\frac{1}{2} Z_{cd} Z^{cd}}} \,  .  \label{Jab} \ee
 This tensor is a non-degenerate symmetric tensor $J^{ij} = \frac{1}{2} J^{ab} \gamma_{ab}{}^{ij} $ in the spinor representation, that determines the preserved supersymmetry as the ones associated to spinor parameters satisfying to
  \be \epsilon_\alpha^i = J^i{}_j  \epsilon_\alpha^j \ . \ee 
 We consider the Euclidean Lagrangian density for which the scalars with negative kinetic terms have been dualised to 4-form potential $B_{IJ}$,
 \be H^\mu_{IJ} = \frac{1}{24 e} \varepsilon^{\mu\nu\sigma\rho\kappa\lambda} \partial_\nu B_{\sigma\rho\kappa\lambda\, IJ} \ ,\ee
 and that reduces to a sum of squares plus a total derivative as follows 
 \bea \frac{1}{e} \cL &=& 20 \partial^\mu \phi \partial_\mu \phi +P_\mu^{ab} P^\mu_{ab} + \frac{1}{2} e^{-8\phi} v_{a}{}^I  v_{b}{}^J H^\mu_{IJ} v^{a K}   v^{b L} H_{\mu KL} \\
 &=& \Scal{ 2 \partial_\mu \phi \, \delta^{ab} - P_\mu^{ab} - e^{-4\phi}  J^{c(a} v^{b)I}  v_{c}{}^J H_{\mu IJ}} \Scal{ 2 \partial^\mu \phi\,  \delta_{ab} - P^\mu_{ab} - e^{-4\phi}   J^c{}_{(a} v_{b)}{}^{I}  v_{c}{}^J H^\mu_{ IJ}} \CR
 && + \frac{1}{2} e^{-8\phi} \scal{ \delta^a_c \delta^b_d -  J^{ae} J_{ce} \delta^b_d- \tfrac{1}{2} J^{ab} J_{cd} } v_a{}^I v_b{}^J v^{c K} v^{d L}  H^\mu_{IJ} H_{\mu KL}  +  \partial_\mu \scal{ e^{-4\phi} J^{ab} v_a{}^I v_b{}^J H^\mu_{IJ}} \nn  \eea
Cancelling the squares gives the equation
\be d\scal{ e^{4\phi} v^\inv_{\; I}{}^a v^\inv_{\; J}{}^b J_{ab}} = \star H_{IJ}  \ . \ee
One obtains the solution 
\bea e^{4\phi} v^\inv_{\; I}{}^a v^\inv_{\; J}{}^b J_{ab} &=&   e^{4\phi_\circ} v^\inv_{\circ \; I}{}^a v^\inv_{\circ \; J}{}^b J_{ab} - \frac{ \kappa^2}{4\pi^3}\frac{q_{IJ}}{r^4} \ ,  \label{EqM}\\ 
H_{IJ} &=& \frac{ \kappa^2}{4\pi^3}  q_{IJ} d\Omega_5 \ ,  \eea
 which action is determined by the total derivative term and gives 
 \be S = e^{-4\phi_\circ} \sqrt{ 2 v_{\circ\, a}{}^I  v_{\circ\, b}{}^J q_{IJ} v_\circ^{a K}   v_\circ^{b L} q_{KL} } \ .  \ee
 The other equations require the scalars to be constant in the directions preserving $J_{ab}$, such that the scalar fields are determined by equation \eqref{EqM} up to constant flat directions.  The Noether charge associated to these solutions satisfies the nilpotency condition 
 \be {\bf Q}_{\bf 10}^{\; 2}=0 \ , \qquad {\bf Q}_{\bf 16}^{\; 2}=0 \ . \ee
 Equation \eqref{12BPS6D} defines a quantised version of these algebraic equations.  Moreover, the form of the associated Fourier mode is characteristic of an instanton correction 
 \be \sim \frac{e^{-10\phi}}{S} e^{-2\pi S + 2\pi  i q_{IJ} a^{IJ}} \ . \ee
  It is therefore legitimate to believe that the next coupling in $\nabla^4 R^4$ will be a function satisfying to differential equations defining a quantisation of the algebraic equations associated to 1/4 BPS instantons. In $\mathfrak{so}(5,5)$, the next to minimal nilpotent orbit is not unique, and there are in fact three disconnected orbits connected to the minimal orbit associated to 1/2 BPS instantons. The two isomorphic smallest orbits are obtained by relaxing the nilpotency condition in the vector representation 
  \be {\bf Q}_{\bf 10}^{\; 3}=0 \ , \qquad {\bf Q}_{\bf 16}^{\; 2}=0 \ . \ee
In this case however, the instanton cannot be defined in the standard Euclidean formulation of the theory, and one must consider a real form of the divisor group that allows for an independent decomposition of the two factors. This is incompatible with the representation of the $SO(5,5)$ symmetry on the 3-form field strengths, and recovering the symmetry would require some analytic continuation of the Euclidean path integral in such a background. One can consider for example the coset $SO(5,5)/(SO(1,4)\times SO(4,1))$ such that only one $Sp(1,1)$ factor decomposes as
\be \sp(1,1) \cong {\bf 3}^\ord{-2} \oplus \scal{ \gl_1 \oplus \mathfrak{su}(2) }^\ord{0} \oplus {\bf 3}^\ord{2}\ .  \ee
In this case the instanton can be described within the scalar fields valued in the Riemannian symmetric space $R_+^* \times SO(4,4)/(SO(4)\times SO(4))$ coupled to eight 4-forms in the ${\bf 8}$ of $SO(4,4)$.

 The two orbits correspond to the choice of $Sp(1,1)$ factor. The coset component then decomposes as 
 \be {\bf 5}\otimes {\bf 5}^\prime \cong {\bf 5^\prime}^\ord{-2} \oplus ( {\bf 3}\otimes {\bf 5}^\prime)^\ord{0} \oplus {\bf 5^\prime}^\ord{2} \ , \ee
 and a representative of the nilpotent orbit is a generic (time-like vector) element of the $ {\bf 5^\prime}^\ord{2}$ component.\footnote{A space-like vector corresponds to a solution that violates the  BPS bound, and which is therefore unphysical.}  The associated solution preserves one half of the chiral (respectively antichiral) supercharges, depending on the choice of $Sp(1,1)$ factor. Note that in the decomposition of the vector representation, with $a$ of $SO(5)^\prime$ and $\hat{a}$ of $SO(5)$, the charge satisfies moreover
 \be Q_a{}^{\hat{c}} Q_{b\hat{c}} = 0 \ , \ee
 although ${\bf Q}_{10}^{\; 2} \ne 0$. 
 
 The third orbit is obtained by relaxing the nilpotency condition in the spinor representation 
  \be {\bf Q}_{\bf 10}^{\; 2}=0 \ , \qquad {\bf Q}_{\bf 16}^{\; 3}=0 \ . \ee
In this case one can consider the standard formulation of the Euclidean theory with coset $SO(5,5)/SO(5,\mathds{C})$ and the decomposition 
 \be \sp(4,\mathds{C}) \cong ( {\bf 1}_\mathds{C})^\ord{-2} \oplus ( {\bf 2}_\mathds{C})^\ord{-1} \oplus \scal{ \mathfrak{gl}_1(\mathds{C}) \oplus \sl_2(\mathds{C})}^\ord{0} \oplus ( {\bf 2}_\mathds{C})^\ord{1}\oplus ( {\bf 1}_\mathds{C})^\ord{2} \ . \ee
 The fundamental representation in which lies the spinor then decomposes as
 \be {\bf 4}_\mathds{C} \cong ( {\bf 1}_\mathds{C})^\ord{-1} \oplus ( {\bf 2}_\mathds{C})^\ord{0}  \oplus ( {\bf 1}_\mathds{C})^\ord{1} \ee
 such that the grad 1 component carries the preserved quarter of supersymmetries. The coset component of $SO(5,5)$ decomposes accordingly such that 
 \be ({\bf 5}\times {\overline{\bf 5}})_\mathds{R} \cong ( {\bf 2} \otimes \overline{\bf 2})_\mathds{R}^\ord{-2} \oplus  ( {\bf 2} \oplus \overline{\bf 2})_\mathds{R}^\ord{-1} \oplus \scal{ \mathds{R} \oplus ({\bf 2}\otimes \overline{\bf 2})_{\mathds{C}}}^\ord{0} \oplus   ( {\bf 2} \oplus \overline{\bf 2})_\mathds{R}^\ord{1} \oplus ( {\bf 2} \otimes \overline{\bf 2})_\mathds{R}^\ord{2}\ee
and a representative of the nilpotent orbit is a generic  (time-like $SO(1,3)$ vector) element of $ ( {\bf 2} \otimes \overline{\bf 2})_\mathds{R}^\ord{2}$. The associated instanton preserves one quarter of supersymmetry (one quarter chiral and one quarter antichiral). 

\subsection{The $\nabla^4 R^4$ type invariants} 
We shall consider in a first place the linearised $\nabla^4 R^4$ invariants. There are three 1/4 BPS measures one can define in the linearised approximation \cite{Bossard:2009sy}, although none of them extends to the non-linear level as one straightforwardly checks using \eqref{Torsion6D}. 
\subsubsection*{The chiral invariant}
The first two  $\nabla^4 R^4$ type invariants are parity conjugate, and we shall only discuss the first. It can be defined in the linearised approximation by considering harmonic variables with respect to one $Sp(2)$ factor only \cite{Bossard:2009sy}, such that the superfield 
\be W^{ij} =  u^{\hat{1}}{}_{\hat{\imath}} u^{\hat{2}}{}_{\hat{\jmath}} L^{ij\hat{\imath}\hat{\jmath}} \  , \ee
satisfies the G-analyticity condition 
\be u^{\hat{r}}{}_{\hat{\imath}} D_\alpha^{\hat{\imath}} W^{ij}  = 0 \ . \ee 
One can again define linearised invariants 
\be \int d^8\theta d^{16}\bar\theta  du F_{\hat{u}}^{[0,n+2k]}  (W^{ij} W_{ij})^{4+k} W^{n [0,n]} \sim    \int d^8\theta_{[0,4]} d^{16}\bar\theta  L^{4+2k\, [0,0],[0,4+2k]} L^{n\, [0,n],[0,n]} \ . \ee
Now there are more representations allowed, and this suggests that one must consider the $(n+2k)^{\rm th}$ derivative of the defining function in all representations $[0,n]\times [0,n+2k]$. This is consistent with the property that 
\be  \cD^2_{[2,0],[2,0]}\cE = 0 \,, \qquad \cD^2_{[0,2],[0,0]} \cE = 0\ ,  \label{SpinorSquare}  \ee
proposed as a quantisation of the corresponding 1/4 BPS condition in the last section. Matching the linearly independent invariant to the independent linearised invariants, one concludes that the complete invariant associated to a function $\cE$ must admit the following expansion 
\bea  \cL[\cE]_\gra{2}{0} &=& \sum_{n=0}^{12} \sum_{k=0}^{8-n/2} \cD^{n+2k}_{[0,n],[0,n+2k]} \cE \, \cL^{[0,n],[0,n+2k]}_\gra{2}{0} \CR
&\sim& \cE \nabla^4 R^4 + \dots + \cD^{16}_{[0,12],[0,16]} \cE \, H^{4 \, [0,0],[0,4]}  \chi^{8\, [0,4],[0,8]} \bar \chi^{8\, [0,8],[0,4]} +\dots \   \label{20Invariant} \eea
Decomposing $d \cL[\cE]_\gra{2}{0} =0$ in the base of  $ \cD^{n+2k}_{[0,n],[0,n+2k]} \cE $, one obtains that 
\begin{multline} d_\omega \cL^{a_1\dots a_{n}}{}_{\hat{a}_1\dots \hat{a}_{n+2k}} = - 2 P^{(a_1}{}_{(\hat{a}_1} \wedge \cL^{a_2\dots a_{n})^\prime}{}_{\hat{a}_2\dots \hat{a}_{n+2k})^\prime} -  \frac{2n+2}{2n+5}  P_{b(\hat{a}_1} \wedge \cL^{a_1\dots a_{n})^\prime b}{}_{\hat{a}_2\dots \hat{a}_{n+2k})^\prime } \\ + 2 A_{n,k} P^{(a_1|\hat{b}} \wedge \cL^{a_2\dots a_{n})^\prime}{}_{\hat{a}_1\dots \hat{a}_{n+2k})^\prime \hat{b}}+ B_{n,k}  P_b{}^{\hat{b}} \wedge  \cL^{a_1\dots a_{n} b}{}_{\hat{a}_1\dots \hat{a}_{n+2k} \hat{b}} \ , \end{multline}
where the two first coefficients are determined by the decomposition 
\bea && \cD_a{}^{\hat{a}}  \, \cD_{(a_1}{}^{(\hat{a}_1} \cD_{a_2}{}^{\hat{a}_2} \cdots \cD_{a_n)^\prime}{}^{\hat{a}_n}  \cD_{b_1}{}^{\hat{a}_{n+1}} \cD^{b_1 \hat{a}_{n+2}}  \cdots   \cD_{b_k}{}^{\hat{a}_{n+2k-1}} \cD^{b_k| \hat{a}_{n+2k})^\prime}    \cE \CR
&=& \cD_{(a}{}^{(\hat{a}}  \, \cD_{a_1}{}^{\hat{a}_1} \cD_{a_2}{}^{\hat{a}_2} \cdots \cD_{a_n)^\prime}{}^{\hat{a}_n}\cD_{b_1}{}^{\hat{a}_{n+1}} \cD^{b_1 \hat{a}_{n+2}}  \cdots   \cD_{b_k}{}^{\hat{a}_{n+2k-1}} \cD^{b_k| \hat{a}_{n+2k})^\prime}  \cE \CR
&& + \frac{n}{2n+3} \delta_{{a}({a}_1} \cD_{a_2}{}^{(\hat{a}_2} \cdots \cD_{a_n)^\prime}{}^{\hat{a}_n} \cD_b{}^{\hat{a}} \cD^{b \hat{a}_1}  \cD_{b_1}{}^{\hat{a}_{n+1}} \cD^{b_1 \hat{a}_{n+2}}  \cdots   \cD_{b_k}{}^{\hat{a}_{n+2k-1}} \cD^{b_k| \hat{a}_{n+2k})^\prime}  \cE  \CR
&&  \hspace{10mm} + \mathcal{O}( \cD^{n+2k-1} \cE) \ . 
\eea
Checking the consistency condition 
\bea &&  d_\omega^{\; 2} \cL^{a_1\dots a_{n}}{}_{\hat{a}_1\dots \hat{a}_{n+2k}} \CR
&=&-n P^{(a_1|\hat{c}} \wedge P_{b\hat{c}} \wedge \cL^{a_2\dots a_{n})b}{}_{\hat{a}_1\dots \hat{a}_{n+2k}} -(n+2k) P_{b(\hat{a}_1} \wedge P^{b\hat{c}}  \wedge \cL^{a_1\dots a_{n}}{}_{\hat{a}_2\dots \hat{a}_{n+2k})\hat{c}} \ , \eea 
one gets the three independent equations 
\bea A_{n,k} &=& \frac{(n+2k+1)(2n+4k+3)}{(n+2k)(2n+4k+5)} A_{n-1,k} \CR
\frac{n+2}{2n+7} B_{n,k} &=& \frac{n+1}{2n+5} \frac{ (n+2k+1)(2n+4k+3)}{(n+2k)(2n+4k+5)} B_{n+1,k-1} \CR
B_{n,k} &=& 2 \frac{n+1}{2n+5} A_{n,k} + \frac{(n+1)(2n+3)(n+2k+1)}{2(2n+5)}\ ,  \eea
that admit the general solution
\bea A_{n,k} &=&  \frac{(n+2k+1)(2k+2s-3)(2k+5-2s)}{4(2n+4k+5)} \ , \CR
 B_{n,k} &=& \frac{2 (n + 1) (n + 2 k + 1)(n + k + s) (n + k + 4 - s) }{(2 n + 
   5) (2 n + 4 k + 5)}  \ . \eea
However assuming the expansion of the invariant \eqref{20Invariant}, $\cL^{[0,n],[0,n+2k]}_\gra{2}{0}$ only exist for $k\ge0$ and therefore $A_{n,-1}$ must vanish by consistency. We get therefore $s=\frac{5}{2}$ or $\frac{3}{2}$, which define the same solutions for  $A_{n,k}$ and $B_{n,k}$. We conclude that the function $\cE$ must satisfy to
\bea && \cD_a{}^{\hat{a}}  \, \cD_{(a_1}{}^{(\hat{a}_1} \cD_{a_2}{}^{\hat{a}_2} \cdots \cD_{a_n)^\prime}{}^{\hat{a}_n}  \cD_{b_1}{}^{\hat{a}_{n+1}} \cD^{b_1 \hat{a}_{n+2}}  \cdots   \cD_{b_k}{}^{\hat{a}_{n+2k-1}} \cD^{b_k| \hat{a}_{n+2k})^\prime}    \cE \CR
&=& \cD_{(a}{}^{(\hat{a}}  \, \cD_{a_1}{}^{\hat{a}_1} \cD_{a_2}{}^{\hat{a}_2} \cdots \cD_{a_n)^\prime}{}^{\hat{a}_n}\cD_{b_1}{}^{\hat{a}_{n+1}} \cD^{b_1 \hat{a}_{n+2}}  \cdots   \cD_{b_k}{}^{\hat{a}_{n+2k-1}} \cD^{b_k| \hat{a}_{n+2k})^\prime}  \cE \CR
&& + \frac{n}{2n+3} \delta_{{a}({a}_1} \cD_{a_2}{}^{(\hat{a}_2} \cdots \cD_{a_n)^\prime}{}^{\hat{a}_n} \cD_b{}^{\hat{a}} \cD^{b \hat{a}_1}  \cD_{b_1}{}^{\hat{a}_{n+1}} \cD^{b_1 \hat{a}_{n+2}}  \cdots   \cD_{b_k}{}^{\hat{a}_{n+2k-1}} \cD^{b_k| \hat{a}_{n+2k})^\prime}  \cE  \CR
&&  -  \frac{k(k-1) (n+2k)}{2n+4k+3} \delta^{\hat{a}(\hat{a}_{n+1}} \cD_{a}{}^{(\hat{a}_{n+2}}  \cD_{(a_1}{}^{(\hat{a}_1}  \cdots \cD_{a_n)^\prime}{}^{\hat{a}_n}  \cD_{b_2}{}^{\hat{a}_{n+3}} \cD^{b_2 \hat{a}_{n+4}}  \cdots   \cD_{b_k}{}^{\hat{a}_{n+2k-1}} \cD^{b_k| \hat{a}_{n+2k})^\prime}    \cE \CR
&& - \frac{n (n+2 k ) (2 n+2k+1) (2n+2k+3)}{4 ( 2 n+3) (2n+4k+3n)}   \CR
&& \hspace{20mm} \times \    \delta_{{a}({a}_{1}}  \delta^{\hat{a}(\hat{a}_1}   \cD_{a_2}{}^{\hat{a}_2} \cdots \cD_{a_n)^\prime}{}^{\hat{a}_n}  \cD_{b_1}{}^{\hat{a}_{n+1}} \cD^{b_1 \hat{a}_{n+2}}  \cdots   \cD_{b_k}{}^{\hat{a}_{n+2k-1}} \cD^{b_k \hat{a}_{n+2k})^\prime}    \cE  \ . 
\label{DiffNextMin} \eea
and in particular 
\be \cD_a{}^{\hat{a}}  \cD_{b}{}^{\hat{b}}  \cE =  \cD_{(a}{}^{(\hat{a}}  \, \cD_{b)^\prime}{}^{\hat{b})^\prime} \cE  + \frac{1}{5} \delta_{{a}{b}} \cD^{c(\hat{a}} \cD_c{}^{\hat{b})^\prime}   \cE - \frac{3}{20} \delta_{ab} \delta^{\hat{a}\hat{b}} \cE \ ,  \ee
such that 
\be \cD_{a}{}^{\hat{c}} \cD_{b\hat{c}} \cE = - \frac{3}{4} \delta_{{a}{b}} \cE \ . \label{VecEq}  \ee
Considering more generally a function $\cE_s$ satisfying to 
\be   {\bf D}_{16}^{\; 2} \cE_s = \frac{s(s-4)}{4}  \mathds{1}_{16} \cE_s \label{EsD5Sol} \ ,  \ee
one computes using the property that  ${\bf D}_{10}$ can be realised from ${\bf D}_{16}$ through a commutator with the $SO(5,5)$ gamma matrices, that 
\be {\bf D}_{10}^{\; 3} \cE_s = (s-1)(s-3) {\bf D}_{10}  \cE_s \ .   \ee
This equation is only consistent with the second equation in \eqref{SpinorSquare} if $s=\frac{5}{2}$ or $\frac{3}{2}$, such that one gets indeed that any non-trivial solution to \eqref{SpinorSquare} must solve \eqref{VecEq}.

The function defining the closed superform \eqref{20Invariant} satisfies therefore to an equation compatible with the function defining the $R^4$ type invariant, consistently with the expected properties of the effective action in type II string theory \cite{Green:2010wi,Green:2010kv}. Solving this spinor differential equation \eqref{EsD5Sol} for $s=\tfrac{5}{2}$ one finds the solution \eqref{SolDS2} for $q_{IJ}=0$, and the complex solution 
\be \cE_q =  \cF(\tau_{ A_1(q)}) \frac{e^{-6\phi}}{\sqrt{Z_2}} e^{-e^{-4\phi} \sqrt{Z_2} +  i q_{IJ} a^{IJ}} ,  \ee
and its complex conjugate, where the upper complex half plan variable $\tau_{ A_1(q)}$ parametrises the $v_{ A_1(q)}$ component of $v_a{}^I$ in the $SL(2)$ subgroup of the stabiliser $SL(2)\times SL(3)\ltimes \mathds{R}^{2\times3} \subset SL(5)$ of $q_{IJ}$. One computes moreover that 
\be \Scal{  {\bf D}_{10}^{\; 2}  + \frac{3}{4} \mathds{1}_{10}} \cE_q =  \left(\begin{array}{cc}\ - \cD^a{}_b \cF(\tau_{ A_1(q)})  \ & \ i J^{(a}{}_{c} \cD^{b)c} \cF(\tau_{ A_1(q)})  \ \\
 \  i J_{(a}{}^{c} \cD_{b)c} \cF(\tau_{ A_1(q)})  \ &\ - \cD_a{}^b \cF(\tau_{ A_1(q)}) \ \end{array}\right) e^{-10\phi} e^{-e^{-4\phi} \sqrt{Z_2}} \ , \ee
with $J_{ab}$ defined as in \eqref{Jab}, which plays the role of a complex structure such that 
\be  i J_{(a}{}^{c} \cD_{b)c} \cF(\tau_{ A_1(q)}) =  \cD_{ab} \cF(\tau_{ A_1(q)})  \ee
for a holomorphic function and \eqref{VecEq} is satisfied. The generic solution to these differential equations is therefore supported on a space of eight variables 
\be \cE[F] = \int_{\mathds{R}^+} dQ \int_{\frac{SL(5)}{SL(2)\times SL(3)\ltimes \mathds{R}^{2\times3}}}  \hspace{-10mm}d\mu(q)\hspace{2mm}F(q,Q) \, e^{i Q   \tau_{ A_1(q)}} \  \frac{e^{-6\phi}}{\sqrt{Z_2}} e^{-e^{-4\phi} \sqrt{Z_2} +  i q_{IJ} a^{IJ}} \ , \label{Fourier20}  \ee
and defines the smallest of the two next to minimal unitary representations of $SO(5,5)$.  The real function $e^{-10\phi} \Evector{\frac{5}{2}} $ does not solve the vector equation \eqref{VecEq}, but one can define the two functions 
\be \cE^\pm_{\frac{5}{2}}(n,m) = e^{-10\phi}\frac{ n^K m_K }{\scal{ n^I M^\inv_{IJ} n^J}^\frac{5}{2}}  \pm e^{-6\phi}\frac{ n^K  M^\inv_{KL} a^{LP}  m_P }{\scal{ n^I M^\inv_{IJ} n^J}^\frac{5}{2}} \ , \ee
which solve both \eqref{EsD5Sol} for $s=\frac{5}{2}$ and $\cE^+_{\frac{5}{2}}(n,m)$ solves  \eqref{VecEq} whereas 
\be \cD_{c\hat{a}} {\cD^c{}}_{\hat{b}} \, \cE^-_{\frac{5}{2}}(n,m)  = - \frac{3}{4}  \delta_{\hat{a}\hat{b}} \, \cE^-_{\frac{5}{2}}(n,m)  \ . \ee
To prove that we note that their sum vanishes for $n^I m_I = 0 $ whereas their difference is then obtained from the character generating  $e^{-6 \phi} \Evector{\frac{3}{2}}$ by an infinitesimal duality transformation of parameter $q_{IJ} =-q_{JI}$ such that  $m_I = q_{IJ} n^J$, \ie
\be \delta \scal{ e^{-4\phi} M^{IJ} } = q_{KL} e^{-4\phi} M^{K(I} a^{J)L} \ , \quad \delta \scal{ n^I e^{4\phi} M^\inv_{IJ} n^J}^{-\frac{3}{2}} = -\frac{3}{2}e^{-6\phi}\frac{ n^I  M^\inv_{IJ} a^{JK} q_{KL} n^L  }{\scal{ n^P M^\inv_{PQ} n^Q}^\frac{5}{2}} \ ,   \ee
and therefore satisfies by construction to \eqref{12BPS6D} as does $e^{-6 \phi} \Evector{\frac{3}{2}}$. When $n^I m_I \ne 0$ the two functions are independent, and one straightforwardly checks that this scalar product is not involved in equation \eqref{EsD5Sol}, and is only relevant in \eqref{VecEq} through
\bea &&  \Scal{  {\bf D}_{10}^{\; 2}  + \frac{3}{4} \mathds{1}_{10}} \cE^\pm_{\frac{5}{2}}(n,m) \CR
& =& \frac{n^I n^J}{2} \left(\begin{array}{cc}\  - 5 v^\inv_{I}{}^a v^\inv_{J b} + \delta^a_{b} M^\inv_{IJ}  \ & \  \pm( 5 v^\inv_{I}{}^a v^\inv_{J}{}^b - \delta^{ab} M^\inv_{IJ} ) \ \\
 \  \pm( 5 v^\inv_{I a} v^\inv_{J b} - \delta_{ab} M^\inv_{IJ})  \ &\ -  5 v^\inv_{I a} v^\inv_{J}{}^{b} + \delta_{a}^{b} M^\inv_{IJ}  \ \end{array}\right) e^{-10\phi} \frac{ n^P m_P}{\scal{ n^K M^\inv_{KL} n^L}^\frac{5}{2}} \ , \label{E5/2Complex} \hspace{5mm} \eea
  which is then only satisfied by $ \cE^+_{\frac{5}{2}}(n,m) $. The linear term in the axion is in contradiction with duality invariance, but the  explicit dependence in the (naked) axion drops out in the real invariant 
 \be \cL[\cE^+_{\frac{5}{2}}(n,m) ]_\gra{2}{0} +  \cL[\cE^-_{\frac{5}{2}}(n,m) ]_\gra{0}{2} \ee
 because the two chiral invariants coincide for a function satisfying to \eqref{12BPS6D}. One checks indeed in the linearised analysis that the superforms $\cL^{[0,n],[0,n]}_\gra{2}{0}$ satisfy to 
 \be \cL^{[0,n],[0,n]}_\gra{2}{0} = \cL^{[0,n],[0,n]}_\gra{0}{2} \ . \ee
 Assuming that they satisfy to the same equation at the non-linear level, one obtains that terms linear in the axion cancel out in the expansion \eqref{20Invariant}. The term in $\cL^{[0,n],[0,n+2k]}_\gra{2}{0}$ for $k\ge 1$ involve the operator \eqref{E5/2Complex} such that they do not depend explicitly on the (naked) axion $a^{IJ}$. This structure is similar to the one associated to the invariant  $\mbox{Re}[\cL[\ln(\eta)]]$ in eight dimensions, for which the axion $a$ only appears polynomially through $\int a \scal{  p_2 - \tfrac{1}{4} p_1^{\; 2}} $. Although in this case there is no topological coupling in the axion, and the supersymmetry invariant only depends on the function $\cE^+_{\frac{5}{2}}(n,m)+\cE^-_{\frac{5}{2}}(n,m) $ and its covariant derivatives. 

The general solution is therefore compatible with the regularised Eisenstein series $\hat{E}{\mbox{\DSOX{\hspace{-0.5mm}\frac{5}{2}}0000}} $ appearing in the $\nabla^4 R^4$ coupling \cite{Green:2010wi,Green:2010kv}, but we should take care however, that the Eisenstein function ${E}{\mbox{\DSOX{\hspace{-0.5mm}s}0000}} $ diverges at $s=\frac{5}{2}$. Note that this function is generated by a specific character, and any covariant differential equation satisfied by the character is also satisfied by the Eisenstein function provided the series converges. Using this property one computes that it satisfies to
 \be   {\bf D}_{16}^{\; 2} {E}_{\mbox{\DSOX s0000}} = \frac{s(s-4)}{4}  \mathds{1}_{16}  {E}_{\mbox{\DSOX s0000}}  \ . \ee
One can use this property to constrains the Fourier modes of this function. Altogether with the constant terms computed in \cite{Green:2010wi}, we conclude that this Eisenstein function admits the expansion
 \begin{multline}  {E}_{\mbox{\DSOX s0000}} = e^{-4s \phi} \Evector{s} + \frac{\pi^{2s-\frac{5}{2}} (s-3)}{\sin( \pi s)\, \Gamma(s) \Gamma(s-\tfrac{3}{2})} \frac{ \zeta(2s-4)}{\zeta(2s-3)} e^{4(s-4)\phi} \Evector{4-s} \\ + 16  \sum_{\ \vspace{-5mm}\begin{array}{c}\vspace{-3mm}  \scriptstyle q\in \mathds{Z}^{10} \\  \scriptstyle q\times q=0\end{array}}\mu_s(q)  E_{[s-\frac{3}{2}]}(v_{ A_1(q)})  \frac{e^{-8\phi}}{\sqrt[4]{Z_2}} K_{s-2}( 2\pi e^{-4\phi}  \sqrt{Z_2}) e^{ 2\pi i q_{IJ} a^{IJ}} 
 \end{multline} 
for some undetermined measure  $\mu_s(q)$. Using this expression, one recovers the singular limit 
\be {E}_{\mbox{\DSOX{\frac{5}{2}{\scriptscriptstyle +}\epsilon}0000}} = \frac{2}{\epsilon} {E}_{\mbox{\DSOX{\frac{3}{2}}0000}}  + \hat{E}_{\mbox{\DSOX{\frac{5}{2}{\scriptscriptstyle +}\epsilon}0000}} + \mathcal{O}(\epsilon) \ , \ee
provided $\mu_\frac{5}{2}(q)= \sum_{n|q_{IJ}} n$ is the same measure as for $s=\frac{3}{2}$.

Using the limit we compute that
\be  {\bf D}_{16}^{\; 2}\hat{E}_{\mbox{\DSOX{\hspace{-0.5mm}\frac{5}{2}}0000}}  = - \frac{15}{16}   \mathds{1}_{16} \hat{E}_{\mbox{\DSOX{\hspace{-0.5mm}\frac{5}{2}}0000}}  + \frac{1}{2} \mathds{1}_{16}{E}_{\mbox{\DSOX{\frac{3}{2}}0000}} \ , \ee
which is not strictly the supersymmetry equation. The logarithms of the moduli appearing in the regularised Eisenstein function are in fact coming from the non-analytic component of the effective action, as we shall discuss at the end of this section. 

\subsubsection*{The parity symmetric invariant}
 The third class of invariants can be obtained in the linearised approximation using harmonic variables parametrising $Sp(2)/(U(1)\times Sp(1))$ with the decomposition ${\bf 4} \cong {\bf 1}^\ord{-2}\oplus{\bf 2}^\ord{0} \oplus {\bf 1}^\ord{2}$ \cite{Bossard:2009sy}. We define  accordingly $u{}_i , \bar u_i , u^r{}_i   $ such that 
\be\Omega^{ij} u{}_i  \bar u_j = 2 \ , \quad    \Omega^{ij} u^r{}_i  u^s{}_{j}   =  \varepsilon^{rs}  \ , \qquad u{}_i \bar u_{j} + \varepsilon_{rs} u^r{}_i  u^s{}_{j} = \Omega_{ij} \ ,\ee 
 and respectively for the second $Sp(2)$ factor. One can then define the G-analytic superfield 
 \be W^{r\hat{s}} = u_i u^r{}_j u_{\hat{\imath}} u^{\hat{s}}{}_{\hat{\jmath}} L^{ij\hat{\imath}\hat{\jmath}} \ , \ee
 that satisfies 
 \be u_i \bar D^{\alpha \, i} W^{r\hat{s}} = 0 \ , \qquad u_{\hat{\imath}} D_\alpha^{\hat{\imath}} W^{r\hat{s}} = 0  \ . \ee
 Using this superfield one can define the class of invariants 
 \bea  &&\int d^{12}\theta d^{12}\bar\theta  du  F^{[2k,0]}_{u}  F^{[2k,0]}_{\hat{u}} (W^{r\hat{s}} W_{r\hat{s}})^{2+k} ( u^s{}_i u^{\hat{r}}{}_{\hat{\jmath}} W_{s\hat{r}} )^{n [n,0],[n,0]} \CR
 &\sim& \int d^{12}\theta_{[4,0]} d^{12}\bar\theta_{[4,0]}\,  L^{4+2k \, [4+2k,0],[4+2k,0]}  L^{n\, [0,n],[0,n]}\ , \eea
where $F^{[n+2k,0]}_{u} $ is the degree $n+2k$ monomial in $\bar u{}_i$ in the corresponding representation. The set of representations involved is again different, and suggests in this case that one must consider the $(n+2k)^{\rm th}$ derivative in all the representations $[2k,n]\times[2k,n]$. 
This is now consistent with the property that 
\be  \cD^2_{[0,2],[0,0]}\cE = 0\ , \qquad  \cD^2_{[0,0],[0,2]}\cE = 0 \ , \label{VectorSquare}  \ee
proposed as a quantisation of the corresponding 1/4 BPS condition in the last section. So such invariant will have the generic form
\bea  \cL[\cE]_\gra{1}{1} &=& \sum_{n=0}^{12} \sum_{k=0}^{8-n/2} \cD^{n+2k}_{[2k,n],[2k,n]} \cE \, \cL^{[2k,n],[2k,n]}_\gra{1}{1} \CR
&=& \cE \nabla^4 R^4 + \dots + \cD^{16}_{[4,12],[4,12]} \cE \, F^{4 \, [4,0],[4,0]}  \chi^{8\, [0,4],[0,8]} \bar \chi^{8\, [0,8],[0,4]} +\dots \label{11superform}  \eea
The form of the linearised invariant therefore strongly suggests that the function $\cE$ must satisfy to equation \eqref{VectorSquare}. In principle one could check this explicitly on the terms multiplying $D^{16}_{[4,12],[4,12]} \cE$, but this computation is rather involved and we shall not carry it out in this paper. Note moreover that the 1/4 BPS condition discussed in the preceding section also requires ${\bf Q}_{16}^{\; 3} = 0$, and considering the expansion \eqref{11superform} requires also that
\be \cD_{[a}{}^{[\hat{a}} \cD_{b}{}^{\hat{b}} \cD_{c]}{}^{\hat{c}]} \, \cE_s =  - \frac{s-2}{24} \varepsilon_{abcde} \varepsilon^{\hat{a}\hat{b}\hat{c}\hat{d}\hat{e}}  \cD^d{}_{\hat{d}} \cD^e{}_{\hat{e}}  \cE_s \ , \label{Spin3} \ee
for some $s$ to be determined, such that there is no new independent term in the gradient expansion of the function $\cE_s$. Using the commutation relation one computes that in general 
\be \varepsilon^{abdef} \cD_{c\hat{c}} \cD_d{}^{\hat{a}} \cD_e{}^{\hat{b}} \cD_f{}^{\hat{c}} = \varepsilon_c{}^{abef} \cD_e{}^{\hat{a}} \cD_f{}^{\hat{b}} +  \varepsilon^{abdef}  \cD_d{}^{\hat{a}} \cD_e{}^{\hat{b}} \cD_f{}^{\hat{c}} \cD_{c\hat{c}} \ , \ee
such that \eqref{Spin3} and \eqref{VectorSquare} are only compatible if 
\be \cD_a{}^{\hat{c}} \cD_{b \hat{c}} \cE_s = \frac{s(s-4)}{4} \delta_{ab} \cE_s \ . \ee
Using these equations and the compatibility condition with \eqref{D5Casimirs} one concludes that
\bea   {\bf D}_{10}^{\; 2} \cE_s &=& \frac{s(s-4)}{4}  \mathds{1}_{10} \cE_s \ , \CR
{\bf D}_{16}^{\; 3} \cE_s &=& - \frac{3(s-2)}{4}  {\bf D}_{16}^{\; 2}   \cE_s + \frac{13 s(s-4) +24}{16}  {\bf D}_{16} \cE_s + \frac{15s(s-2)(s-4)}{64} \mathds{1}_{16} \cE_s   \ .\label{VecSquare}  \eea
It remains now to determine the value of $s$. To do so we note that the linearised invariants in the $[0,n],[0,n]$ representation are all identical  because they have the same 1/2 BPS harmonic integral form 
 \bea \int d^8\theta d^{16}\bar\theta  du F_{\hat{u}}^{[0,n]}  (W^{ij} W_{ij})^{4} W^{n [0,n]} 
 &=&
   \int d^{12}\theta d^{12}\bar\theta  du  F^{[n,0]}_{u}  F^{[n,0]}_{\hat{u}} (W^{r\hat{s}} W_{r\hat{s}})^{4} ( u^s{}_i u^{\hat{r}}{}_{\hat{\jmath}} W_{s\hat{r}} )^{n [n,0],[n,0]}    \CR
   &=&\int d^8\theta d^8\bar\theta  du F_u^{[0,n]} F_{\hat{u}}^{[0,n]} (  \partial_\mu W \partial^\mu W)^2 W^{n}\ .  \eea
This suggests that the associated superforms are also identical at the non-linear level 
\be \cL^{[0,n],[0,n]}_\gra{1}{1}  =  \cL^{[0,n],[0,n]}_\gra{2}{0}  = \cL^{[0,n],[0,n]}_\gra{0}{2} \ . \label{IdenticInvariants} \ee
But this is only possible if the differential equations are compatible and therefore if $s=3$.  The value $s=3$ is indeed consistent with \cite{Green:2010kv}, as we are going to see.

We shall now discuss the solutions of these equations for $s=3$. Solving \eqref{VecSquare} requires the introduction of another class of $SL(5)$ Eisenstein functions satisfying to 
\be \cD_a{}^c \cD_c{}^b \, \Etensor{s} = - \frac{4s-5}{20} \cD_a{}^b \, \Etensor{s}  + \frac{3s(2s-5)}{25} \delta_a^b \, \Etensor{s}  \ . \ee
We checked this equation explicitly on a generating character of these Eisenstein functions. Note that these functions do not satisfy to any quadratic differential equation in the ${\bf 10}$ of $SL(5)$ as does  $\Evector{s}$. This equation is only strictly satisfied by the corresponding Eisenstein functions when they are convergent series. 

 Solving equation \eqref{VecSquare} for a function independent of $a^{IJ}$ one finds the solution
 \be c_1 e^{-30\phi} + e^{-18\phi} \Evector{-\frac{1}{2}} + c_2 e^{-10\phi} +  e^{-10\phi} \Etensor{\frac{5}{2}} +e^{-6\phi} \Etensor{\frac{1}{2}}+e^{-6\phi} \Evector{\frac{3}{2}}\ .   \ee  
All the corresponding Eisenstein functions, but $\Evector{-\frac{1}{2}} $ and $\Etensor{\frac{1}{2}}$, do appear in the decompactification limit of the regularised Eisenstein function $ \hat{E}{\mbox{\DSOX00003}} $ according to \cite{Green:2010kv}. One checks that  $\Evector{-\frac{1}{2}} $ and $\Etensor{\frac{1}{2}}$ only solve  \eqref{VecSquare} for $s=1$. The sign of the terms involving the $\varepsilon$ tensor depend on the chirality, and the corresponding equation in the parabolic gauge also depends on the specific embedding such that the equation $s=1$ corresponds to the M-theory limit of $ \hat{E}{\mbox{\DSOX00003}} $. The Eisenstein functions  $\Evector{-\frac{1}{2}} $ and ${\hat{E}_{\scriptscriptstyle [\mathfrak{0}\mathfrak{2}\mathfrak{0}\mathfrak{0}]}}$ solving the same differential equation as $\Etensor{\frac{1}{2}}$ do indeed appear in the M-theory limit \cite{Green:2010kv}.

 Let us now consider the Fourier modes. Note that the condition $\varepsilon^{IJKLP} q_{IJ} q_{KL}=0$ was coming from the quadratic equation in the spinor representation, and therefore does not hold in this case.  It is therefore convenient to define the two functions 
 \be Z_2 = 2 Z_{ab}(q) Z^{ab}(q) \ , \quad Z_4 =  Z_{ab}(q) Z^{bc}(q) Z_{cd}(q) Z^{da}(q) - \frac{1}{4} \scal{  Z_{ab}(q) Z^{ab}(q)}^2 \ , \ee
 with 
 \be Z_{ab}(q) = v_a{}^I v_b{}^J q_{IJ} \ . \ee
 The off-diagonal equation 
 \be Z_{(a}{}^c(q) \cD_{b)c}  \cE_q = 0 \ \label{offDiag} \ee
 requires that the Fourier modes only depends on the $SL(5)/SO(5)$ scalars through the central charge $Z_{ab}(q)$.  Using these variables, one can rewrite the remaining differential equation as 
 \bea \label{quarterBPSdiff} && Z_{ac}(q) Z^{cd}(q) Z_{de}(q)Z^{eb}(q) \Scal{ - \frac{32}{5} \partial_{Z_4} - \frac{12}{5} Z_4 \partial_{Z_4}^{\; 2} + \frac{4}{5} Z_2 \partial_{Z_2} \partial_{Z_4} + 4 \partial_{Z_2}^{\; 2} - \frac{2}{5} \partial_{Z_4} \partial_\phi } \cE_q \CR
&&  +  Z_{ac}(q) Z^{bc}(q)  \Bigl( \frac{31}{10} Z_2 \partial_{Z_4} - \frac{3}{5} \partial_{Z_2} + \frac{8}{5} Z_2 Z_4 \partial_{Z_4}^{\; 2} + \frac{1}{5} ( 4 Z_4 -Z_2^{\; 2} ) \partial_{ Z_2} \partial_{Z_4}- \frac{8}{5} Z_2 \partial_{Z_2}^{\; 2}\Bigr .  \CR \Bigl . 
&&\hspace{100mm} + \frac{1}{10} Z_2 \partial_{Z_4} \partial_\phi - \frac{1}{5} \partial_{Z_2} \partial_\phi \Bigr) \cE_q \CR
 && + \delta_a^b \Bigl(\frac{56}{25} Z_4 \partial_{Z_4} + \frac{24}{25} Z_2 \partial_{Z_2} + \frac{16}{25}  Z_4^{\; 2}  \partial_{Z_4}^{\; 2} + \frac{16}{25}  Z_2 Z_4 \partial_{ Z_2} \partial_{Z_4}+ \frac{4}{25} Z^{\; 2}_2 \partial_{Z_2}^{\; 2} \Bigr . \CR
 && \Bigl . \hspace{70mm} + \frac{2}{25} Z_4 \partial_{Z_4} \partial_\phi + \frac{1}{25} Z_2 \partial_{Z_2} \partial_\phi + \frac{1}{20^2}  \partial_\phi^{\; 2} + \frac{1}{10} \partial_\phi    \Bigr)  \cE_q \CR
 &=& - \frac{3}{4} \delta_a^b \, \cE_q  \eea  
 This provides three independent second-order equations. One finds the solution 
 \be \cE_q =  \frac{e^{-6\phi}}{ \sqrt{\frac{1}{2} Z_2 +  \sqrt{Z_4}} +\sqrt{\frac{1}{2} Z_2 -  \sqrt{Z_4}}  } \ e^{-e^{-4\phi} \scal{ \sqrt{\frac{1}{2} Z_2 +  \sqrt{Z_4}} +\sqrt{\frac{1}{2} Z_2 -  \sqrt{Z_4}} }} \ . \label{quarterFourier} \ee
 Note that for a 1/2 BPS charge one has 
 \be Z_4 = \frac{1}{4} Z_2^{\; 2} \ , \ee
 and one recovers the same form of the Fourier coefficients as for  ${E}{\mbox{\DSOX{\frac{3}{2}}0000}}  $. The term 
 \be e^{-4\phi} \Scal{ \sqrt{\frac{1}{2} Z_2 +  \sqrt{Z_4}} +\sqrt{\frac{1}{2} Z_2 -  \sqrt{Z_4}} }\ee
 is the action associated to a 1/4 BPS instanton. Considering the central charge in the spinor representation $\frac{1}{2} Z_{ab} \gamma^{ab}{}_i{}^k$, the eigenvalues are 
 \be\pm  \sqrt{\frac{1}{2} Z_2 +  \sqrt{Z_4}} \pm\sqrt{\frac{1}{2} Z_2 -  \sqrt{Z_4}} \ee
 and the BPS bound is defined by the largest. In fact they all define solutions to the equation \eqref{quarterBPSdiff}, but only \eqref{quarterFourier} admits a convergent behaviour in the large radius limit because the others exhibit exponential growth in the asymptotic. The generic solution with a convergent behaviour at infinity  is therefore supported on a set of functions depending on ten variables 
 \be \cE[F] = \int d^{10} q \, F[q] \,  \frac{e^{-6\phi}}{ \sqrt{\frac{1}{2} Z_2 +  \sqrt{Z_4}} +\sqrt{\frac{1}{2} Z_2 -  \sqrt{Z_4}}  } \ e^{-e^{-4\phi} \scal{ \sqrt{\frac{1}{2} Z_2 +  \sqrt{Z_4}} +\sqrt{\frac{1}{2} Z_2 -  \sqrt{Z_4}} } + i q_{IJ} a^{IJ} } \ , \ee
 corresponding the other next to minimal unitary representation of $SO(5,5)$.  
  
 We should also consider the contribution of the 1/2 BPS instantons. But because the solution is then singular by property of the function, one must rather consider the solution for a generic $s$. Because this class of Eisenstein functions is associated to the decomposition of $SO(5,5)$ we use, the generating character of the function $ {E}{\mbox{\DSOX0000s}}$ restricted to the Cartan subgroup is simply $e^{-10 s \phi}$. We computed that $\cE_s = e^{-10 s \phi}$ is a solution to the two equations in \eqref{VecSquare}, and it follows that the Eisenstein function $ {E}{\mbox{\DSOX0000s}}$  also solves them when the series converges. Note that for a rank one Fourier modes (\ie $q\times q=0$), the off-diagonal equation \eqref{offDiag} is not strong enough to impose that the solution only depends on $Z_2$, and the function can also depend on the components of $v_a{}^I$ in the $SL(3)$ subgroup of the stabiliser $SL(2)\times SL(3)\ltimes \mathds{R}^{2\times3} \subset SL(5)$ of $q_{IJ}$, which we shall write  $v_{{A_2(q)}}$. For a 1/2 BPS charge $q_{IJ}$ one finds the solution to the quadratic equation in \eqref{VecSquare}
 \be \cE_q =  e^{-2(7-s) \phi} \EA{s{\scriptscriptstyle \rm -}\frac{3}{2}}(v_{{A_2(q)}}) Z_2^{\; \frac{2s-9}{12}} K_{s-\frac{5}{2}}( e^{-4\phi} \sqrt{Z_2}) + c_1  e^{-2(13-3s) \phi} Z_2^{\; \frac{2s-7}{4}} K_{s-\frac{7}{2}}( e^{-4\phi} \sqrt{Z_2})\ee
together with the conjugate solution obtained by the substitution $s\rightarrow 4-s$. We did not check the cubic equation on these functions, and one cannot determine at this level which of these solutions actually appear in the Fourier expansion of $ {E}{\mbox{\DSOX0000s}}$, but the first solution depending on the $SL(3)$ Eisenstein function admits the appropriate limit to define the singular structure of the regularised Eisenstein function $\hat{E}{\mbox{\DSOX00003}}$ \cite{Green:2010kv}
\be  \hat{E}_{\mbox{\DSOX00003}} =  \lim_{\epsilon\rightarrow 0} \Scal{ {E}_{\mbox{\DSOX0000{3{\scriptscriptstyle+}\epsilon}}} - \frac{45}{4\epsilon} E_{\mbox{\DSOX{\hspace{-0.5mm}\frac{3}{2}}0000}} 
 }  \ . \ee
Indeed 
 \be e^{-2(4-\epsilon) \phi} \EA{\frac{3}{2}{\scriptscriptstyle +}\epsilon\, }(v_{{A_2(q)}}) Z_2^{\; \frac{2\epsilon-3}{12}} K_{\frac{1}{2} + \epsilon}( 2\pi e^{-4\phi} \sqrt{Z_2})  = \frac{\pi}{\epsilon} \frac{e^{-6\phi}}{\sqrt{Z_2}} e^{-2\pi e^{-4\phi} \sqrt{Z_2}} + \mathcal{O}(\epsilon^0) \ . \ee
In particular, we conclude that the 1/2 BPS instanton contributions to the $\nabla^4 R^4$ coupling in string theory combine into 
 \bea&&  \frac{1}{2} \hat{E}_{\mbox{\DSOX{\frac{5}{2}}0000}} + \frac{4}{45} \hat{E}_{\mbox{\DSOX00003}} \CR
 &=& \sum_{\ \vspace{-5mm}\begin{array}{c}\vspace{-3mm}  \scriptstyle q\in \mathds{Z}^{10} \\  \scriptstyle q\times q=0\end{array}} \Scal{ \sum_{n|q_{IJ}}n}  \Scal{ 4 \hat{E}_{[1]}(v_{{A_1(q)}}) + 2 {\hat{E}_{\scriptscriptstyle [\frac{3}{2}\mathfrak{0}]}}(v_{{A_2(q)}})} \frac{e^{-6\phi}}{ \sqrt{Z_2}} e^{-2\pi e^{-4\phi}  \sqrt{Z_2} + 2\pi i q_{IJ} a^{IJ}} + \dots  \qquad \eea
It is rather striking that this combination of $ \hat{E}_{[1]}$ and ${\hat{E}_{\scriptscriptstyle [\frac{3}{2}\mathfrak{0}]}}$ is precisely the one that defines the $R^4$ coupling in eight dimensions \cite{Green:2010wi}, for which the respective $\frac{1}{\epsilon}$ poles cancel out. 

\subsubsection*{The non-analytic terms}
Similarly as  $\hat{E}{\mbox{\DSOX{\hspace{-0.5mm}\frac{5}{2}}0000}} $,  $\hat{E}{\mbox{\DSOX00003}}$ does not strictly satisfy to the supersymmetry equation \eqref{VecSquare}, but rather to
\be {\bf D}_{10}^{\; 2}  \hat{E}_{\mbox{\DSOX00003}} = - \frac{3}{4}  \mathds{1}_{10} \hat{E}_{\mbox{\DSOX00003}}  + \frac{45}{8} \mathds{1}_{10} {E}_{\mbox{\DSOX{\frac{3}{2}}0000}} \ . \ee
A $\nabla^4 R^4$ invariant does not have the right dimension to appear as a counterterm for logarithmic divergences in supergravity, and the non-analytic component of the effective action responsible for these corrections to the differential equations satisfied by the threshold functions must also include massive states contributions. From the supergravity perspective, this comes from the property that $\nabla^4 R^4$ has the correct dimension to be a counterterm for the 1-loop divergence of an $R^4$ invariant operator defined as an insertion.  If we consider the low energy expansion of the effective action, the leading non-analytic components will match the supergravity effective action, but the next order correction will include the insertion of the exact $R^4$ string theory coupling. Schematically, the amplitude is determined by the supergravity path integral of the string theory Wilsonian effective $S$
\be \exp\scal{ i W[J]} = \int \mathcal{D}\varphi \exp\Scal{ \frac{i}{\kappa^2}  \scal{S_0 +  \kappa^3 S_3 + \kappa^{5} S_5 + \dots } + i \int J \varphi } \ ,  \ee
such that the corresponding Legendre transform decomposes as
\be \Gamma[\varphi] = \frac{1}{\kappa^2} S_0 + \Gamma_{\mbox{\tiny 1-loop}} + \kappa  S_3 + \kappa^2 \Gamma_{\mbox{\tiny 2-loop}}  +   \kappa^3 \scal{  S_5 + \bigl[ S_3\cdot \Gamma\, \bigr]_{\mbox{\tiny 1-loop}} } + \dots \  \ee 
 If one considers the perturbative string theory contribution as depicted in \cite{Green:2010kv}, one finds indeed a logarithm correction of the form 
\be  \frac{1}{2} \hat{E}_{\mbox{\DSOX{\frac{5}{2}}0000}} + \frac{4}{45} \hat{E}_{\mbox{\DSOX00003}}  = e^{-3\phi_s} \Scal{ \dots + \phi_s e^{2\phi_s} E_{\mbox{\DSOX{\frac{3}{2}}0000}}  + \dots } \ , \ee
where the overall $e^{-3\phi_s}$ corresponds to the Weyl rescaling to Einstein frame. According to the analysis displayed in \cite{Green:2010sp}, one understands that this logarithm of the dilaton comes from a logarithm of the Mandelstam variable $s$ in the effective action. We see therefore that the tree-level and one-loop corrections to the $R^4$ coupling in string theory contribute respectively to a one-loop and a 2-loop correction to $\ln(s)s^2 R^4$ in the effective action.  In supergravity, this implies that the local operator $\cL[\cE_\frac{3}{2}]_\gra{2}{2}$ defining an arbitrary $R^4$ type invariant, admits a logarithmic divergence at 1-loop, renormalised by a local operator of the form $\cL[\cE_\frac{3}{2}]_\gra{1}{1}$ defining a $\nabla^4 R^4$ type invariant, for the same function $\cE_\frac{3}{2}$. 

 The consistency of this argument requires that the anomalous term in $ {E}{\mbox{\DSOX{\frac{3}{2}}0000}}$ in the two supersymmetry equations associated to the two independent invariants define the same unique  invariant, itself associated to the 1-loop divergence of the corresponding $R^4$ type invariant. Equivalently, the cancelation of the $\frac{1}{\epsilon}$ divergence in the combination 
 \be \lim_{\epsilon\rightarrow 0} \Scal{   \frac{1}{4} \cL\Bigl[{E}^+_{\mbox{\DSOX{\frac{5}{2}{\scriptscriptstyle +}\epsilon}0000}}\Bigr]_\gra{2}{0}+ \frac{4}{45} \cL\Bigl[ {E}_{\mbox{\DSOX0000{3{\scriptscriptstyle -}\epsilon}}}\Bigr]_\gra{1}{1} +   \frac{1}{4} \cL\Bigl[{E}^-_{\mbox{\DSOX{\frac{5}{2}{\scriptscriptstyle +}\epsilon}0000}}\Bigr]_\gra{0}{2}} \ee
requires that for a function $\cE_\frac{3}{2}$ satisfying the 1/2 BPS quadratic equation \eqref{12BPS6D}, the three invariants must be identical, \ie 
\be \cL[\cE_\frac{3}{2}]_\gra{2}{0} =  \cL[\cE_\frac{3}{2}]_\gra{0}{2} = \cL[ \cE_\frac{3}{2}]_\gra{1}{1} \ . \ee 
 The corresponding expansions in derivatives of the function $\cE_\frac{3}{2}$ are indeed of the same form in that case because of the quadratic equations satisfied by  $\cE_\frac{3}{2}$, and these invariants are indeed identical provided \eqref{IdenticInvariants} is satisfied. 
 
\section{$\N=8$ supergravity in four dimensions}

We will now discuss the case of $\N=8$ supergravity in four dimensions \cite{Brink:1979nt,Cremmer:1979up}. The R-symmetry group is then $SU(8)$ and the Lorentz group $SL(2,\mathds{C})$. In this section $i=1$ to $8$ is an $SU(8)$ index. The same construction permits to determine the properties of the function defining the $R^4$ type invariant, and we will propose a conjecture for the equations satisfied by the functions defining the $\nabla^4 R^4$ and $\nabla^6 R^4$ type invariants.
\subsection{The $R^4$ type invariant}
One can define the linearised $R^4$ type invariants in the linearised approximation by using harmonic variables in $SU(8) / S(U(4)\times U(4))$ as in \cite{Hartwell:1994rp}. One obtains that the scalar superfield 
\be W = u^1{}_i u^2{}_j u^3{}_k u^4{}_l W^{ijkl} \ , \ee
is G-analytic with respect to (with $r=1$ to $4$ and $\hat{r}=5$ to $8$)
\be u^r{}_i D^i_\alpha \, W = 0 \ , \qquad u^i{}_{\hat{r}} \bar D_{\adt i} \, W = 0 \ , \ee
such that 
\bea&&  \int d^8\theta d^8 \bar \theta du  F_u^{[0,0,0,n,0,0,0]} \, W^{4+n} \CR
&\sim &W^{n\,  [0,0,0,n,0,0,0]} R^4 + \dots +W^{n-12 \,  [0,0,0,n-12,0,0,0]}   \chi^{8 \, [0,0,0,6,0,0,0]} \bar \chi^{8 \, [0,0,0,6,0,0,0]} \ . \eea
Although the harmonic measure does not extend to the non-linear theory, it suggests strongly that the non-linear invariant admits the expansion 
\be \cL[\cE] = \cE \cL + \cD_{ijkl} \cE \cL^{ijkl} +  \sum_{n=2}^{12} \cD^{n}_{[0,0,0,n,0,0,0]} \cE \, \cL^{[0,0,0,n,0,0,0]}\ . \ee 
As in the preceding section, we will concentrate on the term with the maximal number of derivative carrying the highest weight $SU(8)$ representation. Using representation theory and power counting, one obtains that the maximal weight term can only be the monomial in $\chi^8 \bar \chi^8$ \, because one needs $48$ open indices to get this representation. To show that this monomial exists and is unique, one can use the harmonic projection 
\be \chi_\alpha^{\hat{r}} \equiv  \varepsilon^{\hat{r}\hat{s}\hat{t}\hat{u}} u^i{}_{\hat{s}} u^j{}_{\hat{t}} u^k{}_{\hat{u}} \chi_{\alpha\, ijk} \ , \qquad \bar \chi_{\adt \, r} = \varepsilon_{rstu} u^s{}_i u^t{}_j u^u{}_k \bar \chi_{\adt}^{ijk} \ , \ee 
which define $8+8$ fermionic variables. The maximal monomial is therefore $\chi^8 \bar \chi^8$, and by definition of the harmonic variables, it has maximal $U(1)$ weight such that it is in the $[0,0,0,12,0,0,0]$ representation of $SU(8)$, of Young tableau ${ \Yboxdim3.5pt  {\yng(12,12,12,12)}}$ .  To consider the action of the covariant derivatives on such monomial, we need to consider the independent terms in $\chi^9$ 
\be \chi_\alpha^{[0,0,1,0,0,0,0]}  \chi^{8 \, [0,0,0,6,0,0,0]}   \sim \chi_\alpha^{9 \, [0,1,0,5,1,0,0]} +  \chi_\alpha^{9 \, [1,0,0,5,0,1,0]} +  \chi_\alpha^{9 \, [0,0,0,5,0,0,1]}  \ .  \ee 
Using the first term (of maximal weight), one gets the two possible combinations
\be  \chi_\alpha^{9 \, [0,1,0,5,1,0,0]}  \bar \chi^{8\, [0,0,0,6,0,0,0]} =( \chi_\alpha^9 \bar \chi^8)^{[0,1,0,11,1,0,0]} + ( \chi_\alpha^9 \bar \chi^8)^{[1,0,0,11,0,1,0]} + \dots \ , \ee
 which will both appear in the derivative of $\cD^{12} \cE \chi^8 \bar \chi^8$ as
 \bea&& D_\alpha^i \Scal{ \cD^{12}_{[0,0,0,12,0,0,0]} \cE\, \chi^{8 \, [0,0,0,6,0,0,0]} \bar \chi^{8 \, [0,0,0,6,0,0,0]}} \CR
 &=& \cD^{13}_{[0,1,0,11,0,1,0]} \cE\,  ( \chi_\alpha^{i\, 9} \bar \chi^8)^{[0,1,0,11,1,0,0]} + \cD^{i\, 13}_{[0,1,0,11,0,1,0]} \cE\, ( \chi_\alpha^9 \bar \chi^8)^{[1,0,0,11,0,1,0]}  + \dots \ .  \label{VaryMax24} \eea
The only other way to get $\chi^9$ in  the $[0,1,0,5,1,0,0]$ representation is through 
\be \chi_\alpha^{[0,0,1,0,0,0,0]}  \chi^{8 \, [1,1,0,4,1,0,0]}   \sim \chi_\alpha^{9 \, [0,1,0,5,1,0,0]} +  \chi_\alpha^{9 \, [1,0,0,5,0,1,0]\prime} + \dots   \ ,  \ee 
where the prime states that the $[1,0,0,5,0,1,0]$ is not necessarily the same, because there exists two such combinations of $\chi^9$. Therefore one should also consider terms like
\be \cD^{12}_{[0,1,0,10,0,1,0]} \cE\, \bar\chi^{8 \, [0,0,0,6,0,0,0]} \Scal{ \chi^{8\, [0,0,0,6,0,0,0]}  + \chi^{8\, [1,1,0,4,1,0,0]}} \ . \ee 
However $[0,0,1,0,0,0,0]\times[0,1,0,10,0,1,0]$ does not contain the $[0,1,0,11,1,0,0]$, so such terms can only be used to compensate for the $[1,0,0,11,0,1,0]$ in \eqref{VaryMax24}.

For completeness, less us stress that terms involving bosons  with a maximal number of open $SU(8)$ indices 
\be \cD^{13}_{[0,0,0,11,0,0,0]} \cE \, F_{\alpha\beta}^{[0,1,0,0,0,0,0]} \chi^{6\, \alpha\beta [0,0,0,5,0,0,0]} \bar\chi^{8\, [0,0,0,6,0,0,0]} + {\rm C.C} \ , \ee
and 
\be \cD^{13}_{[1,0,0,11,0,0,1]} \, \cE \, P^{\alpha\bdt\, [0,0,0,1,0,0,0]}  \chi_\alpha^{7\,  [1,0,0,5,0,0,0]} \bar\chi_{\bdt}^{7\, [0,0,0,5,0,0,1]} \ , \ee
could not mix with the terms we have been considering. Moreover the second can be eliminated by the addition of a total derivative, up to the addition of lower derivative terms in $\cD^{12} \cE$.

We conclude that there is nothing that can compensate for the first term in \eqref{VaryMax24}, and the function $\cE$ must therefore satisfy to the equation 
\be \cD^{13}_{[0,1,0,11,0,1,0]} \cE = 0 \ . \ee
Up to lower derivative terms in $\cE$ in lower weight representations, this equation can be reduced to 
\be \cD^{11}_{[0,0,0,11,0,0,0]} \cD^{2}_{[0,1,0,0,0,1,0]} \cE = 0 \ . \ee
The derivative operator $ \cD^{11}_{[0,0,0,11,0,0,0]}$ includes all components $(\cD_{ijkl})^{11}$ (without summation over the indices), and its kernel is the constant tensor. We conclude that the function $\cE$ must satisfy to the quadratic equation 
\be \cD^{2}_{[0,1,0,0,0,1,0]} \cE = 0 \ , \ee
or more explicitly (with the definition $\Delta \equiv 1/3\,  \cD_{ijkl} \cD^{ijkl}$)
\be \frac{1}{24} \varepsilon^{ijpqrstu} \cD_{klpq} \cD_{rstu} \cE = \frac{3}{28} \delta^{ij}_{kl} \Delta \cE \ . \label{Joseph4D} \ee
Using the relation 
\be [ \cD^{ijkl} , \cD_{pqrs} ]  \cD_{tuvw}= - 24 \delta^{ijkl}_{qrs][t} \cD_{uvw][p} + 3 \delta^{ijkl}_{pqrs} \cD_{tuvw} \ , \ee
one obtains the equality between the two quartic invariants 
\be \cD^{ijkl} \cD_{klpq} \cD^{pqrs} \cD_{rsij} = \frac{1}{12}\scal{  \cD_{ijkl} \cD^{ijkl} }^2 + 6 \cD_{ijkl} \cD^{ijkl} \ . \ee
Using this property one can conclude that $\cE$ satisfies 
\be  \Delta^2 \cE = - 42 \Delta \cE \ . \ee
The same argument as for $SO(5,5)$ in the preceding section would permit to show that the only consistent solution satisfies $\Delta \cE =  - 42 \, \cE$, consistently with the analysis of \cite{Green:2010kv}. Using the explicit form of the differential equation of the next section, one computes indeed that there is no non-trivial solution to \eqref{Joseph4D} satisfying to the Laplace equation $\Delta \cE=0$. We conclude that $\cE$ satisfies to
\be \frac{1}{24} \varepsilon^{ijpqrstu} \cD_{klpq} \cD_{rstu} \cE = -\frac{9}{2} \delta^{ij}_{kl}  \cE \ . \label{N8R4Diff}  \ee

\subsection{Minimal unitary representation}
It is convenient to analyse Equation \eqref{N8R4Diff} considering an explicit coset representative in $E_{7(7)}/SU_{\hspace{-0.5mm}\scriptscriptstyle \rm c}(8)$ in the parabolic gauge \DEVII0000001 relevant to the decompactification limit. In this case we have  
\be \cV = \left( \begin{array}{cccc} \ e^{3\phi} \ & \ 0 \ & \ 0 \ & \ 0 \\
0 &\ e^{\phi} V_{ij}{}^I \ &0&0\\
0&0& \ e^{-\phi} V^\inv{}_I{}^{ij} \ &0\\
0&0&0& e^{-3\phi} \end{array}\right) \left( \begin{array}{cccc} \ 1 \ & \ a^J \ & \ \tfrac{1}{2} t_{JKL} a^K a^L  \ & \ \tfrac{1}{3} t_{KLP} a^K a^L a^P \\
0 &\ \delta_I^J \ &t_{IJK} a^K&\tfrac{1}{2} t_{IKL} a^K a^L\\
0&0& \delta_J^I \ &a^I \\
0&0&0& 1 \end{array}\right)  \ , \ee
where $V_{ij}{}^I$ is a representative of $E_{6(6)} / \Sp$ in the fundamental representation, and $t_{IJK}$ is the invariant symmetric tensor of $E_{6(6)}$. 

The decomposition 
\be d\cV \, \cV^{-1} = P + B \ee
in coset and subgroup components gives 
\be P =  \left( \begin{array}{cccc} \ 3 d\phi \ & \ \tfrac{1}{2} e^{2\phi} V^\inv{}_I{}^{kl} da^I  \ & \ 0 \ & \ 0 \\
\ \tfrac{1}{2} e^{2\phi} V^\inv{}_{I\, ij} da^I \ &\ d\phi \delta_{ij}^{kl} + P_{ij}{}^{kl}  \ & \sqrt{2} e^{2\phi} \Omega_{j][k} V^\inv{}_{I\, l][i} &0\\
0&\sqrt{2} e^{2\phi} \Omega^{j][k} V^\inv{}_{I}{}^{\, l][i} & \ - d\phi \delta^{ij}_{kl} - P^{ij}{}_{kl}  \ &\ \tfrac{1}{2} e^{2\phi} V^\inv{}_I{}^{ij} da^I  \\\
0&0&\ \tfrac{1}{2} e^{2\phi} V^\inv{}_{I\, kl} da^I  \ & -3d\phi \end{array}\right) \ , \ee
where all the antisymmetrisations are understood to be projected to the symplectic traceless component 
\be X_{[ij]} = \frac{1}{2} X_{ij} - \frac{1}{2} X_{ji} - \frac{1}{8} \Omega_{ij} \Omega^{kl} X_{kl} \ , \ee
and $\delta_{ij}^{kl} = \delta^{[k}_{[i} \delta^{l]}_{j]} - \frac{1}{8} \Omega_{ij} \Omega^{kl} $. The symplectic matrix $\Omega_{ij}$ satisfies 
\be \Omega^{ik} \Omega_{jk} = \delta^i_j \ee
and we raise and lower $Sp(4)$ indices as 
\be X_i = \Omega_{ij} X^j \ , X^i = X_j \Omega^{ji} \ . \ee 
The metric on the coset space $E_{7(7)} / \SU$ is defined as 
\be ds^2 = \frac{1}{6} \tr P^2 =  12 d\phi^2 + \frac{1}{3} P_{ijkl} P^{ijkl} + e^{4\phi} V^\inv{}_{I\, ij}V^\inv{}_{J}{}^{ij} da^I da^J \ ,  \ee
and its inverse 
\be g^{-1} = \frac{1}{12} \partial_\phi^{\; 2} + \frac{1}{3} \cD_{ijkl} \cD^{ijkl} + e^{-4\phi} V_{ij}{}^I V^{ij\, J} \partial_I \partial_J \ . \ee
Accordingly, we have 
\be \cD_{ijkl}{}^{\upmu} P_{\upmu}{}^{pqrs} = 3 \delta_{ijkl}^{pqrs} \ ,  \qquad \cD_{ijkl}{}^{\upmu} P_{\upnu}{}^{ijkl} = 3 \delta_{\upnu}^{\upmu}\ , \ee
on the symmetric space $E_{6(6)}/ \Sp$. The reader should take care that we use the same notation for the differential operator $\cD_{ijkl}$, that is associated to the 42 variables of $E_{6(6)}/ \Sp$ in this subsection, whereas it was used for the 70 variables of $E_{7(7)}/SU_{\hspace{-0.5mm}\scriptscriptstyle \rm c}(8)$ in the preceding one.

The inverse vielbein on $E_{7(7)} / \SU$ are defined as 
\be {\bf D} =  \left( \begin{array}{cccc} \ \tfrac{1}{4} \partial_\phi  \ & \ \tfrac{1}{2} e^{-2\phi} V^{kl \, I}{}\partial_I   \ & \ 0 \ & \ 0 \\
\ \tfrac{1}{2} e^{-2\phi} V_{ij}{}^I \partial_I   \ &\ \tfrac{1}{12} \partial_\phi  \delta_{ij}^{kl} + \cD_{ij}{}^{kl}  \ & \sqrt{2} e^{-2\phi} \Omega_{j][k} V_{l][i}{}^I \partial_I  &0\\
0&\sqrt{2} e^{-2\phi} \Omega^{j][k} V^{l][i\, I}\partial_I  & \ -\tfrac{1}{12} \partial_\phi  \delta^{ij}_{kl} - \cD^{ij}{}_{kl}  \ &\  \tfrac{1}{2} e^{-2\phi} V^{ij \, I}{}\partial_I   \\\
0&0&\ \tfrac{1}{2} e^{-2\phi} V_{kl}{}^I \partial_I   \ & - \tfrac{1}{4} \partial_\phi  \end{array}\right) \  . \ee
We compute the different components of the differential equation  ${\bf D}^2 \cE = - \tfrac{9}{2} \mathds{1} \cE$ to give 
\bea \Scal{ \tfrac{1}{16} \partial_\phi^{\; 2} + \tfrac{9}{8} \partial_\phi + \tfrac{1}{4} e^{-4\phi} V_{ij}{}^I V^{ij\, J} \partial_I \partial_J  } \cE &=& - \frac{9}{2} \cE  \label{scal} \hspace{40mm}\\
\Scal{  \tfrac{1}{2} e^{-2\phi} V^{pq \, I}{}\partial_I \cD_{pq}{}^{kl} + e^{-2\phi} V^{kl \, I}{}\partial_I \scal{ 2 + \tfrac{1}{6} \partial_\phi}  } \cE &=& 0 \label{Linear}  \\
\tfrac{1}{4} e^{-4\phi} V^\inv{}_I{}^{ij} t^{IJK} \partial_J \partial_K \, \cE &=& 0\label{BPS}  \\
\Bigl( \scal{ \tfrac{1}{12^2} \partial_\phi^{\; 2} + \tfrac{11}{24} \partial_\phi } \delta_{ij}^{kl} + \cD_{ijpq} \cD^{klpq} +\cD_{ij}{}^{kl} \scal{ 1  + \tfrac{1}{6} \partial_\phi} \hspace{20mm}\Bigr .  && \CR
\hspace{30mm} \Bigl . +  e^{-4\phi} \scal{ \delta_{[i}^{[k} V_{j]p}{}^I V^{l]p\, J} + V_{[i}{}^{[k\, I} V_{j]}{}^{l]\, J} } \partial_I \partial_J \Bigr) \cE &=& - \frac{9}{2} \delta_{ij}^{kl} \cE \label{E6} \\
\sqrt{2} e^{-2\phi} \Scal{ V^{p[i\, I} \cD_p{}^{j]kl} -V^{p[k\, I} \cD_p{}^{l]ij} } \partial_I \cE &=& 0  \label{Ortho} 
\eea
The differential operator ${\bf D}$ clearly commutes with $\partial_I$, such that we can decompose the solution into Fourier modes $e^{ i q_I a^I }$. Let us consider in a first place the zero modes $q_I=0$. In this case equation \eqref{scal} implies that 
\be \cE_0(\phi,V) = e^{-6\phi} \cE_5(V) + e^{-12\phi} \cE^\prime_5(V) \ . \ee 
By representation theory, the term in $\cD_{ij}{}^{kl}$ in equation \eqref{E6} cannot mix with the others, such that the function $\cE^\prime_5(V)$ must be a constant. One finds that the function $e^{-12\phi}$ is indeed a solution to the complete differential equation ${\bf D}^2 \cE = - \tfrac{9}{2} \mathds{1} \cE$. In order to define a solution, the other function $\cE_5(V)$ must satisfy to the equation 
\be \cD_{ijpq} \cD^{klpq} \cE_5 = - 2 \delta_{ij}^{kl}  \cE_5 \ , \label{5DJoseph} \ee
which is nothing but the supersymmetry constraint of the five dimensional $R^4$ threshold. Taking its trace, one obtain indeed the Poisson equation \cite{Green:2010kv}
\be \Delta_{E_{6(6)}} \, \cE_5 = \frac{1}{3} \cD_{ijkl} \cD^{ijkl} \cE_5 = - 18\,  \cE_5 \ . \ee
Let us consider now the non-trivial Fourier modes.  Equation \eqref{BPS} implies that 
\be t^{IJK} q_J q_K = 0 \ee
which is the expected equation for a $\frac{1}{2}$-BPS scalar instanton. Equation \eqref{Ortho} is very constraining, and implies that $\cE_q(\phi,V)$ only dependent on the $E_{6(6)}/ \Sp$ coordinates through the invariant mass of the charge $q_I$. So we define 
\be Z_{ij}(q) = V_{ij}{}^I q_I \ , \qquad |Z(q)|^2 = Z_{ij}(q) Z^{ij}(q) \ , \ee
such that $\cE_q(\phi,V) = \cE_q(\phi,|Z(q)|)$. Because $q_I$ is a rank one vector \cite{Ferrara:1997ci}, 
\be  Z_{ik}(q) Z^{jk}(q) = \tfrac{1}{8} \delta_i^j |Z(q)|^2 \ . \ee
Equation \eqref{Linear} determines the dependence in $|Z(q)|$ in terms of the one in $\phi$, such that one obtains an ordinary differential equation. There are two solutions to this system  
\be \cE_q(\phi,V) =   \frac{e^{-6\phi}}{|Z(q)|^3} \scal{ 1 \pm  e^{-2\phi} |Z(q)| } e^{\mp e^{-2\phi} |Z(q)|}  \ . \ee
To check consistency, we use 
\be \cD^{ijkl} Z_{pq}(q) = 3 \Scal{ \delta_{pq}^{[ij} Z^{kl]}(q) - \Omega^{[ij} \delta_{[p}^{k} Z_{q]}{}^{l]}(q) - \frac{1}{4} \Omega_{pq} \Omega^{[ij} Z^{kl]}(q) - \frac{1}{12}    \Omega^{[ij} \Omega^{kl]} Z_{pq}(q) } \ , \ee
to compute that for a function $\cE_q(\phi,|Z|^2)$
\bea \cD_{ijpq} \cD^{klpq} \cE_q(\phi,|Z|^2) &=& \frac{2 }{3} \scal{ Z_{ij}(q) Z^{kl}(q) + Z_{[i}{}^{[k}(q) Z_{j]}{}^{l]}(q)} \Scal{ 2  |Z(q)|^2 \frac{\partial^2 \cE}{\partial {|Z|^2}^2} + 5 \frac{\partial \cE}{\partial {|Z|^2}} } \CR
&& \quad + \frac{1}{36} \delta^{kl}_{ij} |Z(q)|^2 \Scal{ 10  |Z(q)|^2 \frac{\partial^2 \cE}{\partial {|Z|^2}^2} + 73 \frac{\partial \cE}{\partial {|Z|^2}}} \ ,  \eea 
and 
\be \cD_{ij}{}^{kl} \cE_q(\phi,|Z|^2) = \Scal{ 2 Z_{ij}(q) Z^{kl}(q) - 4 Z_{[i}{}^{[k}(q) Z_{j]}{}^{l]}(q) - \frac{1}{6} \delta_{ij}^{kl} |Z(q)|^2} \frac{\partial \cE}{\partial {|Z|^2}} \ . \ee
The generic solution with appropriate boundary conditions is therefore supported by a function of seventeen variables $F(q)$,
 \be \cE[F,G] = \int_{\frac{E_{6(6)}}{Spin(5,5)\ltimes \mathds{R}^{16}}} \hspace{-7mm} d^{17} q  \ F(q) \, \frac{e^{-6\phi}}{|Z(q)|^3} \scal{ 1 + e^{-2\phi} |Z(q)| } e^{- e^{-2\phi} |Z(q)|  + i q_I a^I }  +  e^{-6\phi} \cE_5[G]\ , \ee 
where the additional function $\cE_5[G]$ is a generic solution to \eqref{5DJoseph} supported by a function $G$ of eleven variables. The representation of $E_{7(7)}$ on this space of functions is its minimal unitary representation. 

 We conclude that supersymmetry on its own already constrains the function $\cE$ to have the expected structure for the string theory effective action, and using the explicit coefficients computed in \cite{Green:2010kv} one gets the form of the Eisenstein series 
\bea&& \EiEVII{\frac{3}{2}} \\&=& \frac{2 \pi^2}{3} e^{-12\phi}  + e^{-6\phi} \EiEVI{\frac{3}{2}}  + \sum_{\scriptstyle q \in \mathds{Z}^{27}| q \times q =  0 } \hspace{-5mm} \mu(q) \,   \frac{e^{-6\phi}}{|Z(q)|^3} \Scal{ 1 +2\pi  e^{-2\phi} |Z(q)| } e^{-2\pi e^{-2\phi} |Z(q)| + 2\pi i q_I a^I }  \ .\nn  \eea
The Fourier modes coincide with the analysis of \cite{Pioline:2010kb,Fleig:2013psa}.

 \subsection{$\nabla^4 R^4$ and $\nabla^6 R^4$ type invariants}
In the linearised approximation, the $\nabla^4 R^4$ type invariant can be obtained from a harmonic superspace integral based on $SU(8)/S(U(2) \times U(4)\times U(2))$ harmonic variables \cite{Hartwell:1994rp}, and the G-analytic superfield 
\be  W^{rs} = u^1{}_i u^2{}_j u^r{}_k u^s{}_l W^{ijkl} \ , \ee
with $r=3$ to $6$ of $SU(4)$. $W^{rs}$ is therefore an $SO(6)$ vector, one the most general integrand is a monomial in a symmetric traceless tensor of $SO(6)$ 
\be \int d^{12}\theta d^{12}\bar \theta du \, F_{u\ r_1s_1\dots r_n s_n}^{[0,k,0,n,0,k,0]}\, (\varepsilon_{rstu} W^{rs} W^{tu})^{2+k} W^{(r_1|(s_1} W^{r_2 s_2} \dots W^{r_n)|s_n)} \ee
suggesting that the non-linear invariant admits an expansion 
\be \cL[\cE] = \sum_{n,k} \cD^{n+2k}_{[0,k,0,n,0,k,0]} \cE\, \cL^{[0,k,0,n,0,k,0]} \ . \ee
Consistently with this structure, the function $\cE$ must satisfy to the constraints 
\be \cD^3_{[0,2,0,0,0,0,0]} \cE = 0 \, , \qquad  \cD^3_{[0,0,0,0,0,2,0]} \cE = 0 \, , \qquad  \cD^3_{[1,0,0,0,0,0,1]} \cE = 0 \ . \ee
The two first define a condition on the differential operator to the third power in the fundamental of $E_{7(7)}$, whereas the last corresponds to a constraints on  the differential operator to the third power in the adjoint representation. Indeed, the harmonic decomposition also defines the graded decomposition of $\mathfrak{e}_{7(7)}$ associated to the next to minimal nilpotent orbit, for which the Lie algebra representative satisfies ${\bf Q}_{56}^{\; 3}=0$ and ${\bf Q}_{133}^{\; 3}=0$. 

It turns out that the eigenvalue of the Laplace operator is determined by these equations by consistency. Indeed, assuming that $\cE$ satisfy to the equations 
\be \Delta \cE = \lambda \cE \ , \qquad {\bf D}_{56}^{\; 3} \cE = a  {\bf D}_{56}   \cE \ , \qquad   {\bf D}_{133}^{\; 3} \cE = b  {\bf D}_{133}  \cE \ , \ee
and using the Casimir identities 
\bea \tr {\bf D}_{133}^{\; 2} &=& 3 \tr  {\bf D}_{56}^{\; 2} \ , \qquad  \tr {\bf D}_{133}^{\; 4} = \frac{1}{6}\scal{  \tr {\bf D}_{56}^{\; 2}  }^2  \ ,\CR 
 \tr {\bf D}_{133}^{\; 6} &=&- 2 \tr {\bf D}_{56}^{\; 6} + \frac{5}{288} \scal{  \tr {\bf D}_{56}^{\; 2}  }^3  + \frac{23}{6} \scal{  \tr {\bf D}_{56}^{\; 2}  }^2  + 492   \, \tr {\bf D}_{56}^{\; 2} \ , 
\eea
one computes that the unique solutions are 
\bea \lambda &=& - 42 \, , \quad a = - \frac{9}{2} \ , \quad b = -14\ , \CR
\lambda &=& - 60 \, , \quad a = - 9 \ , \quad b = - 20 \ .  \eea
The first solution corresponds to the constraint satisfied by the $R^4$ threshold, and we conclude that the second solution is the relevant one for the $\nabla^4 R^4$ threshold, consistently with  \cite{Green:2010kv}. So $\cE_\frac{5}{2}$ must satisfy to the Poisson  equation 
\be \Delta   \cE_\frac{5}{2}  = - 60 \cE_\frac{5}{2} \ , \ee
and
\be \cD_{ijpq} \cD^{pqrs} \cD_{rskl} \cE_\frac{5}{2} = - 9 \cD_{ijkl} \cE_\frac{5}{2} \ , \qquad \cD_{t[ijk} \cD^{qtrs} \cD_{l]prs}|_{[1,0,0,1,0,0,1]} \cE_{\frac{5}{2}} = 0 \ , \label{d4R44D} \ee 
for the superform $\cL[\cE_\frac{5}{2}]$ to be closed.

The $\nabla^6 R^4$ type invariant can be defined from a harmonic superspace integral based on $SU(8)/S(U(1) \times U(6)\times U(1))$ harmonic variables \cite{Hartwell:1994rp}, and the G-analytic superfield 
\be  W^{rst} = u^1{}_i u^r{}_j u^s{}_k u^t{}_l W^{ijkl} \ , \ee
with $r=2$ to $7$ of $SU(6)$. In this case the measure extends to the complete theory \cite{Bossard:2011tq}. The number of possible representations of $SU(8)$ becomes rather large, but they are still self-adjoint by construction. It follows that the constraints 
\be \cD^3_{[0,2,0,0,0,0,0]}\cE_\gra{0}{1}  = 0 \, , \qquad  \cD^3_{[0,0,0,0,0,2,0]}\cE_\gra{0}{1} = 0 \ , \ee
still apply, although the second one is not satisfied. Using the closure diagram of $E_{7(7)}$ \cite{E7Djo}, one finds that there is not a unique next to next to minimal nilpotent orbit. 
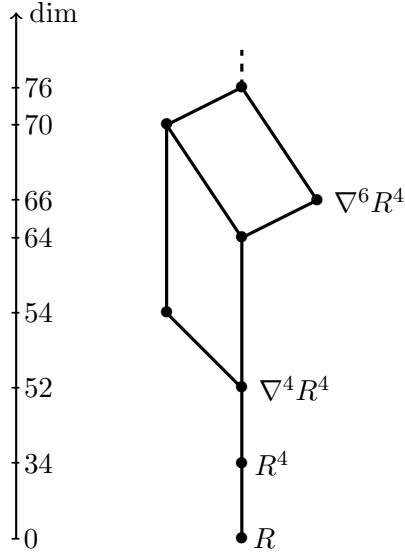
\begin{figure}[htbp]
\begin{center}
 \begin{tikzpicture}
 \draw (\xmin,\ymin) node{\textbullet};
 \draw (\xmin,\ymin - 1) node{\textbullet};
  \draw (\xmin,\ymin - 2)  node{\textbullet};
  \draw (\xmin,\ymin + 2)  node{\textbullet};
  \draw (\xmin,\ymin + 4)  node{\textbullet};
  \draw (\xmin + 1,\ymin + 2.5)  node{\textbullet};
   \draw (\xmin - 1,\ymin + 3.5)  node{\textbullet};
    \draw (\xmin - 1 ,\ymin + 1) node{\textbullet};
   
  \draw (\xmin + 0.3,\ymin - 2) node{$R$};
  \draw (\xmin + 0.4,\ymin - 1) node{$R^4$};
  \draw (\xmin + 0.7,\ymin) node{$ \nabla^4 R^4$};
  \draw (\xmin + 0.7 + 1,\ymin + 2.5) node{$ \nabla^6 R^4$};
  \draw[-,draw=black,very thick](\xmin,\ymin) -- (\xmin,\ymin + 2 );
   \draw[-,draw=black,very thick](\xmin - 1,\ymin + 1) -- (\xmin - 1,\ymin + 3.5);
    \draw[-,draw=black,very thick](\xmin - 1,\ymin + 3.5) -- (\xmin,\ymin + 2);
    \draw[-,draw=black,very thick](\xmin,\ymin + 2) -- (\xmin + 1,\ymin + 2.5);
    \draw[-,draw=black,very thick](\xmin + 1,\ymin + 2.5) -- (\xmin,\ymin + 4);
    \draw[-,draw=black,very thick](\xmin - 1,\ymin + 3.5) -- (\xmin,\ymin + 4);
     \draw[dashed,draw=black,very thick](\xmin,\ymin + 4) -- (\xmin,\ymin + 4.5);
  \draw[-,draw=black,very thick](\xmin,\ymin) -- (\xmin - 1,\ymin + 1);
\draw[-,draw=black,very thick] (\xmin,\ymin - 1) -- (\xmin,\ymin);
\draw[-,draw=black,very thick] (\xmin,\ymin - 2) -- (\xmin,\ymin - 1);
\draw[<-,draw=black,thick] (\xmin - 3,\ymin + 5) -- (\xmin - 3,\ymin - 2);
\draw (\xmin - 3 + 0.2,\ymin - 2) node{$0$};
\draw (\xmin - 3,\ymin - 2) node{-};
\draw (\xmin - 3,\ymin - 1) node{-};
\draw (\xmin - 3,\ymin) node{-};
\draw (\xmin - 3,\ymin + 2) node{-};
\draw (\xmin - 3,\ymin + 1) node{-};
\draw (\xmin - 3,\ymin + 2.5) node{-};
\draw (\xmin - 3,\ymin + 3.5) node{-};
\draw (\xmin - 3,\ymin + 4) node{-};
\draw (\xmin - 3 + 0.3,\ymin - 1) node{$34$};
\draw (\xmin - 3 + 0.3,\ymin) node{$52$};
\draw (\xmin - 3 + 0.3,\ymin + 1) node{$54$};
\draw (\xmin - 3 + 0.3,\ymin + 2) node{$64$};
\draw (\xmin - 3 + 0.3,\ymin + 2.5) node{$66$};
\draw (\xmin - 3 + 0.3,\ymin + 3.5) node{$70$};
\draw (\xmin - 3 + 0.3,\ymin + 4) node{$76$};
\draw (\xmin - 3 + 0.5,\ymin + 5) node{dim};
\end{tikzpicture}
\end{center}
\caption{\small Closure diagram of nilpotent orbits  of $E_{7(7)}$ of  dimension smaller than 76.}
\label{ClosureDiag}
\end{figure}
However the condition ${\bf Q}_{56}^{\; 3}=0$ rules out the dimension 54 orbit. The nilpotent orbit associated to the harmonic decomposition is in fact not the next one of dimension 64 that would also satisfy to ${\bf Q}_{133}^{\; 4} =0$, but the following one of dimension 66. Using harmonic superspace, one finds indeed a non-vanishing integral in the representation $[2,0,0,0,0,0,2]$ by integrating the square of the quartic $SU(6)$ invariant monomial in $W^{rst}$ with the appropriate function of the harmonic variables. Therefore the superform expansion must include terms as
\be \cL[\cE_\gra{0}{1}] = \cE_\gra{0}{1} \cL + \cD_{ijkl} \cE_\gra{0}{1} \cL^{ijkl} + \dots + \cD^4_{[2,0,0,0,0,0,2]} \cE_\gra{0}{1} \cL^{[2,0,0,0,0,0,2]} + \dots \ee
and the corresponding component of ${\bf D}_{133}^{\; 4} \cE_\gra{0}{1}$ acting on the $\su(8)$ adjoint does not vanish.

The determination of the eigenvalue of the Laplace operator does not follow straightforwardly from a group theory argument in that case, and one must moreover consider the corrections to the supersymmetry transformations at this order. Nonetheless, relying on the known Poisson equation satisfied by the function according to  \cite{Green:2010kv}, we find that the function must moreover satisfy to
\be \cD_{ijpq} \cD^{pqrs} \cD_{rskl} \cE_\gra{0}{1} = - 9 \cD_{ijkl} \cE_\gra{0}{1}  - \frac{1}{2} \cE_{\frac{3}{2}} \cD_{ijkl} \cE_{\frac{3}{2}}  \ , \label{d6R44D} \ee
which is consistent with 
\be \Delta   \cE_\gra{0}{1}   = - 60 \, \cE_\gra{0}{1} - \cE_\frac{3}{2}^{\; 2} \ . \label{d6R4}\ee
Let us now analyse these equations in the parabolic gauge as in the preceding section. We shall only analyse the solution for $q_I = 0$, and for the homogenous equation in the fundamental representation. After some computations one obtains 
\bea \Bigl( \frac{1}{64} \partial_\phi^{\; 3} + \frac{21}{32} \partial_\phi^{\; 2} +\frac{9}{2} \partial_\phi - \frac{1}{4} \cD_{ijkl} \cD^{ijkl} \Bigr) \cE_{\frac{5}{2}}    &=& - 9 \times \frac{1}{4} \partial_\phi \cE_\frac{5}{2}\\
 \biggl( \delta_{ij}^{kl} \Bigl( \frac{1}{12^3} \partial_\phi^{\; 3} + \frac{5}{96} \partial_\phi^{\; 2} +\frac{1}{6} \partial_\phi + \frac{1}{6} \cD_{ijkl} \cD^{ijkl} \Bigr) 
 + \cD_{ijpq} \cD^{klpq} \Bigl( \frac{1}{4} \partial_\phi - \frac{15}{4} \Bigr)\hspace{-7mm} && \CR + \cD_{ijpq} \cD^{pq}{}_{rs} \cD^{rskl} + \cD_{ij}{}^{kl} \Bigl( \frac{1}{48} \partial_\phi^{\; 2} + \frac{27}{24} \partial_\phi + \frac{7}{2} \Bigr)   \biggr) \cE_\frac{5}{2}  &=& - 9 \Bigl( \frac{1}{12} \delta_{ij}^{kl} \partial_\phi + \cD_{ij}{}^{kl} \Bigr)  \cE_\frac{5}{2} \nn
 \eea
 where we recall that $\cD_{ijkl}$ states for the covariant derivative on $E_{6(6)}/ Sp_{\scriptscriptstyle \rm c}(4)$ in these expressions. One finds indeed that the decompactification limit of the corresponding Eisenstein series \cite{Green:2010kv}
\be \EiEVII{\frac{5}{2}} = \frac{8 \zeta(8)}{15\pi} e^{-24\phi} + \frac{\pi}{3} e^{-12\phi} \EiEVI{\frac{3}{2}} + \frac{1}{2} e^{-6\phi} \EiEVI{\frac{5}{2}} + \mathcal{O}(e^{-e^{-2\phi}}) \ee
associated to the $\nabla^4 R^4$ correction is a solution provided 
\be \cD_{ijpq} \cD^{pq}{}_{rs} \cD^{rskl}  \EiEVI{\frac{5}{2}} + \frac{4}{3} \cD_{ij}{}^{kl} \EiEVI{\frac{5}{2}} = \frac{25}{4} \Scal{ \cD_{ijpq} \cD^{klpq} + \frac{70}{27} \delta_{ij}^{kl} } \EiEVI{\frac{5}{2}} \ .\ee
The latter equation must therefore define the differential equation satisfied by the function defining the $\nabla^4 R^4$ type invariant in five dimensions, and is indeed consistent with the associated Poisson equation \cite{Green:2010kv}. 

For $q_I\ne 0$, one computes straightforwardly that the equations ${\bf D}^{\; 3}_{56} \cE = - 9 {\bf D}_{56} \cE$ implies moreover 
\be t^{IJK} q_I q_J q_K = 0 \ .\label{FourierCube}  \ee
For $\EiEVIItxt{\frac{5}{2}}$ this is consistent with the property that the next to minimal unitary representation is defined on functions of $26$ variables. Note that the sum of two vectors satisfying to $t^{IJK} q_J q_K=0$ necessarily satisfies \eqref{FourierCube}, such that the complete function $\cE_\gra{0}{1}$ is supported on Fourier modes satisfying to this same constraint \eqref{FourierCube}. Where by $\cE_\gra{0}{1}$ we mean the function appearing in the $\nabla^6 R^4$ type invariant we discuss in this paper, and not the complete function appearing in the four-graviton amplitude. We will explain in another publication that there is in fact a second class of $\nabla^6 R^4$ type invariants associated to the dimension 54 nilpotent orbit, and which admits generic Fourier modes in the decompactification limit. The unitary representation on which $\cE_\gra{0}{1}$ is supported is, however, defined on functions of $33$ variables, therefore the Fourier modes must depend on a non-trivial function of the scalar fields $v_{B_4(q)}$ parametrizing the subgroup $Spin(4,5)\subset E_{6(6)}$ stabilizing $q_I$ \cite{Ferrara:1997ci}. Because $33-26=7$ we expect the function $\cE(v_{B_4(q)})$ to satisfy a differential equation restring effectively its dependence on seven variables. This suggests that the relevant function on $SO(4,5)/(SO(4)\times SO(5))$ should satisfy to the following differential equation associated to a coadjoint $SO(4,5)$ orbit of dimension 14, \ie 
\be {\bf D}_{16}^{\; 2} \, \Evector{s}(v_{B_4(q)}) = \, \frac{s(2s-7)}{8} \, \mathds{1}_{16}  \Evector{s}(v_{B_4(q)}) \ . \ee
We note moreover that the solution to \eqref{d4R44D} is also a solution to the homogeneous equation associated to \eqref{d6R44D}, therefore the restriction of the Fourier mode function to the case in which the function on $SO(4,5)$ is a constant must also be solution. We conclude that the correct value of $s$ must be $s=\frac{7}{2}$. This is precisely the value for which the Eisenstein series diverges in $\frac{1}{2s-7}$, and one concludes that the exact $\nabla^6 R^4$ threshold function $\hat{E}_\gra{0}{1}$ should rather satisfy to a corrected equation of the form 
\be \cD_{ijpq} \cD^{pqrs} \cD_{rskl} \hat{E}_\gra{0}{1}  = - 9 \cD_{ijkl} \hat{E}_\gra{0}{1}  - \frac{1}{4}  \cD_{ijkl} \Scal{ \EiEVII{\frac{3}{2}}}^2    + \xi \,\cD_{ijkl}  \EiEVII{\frac{5}{2}}\ ,  \ee
for some number $\xi$. This implies accordingly that the $\cE_\frac{5}{2} \nabla^4 R^4$ type superform  form factor diverges at 1-loop into  the three-level $\cE_\frac{5}{2} \nabla^6 R^4$ type superform form factor, defined with the same function.

%
%
\section{$\cN=16$ supergravity in three dimensions.} 
In three dimensions the only propagating degrees of freedom are the scalar fields parametrizing the symmetric space $E_{8(8)}/Spin_{\scriptscriptstyle \rm c}(16)$ \cite{Marcus:1983hb}, such that the Maurer--Cartan form 
\be d \cV  \, \cV^{-1} = \left(\begin{array}{cc} - \frac{1}{4} \Gamma^{pq}{}_A{}^B \omega_{pq} \ & \ \frac{1}{4} \Gamma^{kl}{}_{AC} P^C \ \\
\frac{1}{4} \Gamma_{ij}{}^{BC} P_C \ & - 2 \delta_{[i}^{[k} \omega_{j]}{}^{l]} \ \end{array}\right) \ ,  \ee
defines the scalar momentum $P_A$ in the Majorana--Weyl representation of the R-symmetry group $Spin(16)$, whereas the fermion fields $\chi_{\alpha \dot{A}}$ are defined in the opposite chirality Majorana--Weyl representation. Solving the superspace constraints \cite{Greitz:2011vh} the momentum decomposes as 
\be P^A = E^a P_a^A + E^\alpha_i \, \Gamma^{i A\dot{A}}  \chi_{\alpha\, \dot{A}} \ . \ee
The metric on the symmetric space is defined as 
\be ds^2 = \frac{1}{30} \tr \, {\bf P}_{\scriptscriptstyle 248}\big .^{\hspace{-3mm}2\hspace{3mm}}  = P_A P^A \ , \ee
and the covariant derivative satisfies 
\be [ \cD_A , \cD_B ] \cD_C = - \frac{1}{16} \Gamma^{ij}{}_{AB} \Gamma_{ij C}{}^D \cD_D \ . \label{CommE8} \ee

\subsection{The $R^4$ type invariant}
The argumentation proposed in the last section in four dimensions extends to $\cN=16$ supergravity in three dimensions. In this case the equivalent of the $R^4$ type invariant, \ie $(\nabla P)^4$ type invariant in practice, admits a superspace construction in the linearised approximation based on harmonic variables in $SO(16)/U(8)$ \cite{Howe:1994ms}. The linearised superfield $W_A$   as a chiral spinor of $Spin(16)$ decomposes into 
\be {\bf 128}_+ \cong {\bf 1}^\ord{-4} \oplus \overline{\bf 28}^\ord{-2} \oplus {\bf 70}^\ord{0} \oplus {\bf 28}^\ord{2} \oplus {\bf 1}^\ord{4}\ , \ee
and the G-analytic superfield $W$ is in the weight $4$ singlet of $SU(8)$, \ie an $SO(16)$ pure spinor. The Dirac fermion $\chi_{\alpha\dot{A}}$ decomposes accordingly as a Majorana--Weyl spinor of opposite chirality  into 
\be   {\bf 128}_- \cong \overline{\bf 8}^\ord{-3} \oplus \overline{\bf 56}^\ord{-1} \oplus {\bf 56}^\ord{1} \oplus {\bf 8}^\ord{3}  \ ,\ee
and we write $\chi_\alpha^{r}$ the $U(1)$ weight $3$ component, with $r=1$ to $8$ of $SU(8)$. The linearised invariant 
\be \int d^{16} \theta du \, F_u^{\mbox{\WSOXVIn0000000n} } \  W^{4+n}  \sim  (W^n)^{\mbox{\WSOXVIn0000000n} } \, (\nabla P)^4 + \dots +  (W^{n-12})^{\mbox{\WSOXVIn0000000{n\mbox{-}12}} }  (\chi^{16})^{\mbox{\WSOXVIn0000000{12}} }  \ee
suggests the expansion of the non-linear closed superform in 
\be \cL[\cE] = \sum_{n=0}^{12}  \cD^n_{\mbox{\WSOXVI0000000n} } \cE\, \cL^{\mbox{\WSOXVI0000000n} } \ . \label{3DR4} \ee
 The superconformal symmetry $OSp(16|4,\mathds{R})$ of the linearised theory \cite{Chiodaroli:2011pp} suggests that all the supersymmetry invariants are defined by harmonic superspace integrals in the linearised approximation, such that the harmonic superspace integrals are indeed in bijective correspondence with the independent non-linear invariants. One confirms this property by looking at the monomial in the fermions of maximal weight. Using the harmonic decomposition, one gets directly that the $2\times 8$ fermions $\chi_\alpha^r$ to the sixteenth power carries a $U(1)$ weight $48$, just as does $W^{12}$.  Considering the action of the covariant derivative $D_\alpha^i$, one cannot include one more $\chi_\alpha^r$, so the only non-trivial term appears to include instead a weight $1$ fermion $\chi_\alpha^{rst}$. Projecting out the corresponding representations in $D_\alpha^i \cL[\cE]_\gra{3}{0}$ using the harmonic variables, one gets 
\be D^{\ord{1}} _{\alpha r}  \scal{ \cD^{12 \ord{-48} } \cE   ( \chi^{16} )^{\ord{48}}  +  \cD^{12 \ord{-46} }_{st}  \cE   ( \chi^{15} )^{ \beta \ord{45}}_u \chi^{stu \ord{1}}_\beta   + \dots} \sim    \cD^{13 \ord{-48}}_{rstu}  \cE  \, \chi^{stu \ord{1} }      ( \chi^{16} )^{\ord{48}}  + \dots \ee
where the two terms in the first line contribute to two  independent terms in the second $\sim \cD^{ \ord{0}}_{rstu} (\cD^{\ord{-4}})^{12}$ and $\sim \cD^{ \ord{-2}}_{[rs} \cD^{\ord{-2}}_{tu]} (\cD^{\ord{-4}})^{11}$, such that they cannot compensate each other. To deduce the $Spin(16)$ covariant expressions associated to these terms, we note that the rank $p$ antisymmetric tensor representation of $SO(16)$ admits as a highest weight component of weight $p$ the rank $p$ antisymmetric tensor in the anti-fundamental of $SU(8)$. We conclude that $\chi^{stu \ord{1} }      ( \chi^{16} )^{\ord{48}}$ is in the highest weight component of the \WSOXVI0000100{11} representation, whereas $\cD^{13 \ord{-48}}_{rstu}$ is in the lowest weight component of the \WSOXVI0001000{11}, such that this expression corresponds to
\be D^i_\alpha \Scal{ \cD^{12}_{\mbox{\WSOXVI0000000{12}} } \cE\, (\chi^{16})^{\mbox{\WSOXVI0000000{12}} }  } \sim   \cD^{13}_{\mbox{\WSOXVI0001000{11}} } \cE\, (\chi^{17})^{\mbox{\WSOXVI0000100{11}} }  + \dots    \ee
There is no other contribution that could cancel this term, because the next terms of maximal weight are in the ${\mbox{\WSOXVI0001000{10}} } \cE$ and carry a maximal weight component in $\cD^{12\ \ord{-44}}_{rstu} \cE $ whereas 
\be  D^{\ord{1}} _{\alpha r}  \cE \sim \cD^\ord{-2}_{rs} \cE\, \chi_\alpha^{s\, \ord{3}} +  \cD^\ord{0}_{rstu} \cE\, \chi_\alpha^{stu\, \ord{1}} + \dots \ee
and they cannot  contribute to terms in $\cD^{13\ \ord{-48}}_{rstu} \cE $. We conclude similarly as in the preceding sections that the function $\cE$ must satisfy to the differential equation 
\be \cD^{11}_{\mbox{\WSOXVI0000000{11}}} \cD^{2}_{\mbox{\WSOXVI00010000}}  \cE = 0 \ . \ee
Using the property that the $\cD^n$ differential operator of maximal weight in the ${\mbox{\WSOXVI0000000{n}}}$ has no kernel, one obtains that the function $\cE$ must satisfy to the quadratic equation 
\be \Gamma^{ijkl\, AB} \cD_A \cD_B \, \cE = 0 \ . \label{QuadraE8} \ee
Using $SO(16)$ Fierz identities 
\bea \cD_A \cD_B &=& \frac{1}{128} \Scal{ \delta_{AB} (\cD^2) + \frac{1}{4!} \Gamma^{ijkl}_{AB} ( \cD \Gamma_{ijkl} \cD) + \frac{2}{8!} \Gamma^{ijklmnpq}_{AB}  ( \cD \Gamma_{ijklmnpq} \cD) } + \frac{1}{2} [\cD_A, \cD_B]  \ , \CR
\Gamma^{ij} \Gamma_{[n]} \Gamma_{ij} &=& - 4 (n-6)(n-10) \Gamma_{[n]} \ , \quad \Gamma^{ijkl} \Gamma_{[n]} \Gamma_{ijkl} = 16 \scal{ (n-8)^4 - 22 (n-8)^2 + 42} \ , 
\eea
and the commutation relation \eqref{CommE8}, one computes that 
\be \Gamma^{ijkl\, AB} \cD_A \cD_B  \,  \Gamma_{ijkl}{}^{CD} \cD_C \cD_D  = 672\,  \cD_A \cD^A \scal{ \cD_B \cD^B + 120 }  \  . \ee 
Moreover, \eqref{QuadraE8} implies as a consistency condition that the third derivative of the function $\cE$ restricted to the \WSOXVI0100000{1} must also vanish, \ie 
\be \scal{  5 \, \Gamma^{kl\, A(B} \Gamma_{ijkl}{}^{CD)} + 14\,  \Gamma_{ij}{}^{A(B} \delta^{CD)} } \cD_B \cD_C \cD_D \, \cE = 0 \ . \label{CubicE8} \ee 
Using \eqref{QuadraE8} in this equation one obtains 
\be 14\,  \Gamma_{ij}{}^{AB} \cD_B \scal{ \cD_C \cD^C + 120 } \cE = 0 \ , \ee
such that if $\cE$ is canceled by the Laplacian, it must necessarily be a constant, and supersymmetry indeed implies 
\be \Delta \cE = - 120 \cE \ , \ee
consistently with \cite{Green:2010kv}.  

Using these equations, one computes that the covariant derivative in the adjoint representation 
\be {\bf D}_{\scriptscriptstyle 248} = \left(\begin{array}{cc} \ 0 \ & \ \frac{1}{4} \Gamma^{kl}{}_{AC} \cD^C \ \\ 
\frac{1}{4} \Gamma_{ij}{}^{BC} \cD_C \ & \ 0  \ \end{array}\right) \ ,  \ee
satisfies 
\be {\bf D}_{\scriptscriptstyle 248}\big .^{\hspace{-3mm}2\hspace{3mm}} + 15 \, \mathds{1}_{\scriptscriptstyle 248}  = - \left(\begin{array}{cc} \  \cD_A \cD^B \ & \ 0  \ \\ 
 0  \ & \ 0  \ \end{array}\right) \ .  \ee
This equation defines a quantization of the algebraic equation 
\be   {\bf Q}_{\scriptscriptstyle 248}\big .^{\hspace{-3mm}2\hspace{3mm}}   = - \left(\begin{array}{cc} \  Q_A Q^B \ & \ 0  \ \\ 
 0  \ & \ 0  \ \end{array}\right) \ , \ee
 for a Majorana--Weyl pure spinor of $Spin^*(16)$, which is a representative of the minimal nilpotent orbit of $E_{8(8)}$ \cite{Bossard:2009at}. The solutions to the differential equation \eqref{QuadraE8} with appropriate boundary conditions define the minimal unitary representation of $E_{8(8)}$, and are supported on functions depending on 29 variables as explained in \cite{Kazhdan:2001nx,Gunaydin:2001bt}. 

\subsection{The $\nabla^4 R^4$ type invariant}
The $(\nabla^2 P)^4$ type invariant can be defined in harmonic superspace \cite{Howe:1994ms} in the linearised approximation using harmonic variables parametrizing $SO(16)/ ( SO(8) \times U(4))$ such that the Majorana--Weyl representations decomposes as 
\be  {\bf 128}_\pm \cong {\bf 8}_\pm^\ord{-2} \oplus (\overline{\bf 4}\otimes {\bf 8}_\mp)^\ord{-1} \oplus ( {\bf 6} \otimes {\bf 8}_\pm)^\ord{0} \oplus ({ \bf 4} \otimes {\bf 8}_\mp)^\ord{1} \oplus {\bf 8}_\pm^\ord{2} \, \ , \ee
such that the weight $2$ scalar superfield $W^r$ in the chiral spinor representation of $Spin(8)$ is G-analytic. One defines the invariant 
\bea &&  \int d^{24} \theta  du \, F_{u\ r_1r_1\dots r_n} ^{\mbox{\WSOXVInk000k000n} } ( W^r W_r)^{2+k} W^{(r_1} W^{r_2} \dots W^{r_n)}  \CR
& \sim&  (W^{n+2k})^{\mbox{\WSOXVInk000k000n} } \, (\nabla^2  P)^4 + \dots +  (W^{n+2k-16})^{\mbox{\WSOXVInk000k000{n\mbox{-}16}} }  (\chi^{16})^{\mbox{\WSOXVI0000000{12}} }( P^4)^{\mbox{\WSOXVI0000000{4}} } \, ,  \qquad \eea
which suggests the following expansion of the superform defining the invariant at the non-linear level 
\be \cL[\cE] = \sum_{n,k}  \cD^{n+2k}_{\mbox{\WSOXVInk000k000n} } \cE\, \cL^{\mbox{\WSOXVInk000k000n} } \ . \label{3DD4R4} \ee
Assuming that all  $(\nabla^2 P)^4$ type invariants are defined in this way, this shows that the function must have covariant derivatives restricted to these representations. This is the case if and only if he function $\cE$ satisfies the cubic equation \eqref{CubicE8}. Moreover, acting with one more derivative on this equation one obtains using the Fierz rearrangements 
\bea 
&& \frac{1}{72} \scal{\cD \Gamma_{ijklpqrs} \cD} \scal{ \cD \Gamma^{pqrs} \cD} \CR
&=&  \scal{ \cD  \Gamma_{[ij}{}^{pq}  \cD} \scal{ \cD \Gamma_{kl]pq} \cD} +   \scal{ \cD \Gamma_{ijkl} \cD} \scal{ 9 (\cD \cD) +872} \ , \CR
&&  \scal{ \cD \Gamma_{[ij}  \Gamma^{pq}  \cD} \scal{ \cD \Gamma_{kl]pq} \cD} \CR
&=& \scal{ \cD  \Gamma_{[ij}{}^{pq}  \cD} \scal{ \cD \Gamma_{kl]pq} \cD}- 2 \scal{ \cD \Gamma_{ijkl} \cD} \scal{  (\cD \cD) +  360}  \CR
&=& - \frac{1}{2}  \scal{ \cD  \Gamma_{[ij}{}^{pq}  \cD} \scal{ \cD \Gamma_{kl]pq} \cD} - \frac{1}{2}  \scal{ \cD \Gamma_{ijkl} \cD} \scal{ (\cD \cD) -24} - \frac{1}{48} \scal{\cD \Gamma_{ijklpqrs} \cD} \scal{ \cD \Gamma^{pqrs} \cD} \CR
&=& - 6 \scal{ \cD \Gamma_{ijkl} \cD} \scal{ (\cD \cD) +152} \ ,
\eea
that
\bea &&  \Gamma_{[ij}{}^{A}{}_E  \scal{  5 \, \Gamma^{pq\, E(B} \Gamma_{kl]pq}{}^{CD)} + 14\,  \Gamma_{kl]}{}^{E(B} \delta^{CD)} } \cD_A \cD_B \cD_C \cD_D  \CR &=& - 16 \, \Gamma_{ijkl}{}^{AB} \cD_A \cD_B \, \scal{ \cD_C \cD^C  + 180 } \  , \eea
such that the function must then either satisfy to the quadratic equation \eqref{QuadraE8} or to 
\be \Delta \cE = - 180 \cE \ . \label{180Laplace}  \ee
The two equations being incompatible, supersymmetry requires that the function defining the  $(\nabla^2 P)^4$ type invariant satisfies \eqref{180Laplace}, consistently with \cite{Green:2010kv}. Using the latter, \eqref{CubicE8} simplifies to 
\be \Gamma^{kl\, AB}  \Gamma_{ijkl}{}^{CD}  \cD_B \cD_C\cD_D = - 168\,  \Gamma_{ij}{}^{AB} \cD_B \  . \label{3Dd4R4} \ee
Using this equation and \eqref{180Laplace} one computes that 
\be{\bf D}_{\scriptscriptstyle 248}\big .^{\hspace{-3mm}3\hspace{3mm}}   =  \left(\begin{array}{cc} \ 0 \ & \ -3 \Gamma^{kl}{}_{AC} \cD^C   \ \\ 
- \frac{31}{2} \Gamma_{ij}{}^{BC} \cD_C   \ & \ 0  \ \end{array}\right) \ ,  \ee
which defines the quantisation of the algebraic equation ${\bf Q}_{\scriptscriptstyle 248}\big .^{\hspace{-3mm}3\hspace{1mm}}   =0$ defining the next to minimal nilpotent orbit of $E_{8(8)}$ \cite{Bossard:2009at}. We conclude that the solutions to \eqref{3Dd4R4} with appropriate boundary conditions define the next to minimal unitary representation of $E_{8(8)}$ associated to the next to minimal coadjoint orbit.

\section*{Acknowledgment}
We thank Eric D'Hoker, Paul Howe,  Ruben Minasian, Daniel Persson,  Boris Pioline,  K.S. Stelle and Pierre Vanhove for relevant discussions. This work was supported by the French ANR contract 05-BLAN-NT09-573739 and the ERC Advanced Grant no. 226371.


 \begin{appendix}
\section{Conventions in eight dimensions}
The $SU(2)$ invariant tensors $\varepsilon_{ij}$ and $\varepsilon^{ij}$ are defined respectively such that \begin{equation}
 \varepsilon^{i j} \equiv  
\begin{pmatrix}
\ 0 \ & \ 1 \ \\
-1\ & \ 0 \ \\
\end{pmatrix} \ , \qquad  \varepsilon_{i j} \equiv  
\begin{pmatrix}
\ 0 \  & -1 \ \\
\ 1 \ & \ 0 \ \\
\end{pmatrix} \ , \qquad
 \varepsilon_{i j}  \varepsilon^{j k} = \delta_{i}^{k} \ , \qquad
 \varepsilon^{i j}  \varepsilon_{j k} = \delta^{i}_{k}\ . 
\end{equation}
One raises and lower the $SU(2)$ indices according to the rules
\begin{equation}
\chi_{i}{}^{kl\dots } = \varepsilon_{i j} \chi^{jkl\dots}\ ,  \qquad \frac{\partial}{\partial \chi^i} \chi^j = \delta^{j}_{i} \ , \qquad \frac{\partial}{\partial \chi_i} \chi_j = \delta_{j}^{i}\ ,  \qquad \frac{\partial}{\partial \chi_i} = -  \varepsilon^{i j} \frac{\partial}{\partial \chi^j}\ , 
\end{equation}
The conventions for the $SO(1,7)$ invariant tensors are pletely antisymmetric tensor with the local metric are taken to be:
\begin{equation}
 \varepsilon_{01234567} = 1\,  ,     \qquad \eta_{00} = - 1 \ , \quad \eta_{11} = \eta_{22} = \dots = 1 \ , 
\end{equation}
and we define the antisymmetric Kronecker delta tensors 
\begin{equation}
\delta_{a_1 a_2 \dots a_n}^{b_1 b_2 \dots b_n} \equiv \delta_{[a_1}^{[b_1} \delta_{a_2}^{b_2} \dots \delta_{a_n]}^{b_n]} \ . 
\end{equation}
We decompose the spinor representation into the Weyl representation of positive chirality with undotted indices and negative chirality with dotted indices, which are complex conjugate. We use the octonionic representation such that the charge conjugation matrix is the identity, and we have the following relations
\begin{equation}
\begin{split}
(\gamma^a)^{\dot{\alpha} \alpha} & = -\frac{i}{7!}  \varepsilon^a_{\;\;b c d e f g h} (\gamma^{b c d e f g h})^{\dot{\alpha} \alpha} \\
(\gamma^{a b})^{\dot{\alpha} \dot{\beta}} & = -\frac{i}{6!}  \varepsilon^{a b}_{\;\;\;\; c d e f g h} (\gamma^{c d e f g h})^{\dot{\alpha} \dot{\beta}} \\
(\gamma^{a b c})^{\dot{\alpha} \alpha} & = \frac{i}{5!}  \varepsilon^{a b c}_{\;\;\;\;\; d e f g h} (\gamma^{d e f g h})^{\dot{\alpha} \alpha} \\
(\gamma^{a b c d})^{\dot{\alpha} \dot{\beta}} & = \frac{i}{4!}  \varepsilon^{a b c d}_{\;\;\;\; \;\;\;\; e f g h} (\gamma^{e f g h})^{\dot{\alpha} \dot{\beta}} \\
(\gamma^{a b c d e f g h})^{\dot{\alpha} \dot{\beta}} & = i   \varepsilon^{a b c d e f g h} C^{\dot{\alpha}\dot{\beta}}  \\
C^{\alpha \beta} & = \delta^{\alpha\beta} \\
(\gamma^{a b})^{\alpha \beta} & = - (\gamma^{a b})^{\beta \alpha}\\
(\gamma^{a b c d})^{\alpha \beta} & = (\gamma^{a b c d})^{\beta \alpha}
\end{split}
\begin{split}
(\gamma^a)^{\alpha \dot{\alpha}} & = \frac{i}{7!}  \varepsilon^a_{\;\;b c d e f g h} (\gamma^{b c d e f g h})^{\alpha \dot{\alpha}}\\
(\gamma^{a b})^{\alpha \beta} & = \frac{i}{6!}  \varepsilon^{a b}_{\;\;\;\; c d e f g h} (\gamma^{c d e f g h})^{\alpha \beta}\\
(\gamma^{a b c})^{\alpha \dot{\alpha}} & = - \frac{i}{5!}  \varepsilon^{a b c}_{\;\;\;\;\; d e f g h} (\gamma^{d e f g h})^{\alpha \dot{\alpha}}\\
(\gamma^{a b c d})^{\alpha \beta} & = - \frac{i}{4!}  \varepsilon^{a b c d}_{\;\;\;\; \;\;\;\; e f g h} (\gamma^{e f g h})^{\alpha \beta}\\
(\gamma^{a b c d e f g h})^{\alpha \beta} & = - i  \varepsilon^{a b c d e f g h} C^{\alpha \beta}  \\
C^{\dot \alpha \dot \beta} & = \delta^{ \dot \alpha\dot \beta} \\
(\gamma^{a b})^{\dot \alpha \dot \beta} & = - (\gamma^{a b})^{\dot \beta \dot \alpha}\\
(\gamma^{a b c d})^{\dot \alpha \dot \beta} & = (\gamma^{a b c d})^{\dot \beta \dot \alpha}
\end{split}
\end{equation}
\section{Dimension 1 solution to the superspace Bianchi identities} \label{appendix1Dim}
In this appendix we give the dimension 1 Bianchi identities of $\cN=2$ supergravity in eight dimensions and solve them. The result are ordered in function of the $U(1)$ weight. 
\subsection{Dimension 1 Bianchi identities}
The components of $d _{\omega} T = R$  of dimension 1 and $U(1)$ weight $4, 3$ and $0$ are
\bea
3 D_{(\alpha}^{\;i} T_{\beta \gamma)}^{j k\;\;\dot\delta l} + 3 T_{(\alpha \beta}^{\;i j} {}^{\varepsilon}_{m} T_{ \varepsilon \gamma)}^{m k\;\dot\delta l} + 3 T_{(\alpha \beta}^{\;i j\; \bdt m} T_{\bdt m \gamma)}^{\;\;\;\;k\;\dot\delta l} & =& 0\CR
2 T_{c (\alpha}^{\;\;\;i\;\; \dot{\alpha}k} T_{\dot{\alpha}k \beta)}^{\;\;\;\;j\;\; d} & =& R_{\alpha \beta c}^{ij\;\;\;\;\;d} \CR
3 D_{(\alpha}^{\;i} T_{\beta \gamma)\;l}^{j k\;\;\delta} + 3 T_{(\alpha \beta\;m}^{\;i j\; \varepsilon} T_{ \varepsilon \gamma)\;l}^{m k\;\delta} + 3 T_{(\alpha \beta}^{\;i j\; \bdt m} T_{\bdt m \gamma)\;l}^{\;\;\;\;k\;\delta} & = &\frac{3}{4} R_{(\alpha \beta c d}^{\;i j} 
(\gamma^{c d})_{\gamma)}^{\;\;\;\delta} \delta_{l}^k + 3 R_{(\alpha \beta \;\;\; l}^{\;i j\;\; k} \delta_{\gamma)}^{\delta}\CR
 \bar D_{\dot \alpha i} T_{\beta \gamma}^{jk\;\dot \delta l} + 2 D_{(\beta}^{\;j} T_{\gamma) \dot \alpha i}^{k \;\;\;\;\;\dot \delta l} + T_{\beta \gamma}^{jk\;\bdt m} T_{\bdt m \dot \alpha i}^{\;\;\;\;\;\;\;\;\dot \delta l} +  T_{\beta \gamma\;m}^{jk\; \varepsilon} T_{\; \varepsilon\; \dot \alpha i}^{m\;\;\;\;\dot \delta l}  \hspace{0mm} && \CR+ 2 T_{\dot \alpha i (\beta}^{\;\;\;\;\;j\;\bdt m} T_{\dot \varepsilon m\gamma)}^{\;\;\;\;k\;\dot \delta l}  + 2 T_{\dot \alpha i (\beta \;m}^{\;\;\;\;\;j\; \varepsilon} T_{\; \varepsilon\;\gamma)}^{mk\;\;\dot \delta l} 
+ T_{\dot \alpha i (\beta}^{\;\;\;\;\;j\;e} T_{e\;\gamma)}^{\;\;k\;\;\dot \delta l} &=&  \frac{1}{4} R_{\beta \gamma\;a b}^{jk} (\gamma^{ab})_{\dot \alpha}^{\;\;\dot \delta}\delta_{i}^{l} + R_{\beta \gamma i}^{jk\;\;\;l} \delta_{\dot \alpha}^{\dot \delta} \CR
T_{c \alpha\;k}^{\;\;i\; \gamma} T_{\gamma \dot{\beta} j}^{k\;\;\;\;d} + T_{c \dot{\beta} j}^{\;\;\;\;\; \dot{\gamma} k} T_{\dot{\gamma} k \alpha}^{\;\;\;\;i\;d} &=& R_{\alpha \dot{\beta}j c}^{i\;\;\;\;\;\;\;d} 
\eea
\vspace{-9mm}
\begin{multline}
 2 D_{(\alpha}^{\;i} T_{\beta) \dot \gamma k\;l}^{j\;\;\;\;\;\;\delta} + \bar D_{\dot \gamma k} T_{\alpha \beta \;l}^{i j\;\;\delta} +
T_{\alpha \beta\;m}^{\;i j\; \varepsilon} T_{\;  \varepsilon \;\dot \gamma k\;l}^{m \;\;\;\;\delta}+ T_{\alpha \beta}^{\;i j\;\bdt m} T_{\bdt m \dot \gamma k\;l}^{\;\;\;\;\;\;\;\;\;\delta} \\ + 
2 T_{\dot \gamma k (\alpha\;m}^{\;\;\;\;\; i \;\; \varepsilon} T_{\;  \varepsilon \;\beta) \;l}^{m j\;\;\delta} 
 + 2 T_{\dot \gamma k (\alpha}^{\;\;\;\;\; i \;\;\bdt m} T_{\bdt m \beta) \;l}^{\;\;\;\;j\;\;\delta} + 2 T_{\dot \gamma k (\alpha}^{\;\;\;\;\; i \;\;e} T_{e \beta) \;l}^{\;\;j\;\;\delta} \\
  =
\frac{1}{2} R_{\dot \gamma k (\alpha \;c d}^{\;\;\;\;\;i} (\gamma^{cd})_{\beta)}^{\;\;\;\delta} \delta_l^j +
2 R_{\dot \gamma k (\alpha}^{\;\;\;\;\;i} \delta_{\beta)}^{\delta} \delta_l^j + 2 R_{\dot \gamma k (\alpha\;\;l}^{\;\;\;\;\;i\;j} \delta_{\beta)}^{\delta}
\end{multline} 
The Bianchi identity for the 2-form field strength  $\bar F$ decomposes in components of $U(1)$ weight $4, 2$ and $0$ as
\bea
2 D_{(\alpha}^i \bar F_{\beta) a}^{j\;mn} + T_{\alpha \beta\;l}^{ij\; \;  \varepsilon}\bar F_{ \varepsilon a}^{l\;mn}+ T_{\alpha \beta}^{ij\; \; \dot \varepsilon l}\bar F_{\dot  \varepsilon l a}^{mn} &=& 2 P_{(\alpha}^{\;i\;mnpq} \bar F_{\beta) a\;pq}^{j} + \bar P_{a} F_{\alpha \beta}^{ij\;mn} \CR
D_{\alpha}^i \bar F_{\dot \beta j a}^{mn} + \bar D_{\dot \beta j} \bar F_{\alpha a}^{i\;mn} + T_{\alpha \dot \beta j\;l}^{i\;\;\;\;\;  \varepsilon}\bar F_{ \varepsilon a}^{l\;mn}  && \hspace{-4mm}+ \ T_{\alpha \dot \beta j}^{i\;\;\;\;\; \dot \varepsilon l}\bar F_{\dot \varepsilon l a}^{mn} +T_{\alpha \dot \beta j}^{i\;\;\;\;\; e}\bar F_{e a}^{mn} + T_{a \alpha}^{\;\;i\; \dot  \varepsilon l}\bar F_{\dot  \varepsilon l \dot \beta j}^{mn}\CR
 &=& P_{\alpha}^{i\;mnpq} \bar F_{\dot \beta j a\;pq} + P_{\dot \beta j}^{mnpq} \bar F_{\alpha a\;pq}^i +\bar P_{\dot \beta j} F_{\alpha a}^{i\;mn} \CR
2 \bar D_{(\dot \alpha i} \bar F_{\dot \beta j) a}^{mn} + T_{\dot \alpha i \dot \beta j\;l}^{\;\;\;\;\;\;\;\;  \varepsilon}\bar F_{ \varepsilon a}^{l\;mn}&&  \hspace{-4mm} + \  T_{\dot \alpha i \dot \beta j}^{\;\;\;\;\;\;\; \dot \varepsilon l}\bar F_{\dot \varepsilon l a}^{mn}  + 2 T_{a ( \dot \alpha i}^{\;\;\;\;\;\;\; \dot  \varepsilon l}\bar F_{\dot  \varepsilon l \dot \beta j)}^{mn}  \CR
&=& 2 P_{(\dot \alpha i}^{mnpq} \bar F_{\dot \beta j) a pq} + P_{a}^{mnpq} \bar F_{\dot \alpha i \dot \beta j pq} + 2 \bar P_{(\dot \alpha i} F_{\dot \beta j) a}^{mn} \hspace{5mm} \eea
The Bianchi identity for the 3-form field strength decomposes in components of $U(1)$ weight $2$ and $0$ as
\begin{multline} 
2 D_{(\alpha}^{\;i} H_{\beta) a b}^{j\;mn} + 4 T_{a] (\alpha}^{\;\;\;\;i\;\dot  \varepsilon l} H_{\dot  \varepsilon l \beta) [b}^{\;\;\;j\;mn} + T_{\alpha \beta\;l}^{ij\;\; \varepsilon} H_{ \varepsilon a b}^{l\;mn}   + T_{\alpha \beta}^{ij\;\dot  \varepsilon l} H_{\dot  \varepsilon l a b}^{mn}  \\
= - 2 P_{(\alpha}^{\;i\;mnpq} H_{\beta) a b\;pq}^{j}  + F_{\alpha \beta}^{ij\;p(m} \bar F_{a b\;p}^{\;n)} + 4 F_{a] (\alpha}^{\;\;\;\;i\;p(m} \bar F_{\beta) [b\;\;p}^{j\;\;\;n)} 
\end{multline}
\begin{multline}
D_{\alpha}^{i} H_{\dot \beta j a b}^{mn} + \bar D_{\dot \beta j} H_{\alpha j a b}^{i\;mn} + T_{\alpha \dot \beta j\;l}^{i\;\;\;\;\; \varepsilon} H_{ \varepsilon a b}^{l\;mn}  + T_{\alpha \dot \beta j}^{i\;\;\;\;\dot  \varepsilon l} H_{\dot  \varepsilon l a b}^{mn}  + T_{\alpha \dot \beta j}^{i\;\;\;\;\dot  \varepsilon l} H_{\dot  \varepsilon l a b}^{mn} + T_{\alpha \dot \beta j}^{i\;\;\;\;e} H_{e a b}^{mn}   \\
 + 2 T_{a] \alpha\;l}^{\;\;\;i\;  \varepsilon} H_{ \varepsilon \dot \beta j [b}^{l \;mn}  + 2 T_{a] \dot \beta j}^{\;\;\;\;\;\;\dot  \varepsilon l} H_{\dot  \varepsilon l  \alpha [b}^{\;\;\;i \;mn} \\= - P_{\alpha}^{i\;mnpq} H_{\dot \beta j a b\;pq}  - P_{\dot \beta j}^{mnpq} H_{\alpha a b\;pq}^i - 2 P_{[a}^{mnpq} H_{\alpha \dot \beta j b]\;pq}^i   
+ 2 F_{[a \alpha}^{\;\;\;i\;p(m} \bar F_{\dot \beta j b]\;\;p}^{\;\;\;\;\;n)} + 2 F_{[a \dot \beta j }^{p(m} \bar F_{\alpha b]\;\;p}^{i\;\;\;n)} 
\end{multline}
The Bianchi identity for the 4-form field strength $\bar G$ decomposes in components of $U(1)$ weight $4, 2$ and $0$ as follows 
\bea
2 D_{(\alpha}^{\;i} \bar G_{\beta) a b c}^{j} + T_{\alpha \beta\;l}^{ij\;\; \varepsilon}  \bar G_{ \varepsilon a b c}^{l} &=& 3 \bar P_{[a} G_{\alpha \beta b c]}^{ij} + 6 H_{a b] (\alpha}^{\;\;\;\;\;\;i\;pq} \bar F_{\beta) [c\;pq}^{j} \CR
\bar D_{\dot \beta j} \bar G_{\alpha a b c}^{i} + T_{\alpha \dot \beta j\;l}^{i\;\;\;\;\; \varepsilon}  \bar G_{ \varepsilon a b c}^{l}  && \hspace{-4mm}+ \   T_{\alpha \dot \beta j}^{i\;\;\;\;e}  \bar G_{e a b c}   +3 T_{[a \alpha}^{\;\;i\;\dot  \varepsilon l}  \bar G_{\dot  \varepsilon l \dot \beta j b c]} \CR
&=&  3 H_{\dot \beta j [a b}^{pq} \bar F_{\alpha c]\;pq}^{i}  + 3 H_{\alpha [a b}^{i\;pq} \bar F_{\dot \beta j c]\;pq} + 3 H_{\alpha \dot \beta j [a}^{i\;pq} \bar F_{bc]\;pq} \CR
T_{\dot \alpha i \dot \beta j\;l}^{\;\;\;\;\;\;\;\; \varepsilon}  \bar G_{ \varepsilon a b c}^{l} + 6 T_{a] (\dot \alpha i}^{\;\;\;\;\;\;\;\dot  \varepsilon l}  \bar G_{\dot  \varepsilon l \dot \beta j) [b c} &=& 2 \bar P_{(\dot \alpha i} G_{\dot \beta j) a b c}  + H_{a b c}^{pq} \bar F_{\dot \alpha i \dot \beta j\;pq}  + 6 H_{a b] (\dot \alpha i }^{pq} \bar F_{\dot \beta j) [c\;pq}
\eea
The dimension 1 components of $d_{\omega} \bar P = 0$  of respective $U(1)$ weight $6,4$ and $2$ read 
\bea
T_{\alpha \beta}^{i j\;\dot  \varepsilon l } \bar P_{\dot  \varepsilon l} = 0 \CR
D_{\alpha}^i \bar P_{\dot \beta j} + T_{\alpha \dot \beta j}^{i\;\;\;\;e} \bar P_{e} + T_{\alpha \dot \beta j}^{i\;\;\;\;\dot  \varepsilon l} \bar P_{\dot  \varepsilon l} = 0 \CR
2 \bar D_{(\dot \alpha i} \bar P_{\dot \beta j)} + T_{\dot \alpha i \dot \beta j}^{i\;\;\;\;\;\;\dot  \varepsilon l} \bar P_{\dot  \varepsilon l} = 0
\eea
And similarly the components of $d_{\omega} P^{ijkl} = 0$ of dimension 1 and respective $U(1)$ weight $2$ and $ 0$ are 
\bea
2 D_{(\alpha}^{\;i} P_{\beta)}^{j\;pqrs} +  T_{\alpha \beta}^{ij\;\bdt l} P_{\bdt l}^{pqrs} +  T_{\alpha \beta\;l}^{ij\;\; \varepsilon} P_{ \varepsilon}^{l\;pqrs} &=& 0 \CR
D_{\alpha}^{i} P_{\dot \beta j}^{pqrs} + \bar D_{\dot \beta j} P_{\alpha}^{i\;pqrs} + T_{\alpha \dot\beta j}^{i\;\;\;\;\bdt l} P_{\bdt l}^{pqrs} +  T_{\alpha \dot \beta j\;l}^{i\;\;\;\;\; \varepsilon} P_{ \varepsilon}^{l\;pqrs} +  T_{\alpha \dot \beta j}^{i\;\;\;\;\;e} P_{e}^{pqrs} &=& 0
\eea

\subsection{Dimension 1 solution} \label{sec:dim1_solutions}
The only component of $U(1)$ weight $4$ is the covariant derivative of the fermion field $\bar \chi$ 
\begin{equation} \label{eq:DChiBar}
 D_{\alpha}^i \bar \chi_{\dot \beta}^j =  \frac{1}{2} \left(\gamma^a\right)_{\alpha \dot \beta} \scal{ i  \varepsilon^{i j}  \bar P_a +  (\bar \chi_k \gamma_a \lambda^{ijk} )}  + \frac{3}{8}  \lambda_\alpha^{ijk} \bar \chi_{\bdt\, k}\  .  \end{equation}
From weight $2$ and above, there are more components, and for convenience we will define the following basis of bilinear in the fermions in irreducible representations of $SU(2)\times Spin(1,7)$
\begin{equation} \label{eq:D1U2Basis}
\begin{split}
\left(\lambda \lambda\right) &\equiv \lambda^{i j k} \lambda_{i j k} \ , \\
\left(\lambda \lambda\right)^{ijkl} & \equiv \lambda^{m (ij} \lambda^{kl)}_{\;\;\;\;m} \ , \\
\left(\bar \chi \bar \lambda\right)_{ab}^{ij} & \equiv \bar \chi^{k} \gamma_{a b} \bar \lambda_{k}^{ij} \ , \\
 \left(\bar \chi \bar \lambda\right)^{ijkl}_{ab} & \equiv \bar \chi^{(i} \gamma_{ab}\bar \lambda^{jkl)}\ , 
\end{split} \hspace{10mm}
\begin{split}
\left(\lambda \lambda\right)_{abcd} & \equiv \lambda^{i j k}\gamma_{abcd} \lambda_{i j k}\ ,  \\
\left(\lambda \lambda\right)_{abcd}^{ijkl} & \equiv \lambda^{m (ij} \gamma_{abcd} \lambda^{kl)}_{\;\;\;\;m}\ ,  \\
\left(\bar \chi \bar \lambda\right)_{abcd}^{ij} & \equiv \bar \chi^{k} \gamma_{a b c d} \bar \lambda_{k}^{ij} \ , 
\end{split} \hspace{10mm}
\begin{split}
\left(\lambda \lambda\right)_{ab}^{ij} & \equiv \lambda^{i k l}\gamma_{ab} \lambda^{j}_{\;\;k l} \ , \\
\left(\bar \chi \bar \lambda\right)^{ij} & \equiv \bar \chi^{k} \bar \lambda_{k}^{ij}\ ,  \\
\left(\bar \chi \bar \lambda\right)^{ijkl} & \equiv \bar \chi^{(i} \bar \lambda^{jkl)}\ . 
\end{split}
\end{equation}
The corresponding torsion component is
\begin{multline} \label{eqn:D1Torsion}
T_{a \alpha}^{\;\;i\;\;\dot\beta j} =  (\gamma^{bcd})_{\alpha}^{\;\;\dot \beta}  \varepsilon^{ij} \left(\frac{i}{24} \bar G^{-}_{a b c d} - \frac{i}{576}  \left(\lambda \lambda\right)_{abcd} \right) \\
 + (\gamma_{a}^{\;\;bc})_{\alpha}^{\;\;\dot \beta}  \left( \frac{i}{24}  \bar{F}_{cd}^{ij} + \frac{i}{48}   \left(\bar \chi \bar \lambda\right)_{cd}^{ij} + \frac{i}{12} \left(\lambda \lambda\right)_{cd}^{ij}\right) \\
 + (\gamma^{b})_{\alpha}^{\;\;\dot \beta}  \left(\frac{5i}{12}  \bar{F}_{ab}^{ij} + \frac{i}{12} \left(\bar \chi \bar \lambda\right)_{ab}^{ij} + \frac{i}{3} \left(\lambda \lambda\right)_{ab}^{ij}\right) \ . 
\end{multline}
The $(0,2,0)$ Riemann curvature component decomposes into the $\so(1,7)$ part
\begin{multline}
 R_{\alpha}^i{}_{\beta}^{j}{}_c{}^d = C_{\alpha \beta} \left(\frac{5}{6}\bar F_{c}^{\;\;d\;ij} + \frac{2}{3}\left(\bar \chi \bar \lambda\right)_{c}^{\;\;d\;ij}  + \frac{1}{6} \left(\lambda \lambda\right)_{c}^{\;\;d\;ij} \right) \\
 +  (\gamma_{c}^{\;\;d a b})_{\alpha \beta} \left(\frac{1}{12}\bar F_{ab}^{ij} + \frac{1}{6}\left(\bar \chi \bar \lambda\right)_{ab}^{ij}  + \frac{1}{24} \left(\lambda \lambda\right)_{ab}^{ij} \right)  
 \\
+  (\gamma^{a b})_{\alpha \beta}  \varepsilon^{ij} \left(\frac{1}{4} \bar G^{-}_{abc}{}^{d} - \frac{1}{96} \left(\lambda \lambda\right)_{abc}{}^{d} \right) \ , 
\end{multline}
the $\su(2)$ part
\begin{equation}
  R_{\alpha \beta\;\;\;l}^{i \;j \;k} = P_{\alpha}^{i\;k m n p} P_{\beta\;l m n p}^{j} - \frac{1}{2} \delta^k_l P_{\alpha}^{i\;m n p q} P_{\beta\;m n p q}^{j} \ , \qquad P_{\alpha}^{i\;jklm} = -  \varepsilon^{i (j} \lambda_{\alpha}^{klm)} \ ,
\end{equation}
and the $\mathfrak{u}(1)$ part that vanishes. The covariant derivative of the fermion fields of $U(1)$ weight $2$ are 
\begin{multline}
 D^{i}_{\alpha} \lambda^{jkl}_{\beta} = 
C_{\alpha \beta} \left(- \left(\bar \chi \bar \lambda\right)^{ijkl} -\frac{15}{32} \left(\lambda \lambda\right)^{ijkl} + \frac{3}{4} \varepsilon^{i ( j} \left(\bar \chi \bar \lambda\right)^{k l)} \right) +  \frac{1}{1536} (\gamma^{a b c d})_{\alpha \beta} \left(\lambda \lambda\right)_{abcd}^{ijkl}\\
+  (\gamma^{ab})_{\alpha \beta} \left(- \frac{1}{4}  \varepsilon^{i (j}  \bar{F}^{kl)}_{ab} - \frac{1}{128} 
  \varepsilon^{i (j} \left(\lambda \lambda \right)^{kl)}_{ab} + \frac{1}{16}  \varepsilon^{i (j} \left(\bar \chi \bar \lambda \right)^{kl)}_{ab} + \frac{1}{4} \left(\bar \chi \bar \lambda\right)^{ijkl}_{ab} \right)\ , 
\end{multline}
\begin{multline} \label{eq:DBarChiBar}
 \bar D_{\dot \alpha i} \bar \chi_{\dot \beta}^j = C_{\dot \alpha \dot \beta}\left(- \frac{3}{32} \delta_i^j \left(\lambda \lambda\right) - \frac{15}{32}  \left(\bar \chi \bar \lambda\right)_{i}^{\;\;j} \right)+ (\gamma^{abcd})_{\dot \alpha \dot \beta}\left(\frac{1}{192} \delta_i^j \bar G^{-}_{a b c d} + \frac{1}{1536} \left(\bar \chi \bar \lambda \right)_{abcd\;i}^{\;\;\;\;\;\;\;\;\;j} \right) \\
+ (\gamma^{ab})_{\dot \alpha \dot \beta}\left(-\frac{1}{8} \bar F^{\;\;\;\;\;\;j}_{ab\;i} - \frac{1}{64} \left(\bar \chi \bar \lambda \right)^{\;\;\;\;\;\;j}_{ab\;i} + \frac{1}{32} \left(\lambda \lambda \right)^{\;\;\;\;\;\;j}_{ab\;i}\right)\ . 
\end{multline}
In our notations, the field $\bar F_{ab}$ and $H_{abc}$ coincide with the corresponding  (respectively $(2,0,0)$ and $(3,0,0)$) components of their associated superforms, whereas the $(4,0,0)$ component of the 4-form superform decomposes into a complex selfdual part  $\bar G_{abcd}$ and a complex antiselfdual part bilinear in the fermions, \ie 
\begin{equation}
\begin{split}
\bar{G}_{abcd} = \bar{G}_{abcd}^{-} - \frac{1}{8} \left(\lambda^{ijk} \gamma_{abcd} \lambda_{ijk}\right)\ . 
\end{split}
\end{equation}
We now consider the $U(1)$ invariant components, with the following basis of bilinear in the fermions in irreducible representations of $SU(2)\times Spin(1,7)$
\begin{equation}
\begin{split}
\left(\lambda \bar \lambda\right)^{ijkl}_{abc} &\equiv \lambda^{m (i j} \gamma_{abc} \bar{\lambda}^{kl)}_{\;\;\;\;m} \ , \\
\left(\lambda \bar \lambda\right)^{ij}_{a} &\equiv \lambda^{k l (i} \gamma_{a} \bar{\lambda}^{j)}_{\;\;kl} \ , \\
\left(\chi \bar{\chi}\right)^{ij}_{abc} &\equiv \chi^{(i} \gamma_{abc} \bar{\chi}^{j)} \ , \\
\left(\chi \bar{\chi}\right)_{a} & \equiv \chi^{i} \gamma_{a} \bar{\chi}_{i} \ , 
\end{split}\hspace{10 mm}
\begin{split}
\left(\lambda \bar \lambda\right)^{ijkl}_{a} & \equiv \lambda^{m (i j} \gamma_{a} \bar{\lambda}^{kl)}_{\;\;\;\;m} \ , \\
\left(\lambda \bar \lambda\right)_{abc} & \equiv \lambda^{i j k} \gamma_{abc} \bar{\lambda}_{ijk} \ , \\
\left(\chi \bar{\chi}\right)^{ij}_{a} & \equiv \chi^{(i} \gamma_{a} \bar{\chi}^{j)}  \ , \end{split}\hspace{10 mm}
\begin{split}
\left(\lambda \bar \lambda\right)^{ij}_{abc} & \equiv \lambda^{k l (i} \gamma_{abc} \bar{\lambda}^{j)}_{\;\;kl}\ ,  \\
\left(\lambda \bar \lambda\right)_{a} & \equiv \lambda^{i j k} \gamma_{a} \bar{\lambda}_{ijk}\ , \\
\left(\chi \bar{\chi}\right)_{abc} & \equiv \chi^{i} \gamma_{abc} \bar{\chi}_{i}\ . 
\end{split}
\end{equation}
The corresponding component of the torsion is\begin{multline}
T_{a \alpha\;j}^{\;i\;\;\beta} = (\gamma^{bc})_{\alpha}^{\;\;\beta} \delta^i_j \left(- \frac{i}{64} \left(\lambda \bar{\lambda}\right)_{abc} + \frac{i}{16}\left(\chi \bar{\chi}\right)_{abc}\right) + (\gamma_{a}^{\;\;bcd})_{\alpha}^{\;\;\beta}  \delta^i_j  \left(- \frac{i}{192}\left(\lambda \bar{\lambda}\right)_{bcd} + \frac{i}{48} \left(\chi \bar{\chi}\right)_{bcd}\right) \\
+ (\gamma_a^{\;\;b})_{\alpha}^{\;\;\beta} \left( - \frac{i}{16}  \left(\lambda \bar{\lambda}\right)_{b\;j}^{i} +  \frac{i}{12} \left(\chi \bar{\chi}\right)_{b\;j}^{i}\right) + (\gamma_a^{\;\;b})_{\alpha}^{\;\;\beta} \delta_j^i \left( - \frac{i}{96} \left(\lambda \bar{\lambda}\right)_{b} + \frac{i}{8} \left(\chi \bar{\chi}\right)_{b}\right)\\
 \hspace{10mm} + \delta_{\alpha}^{\beta} \left( - \frac{5i}{16} \left(\lambda \bar{\lambda}\right)_{a\;j}^{i} + \frac{5i}{12} \left(\chi \bar{\chi}\right)_{a\;j}^{i}\right) + \delta_{\alpha}^{\beta} \delta_j^i \left( - \frac{5i}{96}  \left(\lambda \bar{\lambda}\right)_{a} - \frac{5i}{8} \left(\chi \bar{\chi}\right)_{a}\right) \\ 
\hspace{30mm} + (\gamma_{a}^{\;\;bcd})_{\alpha}^{\;\;\beta}\left(- \frac{1}{36} H_{bcd\;j}^{i} +  \frac{i}{96}\left(\lambda \bar{\lambda}\right)^{i}_{bcd\;j} +  \frac{i}{24}\left(\chi \bar{\chi}\right)^{i}_{bcd\;j}\right) \\
+ (\gamma^{bc})_{\alpha}^{\;\;\beta}\left(- \frac{1}{6} H_{a b c\;j}^{i} + \frac{i}{32}\left(\lambda \bar{\lambda}\right)^{i}_{abc\;j} + \frac{1}{8} \left(\chi \bar{\chi}\right)^{i}_{abc\;j}\right) \ ,
 \end{multline}
and the $(0,1,1)$ component of the Riemman curvature in $\so(1,7)$ is
\begin{multline}
R_{\alpha \dot{\beta} j c}^{i\;\;\;\;\;\;\;\;d} =  (\gamma^a)_{\alpha \dot \beta} \delta_j^i \left(\frac{1}{16}\left(\lambda \bar{\lambda}\right)_{a c}^{\;\;\;\;d} - \frac{1}{4} \left(\chi \bar{\chi}\right)_{a c}^{\;\;\;\;d} \right) + 
 (\gamma^{a\;\;d}_{\;\;c})_{\alpha \dot \beta} \left(\frac{1}{8} \left(\lambda \bar{\lambda}\right)_{a\;j}^{i}- \frac{1}{6}  \left(\chi \bar{\chi}\right)_{a\;j}^{i}\right) \\
 + (\gamma^{a b e})_{\alpha \dot \beta}  \varepsilon_{a b e}{}^{f g h}{}_{c}^{\;\;d} \left( - \frac{1}{108}H_{f g h\;j}^{i} +\frac{i}{288} \left(\lambda \bar{\lambda}\right)_{f g h\;j}^{i} + \frac{i}{72}  \left(\chi \bar{\chi}\right)_{f g h\;j}^{i} \right) \\
 + (\gamma^{a b e})_{\alpha \dot \beta} \varepsilon_{a b e}{}^{f g h}{}_{c}^{\;\;d}  \delta_j^i \left(- \frac{i}{576} \left(\lambda \bar{\lambda}\right)_{f g h} + \frac{i}{144} \left(\chi \bar{\chi}\right)_{f g h} \right) \\
 \hspace{30 mm}+ (\gamma^a)_{\alpha \dot \beta} \left(- \frac{2 i}{3}H_{a c\;\;\;\;\;j}^{\;\;\;\;d\;i} - \frac{1}{8} \left(\lambda \bar{\lambda}\right)_{a c\;\;\;\;\;j}^{\;\;\;\;d\;i} - \frac{1}{2} \left(\chi \bar{\chi}\right)_{a c\;\;\;\;\;j}^{\;\;\;\;d\;i} \right) \\
  + (\gamma^{a\;\;d}_{\;\;c})_{\alpha \dot \beta} \delta_j^i \left( \frac{1}{48}\left(\lambda \bar{\lambda}\right)_{a} + \frac{1}{4}\left(\chi \bar{\chi}\right)_{a}\right) \ . 
\end{multline}
whereas its component in $\mathfrak{u}(1)$ and $\su(2)$ are
\bea
R_{\alpha \dot \beta j}^{i} & = &- 2 \chi_\alpha^i \bar \chi_{\bdt j}  \ ,   \CR
R_{\alpha \dot \beta j\;\;\;\;l}^{i\;\;\;\;\;\;k} & =& P_{\alpha}^{i\;kmnp} P_{\dot \beta j\;lmnp} - \frac{1}{2} \delta^k_l P_{\alpha}^{i\;mnpq} P_{\dot \beta j\;mnpq} \ , \qquad P_{\alpha}^{i\;jklm} = -  \varepsilon^{i (j} \lambda_{\alpha}^{klm)}\ . 
\eea
The covariant derivative of the fermion $\lambda$ is
\begin{multline}
\bar D_{i \dot \alpha} \lambda_{\alpha}^{jkl} = (\gamma^a)_{\dot \alpha \alpha} \left(i P_{ai}^{\;\;\;jkl} - \frac{13}{32} \left(\lambda \bar \lambda\right)_{ai}^{\;\;\;jkl} + \frac{3}{64} \delta_{i}^{(j} \left(\lambda \bar \lambda\right)_{a}^{kl)} + \delta_{i}^{(j} \left(\chi \bar \chi\right)_{a}^{kl)}\right) \\
+ (\gamma^{abc})_{\dot \alpha \alpha} \left( - \frac{1}{64} \left(\lambda \bar \lambda\right)_{abci}^{\;\;\;\;\;\;\;jkl} - \frac{i}{12} \delta_i^{(j} H^{kl)}_{abc} - \frac{1}{128} \delta_i^{(j} \left(\lambda \bar \lambda\right)_{abc}^{kl} \right)\ . 
\end{multline}

\section{Dimension 3/2 solution to the superspace Bianchi identities} \label{appendix3/2Dim}
In the core of the paper we use the dimension 1/2 covariant derivative of the dimension 1 fields and the equation of motion of the fermion field $\bar \chi$, which we derive from the dimension 3/2 Bianchi identities and the algebra of the covariant derivatives in this appendix. We do not derive the expression of the dimension $3/2$ Riemann curvature that we do not need in this paper. 
\subsection{Dimension 3/2 Bianchi identities}\label{sec:dim32Id}
The components of dimension 3/2 of  $d_{\omega} \bar P = 0$  of respective $U(1)$ weight $5$ and $3$ are 
\bea \label{eq:DPBar}
D_{\alpha}^{i} \bar{P}_{a} + T_{\alpha a}^{i\;\;\bdt l} \bar{P}_{\bdt l} &=& 0 \ , \CR
\bar D_{\dot \alpha i} \bar{P}_{a} - D_a \bar{P}_{\dot \alpha i} + T_{\dot \alpha i a}^{\;\;\;\;\;\bdt l} \bar{P}_{\bdt l} &=& 0 \ ,
\eea
whereas the dimension 3/2 component of $d_{\omega} P^{ijkl} = 0$ is
\be
D_{\alpha}^{i} \bar{P}^{jklm}_{a} - D_a \bar{P}^{i\;jklm}_{\alpha} +  T_{\alpha a}^{i\;\;\bdt p} P_{\bdt p}^{jklm}  +  T_{\alpha a\;p}^{i\;\;\; \beta} P_{ \beta}^{p\;jklm}  = 0 \ . 
\ee
The Bianchi identity for the 2-form field strength $\bar F$ gives at this dimension  the following equations of $U(1)$ weight $3$ and $1$
\bea
 D_{\alpha}^{i} \bar{F}_{ab}^{jk} + 2 D_{[a} \bar{F}_{b]\alpha}^{\;\;\;i\;jk} + 2 T_{\alpha [a \;l}^{i\;\;\;\; \gamma} \bar{F}_{ \gamma b]}^{l\;jk} && \hspace{-4mm}+\  2 T_{\alpha [a }^{i\;\;\;\bdt l} \bar{F}_{\bdt l b]}^{jk} \CR
 & = &  2 P_{[a}^{jklm} \bar{F}_{b]\alpha\;lm}^{\;\;\;i}  +  P_{\alpha}^{i\;jklm} \bar{F}_{ab\;lm} + 2 \bar{P}_{[a} F^{\;\;i\;jk}_{b] \alpha} \CR
\bar D_{\dot \alpha i}  \bar{F}_{ab}^{jk}  + 2 D_{[a} \bar{F}_{b]\dot \alpha i}^{jk}  + 2 T_{\dot \alpha i [a \;l}^{\;\;\;\;\;\;\; \gamma} \bar{F}_{ \gamma b]}^{l\;jk} && \hspace{-4mm}+\   2 T_{\dot \alpha i [a }^{\;\;\;\;\;\;\bdt l} \bar{F}_{\bdt l b]}^{jk} + T_{a b}^{\;\;\bdt l} \bar{F}_{\bdt l \dot \alpha i}^{jk} \CR
&=& 2 \bar{P}_{[a} F^{jk}_{b] \dot \alpha i} + \bar{P}_{\dot \alpha i} F^{jk}_{a b} + 2 P_{[a}^{jklm} \bar{F}_{b]\dot \alpha i\;lm} + P_{\dot \alpha i}^{jklm} \bar{F}_{ab\;lm} \hspace{8mm} 
\eea
The Bianchi identity for the 4-form field strength $\bar G$ gives the following equations of respective  $U(1)$ weight $3$ and $1$
\bea
 D_{\alpha}^{i} \bar{G}_{abcd} + 4 D_{[a} \bar{G}_{bcd] \alpha}^{\;\;\;\;\;\;i} + 4 T_{\alpha [a\;l}^{i\;\;\;\; \gamma} \bar{G}_{ \gamma b c d]}^{l}  &=& 4 H_{[abc}^{jk} \bar F_{d] \alpha\;jk}^{\;\;i} + 6 H_{\alpha[ab}^{i\;jk} \bar F_{cd]\;jk} \CR
\bar D_{\dot \alpha i} \bar{G}_{abcd} + 4 T_{\dot \alpha i [a\;l}^{\;\;\;\;\;\;\; \gamma}  \bar{G}_{ \gamma bcd]}^l + 6 T_{[ab}^{\;\;\;\;\bdt l} \bar{G}_{\bdt l \dot \alpha i cd]} &=& \bar P_{\dot \alpha i} G_{a b c d} + 4 \bar P_{[a} G_{b c d] \dot \alpha i} \CR
&& + 4 H_{[ab c}^{jk} \bar F_{d] \dot \alpha i\;jk} + 6 H_{\dot \alpha i [a b}^{jk} \bar F_{cd]\;jk}
\eea
The Bianchi identity for the 3-form field strength gives the following equation of $U(1)$ weight 1
\begin{multline}
D_{\alpha}^i H_{a b c}^{j k} - 3 D_{[a} H_{b c] \alpha}^{\;\;\;\;i\;j k} + 3 T_{\alpha [a\;l}^{i\;\;\;\; \gamma} H_{ \gamma b c]}^{l\;jk}  + 3 T_{\alpha [a}^{i\;\;\;\bdt l} H_{\bdt l b c]}^{jk} + 3 T_{[a b}^{\;\;\;\;\bdt l} H_{\bdt l \alpha c]}^{\;\;\;i} \\
= - P_{\alpha}^{i\;jkpq} H_{a b c\;pq} - 3 P_{[a}^{jkpq} H_{b c]\alpha\;pq}^{\;\;\;\;i} - 6 F_{[a b}^{p(j} \bar F_{c] \alpha\;\;p}^{\;\;i\;k)} + 6 F_{\alpha [a}^{i\;p(j} \bar F_{b c]\;p}^{\;k)}
\end{multline}
We will also make use of the following commutation relations between the covariant derivatives acting of the fermions, ordered with respect to their $U(1)$ weight from 5 to 1
\bea
\left\{D_{\alpha}^{i}, D_{\beta}^{j}\right\} \bar \chi_{\dot \gamma}^k &=& - T_{\alpha \beta\;l}^{i j\;\; \varepsilon}  D_{ \varepsilon}^{l} \bar \chi_{\dot \gamma}^k - T_{\alpha \beta}^{i j\;\bdt l} \bar D_{\bdt l} \bar \chi_{\dot \gamma}^k 
- \frac{1}{4} R_{\alpha \beta c d}^{i j} (\gamma^{cd} )_{\dot \gamma}^{\;\;\dot \delta} \bar \chi_{\dot \delta}^k - R_{\alpha \beta\;\;\;l}^{i j\;\;k} \bar \chi_{\dot \delta}^l\nn \\  
\left\{D_{\alpha}^{i}, D_{\beta}^{j}\right\} \lambda_{\gamma}^{pqm} &=& - T_{\alpha \beta\;l}^{i j\;\; \varepsilon}  D_{ \varepsilon}^{l}  \lambda_{\gamma}^{pqm} - T_{\alpha \beta}^{i j\;\bdt l} \bar D_{\bdt l} \lambda_{\gamma}^{pqm}
- \frac{1}{4} R_{\alpha \beta c d}^{i j} (\gamma^{cd} )_{\gamma}^{\;\;\delta} \lambda_{\delta}^{pqm} - 3 R_{\alpha \beta\;\;\;\;l}^{i j\;\;(p} \bar \lambda_{\dot \delta}^{qm)l} \CR
\left\{D_{\alpha}^{i}, \bar D_{\dot \beta j}\right\} \bar \chi_{\dot \gamma}^k &=& - T_{\alpha \dot \beta j\;l}^{i \;\;\;\;\; \varepsilon} D_{ \varepsilon}^{l} \bar \chi_{\dot \gamma}^k - T_{\alpha \dot \beta j}^{i \;\;\;\;\ddt l} \bar D_{\ddt l} \bar \chi_{\dot \gamma}^k
- T_{\alpha \dot \beta j}^{i \;\;\;\;e} D_e \bar \chi_{\dot \gamma}^k  \CR
&& - \frac{1}{4} R_{\alpha \dot \beta j c d}^{i} (\gamma^{cd} )_{\dot \gamma}^{\;\;\dot \delta} \bar \chi_{\dot \delta}^k - R_{\alpha \dot \beta j\;\;\;l}^{i\;\;\;\;\;k} \bar \chi_{\dot \gamma}^l  - 3 R_{\alpha \dot \beta j}^{i} \bar \chi_{\dot \gamma}^k \CR
 \left\{\bar D_{\dot \alpha i},\bar D_{\dot \beta j}\right\} \bar \chi_{\dot \gamma}^k &=&  - T_{\dot \alpha i \dot  \beta j \;l}^{\;\;\;\;\;\;\; \varepsilon}  D_{ \varepsilon}^{l} \bar \chi_{\dot \gamma}^k - T_{\dot \alpha i \dot  \beta j}^{\;\;\;\;\;\;\;\ddt l} \bar D_{\ddt l} \bar \chi_{\dot \gamma}^k 
- \frac{1}{4} R_{\dot \alpha i \dot \beta  j c d} (\gamma^{cd} )_{\dot \gamma}^{\;\;\dot \delta} \bar \chi_{\dot \delta}^k - R_{\dot \alpha i \dot \beta  j\;\;\;l}^{\;\;\;\;\;\;\;\;k} \bar \chi_{\dot \delta}^l \CR
\left\{D_{\alpha}^{i}, \bar D_{\dot \beta j}\right\} \lambda_{\gamma}^{pqm} &=& - T_{\alpha \dot \beta j\;l}^{i \;\;\;\;\; \varepsilon} D_{ \varepsilon}^{l} \lambda_{\gamma}^{pqm} - T_{\alpha \dot \beta j}^{i \;\;\;\;\ddt l} \bar D_{\ddt l} \lambda_{\gamma}^{pqm}
- T_{\alpha \dot \beta j}^{i \;\;\;\;e} D_e \lambda_{\gamma}^{pqm} \CR
&& - \frac{1}{4} R_{\alpha \dot \beta j c d}^{i} (\gamma^{cd} )_{\gamma}^{\;\;\delta} \lambda_{\delta}^{pqm}  - 3 R_{\alpha \dot \beta j\;\;\;l}^{i\;\;\;\;\;(p} \lambda_{\gamma}^{qm)l} - R_{\alpha \dot \beta j}^{i} \lambda_{\gamma}^{pqm}\CR
 \left\{D_{\alpha}^{i}, D_{\beta}^{j}\right\} \bar \lambda_{\dot \gamma}^{pqm} &=& - T_{\alpha \beta\;l}^{i j\;\; \varepsilon}  D_{ \varepsilon}^{l} \bar \lambda_{\dot \gamma}^{pqm}  - T_{\alpha \beta}^{i j\;\ddt l} \bar D_{\ddt l} \bar \lambda_{\dot \gamma}^{pqm}  
- \frac{1}{4} R_{\alpha \beta c d}^{i j} (\gamma^{cd} )_{\dot \gamma}^{\;\;\dot \delta} \bar \lambda_{\dot \delta}^{pqm}  - 3 R_{\alpha \beta\;\;\;l}^{i j\;\;(p} \bar \lambda_{\dot \gamma}^{qm)l} \CR
\eea

\subsection{Dimension 3/2 solution}
The number of linearly independent dimension 3/2 monomials in the fields is rather large, and we find it convenient to define the following basis in irreducible representations of $SU(2)$, and filtrated with respect to $Spin(1,7)$ irreducible representations, such that the larger representations are not irreducible. It is indeed convenient to keep the gamma traces rather than to remove them systematically. The elements of $U(1)$ weight 5 are 
\begin{equation}
\begin{split}
 \bigl(\bar G \bar \chi\bigr)^{i}_{\dot{\alpha}} & \equiv \bar G^{-}_{a b c d} \bigl(\gamma^{abcd}\bar{\chi}^i \bigr)_{\dot \alpha} \ , \\
 \bigl(\bar{F}\bar{\chi}\bigr)^i_{\dot{\alpha}} & \equiv \bar{F}^{ij}_{ab} \bigl(\gamma^{ab} \bar{\chi}_j\bigr)_{\dot{\alpha}} \ , \\
 \bigl(\bar \chi \bar \chi \bar \lambda\bigr)^i_{\dot\alpha} & \equiv \bigl(\bar \chi \bar \chi\bigr)_{ab}^{jk} \bigl(\gamma^{ab}\bar\lambda^i_{jk}\bigr)_{\dot \alpha}\ ,  \\
 \bigl(\bar \chi \lambda \lambda \bigr)^i_{\dot \alpha} & \equiv \bigl(\gamma^{ab}\bar \chi^j\bigr)_{\dot{\alpha}}  \bigl(\lambda \lambda\bigr)^{i}_{ab\;j} \ , \\
 \bigl(\bar \chi \lambda \lambda \bigr)^i_{a \alpha} & \equiv \bigl(\gamma^{b}\bar \chi^j\bigr)_{\alpha}  \bigl(\lambda \lambda\bigr)^{i}_{ab\;j} \ , \end{split} \hspace{20 mm}
 \begin{split}
\bigl(\bar{F}\bar{\chi}\bigr)^i_{a \alpha} & \equiv \bar{F}^{ij}_{ab} \bigl(\gamma^{b} \bar{\chi}_j\bigr)_{\alpha}\ ,  \\
\bigl(\bar \chi \bar \chi \bar \lambda\bigr)^i_{a \alpha} & \equiv \bigl(\bar \chi \bar \chi\bigr)_{ab}^{jk} \bigl(\gamma^{b}\bar\lambda^i_{jk}\bigr)_{\dot \alpha}\ , \\
\bigl(\bar \chi \lambda \lambda \bigr)^i_{A\; \dot \alpha} & \equiv \bar \chi^i_{\dot \alpha} \bigl(\lambda \lambda\bigr) \ , \\
\bigl(\bar \chi \lambda \lambda \bigr)^i_{A\; a \alpha} & \equiv \bigl( \gamma^{bcd} \bar \chi^i_{\dot \alpha}\bigr)_{\alpha} \bigl(\lambda \lambda\bigr)_{abcd}\ . 
 \end{split}
\end{equation}
where we use the bilinear in the fermions defined in \eqref{eq:D1U2Basis}. Solving equation \eqref{eq:DPBar}  one gets
\begin{multline}
 D_{\alpha}^i \bar{P}_{a} =  (\gamma_{a})_{\alpha}^{\;\;\dot\alpha}\left(\frac{i}{12}\bigl(\bar{F}\bar{\chi}\bigr)^i_{\dot{\alpha}} +  \frac{i}{96} \bigl(\bar G \bar \chi\bigr)^{i}_{\dot{\alpha}} + \frac{7i}{48} \bigl(\bar \chi \bar \chi \bar \lambda\bigr)^i_{\dot\alpha} - \frac{i}{24} \bigl(\bar \chi \lambda \lambda \bigr)^i_{\dot \alpha} \right) \\ 
 +\frac{2 i}{3} \bigl(\bar{F}\bar{\chi}\bigr)^i_{a \alpha} - \frac{i}{3}  \bigl(\bar \chi \bar \chi \bar \lambda\bigr)^i_{a \alpha} - \frac{i}{12}  \bigl(\bar \chi \lambda \lambda \bigr)^i_{a \alpha} - \frac{i}{288} \bigl(\bar \chi \lambda \lambda \bigr)^i_{A\;a \alpha} 
\end{multline}
From $U(1)$ weight 3 and below the number of monomials increases considerably, and we shall display them in increasing order of the number of fields. At the linear level we have the covariant derivative of the fermion field $\bar \chi$, but because it satisfies the Dirac equation, we distinguish its irreducible component $ (D_{a} \bar{\chi}^i_{\dot \alpha} )' $ from the gamma trace that is equal to a sum of monomials in the fields. Here the prime states for the projection to the \DSOVIII1010 irreducible representation of $Spin(1,7)$. The list of bilinear in the fields is
\begin{equation}
\begin{split}
 \bigl(\bar P \bar \lambda \bigr)_{\alpha}^{ijk} & \equiv \bar P_a \bigl(\gamma^a \bar \lambda^{ijk}\bigr)_{\alpha} \ , \\
 \bigl(\bar F \lambda \bigr)^i_{\alpha} & \equiv \bar{F}^{jk}_{ab} \bigl(\gamma^{ab}\lambda^i_{jk}\bigr)_{\alpha} \ , \\
 \bigl(\bar F \lambda \bigr)^{ijk}_{\alpha} & \equiv \bar{F}^{l(i}_{ab} \bigl(\gamma^{ab}\lambda^{jk)}_{l}\bigr)_{\alpha} \ , \\
 \bigl(\bar G \lambda \bigr)^{ijk}_{a \dot \alpha} & \equiv \bar G^{-}_{abcd} \bigl(\gamma^{bcd} \lambda^{ijk} \bigr)_{\dot \alpha} \ , \\
 \bigl(P^4 \bar \chi \bigr)^{ijk}_{\alpha} & \equiv P_a^{ijkl} \bigl(\gamma^a\bar\chi_l\bigr)_{\alpha} \ , \\
 \bigl(H \bar \chi \bigr)_{\alpha}^{i} & \equiv H^{ij}_{abc}\bigl(\gamma^{abc} \bar \chi_j\bigr)_{\alpha} \ , \\
 \bigl(H \bar \chi \bigr)_{a b c \dot \alpha}^{i} & \equiv H^{ij}_{abc} \bar \chi_{\dot \alpha\;j} \ , \\
 \bigl(H \bar \chi \bigr)_{a b \alpha}^{ijk} & \equiv H^{(ij}_{abc}\bigl(\gamma^{c} \bar \chi^{k)}\bigr)_{\alpha} \ , 
 \end{split} \hspace{5 mm}
 \begin{split}
 \bigl(\bar P \bar \lambda \bigr)_{a \dot\alpha}^{ijk} & \equiv \bar P_a  \bar  \lambda^{ijk}_{\dot \alpha} \ , \\
 \bigl(\bar F \lambda \bigr)^i_{a \dot \alpha} & \equiv \bar{F}^{jk}_{ab} \bigl(\gamma^{b}\lambda^i_{jk}\bigr)_{\dot \alpha} \ , \\
 \bigl(\bar F \lambda \bigr)^{ijk}_{a \dot \alpha} & \equiv \bar{F}^{l(i}_{ab} \bigl(\gamma^{b}\lambda^{jk)}_{l}\bigr)_{\dot \alpha} \ , \\
 \bigl(\bar G \lambda \bigr)^{ijk}_{abcd \alpha} & \equiv \bar G^{-}_{abcd} \lambda^{ijk}_{\alpha} \ , \\
 \bigl(P^4 \bar \chi \bigr)^{ijk}_{a \dot\alpha} & \equiv P_a^{ijkl} \bar\chi_{\dot \alpha\;l} \ , \\
 \bigl(H \bar \chi \bigr)_{a \dot\alpha}^{i} & \equiv H^{ij}_{abc}\bigl(\gamma^{bc} \bar \chi_j\bigr)_{\dot \alpha} \ , \\
 \bigl(H \bar \chi \bigr)_{\alpha}^{ijk} & \equiv H^{(ij}_{abc}\bigl(\gamma^{abc}\bar \chi^{k)}\bigr)_{\alpha} \ , 
 \end{split} \hspace{5 mm}
 \begin{split}
 \bigl(\bar F \lambda \bigr)^i_{a b \alpha} & \equiv \bar{F}^{jk}_{ab} \lambda^i_{\alpha\;jk}\ , \\
 \bigl(\bar F \lambda \bigr)^{ijk}_{a b \alpha} & \equiv \bar{F}^{l(i}_{ab} \lambda^{jk)}_{\alpha\;l}\ , \\
 \bigl(H \bar \chi \bigr)_{a b \alpha}^{i} & \equiv H^{ij}_{abc}\bigl(\gamma^{c} \bar\chi_j\bigr)_{\alpha} \ , \\
 \bigl(H \bar \chi \bigr)_{a \dot\alpha}^{ijk} & \equiv H^{(ij}_{abc}\bigl(\gamma^{bc} \bar \chi^{k)}\bigr)_{\dot \alpha} \ . 
 \end{split}
\end{equation}
Finally we must also consider the cubic terms in the fermions. We list in a first place the monomials in $\chi \bar \chi^2$
\begin{equation}
\begin{split}
 \bigl(\chi \bar \chi \bar \chi \bigr)_{\alpha}^i & \equiv \chi^i_{\alpha} \bigl(\bar\chi \bar\chi\bigr) \ , \\
 \bigl(\chi \bar \chi \bar \chi \bigr)_{a \dot \alpha}^i & \equiv \bigl(\gamma^{bcd}\chi^i\bigr)_{\alpha} \bigl(\bar\chi \bar\chi\bigr)_{abcd} \ , \\
 \bigl(\chi \bar \chi \bar \chi \bigr)_{a b \alpha}^i & \equiv \chi^j_{\alpha} \bigl(\bar\chi \bar\chi\bigr)_{ab\;j}^i \ , \\
 \bigl(\chi \bar \chi \bar \chi \bigr)_{\alpha}^{ijk} & \equiv \bigl(\gamma^{ab}\chi^{(i}\bigr)_{\alpha} \bigl(\bar\chi \bar\chi\bigr)_{ab}^{jk)} \ , \\
 \bigl(\chi \bar \chi \bar \chi \bigr)_{a b \alpha}^{ijk} & \equiv \chi^{(i}_{\dot\alpha} \bigl(\bar\chi \bar\chi\bigr)_{ab}^{jk)}\ . 
\end{split} \hspace{10 mm}
\begin{split}
 \bigl(\chi \bar \chi \bar \chi \bigr)_{A \; \alpha}^i & \equiv \bigl(\gamma^{ab}\chi^j\bigr)_{\alpha} \bigl(\bar\chi \bar\chi\bigr)_{ab\;j}^{i} \ , \\
 \bigl(\chi \bar \chi \bar \chi \bigr)_{A \; a \dot \alpha}^i & \equiv \bigl(\gamma^{b}\chi^j\bigr)_{\alpha} \bigl(\bar\chi \bar\chi\bigr)_{ab\;j}^i\ , \\
 \bigl(\chi \bar \chi \bar \chi \bigr)_{a b c d \alpha}^i & \equiv \chi^i_{\alpha} \bigl(\bar\chi \bar\chi\bigr)_{abcd} \ , \\
 \bigl(\chi \bar \chi \bar \chi \bigr)_{a \dot\alpha}^{ijk} & \equiv \bigl(\gamma^{b}\chi^{(i}\bigr)_{\dot\alpha} \bigl(\bar\chi \bar\chi\bigr)_{ab}^{jk)}\ , 
\end{split}
\end{equation}
where we use definition \eqref{eq:D1U2Basis} for the bilinear in $\bar \chi$ as well as
\begin{equation} \label{eq:D1U2Basis_1}
 \bigl(\bar \chi \bar \chi \bigr)_{a b c d} \equiv \bar \chi^i \gamma_{abcd} \bar \chi_i \ , \qquad  \bigl(\bar \chi \bar \chi \bigr) \equiv \bar \chi^i \bar \chi_i\ .  
\end{equation}
Then we define the basis of three-linear in  $\bar \chi \lambda \bar \lambda$
\begin{equation}
\begin{split}
 \bigl(\bar \chi \lambda \bar \lambda \bigr)_{\alpha}^i & \equiv \bigl(\gamma^a \bar \chi^i\bigr)_{\alpha} \bigl(\lambda \bar \lambda \bigr)_a \ , \\
 \bigl(\bar \chi \lambda \bar \lambda \bigr)_{B\;\alpha}^i & \equiv \bigl(\gamma^{abc} \bar \chi^i\bigr)_{\alpha} \bigl(\lambda \bar \lambda \bigr)_{abc} \ , \\
 \bigl(\bar \chi \lambda \bar \lambda \bigr)_{a \dot \alpha}^i & \equiv  \bar \chi^i_{\dot \alpha} \bigl(\lambda \bar \lambda \bigr)_a  \ , \\
 \bigl(\bar \chi \lambda \bar \lambda \bigr)_{B\; a \dot \alpha}^i & \equiv \bigl(\gamma^{bc} \bar \chi^i\bigr)_{\dot \alpha} \bigl(\lambda \bar \lambda \bigr)_{abc} \ , \\
 \bigl(\bar \chi \lambda \bar \lambda \bigr)_{a b \alpha}^i & \equiv  \bigl(\gamma^{c} \bar \chi^i\bigr)_{\alpha} \bigl(\lambda \bar \lambda \bigr)_{abc} \ , \\
 \bigl(\bar \chi \lambda \bar \lambda \bigr)_{a b c \dot \alpha}^i & \equiv  \bar \chi^i_{\dot \alpha} \bigl(\lambda \bar \lambda \bigr)_{abc} \ , \\
 \bigl(\bar \chi \lambda \bar \lambda \bigr)_{\alpha}^{ijk} & \equiv  \bigl(\gamma^a \bar \chi^{(i}\bigr)_{\alpha} \bigl(\lambda \bar \lambda \bigr)^{jk)}_a \ , \\
 \bigl(\bar \chi \lambda \bar \lambda \bigr)_{B\;\alpha}^{ijk} & \equiv  \bigl(\gamma^{abc} \bar \chi^{(i}\bigr)_{\alpha} \bigl(\lambda \bar \lambda \bigr)^{jk)}_{abc} \ , \\
 \bigl(\bar \chi \lambda \bar \lambda \bigr)_{a \dot\alpha}^{ijk} & \equiv  \bar \chi^{(i}_{\dot\alpha} \bigl(\lambda \bar \lambda \bigr)^{jk)}_a \ , \\
 \bigl(\bar \chi \lambda \bar \lambda \bigr)_{B\;a \dot\alpha}^{ijk} & \equiv  \bigl(\gamma^{bc} \bar \chi^{(i}\bigr)_{\dot\alpha} \bigl(\lambda \bar \lambda \bigr)^{jk)}_{abc} \ , \\
 \bigl(\bar \chi \lambda \bar \lambda \bigr)_{ab \alpha}^{ijk} & \equiv  \bigl(\gamma^{c} \bar \chi^{(i}\bigr)_{\alpha} \bigl(\lambda \bar \lambda \bigr)^{jk)}_{abc}  \ . 
 \end{split}  \hspace{20 mm}
\begin{split}
 \bigl(\bar \chi \lambda \bar \lambda \bigr)_{A \; \alpha}^i & \equiv \bigl(\gamma^a \bar \chi^j\bigr)_{\alpha} \bigl(\lambda \bar \lambda \bigr)_{a\;j}^i \ , \\
 \bigl(\bar \chi \lambda \bar \lambda \bigr)_{C \; \alpha}^i & \equiv \bigl(\gamma^{abc} \bar \chi^j\bigr)_{\alpha} \bigl(\lambda \bar \lambda \bigr)_{abc\;j}^i \ , \\
 \bigl(\bar \chi \lambda \bar \lambda \bigr)_{A \; a \dot \alpha}^i & \equiv \bar \chi^j_{\dot \alpha}\bigl(\lambda \bar \lambda \bigr)_{a\;j}^i \ , \\
 \bigl(\bar \chi \lambda \bar \lambda \bigr)_{C \; a \dot \alpha}^i & \equiv \bigl(\gamma^{bc} \bar \chi^j\bigr)_{\dot \alpha} \bigl(\lambda \bar \lambda \bigr)_{abc\;j}^i\ , \\
 \bigl(\bar \chi \lambda \bar \lambda \bigr)_{A \; a b \alpha}^i & \equiv \bigl(\gamma^{c} \bar \chi^j\bigr)_{\alpha} \bigl(\lambda \bar \lambda \bigr)_{abc\;j}^i\ , \\
 \bigl(\bar \chi \lambda \bar \lambda \bigr)_{A \; a b c \dot \alpha}^i & \equiv \bar \chi^j_{\dot \alpha} \bigl(\lambda \bar \lambda \bigr)_{abc\;j}^i\ , \\
 \bigl(\bar \chi \lambda \bar \lambda \bigr)_{A \alpha}^{ijk} & \equiv \bigl(\gamma^a \bar \chi^{l}\bigr)_{\alpha} \bigl(\lambda \bar \lambda \bigr)^{ijk}_{a\;l} \ , \\
 \bigl(\bar \chi \lambda \bar \lambda \bigr)_{C\;\alpha}^{ijk} & \equiv \bigl(\gamma^{abc} \bar \chi^{l}\bigr)_{\alpha} \bigl(\lambda \bar \lambda \bigr)^{ijk}_{abc\;l}\ , \\
 \bigl(\bar \chi \lambda \bar \lambda \bigr)_{A\;a \dot\alpha}^{ijk} & \equiv \bar \chi^{l}_{\dot\alpha} \bigl(\lambda \bar \lambda \bigr)^{ijk}_{a\;l} \ , \\
 \bigl(\bar \chi \lambda \bar \lambda \bigr)_{C\;a \dot\alpha}^{ijk} & \equiv \bigl(\gamma^{bc} \bar \chi^{l}\bigr)_{\dot\alpha} \bigl(\lambda \bar \lambda \bigr)^{ijk}_{abc\;l}\ , \\
 \bigl(\bar \chi \lambda \bar \lambda \bigr)_{A\;a b \alpha}^{ijk} & \equiv \bigl(\gamma^{c} \bar \chi^{l}\bigr)_{\alpha} \bigl(\lambda \bar \lambda \bigr)^{ijk}_{abc\;l}\ , 
 \end{split} 
\end{equation}
where we use the following definitions
\begin{equation}
\begin{split}
 \bigl(\lambda \bar\lambda \bigr)_{a} & \equiv \lambda^{ijk} \gamma_{a} \bar \lambda_{ijk} \ , \\
 \bigl(\lambda \bar\lambda \bigr)^{ij}_{a} & \equiv \lambda^{kl(i} \gamma_{a} \bar \lambda_{kl}^{j)} \ , \\
 \bigl(\lambda \bar\lambda \bigr)^{ijkl}_{a} & \equiv \lambda^{m(ij} \gamma_{a} \bar \lambda_{m}^{kl)}  \ , 
\end{split} \hspace{20 mm}
\begin{split}
 \bigl(\lambda \bar\lambda \bigr)_{a b c} & \equiv \lambda^{ijk} \gamma_{a b c} \bar \lambda_{ijk} \ , \\
 \bigl(\lambda \bar\lambda \bigr)^{ij}_{abc} & \equiv \lambda^{kl(i} \gamma_{abc} \bar \lambda_{kl}^{j)} \ , \\
 \bigl(\lambda \bar\lambda \bigr)^{ijkl}_{abc} & \equiv \lambda^{m(ij} \gamma_{abc} \bar \lambda_{m}^{kl)} \  .
\end{split}
\end{equation}
Finally, the list of three-linear in $\lambda^3$ is
\begin{equation}
\begin{split}
 \bigl(\lambda \lambda \lambda\bigr)_{\alpha}^i & \equiv \lambda_{\beta}^{ikj} \lambda^{\beta\, lm}_{j} \lambda_{\alpha \, klm} \ , \\
 \bigl(\lambda \lambda \lambda\bigr)_{\alpha}^{ijk} & \equiv \lambda^{ijk}_{\alpha} \bigl(\lambda \lambda\bigr)  \ , \\
 \bigl(\lambda \lambda \lambda\bigr)_{a b\alpha}^{ijk} & \equiv \lambda^{l(ij}_{\alpha} \bigl(\lambda \lambda\bigr)_{ab\;l}^{k)} \ , 
\end{split} \hspace{20 mm}
\begin{split}
 \bigl(\lambda \lambda \lambda\bigr)_{ab\alpha}^i & \equiv \lambda_{\beta}^{ijk} \bigl(\lambda_{j}^{lm}\gamma_{ab}\lambda_{klm}\bigr)\ , \\
 \bigl(\lambda \lambda \lambda\bigr)_{A\;\alpha}^{ijk} & \equiv \bigl(\gamma^{abcd}\lambda^{ijk}\bigr)_{\alpha} \bigl(\lambda \lambda\bigr)_{abcd}\ , \\
 \bigl(\lambda \lambda \lambda\bigr)_{a \dot\alpha}^{ijk} & \equiv \bigl(\gamma^b\lambda^{l(ij}\bigr)_{\dot\alpha} \bigl(\lambda \lambda\bigr)_{ab\;l}^{k)}\ , 
\end{split}
\end{equation}
where we use again $\eqref{eq:D1U2Basis}$.\\[0.5 cm]
Within this basis, one computes the Dirac equation for the fermion field  $\bar \chi$,  solving the Bianchi identities displayed in section \ref{sec:dim32Id}, such that 
\begin{multline} \label{eq:PartialChiBar}
 D_{a}\bar \chi_{\dot \alpha}^i = (D_{a}\bar \chi_{\dot \alpha}^i )' + \bigl(\gamma_{a}\bigr)_{\dot \alpha}^{\;\;\;\alpha} \biggl(- \frac{i}{64} \bigl(\bar F \lambda \bigr)^i_{\alpha} +  \frac{1}{96} \bigl(H \bar \chi \bigr)_{\alpha}^{i} - \frac{3i}{16} \bigl(\chi \bar \chi \bar \chi \bigr)_{\alpha}^i - \frac{i}{32} \bigl(\chi \bar \chi \bar \chi \bigr)_{A\;\alpha}^i \\
 \hspace{20 mm} +  \frac{5i}{256} \bigl(\bar \chi \lambda \bar \lambda \bigr)_{\alpha}^i +  \frac{5i}{128}  \bigl(\bar \chi \lambda \bar \lambda \bigr)_{A \; \alpha}^i + \frac{i}{1536}   \bigl(\bar \chi \lambda \bar \lambda \bigr)_{B\;\alpha}^i + \frac{i}{768} \bigl(\bar \chi \lambda \bar \lambda \bigr)_{C \; \alpha}^i + \frac{i}{32} \bigl(\lambda \lambda \lambda\bigr)_{\alpha}^i \biggr) 
\end{multline}
The covariant derivative of the scalar momentum gives 
\begin{multline}
 \bar D_{i \dot \alpha} \bar P_a = \biggl( 2 (D_{a}\bar \chi_{\dot \alpha i})' - \frac{1}{6} \bigl(H \bar \chi \bigr)_{a \dot\alpha i} +  \frac{i}{48} \bigl(\chi \bar \chi \bar \chi \bigr)_{a \dot\alpha i} +   \frac{i}{6}  \bigl(\chi \bar \chi \bar \chi \bigr)_{A \; a \dot \alpha i} + \frac{i}{12}  \bigl(\bar \chi \lambda \bar \lambda \bigr)_{a \dot \alpha i} \\
  -  \frac{i}{2}  \bigl(\bar \chi \lambda \bar \lambda \bigr)_{A\; a \dot \alpha i} \biggr)  + \bigl(\gamma_{a}\bigr)_{\dot \alpha}^{\;\;\;\alpha} \biggl(- \frac{i}{32} \bigl(\bar F \lambda \bigr)_{\alpha i}  - \frac{5}{144} \bigl(H \bar \chi \bigr)_{\alpha i} + i \bigl(\chi \bar \chi \bar \chi \bigr)_{\alpha i} \\
\hspace{10 mm}- \frac{i}{48} \bigl(\chi \bar \chi \bar \chi \bigr)_{A\;\alpha i} +  \frac{23i}{384}  \bigl(\bar \chi \lambda \bar \lambda \bigr)_{\alpha i} - \frac{3i}{64} \bigl(\bar \chi \lambda \bar \lambda \bigr)_{A \; \alpha i} - \frac{7i}{768}  \bigl(\bar \chi \lambda \bar \lambda \bigr)_{B\;\alpha i} \\
- \frac{7i}{384}  \bigl(\bar \chi \lambda \bar \lambda \bigr)_{C \; \alpha i} + \frac{i}{16}   \bigl(\lambda \lambda \lambda\bigr)_{\alpha i} \biggr)\ , 
\end{multline}
whereas the covariant derivative of  $ \bar F$ is
\begin{multline} \label{eq:DFBar}
 \hspace{2mm} D_{\alpha}^i \bar F_{ab}^{jk}
\\ 
\hspace{-18mm}= \varepsilon^{i (j} (\gamma_{ab})_{\alpha}^{\;\;\beta}\biggl(- \frac{1}{144} \bigl(\bar F \lambda \bigr)^{k)}_{\beta} - \frac{i}{216}\bigl(H \bar \chi \bigr)_{\beta}^{k)} - \frac{1}{3}  \bigl(\chi \bar \chi \bar \chi \bigr)_{\beta}^{k)} + \frac{7}{24} \bigl(\chi \bar \chi \bar \chi \bigr)_{A\;\beta}^{k)} - \frac{35}{576}  \bigl(\bar \chi \lambda \bar \lambda \bigr)_{\beta}^{k)}  \biggr . \\
 \biggl . + \frac{13}{288} \bigl(\bar \chi \lambda \bar \lambda \bigr)_{A \; \beta}^{k)} -\frac{23}{3456} \bigl(\bar \chi \lambda \bar \lambda \bigr)_{B\;\beta}^{k)} + \frac{25}{1728} \bigl(\bar \chi \lambda \bar \lambda \bigr)_{C \; \beta}^{k)}  -  \frac{5}{24} \bigl(\lambda \lambda \lambda\bigr)_{\beta}^{k)}\biggr)\\
  +  \varepsilon^{i (j} (\gamma_{[a})_{\alpha}^{\;\;\dot\alpha} \biggl(-4 i (D_{b]}\bar \chi_{\dot \alpha}^{k)})' + \frac{2}{9} \bigl(\bar F \lambda \bigr)^{k)}_{b] \dot \alpha} - \frac{i}{9} \bigl(H \bar \chi \bigr)_{b] \dot\alpha}^{k)} - \bigl. \frac{1}{72} \bigl(\chi \bar \chi \bar \chi \bigr)_{b] \dot \alpha}^{k)} - \frac{17}{9}  \bigl(\chi \bar \chi \bar \chi \bigr)_{A \; b] \dot \alpha}^{k)} \bigr. \\
  - \frac{13}{36}  \bigl(\bar \chi \lambda \bar \lambda \bigr)_{b] \dot \alpha}^{k)} - \frac{7}{18} \bigl(\bar \chi \lambda \bar \lambda \bigr)_{A\; b] \dot \alpha}^{k)}  + \bigl.\frac{1}{24}  \bigl(\bar \chi \lambda \bar \lambda \bigr)_{B\; b] \dot \alpha}^{k)} - \frac{1}{12}  \bigl(\bar \chi \lambda \bar \lambda \bigr)_{C\; b] \dot \alpha}^{k)} \biggr)\\
  +  \varepsilon^{i (j} \biggl(\frac{1}{18}\bigl(\bar F \lambda \bigr)^{k)}_{ab \alpha} + \frac{4i}{9}\bigl(H \bar \chi \bigr)_{ab \alpha}^{k)} - \frac{1}{9}\bigl(\bar \chi \lambda \bar \lambda \bigr)_{ab \alpha}^{k)} -  \frac{2}{9} \bigl(\bar \chi \lambda \bar \lambda \bigr)_{A\;ab \alpha}^{k)} + \bigl. \frac{1}{9} \bigl(\lambda \lambda \lambda\bigr)_{ab\alpha}^{k)} \biggr)\\
  + (\gamma_{ab})_{\alpha}^{\;\;\beta} \biggl(-\frac{1}{12} \bigl(\bar F \lambda \bigr)^{ijk}_{\beta}- \frac{i}{9} \bigl(H \bar \chi \bigr)_{\beta}^{ijk}  - \frac{1}{4}\bigl(\chi \bar \chi \bar \chi \bigr)_{\beta}^{ijk} +  \frac{5}{48} \bigl(\bar \chi \lambda \bar \lambda \bigr)_{\beta}^{ijk} 
  - \frac{7}{24}  \bigl(\bar \chi \lambda \bar \lambda \bigr)_{A \; \beta}^{ijk}\\  - \frac{7}{288}  \bigl(\bar \chi \lambda \bar \lambda \bigr)_{B\;\beta}^{ijk}+\frac{5}{144} \bigl(\bar \chi \lambda \bar \lambda \bigr)_{C \; \beta}^{ijk}- \frac{1}{8} \bigl(\lambda \lambda \lambda\bigr)_{\beta}^{ijk} \bigl.  + \frac{1}{2304} \bigl(\lambda \lambda \lambda\bigr)_{A\;\beta}^{ijk}\biggr)\\
  + (\gamma_{[a})_{\alpha}^{\;\;\dot\alpha} \biggl(- 2 i \bigl(\bar P \bar \lambda \bigr)_{b] \dot\alpha}^{ijk}- \frac{1}{3} \bigl(\bar F \lambda \bigr)^{ijk}_{b] \dot \alpha} - 4 i \bigl(P^4 \bar \chi \bigr)^{ijk}_{b] \dot \alpha} +\frac{1}{12}\bigl(\bar G \lambda \bigr)^{ijk}_{b] \dot \alpha} + \frac{i}{3}\bigl(H \bar \chi \bigr)_{b] \dot\alpha}^{ijk} +  \frac{7}{3} \bigl(\chi \bar \chi \bar \chi \bigr)_{b] \dot \alpha}^{ijk} \\
 \hspace{10mm}  + \frac{1}{12} \bigl(\bar \chi \lambda \bar \lambda \bigr)_{b] \dot \alpha}^{ijk} - \frac{5}{6} \bigl(\bar \chi \lambda \bar \lambda \bigr)_{A\;b] \dot \alpha}^{ijk} +   \frac{1}{8} \bigl(\bar \chi \lambda \bar \lambda \bigr)_{B\; b] \dot \alpha}^{ijk} -  \frac{1}{4}  \bigl(\bar \chi \lambda \bar \lambda \bigr)_{C\;b] \dot \alpha}^{ijk}  - \bigl. \frac{1}{4}  \bigl(\lambda \lambda \lambda\bigr)_{b] \dot\alpha}^{ijk} \biggr) \\
 - \frac{5}{6} \bigl(\bar F \lambda \bigr)^{ijk}_{ab \alpha} - \frac{4i}{3} \bigl(H \bar \chi \bigr)_{ab \alpha}^{ijk} -\frac{1}{6} \bigl(\bar \chi \lambda \bar \lambda \bigr)_{ab \alpha}^{ijk} - \frac{1}{3} \bigl(\bar \chi \lambda \bar \lambda \bigr)_{A\;ab \alpha}^{ijk} 
 -\frac{1}{3} \bigl(\lambda \lambda \lambda\bigr)_{a b\alpha}^{ijk}
 \end{multline}
 and the one of $\bar G^{-}$ 
 \begin{multline} \label{eq:DGBar}
 D_{\alpha}^i \bar G^{-}_{a b c d} = (\gamma_{[a})_{\alpha}^{\;\;\dot\alpha}\biggl(- \frac{16i}{3}\bigl(H \bar \chi \bigr)_{bcd]\dot \alpha}^{i} - \bigl(\bar \chi \lambda \bar \lambda \bigr)_{bcd] \dot \alpha}^i + 2 \bigl(\bar \chi \lambda \bar \lambda \bigr)_{A\;bcd] \dot \alpha}^{i} \biggr) - 2 \bigl(\chi \bar \chi \bar \chi \bigr)_{a b c d \alpha}^i \\
\hspace{13 mm} +  (\gamma_{[abc})_{\alpha}^{\;\;\dot\alpha} \biggl(- 4 i (D_{d]}\bar \chi_{\dot \alpha}^i)'+  \bigl(\bar F \lambda \bigr)^i_{d] \dot \alpha} + \frac{5 i}{3} \bigl(H \bar \chi \bigr)_{d] \dot\alpha}^{i} +  \frac{1}{24} \bigl(\chi \bar \chi \bar \chi \bigr)_{d] \dot \alpha}^i + \frac{1}{3} \bigl(\chi \bar \chi \bar \chi \bigr)_{A \; d] \dot \alpha}^i \bigr.\\
 +  \frac{1}{12} \bigl(\bar \chi \lambda \bar \lambda \bigr)_{d] \dot \alpha}^i \bigl. + \frac{1}{2}  \bigl(\bar \chi \lambda \bar \lambda \bigr)_{A\; d] \dot \alpha}^i + \frac{1}{8}  \bigl(\bar \chi \lambda \bar \lambda \bigr)_{B\; d] \dot \alpha}^i - \frac{1}{4}  \bigl(\bar \chi \lambda \bar \lambda \bigr)_{C\; d] \dot \alpha}^i \biggr)  \\ 
 \hspace{20 mm}  + (\gamma_{abcd})_{\alpha}^{\;\;\beta}\biggl(- \frac{1}{8} \bigl(\bar F \lambda \bigr)^i_{\beta} \bigr. - \frac{11i}{72} \bigl(H \bar \chi \bigr)^i_{\beta} - \frac{1}{24} \bigl(\chi \bar \chi \bar \chi \bigr)^i_{A\;\beta} - \frac{1}{96} \bigl(\bar \chi \lambda \bar \lambda \bigr)^i_{\beta} \\
  - \frac{1}{16} \bigl(\bar \chi \lambda \bar \lambda \bigr)^i_{A\;\beta} - \frac{1}{192} \bigl(\bar \chi \lambda \bar \lambda \bigr)^i_{B\;\beta} + \bigl. \frac{1}{96} \bigl(\bar \chi \lambda \bar \lambda \bigr)^i_{C\;\beta} \biggr) \\
 + (\gamma_{[a b})_{\alpha}^{\;\;\beta} \biggl(4 i \bigl(H \bar \chi \bigr)_{cd]\beta}^{i} + \frac{3}{4} \bigl(\bar \chi \lambda \bar \lambda \bigr)_{cd]\beta}^{i} - \frac{3}{2} \bigl(\bar \chi \lambda \bar \lambda \bigr)_{A\;cd]\beta}^{i}\biggr) \ . 
\end{multline}

We shall finally consider the components of $U(1)$ weight $1$, for which the number of independent elements is the largest. Similarly as for $\bar \chi$ we define $(D_{a} \lambda_{\alpha}^{ijk})'$ as the irreducible representation component of the covariant derivative of the fermion $\lambda$  in the \DSOVIII1001, and $ \bar{\rho}_{a b \dot \alpha}^{i}$ as the component of the Rarita--Schwinger field strength in the irreducible representation \DSOVIII0110, all other components of the Rarita--Schwinger field strength being equal to monomials in the other fields through the Rarita--Schwinger equation.  We define in a first place the bilinear combinations 
\begin{equation}
\begin{split}
 \bigl(\bar P \chi \bigr)_{\dot \alpha}^i & \equiv  \bar P_{a}  \bigl(\gamma^{a} \chi^i\bigr)_{\dot \alpha} \ , \\
 \bigl(\bar F \bar \lambda \bigr)_{\dot \alpha}^i & \equiv \bar F_{ab}^{jk} \bigl(\gamma^{ab} \bar \lambda_{\;jk}^i \bigr)_{\dot \alpha}\ , \\
 \bigl(\bar F \bar \lambda \bigr)_{a b \dot \alpha}^i & \equiv \bar F_{ab}^{jk} \bar \lambda_{\dot \alpha\;jk}^i \ , \\
 \bigl(\bar F \bar \lambda \bigr)_{a \alpha}^{ijk} & \equiv \bar F_{ab}^{l(i} \bigl(\gamma^{b} \bar \lambda_{l}^{\;jk)} \bigr)_{\alpha} \ , \\
 \bigl(\bar F \bar \lambda \bigr)_{\dot \alpha}^{ijklm} & \equiv \bar F_{ab}^{(ij} \bigl(\gamma^{ab} \bar \lambda^{klm)}\bigr)_{\dot \alpha} \ , \\
 \bigl(\bar G \bar \lambda \bigr)_{\dot \alpha}^{ijk} & \equiv \bar G^{-}_{abcd} \bigl(\gamma^{abcd} \bar \lambda^{ijk} \bigr)_{\dot \alpha} \ , \\
 \bigl(H \lambda \bigr)_{\dot \alpha}^{i} & \equiv H_{abc}^{jk} \bigl(\gamma^{abc} \lambda_{\;jk}^{i} \bigr)_{\dot \alpha} \ , \\
 \bigl(H \lambda \bigr)_{a b \dot \alpha}^{i} & \equiv H_{abc}^{jk}  \bigl(\gamma^{c} \lambda_{\;jk}^{i} \bigr)_{\dot \alpha} \ , \\
 \bigl(H \lambda \bigr)_{\dot \alpha}^{ijk} & \equiv H_{abc}^{l(i} \bigl(\gamma^{abc} \lambda_{l}^{\;jk)} \bigr)_{\dot \alpha} \ , \\
 \bigl(H \lambda \bigr)_{a b \dot \alpha}^{ijk} & \equiv H_{abc}^{l(i}  \bigl(\gamma^{c} \lambda_{l}^{\;jk)} \bigr)_{\dot \alpha} \ , \\
 \bigl(H \lambda \bigr)_{\dot \alpha}^{ijklm} & \equiv H_{abc}^{(ij} \bigl(\gamma^{abc} \lambda^{klm)} \bigr)_{\dot \alpha} \ , \\
 \bigl(P^4 \lambda \bigr)_{\dot \alpha}^{i} & \equiv P_{a}^{i j k l} \bigl(\gamma^a \lambda_{ijk} \bigr)_{\dot \alpha} \ , \\
 \bigl(P^4 \lambda \bigr)_{\dot \alpha}^{ijk} & \equiv P_{a}^{l m (i j} \bigl(\gamma^a \lambda_{l m}{}^{k)} \bigr)_{\dot \alpha} \ , \\
 \bigl(P^4 \lambda \bigr)_{\dot \alpha}^{ijklm} & \equiv P_{a}^{n (i j k} \bigl(\gamma^a \lambda_{n}{}^{lm)} \bigr)_{\dot \alpha}\ , \\
 \bigl(F \bar \chi \bigr)^i_{\dot \alpha} & \equiv F_{ab}^{i j} \bigl(\gamma^{ab} \bar \chi_{j} \bigr)_{\dot \alpha} \ , \\
 \bigl(F \bar \chi \bigr)^i_{a b \dot \alpha} & \equiv F_{ab}^{i j} \bar \chi_{\dot \alpha \; j} \ , \\
 \bigl(F \bar \chi \bigr)^{ijk}_{a \alpha} & \equiv F_{ab}^{(i j} \bigl(\gamma^{b} \bar \chi^{k)} \bigr)_{\alpha} \ , \\
 \bigl(G \bar \chi \bigr)_{a \alpha}^i & \equiv G^{+}_{abcd} \bigl(\gamma^{bcd} \bar \chi^i \bigr)_{\alpha} 
  \end{split} 
  \hspace{20 mm}
  \begin{split}
 \bigl(\bar P \chi \bigr)_{a \alpha}^i & \equiv \bar P_{a} \chi^i_{\alpha} \ , \\
 \bigl(\bar F \bar \lambda \bigr)_{a \alpha}^i & \equiv \bar F_{ab}^{jk} \bigl(\gamma^{b} \bar \lambda_{\;jk}^i \bigr)_{\alpha} \ , \\
 \bigl(\bar F \bar \lambda \bigr)_{\dot \alpha}^{ijk} & \equiv \bar F_{ab}^{l(i} \bigl(\gamma^{ab} \bar \lambda_{l}{}^{jk)} \bigr)_{\dot \alpha} \ , \\
 \bigl(\bar F \bar \lambda \bigr)_{a b \dot \alpha}^{ijk} & \equiv \bar F_{ab}^{l(i} \bar \lambda_{\dot \alpha\;l}^{jk)} \ , \\
 \bigl(\bar F \bar \lambda \bigr)_{a \alpha}^{ijklm} & \equiv \bar F_{ab}^{(ij} \bigl(\gamma^{b} \bar \lambda^{klm)}\bigr)_{\alpha} \ , \\
 \bigl(\bar G \bar \lambda \bigr)_{ab \dot \alpha}^{ijk} & \equiv \bar G^{-}_{abcd} \bigl(\gamma^{cd} \bar \lambda^{ijk} \bigr)_{\dot \alpha} \ , \\
 \bigl(H \lambda \bigr)_{a \alpha}^{i} & \equiv H_{abc}^{jk} \bigl(\gamma^{bc} \lambda_{\;jk}^{i} \bigr)_{\alpha} \ , \\
 \bigl(H \lambda \bigr)_{a b c \alpha}^{i} & \equiv H_{abc}^{jk} \lambda_{\alpha\;jk}^{i} \ , \\
 \bigl(H \lambda \bigr)_{a \alpha}^{ijk} & \equiv H_{abc}^{l(i}  \bigl(\gamma^{bc} \lambda_{l}^{\; jk)}  \bigr)_{\alpha} \ , \\
 \bigl(H \lambda \bigr)_{a b c \alpha}^{ijk} & \equiv H_{abc}^{l(i} \lambda_{\alpha\;l}^{jk)} \ , \\
 \bigl(H \lambda \bigr)_{a \alpha}^{ijklm} & \equiv H_{abc}^{(ij} \bigl(\gamma^{bc} \lambda^{klm)}  \bigr)_{\alpha} \ , \\
 \bigl(P^4 \lambda \bigr)_{a \alpha}^{i} & \equiv P_{a}^{i j k l} \lambda_{\alpha \; ijk} \ , \\
 \bigl(P^4 \lambda \bigr)_{a \alpha}^{ijk} & \equiv P_{a}^{l m (i j} \lambda_{\alpha \; l m}^{k)}\ , \\
 \bigl(P^4 \lambda \bigr)_{a \alpha}^{ijklm} & \equiv P_{a}^{n (i j k} \lambda_{\alpha \; n}^{lm)} \ , \\
\bigl(F \bar \chi \bigr)^i_{a \alpha} & \equiv F_{ab}^{i j} \bigl(\gamma^{b} \bar \chi_{j} \bigr)_{\alpha}  \\
\bigl(F \bar \chi \bigr)^{ijk}_{\dot \alpha} & \equiv F_{ab}^{(i j} \bigl(\gamma^{ab} \bar \chi^{k)}  \bigr)_{\dot \alpha} \\
\bigl(F \bar \chi \bigr)^{ijk}_{a b \dot \alpha} & \equiv F_{ab}^{(i j} \bar \chi_{\dot \beta}^{k)} \\
\bigl(G \bar \chi \bigr)_{a b c d \dot \alpha}^i & \equiv G^{+}_{abcd} \bar \chi_{\dot \alpha}^i
\end{split}
\end{equation}
Then comes the base of three-linear in the fermions, starting with the terms in $\chi \bar \chi \lambda$
\begin{equation}
\begin{split}
 \bigl(\chi \bar \chi \lambda \bigr)_{\dot \alpha}^i & \equiv \bigl(\chi \bar \chi \bigr)_{a}^{jk} \bigl(\gamma^a \lambda_{jk}{}^{i} \bigr)_{\dot \alpha} \ , \\
 \bigl(\chi \bar \chi \lambda \bigr)_{a \alpha}^i & \equiv \bigl(\chi \bar \chi \bigr)_{a}^{jk} \lambda_{\alpha\;jk}^{i} \ , \\
 \bigl(\chi \bar \chi \lambda \bigr)_{a b \dot \alpha}^i & \equiv \bigl(\chi \bar \chi \bigr)_{abc}^{jk} \bigl(\gamma^{c} \lambda_{jk}{}^{i}\bigr)_{\dot \alpha} \ , \\
 \bigl(\chi \bar \chi \lambda \bigr)_{\dot \alpha}^{ijk} & \equiv \bigl(\chi \bar \chi \bigr)_{a} \bigl(\gamma^a \lambda^{ijk} \bigr)_{\dot \alpha} \ , \\
  \bigl(\chi \bar \chi \lambda \bigr)_{B\;\dot \alpha}^{ijk} & \equiv \bigl(\chi \bar \chi \bigr)_{abc} \bigl(\gamma^{abc} \lambda^{ijk}  \bigr)_{\dot \alpha} \ , \\
 \bigl(\chi \bar \chi \lambda \bigr)_{a \alpha}^{ijk} & \equiv  \bigl(\chi \bar \chi \bigr)_{a} \lambda_{\alpha}^{ijk}  \ , \\
 \bigl(\chi \bar \chi \lambda \bigr)_{B\;a \alpha}^{ijk} & \equiv \bigl(\chi \bar \chi \bigr)_{abc} \bigl(\gamma^{bc} \lambda^{ijk} \bigr)_{\alpha}\ , \\
 \bigl(\chi \bar \chi \lambda \bigr)_{ab \dot \alpha}^{ijk} & \equiv \bigl(\chi \bar \chi \bigr)_{abc} \bigl(\gamma^{c}  \lambda^{ijk} \bigr)_{\dot \alpha} \ , \\
 \bigl(\chi \bar \chi \lambda \bigr)_{ab c \alpha}^{ijk} & \equiv \bigl(\chi \bar \chi \bigr)_{abc} \lambda_{\alpha}^{ijk} \ , \\
 \bigl(\chi \bar \chi \lambda \bigr)_{\dot \alpha}^{ijklm} & \equiv \bigl(\chi \bar \chi \bigr)_{a}^{(ij} \bigl(\gamma^a \lambda^{klm)} \bigr)_{\dot \alpha} \ , \\
 \bigl(\chi \bar \chi \lambda \bigr)_{a \alpha}^{ijklm} & \equiv \bigl(\chi \bar \chi \bigr)_{a}^{(ij} \lambda_{\alpha}^{klm)} \ ,   \end{split} \hspace{20 mm}
  \begin{split}
 \bigl(\chi \bar \chi \lambda \bigr)_{A\; \dot \alpha}^i & \equiv \bigl(\chi \bar \chi \bigr)_{abc}^{jk} \bigl(\gamma^{abc} \lambda_{jk}{}^{i} \bigr)_{\dot \alpha} \ , \\
 \bigl(\chi \bar \chi \lambda \bigr)_{A\; a \alpha}^i & \equiv \bigl(\chi \bar \chi \bigr)_{abc}^{jk} \bigl(\gamma^{bc} \lambda_{jk}{}^{i} \bigr)_{\alpha} \ , \\
 \bigl(\chi \bar \chi \lambda \bigr)_{a b c \alpha}^i & \equiv  \bigl(\chi \bar \chi \bigr)_{abc}^{jk} \lambda_{\alpha\;jk}^{i} \ , \\
 \bigl(\chi \bar \chi \lambda \bigr)_{A\;\dot \alpha}^{ijk} & \equiv \bigl(\chi \bar \chi \bigr)_{a}^{l(i} \bigl(\gamma^a \lambda_{l}{}^{jk)}\bigr)_{\dot \alpha} \ , \\
 \bigl(\chi \bar \chi \lambda \bigr)_{C\;\dot \alpha}^{ijk} & \equiv \bigl(\chi \bar \chi \bigr)_{abc}^{l(i} \bigl(\gamma^{abc} \lambda_{l}{}^{jk)} \bigr)_{\dot \alpha} \ , \\
 \bigl(\chi \bar \chi \lambda \bigr)_{A\;a \alpha}^{ijk} & \equiv \bigl(\chi \bar \chi \bigr)_{a}^{l(i} \lambda_{\alpha\;l}^{jk)} \ , \\
 \bigl(\chi \bar \chi \lambda \bigr)_{C\;a  \alpha}^{ijk} & \equiv \bigl(\chi \bar \chi \bigr)_{abc}^{l(i} \bigl(\gamma^{bc} \lambda_{l}{}^{jk)}  \bigr)_{\alpha} \ , \\
 \bigl(\chi \bar \chi \lambda \bigr)_{A\;a b \dot \alpha}^{ijk} & \equiv \bigl(\chi \bar \chi \bigr)_{abc}^{l(i} \bigl(\gamma^{c} \lambda_{l}{}^{jk)}  \bigr)_{\dot \alpha}\ , \\
 \bigl(\chi \bar \chi \lambda \bigr)_{A\;a b c \alpha}^{ijk} & \equiv  \bigl(\chi \bar \chi \bigr)_{abc}^{l(i} \lambda_{\alpha\;l}^{jk)} \ , \\
 \bigl(\chi \bar \chi \lambda \bigr)_{A\;\dot \alpha}^{ijklm} & \equiv \bigl(\chi \bar \chi \bigr)_{abc}^{(ij} \bigl(\gamma^{abc} \lambda^{klm)}\bigr)_{\dot \alpha} \ , \\
 \bigl(\chi \bar \chi \lambda \bigr)_{A\;a \alpha}^{ijklm} & \equiv \bigl(\chi \bar \chi \bigr)_{abc}^{(ij} \bigl(\gamma^{bc} \lambda^{klm)}\bigr)_{\alpha}\ , 
 \end{split}
\end{equation}
where we have used the basis of bilinear defined in section $\ref{sec:dim1_solutions}$. For the terms in $\bar \chi \bar \lambda^{2}$ we give the following basis 
\begin{equation}
\begin{split}
 \bigl(\bar \chi \bar \lambda \bar\lambda \bigr)_{\dot \alpha}^i & \equiv \bar \chi^i_{\dot \alpha}  \bigl(\bar \lambda \bar \lambda \bigr)  \ , \\
 \bigl(\bar \chi \bar \lambda \bar\lambda \bigr)_{B\; \dot \alpha}^i & \equiv \bigl(\gamma^{ab} \bar \chi^j \bigr)_{\dot \alpha} \bigl(\bar \lambda \bar \lambda \bigr)_{ab\;j}^i \ , \\
 \bigl(\bar \chi \bar \lambda \bar\lambda \bigr)_{ab \dot \alpha}^i & \equiv \bigl(\gamma^{cd} \bar \chi^i \bigr)_{\dot \alpha} \bigl(\bar \lambda \bar \lambda \bigr)_{abcd} \ , \\
 \bigl(\bar \chi \bar \lambda \bar\lambda \bigr)_{abcd \dot \alpha}^i & \equiv  \bar \chi^i_{\dot \alpha} \bigl(\bar \lambda \bar \lambda \bigr)_{abcd} \ , \\
 \bigl(\bar \chi \bar \lambda \bar\lambda \bigr)_{A\;\dot \alpha}^{ijk} & \equiv  \bigl(\gamma^{abcd} \bar \chi^l \bigr)_{\dot \alpha} \bigl(\bar \lambda \bar \lambda \bigr)_{abcd\;l}^{ijk} \ , \\
 \bigl(\bar \chi \bar \lambda \bar\lambda \bigr)_{a \alpha}^{ijk} & \equiv  \bigl(\gamma^{b} \bar \chi^{(i} \bigr)_{\alpha} \bigl(\bar \lambda \bar \lambda \bigr)_{ab}^{jk)} \ , \\
 \bigl(\bar \chi \bar \lambda \bar\lambda \bigr)_{A\;a b \dot \alpha}^{ijk} & \equiv   \bar \chi^{(i}_{\dot \alpha} \bigl(\bar \lambda \bar \lambda \bigr)_{ab}^{jk)} \ , \\
 \bigl(\bar \chi \bar \lambda \bar\lambda \bigr)_{A\; \dot \alpha}^{ijklm} & \equiv  \bigl(\gamma^{abcd} \bar \chi^{(i} \bigr)_{\dot \alpha} \bigl(\bar \lambda \bar \lambda \bigr)^{jklm)}_{abcd} \ , \\
 \bigl(\bar \chi \bar \lambda \bar\lambda \bigr)_{a \alpha}^{ijklm} & \equiv  \bigl(\gamma^{b} \bar \chi^n \bigr)_{\alpha} \bigl(\bar \lambda \bar \lambda \bigr)_{ab}^{ijklm}{}_{n} \ , 
 \end{split} \hspace{20 mm}
 \begin{split}
 \bigl(\bar \chi \bar \lambda \bar\lambda \bigr)_{A\; \dot \alpha}^i & \equiv  \bigl(\gamma^{abcd}\bigr)_{\dot \alpha}^{\;\;\;\dot \beta} \bar \chi^i_{\dot \beta} \bigl(\bar \lambda \bar \lambda \bigr)_{abcd} \ , \\
 \bigl(\bar \chi \bar \lambda \bar\lambda \bigr)_{a \alpha}^i & \equiv  \bigl(\gamma^{b}  \bar \chi^j\bigr)_{\alpha} \bigl(\bar \lambda \bar \lambda \bigr)_{ab\;j}^i \ , \\
 \bigl(\bar \chi \bar \lambda \bar\lambda \bigr)_{A\;a b \dot \alpha}^i & \equiv   \bar \chi^j_{\dot \alpha} \bigl(\bar \lambda \bar \lambda \bigr)_{ab\;j}^i \ , \\
 \bigl(\bar \chi \bar \lambda \bar\lambda \bigr)_{\dot \alpha}^{ijk} & \equiv  \bar \chi^l_{\dot \alpha} \bigl(\bar \lambda \bar \lambda \bigr)^{ijk}_{l}\ , \\
 \bigl(\bar \chi \bar \lambda \bar\lambda \bigr)_{B\; \dot \alpha}^{ijk} & \equiv  \bigl(\gamma^{ab} \bar \chi^{(i}\bigr)_{\dot \alpha}  \bigl(\bar \lambda \bar \lambda \bigr)_{ab}^{jk)} \ , \\
 \bigl(\bar \chi \bar \lambda \bar\lambda \bigr)_{a b \dot \alpha}^{ijk} & \equiv   \bigl(\gamma^{cd} \bar \chi^l\bigr)_{\dot \alpha} \bigl(\bar \lambda \bar \lambda \bigr)_{abcd\;l}^{ijk}\ , \\
 \bigl(\bar \chi \bar \lambda \bar\lambda \bigr)_{\dot \alpha}^{ijklm} & \equiv  \bar \chi^{(i}_{\dot \alpha} \bigl(\bar \lambda \bar \lambda \bigr)^{jklm)} \ , \\
 \bigl(\bar \chi \bar \lambda \bar\lambda \bigr)_{B\; \dot \alpha}^{ijklm} & \equiv  \bigl(\gamma^{ab} \bar \chi^n_{\dot \beta} \bigr)_{\dot \alpha} \bigl(\bar \lambda \bar \lambda \bigr)_{ab}^{ijklm}{}_n \ ,
 \end{split}
\end{equation}
where we use
\begin{equation}
\begin{split}
 \bigl(\bar \lambda \bar \lambda \bigr) & \equiv \bar \lambda^{i j k} \bar \lambda_{i j k} \ , \\
 \bigl(\bar \lambda \bar \lambda \bigr)^{i j k l} & \equiv \bar \lambda^{m (i j} \bar \lambda^{k l)}{}_{m} \ , 
\end{split} \hspace{10 mm}
\begin{split}
 \bigl(\bar \lambda \bar \lambda \bigr)_{abcd} & \equiv \bar \lambda^{i j k}\gamma_{abcd} \bar \lambda_{i j k} \ , \\
 \bigl(\bar \lambda \bar \lambda \bigr)_{abcd}^{i j k l} & \equiv \bar \lambda^{m (i j}\gamma_{abcd} \bar \lambda^{k l)}{}_{m} \ , 
\end{split} \hspace{10 mm}
\begin{split}
 \bigl(\bar \lambda \bar \lambda \bigr)_{ab}^{i j} & \equiv \bar \lambda^{k l (i} \gamma_{a b} \bar \lambda^{j)}{}_{k l}  \ , \\
 \bigl(\bar \lambda \bar \lambda \bigr)_{ab}^{i j k l m n} & \equiv \bar \lambda^{(i j k} \gamma_{a b} \bar \lambda^{l m n)} \ . 
 \end{split}
\end{equation}
Finally we define the basis of three-linear in $\lambda^{2} \bar \lambda$ to be
\begin{equation}
\begin{split}
 \bigl(\lambda \lambda \bar \lambda \bigr)_{\dot \alpha}^i & \equiv \bigl(\lambda \lambda \bigr)^{ijkl} \bar \lambda_{\dot \alpha\; jkl} \ , \\
 \bigl(\lambda \lambda \bar \lambda \bigr)_{a \alpha}^i & \equiv \bigl(\lambda \lambda \bigr)^{ijkl}_{abcd} \bigl(\gamma^{bcd} \bar \lambda_{jkl} \bigr)_{\alpha} \ , \\
 \bigl(\lambda \lambda \bar \lambda \bigr)_{a b\dot \alpha}^i & \equiv \bigl(\lambda \lambda \bigr)^{jk}_{ab} \bar \lambda_{\dot \alpha\; jk}^i \ , \\
 \bigl(\lambda \lambda \bar \lambda \bigr)_{\dot \alpha}^{ijk} & \equiv \bigl(\lambda \lambda \bigr) \bar \lambda_{\dot \alpha}^{ijk} \ , \\
 \bigl(\lambda \lambda \bar \lambda \bigr)_{B\;\dot \alpha}^{ijk} & \equiv \bigl(\lambda \lambda \bigr)^{l(i}_{ab}  \bigl(\gamma^{ab} \bar \lambda_{l}{}^{jk)} \bigr)_{\dot \alpha} \ , \\
 \bigl(\lambda \lambda \bar \lambda \bigr)_{a \alpha}^{ijk} & \equiv \bigl(\lambda \lambda \bigr)_{abcd} \bigl(\gamma^{bcd} \bar \lambda^{ijk}\bigr)_{\alpha} \ , \\
 \bigl(\lambda \lambda \bar \lambda \bigr)_{B\;a \alpha}^{ijk} & \equiv \bigl(\lambda \lambda \bigr)^{l(i}_{ab} \bigl(\gamma^{b} \bar \lambda_{l}{}^{jk)} \bigr)_{\alpha} \ , \\
 \bigl(\lambda \lambda \bar \lambda \bigr)_{a b \dot \alpha}^{ijk} & \equiv \bigl(\lambda \lambda \bigr)^{l(i}_{ab} \bar \lambda_{\dot \alpha\; l}^{jk)} \ , \\
 \bigl(\lambda \lambda \bar \lambda \bigr)_{abcd\dot \alpha}^{ijk} & \equiv \bigl(\lambda \lambda \bigr)_{abcd} \bar \lambda_{\dot \alpha}^{ijk} \ , \\
 \bigl(\lambda \lambda \bar \lambda \bigr)_{\dot \alpha}^{ijklm} & \equiv  \bigl(\lambda \lambda \bigr)^{n(ijk} \bar \lambda_{\dot \alpha\; n}^{lm)} \ , \\
 \bigl(\lambda \lambda \bar \lambda \bigr)_{B\;\dot \alpha}^{ijklm} & \equiv \bigl(\lambda \lambda \bigr)^{np(ijkl}_{ab} \bigl(\gamma^{ab} \bar \lambda_{np}{}^{m)}\bigr)_{\dot \alpha} \ , \\
 \bigl(\lambda \lambda \bar \lambda \bigr)_{A\;a \alpha}^{ijklm} & \equiv  \bigl(\lambda \lambda \bigr)^{(ij}_{ab} \bigl(\gamma^{b} \bar \lambda_{\dot \beta}^{klm)}\bigr)_{\alpha}\ , \\
 \end{split} \hspace{20 mm}
 \begin{split}
 \bigl(\lambda \lambda \bar \lambda \bigr)_{A\;\dot \alpha}^i & \equiv \bigl(\lambda \lambda \bigr)^{jk}_{ab} \bigl(\gamma^{ab} \bar \lambda_{jk}{}^i \bigr)_{\dot \alpha} \ , \\
 \bigl(\lambda \lambda \bar \lambda \bigr)_{A\;a \alpha}^i & \equiv \bigl(\lambda \lambda \bigr)^{jk}_{ab} \bigl(\gamma^{b}\bar \lambda_{jk}{}^i\bigr)_{\alpha}  \ , \\
 \bigl(\lambda \lambda \bar \lambda \bigr)_{abcd\dot \alpha}^i & \equiv \bigl(\lambda \lambda \bigr)^{ijkl}_{abcd} \bar \lambda_{\dot \alpha\; jkl} \ , \\
 \bigl(\lambda \lambda \bar \lambda \bigr)_{A\;\dot \alpha}^{ijk} & \equiv  \bigl(\lambda \lambda \bigr)^{lm(ij} \bar \lambda_{\dot \alpha\; lm}^{k)} \ , \\
 \bigl(\lambda \lambda \bar \lambda \bigr)_{C\;\dot \alpha}^{ijk} & \equiv \bigl(\lambda \lambda \bigr)^{ijklmn}_{ab} \bigl(\gamma^{ab} \bar \lambda_{lmn}\bigr)_{\dot \alpha} \ , \\
 \bigl(\lambda \lambda \bar \lambda \bigr)_{A\;a \alpha}^{ijk} & \equiv \bigl(\lambda \lambda \bigr)^{lm(ij}_{abcd} \bigl(\gamma^{bcd}\bar \lambda_{lm}^{k)}\bigr)_{\alpha} \ , \\
 \bigl(\lambda \lambda \bar \lambda \bigr)_{C\;a \alpha}^{ijk} & \equiv \bigl(\lambda \lambda \bigr)^{ijklmn}_{ab} \bigl(\gamma^{b} \bar \lambda_{lmn}\bigr)_{\alpha} \ , \\
 \bigl(\lambda \lambda \bar \lambda \bigr)_{A\;ab\dot \alpha}^{ijk} & \equiv \bigl(\lambda \lambda \bigr)^{ijklmn}_{ab} \bar \lambda_{\dot \alpha\; lmn} \ , \\
 \bigl(\lambda \lambda \bar \lambda \bigr)_{A\;abcd\dot \alpha}^{ijk} & \equiv \bigl(\lambda \lambda \bigr)^{lm(ij}_{abcd} \bar \lambda_{\dot \alpha\;lm}^{k)} \ , \\
 \bigl(\lambda \lambda \bar \lambda \bigr)_{A\;\dot \alpha}^{ijklm} & \equiv \bigl(\lambda \lambda \bigr)^{(ij}_{ab}  \bigl(\gamma^{ab} \bar \lambda^{klm)}\bigr)_{\dot \alpha} \ , \\
 \bigl(\lambda \lambda \bar \lambda \bigr)_{a \alpha}^{ijklm} & \equiv \bigl(\lambda \lambda \bigr)^{n(ijk}_{abcd} \bigl(\gamma^{bcd} \bar \lambda_{n}{}^{lm)} \bigr)_{\alpha} \ , \\
 \bigl(\lambda \lambda \bar \lambda \bigr)_{B\;a \alpha}^{ijklm} & \equiv \bigl(\lambda \lambda \bigr)^{np(ijkl}_{ab} \bigl(\gamma^{b} \bar \lambda_{np}{}^{m)} \bigr)_{\alpha} \ , \\
 \end{split}
\end{equation}
where we use definition $\eqref{eq:D1U2Basis}$  together with the following ones 
\begin{equation}
 \bigl(\lambda \lambda\bigr)_{abcd}^{i j k l} \equiv \lambda^{m (i j}\gamma_{abcd} \lambda^{k l)}{}_{m}\ ,   \qquad \bigl(\lambda \lambda \bigr)_{ab}^{i j k l m n}  \equiv \lambda^{(i j k} \gamma_{a b} \lambda^{l m n)} \ . 
\end{equation}
Now we can use this basis to write down the solution to the Bianchi identities. The Dirac equation of $\lambda$ gives the following decomposition
\begin{multline}
 D_{a} \lambda_{\alpha}^{ijk}  = (D_{a}\lambda_{\alpha}^{ijk})' + \bigl(\gamma_{a}\bigr)_{\alpha}^{\;\;\;\dot \beta}\biggl(- \frac{i}{32} \bigl(\bar{F} \bar \lambda \bigr)_{\dot \beta}^{ijk} + \frac{i}{768} \bigl(\bar G \bar \lambda \bigr)_{\dot \beta}^{ijk} + \frac{1}{96} \bigl(H \lambda \bigr)_{\dot \beta}^{ijk} \bigr.\\
  - \frac{i}{192}\bigl(\chi \bar{\chi} \lambda \bigr)_{C\;\dot \beta}^{ijk} - \frac{7}{64}  \bigl(\bar \chi \bar \lambda \bar \lambda \bigr)_{\dot \beta}^{ijk} + \frac{i}{3072}  \bigl(\bar \chi \bar \lambda \bar \lambda \bigr)_{A\;\dot \beta}^{ijk}-\frac{3i}{256}  \bigl(\bar \chi \bar \lambda \bar \lambda \bigr)_{B\;\dot \beta}^{ijk} \\
  + \frac{i}{16} \bigl(F \bar \chi \bigr)_{\dot \beta}^{ijk} + \frac{5i}{64} \bigl(\chi \bar{\chi} \lambda \bigr)_{\dot \beta}^{ijk} + \frac{5i}{32}  \bigl(\chi \bar{\chi} \lambda \bigr)_{A\;\dot \beta}^{ijk} - \frac{i}{384} \bigl(\chi \bar{\chi} \lambda \bigr)_{B\;\dot \beta}^{ijk} \\
  + \frac{3i}{64} \bigl(\lambda \lambda \bar \lambda\bigr)_{A\;\dot \beta}^{ijk} - \frac{11 i}{1280} \bigl(\lambda \lambda \bar \lambda\bigr)_{B\;\dot \beta}^{ijk} + \bigl.  \frac{i}{128} \bigl(\lambda \lambda \bar \lambda\bigr)_{C\;\dot \beta}^{ijk} \biggr) 
\end{multline}
The covariant derivative of the scalar momentum $P^{ijkl}$ yields
\begin{multline}
 D_{\alpha}^i P_{a}^{jklm} \\
 =   \varepsilon^{i(j}\bigl(\gamma_a\bigr)_{\alpha}^{\;\;\dot \beta} \biggl( \frac{i}{160} \bigl(\bar{F} \bar \lambda \bigr)_{\dot \beta}^{ijk)} + \frac{i}{256} \bigl(\bar G \bar \lambda \bigr)_{\dot \beta}^{ijk)} + \frac{1}{160} \bigl(H \lambda \bigr)_{\dot \beta}^{ijk)} - \frac{i}{16} \bigl(F \bar \chi \bigr)_{\dot \beta}^{ijk)} + \frac{3i}{64} \bigl(\chi \bar{\chi} \lambda \bigr)_{\dot \beta}^{ijk)} \biggr . \\ 
 \biggl . - \frac{33i}{160}  \bigl(\chi \bar{\chi} \lambda \bigr)_{A\;\dot \beta}^{ijk)} - \frac{7i}{384} \bigl(\chi \bar{\chi} \lambda \bigr)_{B\;\dot \beta}^{ijk)}  - \frac{19i}{960}\bigl(\chi \bar{\chi} \lambda \bigr)_{C\;\dot \beta}^{ijk)} - \frac{21i}{64}  \bigl(\bar \chi \bar \lambda \bar \lambda \bigr)_{\dot \beta}^{ijk)} + \frac{i}{1024}  \bigl(\bar \chi \bar \lambda \bar \lambda \bigr)_{A\;\dot \beta}^{ijk)}\biggr . \\ \biggl .  \hspace{15mm} -\frac{13i}{1280}  \bigl(\bar \chi \bar \lambda \bar \lambda \bigr)_{B\;\dot \beta}^{ijk)}  \frac{i}{64} \bigl(\lambda \lambda \bar \lambda\bigr)_{\dot \beta}^{ijk)} + \frac{57i}{320} \bigl(\lambda \lambda \bar \lambda\bigr)_{A\;\dot \beta}^{ijk)} + \frac{83 i}{6400} \bigl(\lambda \lambda \bar \lambda\bigr)_{B\;\dot \beta}^{ijk)} +   \frac{i}{1920} \bigl(\lambda \lambda \bar \lambda\bigr)_{C\;\dot \beta}^{ijk)} \biggr) \\
+  \varepsilon^{i(j} \biggl(- \frac{i}{5} \bigl(\bar{F} \bar \lambda \bigr)^{klm)}_{a \alpha} +  \frac{1}{20} \bigl(H \lambda \bigr)^{klm)}_{a \alpha}  \bigr. + \frac{i}{2} \bigl(\chi \bar{\chi} \lambda \bigr)^{klm)}_{a \alpha} - \frac{i}{5} \bigl(\chi \bar{\chi} \lambda \bigr)^{klm)}_{A\;a \alpha}+ \frac{i}{20} \bigl(\chi \bar{\chi} \lambda \bigr)^{klm)}_{a \alpha} 
  \biggr . \\ \biggl . 
 \hspace{10mm}  - \frac{i}{1152} \bigl(\lambda \lambda \bar \lambda\bigr)^{klm)}_{a \alpha} +  \frac{i}{480} \bigl(\lambda \lambda \bar \lambda\bigr)^{klm)}_{A\;a \alpha} -  \frac{11i}{200} \bigl(\lambda \lambda \bar \lambda\bigr)^{klm)}_{B\;a \alpha}  +  \frac{i}{120} \bigl(\lambda \lambda \bar \lambda \bigr)^{klm)}_{C\;a \alpha} - (D_{a} \lambda_{\alpha}^{k l m)})' \biggr)  \\
  + \bigl(\gamma_a\bigr)_{\alpha}^{\;\;\dot \beta} \biggl(\frac{i}{24} \bigl(\bar{F} \bar \lambda \bigr)^{ijklm}_{\dot \beta} - \frac{1}{36} \bigl(H \lambda \bigr)^{ijklm}_{\dot \beta} + \frac{i}{12} \bigl(\chi \bar{\chi} \lambda \bigr)^{ijklm}_{\dot \beta} + \frac{i}{24} \bigl(\chi \bar{\chi} \lambda \bigr)^{ijklm}_{A\;\dot \beta}  - \frac{35i}{96} \bigl(\bar \chi \bar \lambda \bar \lambda \bigr)^{ijklm}_{\dot \beta} \biggr . \\
  \biggl .  + \frac{5i}{4608} \bigl(\bar \chi \bar \lambda \bar \lambda \bigr)^{ijklm}_{A\;\dot \beta} - \frac{7i}{96} \bigl(\bar \chi \bar \lambda \bar \lambda \bigr)^{ijklm}_{B\;\dot \beta}  + \frac{17 i}{32} \bigl(\lambda \lambda \bar \lambda\bigr)^{ijklm}_{\dot \beta}   + \frac{31i}{1920} \bigl(\lambda \lambda \bar \lambda\bigr)^{ijklm}_{A\;\dot \beta}  - \bigl.\frac{i}{64} \bigl(\lambda \lambda \bar \lambda\bigr)^{ijklm}_{B\;\dot \beta}\biggr)  \\
  + \frac{i}{3} \bigl(\bar{F} \bar \lambda \bigr)^{ijklm}_{a \alpha}  - \frac{1}{12} \bigl(H \lambda \bigr)^{ijklm}_{a \alpha} + \frac{i}{3} \bigl(\chi \bar{\chi} \lambda \bigr)^{ijklm}_{a \alpha} + \frac{i}{6} \bigl(\bar \chi \bar \lambda \bar \lambda \bigr)^{ijklm}_{a \alpha}  \bigr.\\
  - \frac{i}{192}  \bigl(\lambda \lambda \bar \lambda\bigr)^{ijklm}_{a \alpha} + \frac{29i}{480}  \bigl(\lambda \lambda \bar \lambda\bigr)^{ijklm}_{A\;a \alpha} + \bigl.\frac{i}{16} \bigl(\lambda \lambda \bar \lambda\bigr)^{ijklm}_{B\;a \alpha} \ . 
\end{multline}
The $3/2$ dimensional component of the torsion is
\begin{multline}
 T_{a b}^{\;\;\;\dot \alpha i} = \bigl(\gamma_{ab}\bigr)^{\dot \alpha \dot \beta}\biggl(\frac{i}{21}  \bigl(\bar{P} \chi \bigr)_{\dot \beta}^{i} - \frac{1}{56} \bigl(\bar{F} \bar \lambda \bigr)_{\dot \beta}^{i} + \frac{5 i}{504} \bigl(H \lambda \bigr)_{\dot \beta}^{i} + \frac{i}{24}\bigl(P^4 \lambda \bigr)_{\dot \beta}^{i} \bigr.\\
  \hspace{-25mm}- \frac{1}{28} \bigl(F \bar \chi \bigr)_{\dot \beta}^{i} - \frac{1}{56} \bigl(\chi \bar{\chi} \lambda \bigr)_{\dot \beta}^{i} + \frac{11}{1008} \bigl(\chi \bar{\chi} \lambda \bigr)_{A\;\dot \beta}^{i} + \frac{1}{672}  \bigl(\bar \chi \bar \lambda \bar \lambda \bigr)_{\dot \beta}^{i} \\
 - \frac{1}{32256}  \bigl(\bar \chi \bar \lambda \bar \lambda \bigr)_{A\;\dot \beta}^{i} - \frac{1}{96}  \bigl(\bar \chi \bar \lambda \bar \lambda \bigr)_{B\;\dot \beta}^{i} - \frac{1}{224} \bigl(\lambda \lambda \bar \lambda\bigr)_{\dot \beta}^{i} - \bigl. \frac{11}{2688} \bigl(\lambda \lambda \bar \lambda\bigr)_{A\;\dot \beta}^{i}\biggr) \\
+ \bigl(\gamma_{[a}\bigr)^{\dot \alpha \beta} \biggl(- \frac{2 i}{3} \bigl(\bar{P} \chi \bigr)_{b] \beta}^{i} + \frac{1}{6} \bigl(\bar{F} \bar \lambda \bigr)_{b] \beta}^{i} - \frac{i}{12} \bigl(H \lambda \bigr)_{b] \beta}^{i} \bigr.  -  \frac{i}{3}  \bigl(P^4 \lambda \bigr)_{b] \beta}^{i} + \frac{1}{3} \bigl(F \bar \chi \bigr)_{b] \beta}^{i}  \\
 + \frac{1}{36} \bigl(G \bar \chi \bigr)_{b] \beta}^{i} + \frac{1}{6} \bigl(\chi \bar{\chi} \lambda \bigr)_{b] \beta}^{i} - \frac{1}{12} \bigl(\chi \bar{\chi} \lambda \bigr)_{A\;b] \beta}^{i} +\frac{1}{12} \bigl(\bar \chi \bar \lambda \bar \lambda \bigr)_{b] \beta}^{i} \\
 +  \frac{1}{576}   \bigl( \lambda \lambda \bar \lambda \bigr)_{b] \beta}^{i} + \bigl. \frac{1}{32}   \bigl( \lambda \lambda \bar \lambda \bigr)_{A\;b] \beta}^{i}\biggr) + \bar \rho_{ab}^{\dot \alpha i}
\end{multline}
where we have defined the projection to the irreducible representation \DSOVIII0110 to be $\bar \rho$. \\[0.5 cm]
We will now give the fermionic covariant derivative of the field strength $\bar G^{-}$, $\bar F$ and $H$ having $U(1)$ weight $1$. We get
\begin{multline}
 \bar D_{\dot \alpha i} \bar G^{-}_{abcd} = \bigl(\gamma_{abcd}\bigr)_{\dot \alpha}^{\;\;\;\dot \beta} \biggl(-\frac{2i}{7}  \bigl(\bar{P} \chi \bigr)_{\dot \beta i} + \frac{3}{28} \bigl(\bar{F} \bar \lambda \bigr)_{\dot \beta i} - \frac{9 i}{112} \bigl(H \lambda \bigr)_{\dot \beta i} + \frac{3i}{28}\bigl(P^4 \lambda \bigr)_{\dot \beta i} \bigr.\\
  \hspace{- 5 mm} + \frac{1}{21} \bigl(F \bar \chi \bigr)_{\dot \beta i} - \frac{5}{84} \bigl(\chi \bar{\chi} \lambda \bigr)_{\dot \beta i} - \frac{13}{252} \bigl(\chi \bar{\chi} \lambda \bigr)_{A\;\dot \beta i} - \frac{1}{112}  \bigl(\bar \chi \bar \lambda \bar \lambda \bigr)_{\dot \beta i} \\
 + \frac{73}{16128}  \bigl(\bar \chi \bar \lambda \bar \lambda \bigr)_{A\;\dot \beta i} - \frac{1}{48}  \bigl(\bar \chi \bar \lambda \bar \lambda \bigr)_{B\;\dot \beta i} + \frac{3}{112} \bigl(\lambda \lambda \bar \lambda\bigr)_{\dot \beta i} + \frac{1}{896} \bigl(\lambda \lambda \bar \lambda\bigr)_{A\;\dot \beta i}\biggr)  \\ + \bigl(\gamma_{[abc}\bigr)_{\dot \alpha}^{\;\;\;\beta} \biggl(- \bigl(\bar{F} \bar \lambda \bigr)_{d] \beta i} + i \bigl(H \lambda \bigr)_{d] \beta i} \bigr. - \frac{4}{3} \bigl(F \bar \chi \bigr)_{d] \beta i}  + \frac{2}{3} \bigl(\chi \bar{\chi} \lambda \bigr)_{A\;d] \beta i} + \bigl. \frac{1}{3} \bigl(\bar \chi \bar \lambda \bar \lambda \bigr)_{d] \beta i}\biggr)\\
  + \bigl(\gamma_{[ab}\bigr)_{\dot \alpha}^{\;\;\;\dot \beta} \biggl(- 6 \bar \rho_{cd ]  \dot \beta i} -3 \bigl(\bar{F} \bar \lambda \bigr)_{cd ]  \dot \beta i} \bigr. + 3 i \bigl(H \lambda \bigr)_{cd ]  \dot \beta i} -  4 \bigl(F \bar \chi \bigr)_{cd ]  \dot \beta i}  + 2 \bigl(\chi \bar{\chi} \lambda \bigr)_{cd ]  \dot \beta i}  \biggr .  \\ \biggl  . + \frac{1}{4} \bigl(\bar \chi \bar \lambda \bar \lambda \bigr)_{cd ]  \dot \beta i} + \bigl(\bar \chi \bar \lambda \bar \lambda \bigr)_{A\;cd ]  \dot \beta i} \biggr) + \frac{1}{12}  \bigl(\bar \chi \bar \lambda \bar \lambda \bigr)_{abcd \dot \alpha i}\ , 
\end{multline}
and
\begin{multline}
 \bar D_{i \dot \alpha} \bar F_{ab}^{jk} \\
 =  \delta_{i}^{(j} \biggl(2 \bar \rho_{ab \dot \alpha}^{k)}+\frac{5}{6} \bigl(\bar{F} \bar \lambda \bigr)_{ab \dot \alpha}^{k)} - \frac{4 i}{9} \bigl(H \lambda \bigr)_{ab \dot \alpha}^{k)} +  \frac{20}{9} \bigl(F \bar \chi \bigr)_{ab \dot \alpha}^{k)}\bigr. + \frac{2}{9} \bigl(\chi \bar{\chi} \lambda \bigr)_{ab \dot \alpha}^{k)} -\frac{1}{24} \bigl(\bar \chi \bar \lambda \bar \lambda \bigr)_{ab \dot \alpha}^{k)} + \frac{1}{9} \bigl(\bar \chi \bar \lambda \bar \lambda \bigr)_{A\; ab \dot \alpha}^{k)} \biggr)\\
   - \frac{1}{2}  \bigl(\bar{F} \bar \lambda \bigr)_{ab \dot \alpha\;i}^{jk} + \frac{2i}{3}  \bigl(H \lambda \bigr)_{ab \dot \alpha\;i}^{jk} - \frac{2}{3} \bigl(F \bar \chi \bigr)_{ab \dot \alpha\;i}^{jk} + \frac{1}{3} \bigl(\chi \bar{\chi} \lambda \bigr)_{ab \dot \alpha\;i}^{jk} + \frac{2}{3} \bigl(\chi \bar{\chi} \lambda \bigr)_{A\;ab \dot \alpha\;i}^{jk} + \frac{1}{3} \bigl(\bar \chi \bar \lambda \bar \lambda \bigr)_{ab \dot \alpha\;i}^{jk}  \\
 + \delta_{i}^{(j} \bigl(\gamma_{[a}\bigr)_{\dot \alpha}^{\;\;\;\beta} \biggl( \frac{8 i}{3} \bigl(\bar{P} \chi \bigr)_{b] \beta}^{k)} + \frac{1}{3} \bigl(\bar{F} \bar \lambda \bigr)_{b] \beta}^{k)} - \frac{i}{18} \bigl(H \lambda \bigr)_{b] \beta}^{k)} \bigr. + \frac{2 i}{3}  \bigl(P^4 \lambda \bigr)_{b] \beta}^{k)}  +  \frac{8}{9} \bigl(F \bar \chi \bigr)_{b] \beta}^{k)} - \frac{1}{9} \bigl(G \bar \chi \bigr)_{b] \beta}^{k)}  \biggr . \\ 
 \biggl . + \frac{1}{3} \bigl(\chi \bar{\chi} \lambda \bigr)_{b] \beta}^{k)} + \frac{1}{9} \bigl(\bar \chi \bar \lambda \bar \lambda \bigr)_{b] \beta}^{k)}  - \frac{1}{288}   \bigl( \lambda \lambda \bar \lambda \bigr)_{b] \beta}^{k)}  - \bigl. \frac{1}{16}   \bigl( \lambda \lambda \bar \lambda \bigr)_{A\;b] \beta}^{k)}\biggr) \\
 + \bigl(\gamma_{[a}\bigr)_{\dot \alpha}^{\;\;\;\beta} \biggl(- \frac{i}{6} \bigl(H \lambda \bigr)_{b] \beta\;i}^{jk} + 2i \bigl(P^4 \lambda \bigr)_{b] \beta\;i}^{jk} - \frac{2}{3}\bigl(F \bar \chi \bigr)_{b] \beta\;i}^{jk}
  - \frac{1}{3} \bigl(\chi \bar{\chi} \lambda \bigr)_{b] \beta\;i}^{jk} - 2 \bigl(\chi \bar{\chi} \lambda \bigr)_{A\;b] \beta\;i}^{jk} 
  \biggr . \\ \biggl .  - \frac{1}{6} \bigl(\lambda \lambda \bar \lambda\bigr)_{b] \beta\;i}^{jk} + \frac{1}{96} \bigl(\lambda \lambda \bar \lambda\bigr)_{A\;b] \beta\;i}^{jk}  - \frac{3}{40} \bigl(\lambda \lambda \bar \lambda\bigr)_{B\;b] \beta\;i}^{jk} +\bigl. \frac{1}{3} \bigl(\lambda \lambda \bar \lambda \bigr)_{C\;b] \beta\;i}^{jk} - 2 i (D_{b]} \lambda_{\beta\;i}^{jk})' \biggr) \\
 \hspace{-25mm}+ \delta_i^{(j} \bigl(\gamma_{ab}\bigr)_{\dot \alpha}^{\;\;\; \dot \beta} \biggl(\frac{2 i}{21}  \bigl(\bar{P} \chi \bigr)_{\dot \beta}^{k)} - \frac{1}{28} \bigl(\bar{F} \bar \lambda \bigr)_{\dot \beta}^{k)} - \frac{13 i}{756} \bigl(H \lambda \bigr)_{\dot \beta}^{k)} + \frac{i}{21}\bigl(P^4 \lambda \bigr)_{\dot \beta}^{k)} \bigr.\\
  - \frac{1}{63} \bigl(F \bar \chi \bigr)_{\dot \beta}^{k)} - \frac{5}{42} \bigl(\chi \bar{\chi} \lambda \bigr)_{\dot \beta}^{k)} - \frac{43}{756} \bigl(\chi \bar{\chi} \lambda \bigr)_{A\;\dot \beta}^{k)} + \frac{1}{366}  \bigl(\bar \chi \bar \lambda \bar \lambda \bigr)_{\dot \beta}^{k)} \\
  \hspace{20mm} - \frac{5}{5376}  \bigl(\bar \chi \bar \lambda \bar \lambda \bigr)_{A\;\dot \beta}^{k)} + \frac{1}{144}  \bigl(\bar \chi \bar \lambda \bar \lambda \bigr)_{B\;\dot \beta}^{k)} + \frac{11}{336} \bigl(\lambda \lambda \bar \lambda\bigr)_{\dot \beta}^{k)} +\bigl. \frac{31}{1344} \bigl(\lambda \lambda \bar \lambda\bigr)_{A\;\dot \beta}^{k)}\biggr) \\
   + \bigl(\gamma_{ab}\bigr)_{\dot \alpha}^{\;\;\; \dot \beta} \biggl(- \frac{1}{16} \bigl(\bar{F} \bar \lambda \bigr)_{\dot \beta\;i}^{jk} + \frac{1}{384} \bigl(\bar G \bar \lambda \bigr)_{\dot \beta\;i}^{jk} + \frac{5 i}{144} \bigl(H \lambda \bigr)_{\dot \beta\;i}^{jk} \bigr. - \frac{1}{24} \bigl(F \bar \chi \bigr)_{\dot \beta\;i}^{jk}  + \frac{11}{96} \bigl(\chi \bar{\chi} \lambda \bigr)_{\dot \beta\;i}^{jk} \biggr . \\
   \biggl . - \frac{7}{16} \bigl(\chi \bar{\chi} \lambda \bigr)_{A\;\dot \beta\;i}^{jk}  + \frac{1}{576} \bigl(\chi \bar{\chi} \lambda \bigr)_{B\;\dot \beta\;i}^{jk} +\frac{1}{288} \bigl(\chi \bar{\chi} \lambda \bigr)_{C\;\dot \beta\;i}^{jk}   - \frac{7}{32}  \bigl(\bar \chi \bar \lambda \bar \lambda \bigr)_{\dot \beta\;i}^{jk} + \frac{1}{1536}  \bigl(\bar \chi \bar \lambda \bar \lambda \bigr)_{A\;\dot \beta\;i}^{jk}\\+ \frac{23}{384}  \bigl(\bar \chi \bar \lambda \bar \lambda \bigr)_{B\;\dot \beta\;i}^{jk} - \frac{1}{48} \bigl(\lambda \lambda \bar \lambda\bigr)_{\dot \beta\;i}^{jk} 
   - \frac{3}{32} \bigl(\lambda \lambda \bar \lambda\bigr)_{A\;\dot \beta\;i}^{jk} + \frac{13}{640} \bigl(\lambda \lambda \bar \lambda\bigr)_{B\;\dot \beta\;i}^{jk} - \bigl.  \frac{3}{64} \bigl(\lambda \lambda \bar \lambda\bigr)_{C\;\dot \beta\;i}^{jk} \biggr)\ , 
\end{multline}
and finally 
\begin{multline}
 D_{\alpha}^i H_{abc}^{jk} \\
\hspace{-22mm} = \bigl(\gamma_{abc}\bigr)_{\alpha}^{\;\;\;\dot \beta}\biggl(- \frac{i}{64} \bigl(\bar{F} \bar \lambda \bigr)_{\dot \beta}^{ijk} + \frac{3i}{512} \bigl(\bar G \bar \lambda \bigr)_{\dot \beta}^{ijk} - \frac{11}{192} \bigl(H \lambda \bigr)_{\dot \beta}^{ijk} \bigr. - \frac{3i}{32} \bigl(F \bar \chi \bigr)_{\dot \beta}^{ijk}  + \frac{9i}{128} \bigl(\chi \bar{\chi} \lambda \bigr)_{\dot \beta}^{ijk}\biggr . \\
 \biggl  .  - \frac{7i}{64} \bigl(\chi \bar{\chi} \lambda \bigr)_{A\;\dot \beta}^{ijk}  -  \frac{7i}{256} \bigl(\chi \bar{\chi} \lambda \bigr)_{B\;\dot \beta}^{ijk} + \frac{9 i}{128} \bigl(\chi \bar{\chi} \lambda \bigr)_{C\;\dot \beta}^{ijk} 
   - \frac{7 i}{128}  \bigl(\bar \chi \bar \lambda \bar \lambda \bigr)_{\dot \beta}^{ijk}  + \frac{i}{6144}  \bigl(\bar \chi \bar \lambda \bar \lambda \bigr)_{A\;\dot \beta}^{ijk}  \\
 \hspace{15mm}   + \frac{29 i}{512}  \bigl(\bar \chi \bar \lambda \bar \lambda \bigr)_{B\;\dot \beta}^{ijk}
    - \frac{15 i}{128} \bigl(\lambda \lambda \bar \lambda\bigr)_{A\;\dot \beta}^{ijk} 
  - \frac{71 i}{2560} \bigl(\lambda \lambda \bar \lambda\bigr)_{B\;\dot \beta}^{ijk} + \bigl. \frac{i}{256} \bigl(\lambda \lambda \bar \lambda\bigr)_{C\;\dot \beta}^{ijk} \biggr)\\
   + \bigl(\gamma_{abc}\bigr)_{\alpha}^{\;\;\;\dot \beta}  \varepsilon^{i(j} \biggl(-\frac{1}{7}  \bigl(\bar{P} \chi \bigr)_{\dot \beta}^{k)}  - \frac{i}{84} \bigl(\bar{F} \bar \lambda \bigr)_{\dot \beta}^{k)} - \frac{1}{504} \bigl(H \lambda \bigr)_{\dot \beta}^{k)} - \frac{1}{14}\bigl(P^4 \lambda \bigr)_{\dot \beta}^{k)} - \frac{3i}{28} \bigl(F \bar \chi \bigr)_{\dot \beta}^{k)} - \frac{23 i}{168} \bigl(\chi \bar{\chi} \lambda \bigr)_{\dot \beta}^{k)} \\
    \hspace{10mm} - \frac{i}{112} \bigl(\chi \bar{\chi} \lambda \bigr)_{A\;\dot \beta}^{k)} - \frac{23 i}{336}  \bigl(\bar \chi \bar \lambda \bar \lambda \bigr)_{\dot \beta}^{k)} +  \frac{i}{8064} \bigl(\bar \chi \bar \lambda \bar \lambda \bigr)_{A\;\dot \beta}^{k)} + \frac{i}{48}  \bigl(\bar \chi \bar \lambda \bar \lambda \bigr)_{B\;\dot \beta}^{k)} 
  + \frac{i}{56} \bigl(\lambda \lambda \bar \lambda\bigr)_{\dot \beta}^{k)} +\bigl. \frac{i}{1344} \bigl(\lambda \lambda \bar \lambda\bigr)_{A\;\dot \beta}^{k)}\biggr)\\
  + \bigl(\gamma_{[ab}\bigr)_{\alpha}^{\;\;\;\beta}\biggl(\frac{3}{8} \bigl(H \lambda \bigr)_{c] \beta}^{ijk}  - \frac{3}{2} \bigl(P^4 \lambda \bigr)_{c] \beta}^{ijk}+ \frac{3i}{8}  \bigl(\chi \bar{\chi} \lambda \bigr)_{B\;c] \beta}^{ijk} -\frac{3i}{4} \bigl(\chi \bar{\chi} \lambda \bigr)_{C\;c] \beta}^{ijk}-\frac{3i}{8} \bigl(\bar \chi \bar \lambda \bar \lambda\bigr)_{c] \beta}^{ijk} \\
    \hspace{15mm} - \frac{i}{96} \bigl(\lambda \lambda \bar \lambda\bigr)_{c] \beta}^{ijk} - \frac{i}{128}\bigl(\lambda \lambda \bar \lambda\bigr)_{A\;c] \beta}^{ijk} + \frac{9i}{32} \bigl(\lambda \lambda \bar \lambda\bigr)_{B\;c] \beta}^{ijk} - \bigl. \frac{3}{2} (D_{c]} \lambda_{\beta}^{ijk})' \biggr) \\
\hspace{-3mm}+ \bigl(\gamma_{[ab}\bigr)_{\alpha}^{\;\;\;\beta}   \varepsilon^{i(j} \biggl(2 \bigl(\bar{P} \chi \bigr)_{c] \beta}^{k)} + \frac{i}{2} \bigl(\bar{F} \bar \lambda \bigr)_{c] \beta}^{k)} + i  \bigl(F \bar \chi \bigr)_{c] \beta}^{k)} \bigr. + \frac{i}{12} \bigl(G \bar \chi \bigr)_{c] \beta}^{k)} + \frac{i}{2} \bigl(\chi \bar{\chi} \lambda \bigr)_{c] \beta}^{k)}  + \frac{i}{4} \bigl(\chi \bar{\chi} \lambda \bigr)_{A\;c] \beta}^{k)} - \frac{i}{4} \bigl. \bigl(\bar \chi \bar \lambda \bar \lambda \bigr)_{c] \beta}^{k)} \biggr)  \\
+ \bigl(\gamma_{[a}\bigr)_{\alpha}^{\;\;\;\dot\beta} \biggl( \frac{3 i}{2}  \bigl(\bar{F} \bar \lambda \bigr)_{bc] \dot \beta}^{ijk} + \bigr. \frac{3 i}{8} \bigl(\bar G \bar \lambda \bigr)_{bc] \dot \beta}^{ijk} -  3 i \bigl(F \bar \chi \bigr)_{bc] \dot \beta}^{ijk} + \frac{3 i}{4} \bigl(\chi \bar{\chi} \lambda \bigr)_{bc] \dot \beta}^{ijk}  - \frac{3 i}{2} \bigl(\chi \bar{\chi} \lambda \bigr)_{A\;bc] \dot \beta}^{ijk} + \bigl. \frac{3 i}{4} \bigl(\bar \chi \bar \lambda \bar \lambda \bigr)_{bc] \dot \beta}^{ijk} \biggr) \\
+ \bigl(\gamma_{[a}\bigr)_{\alpha}^{\;\;\;\dot\beta}  \varepsilon^{i(j} \biggl(2 i \bigl(\bar{F} \bar \lambda \bigr)_{bc] \dot \beta}^{k)} + 4 i \bigl(F \bar \chi \bigr)_{bc] \dot \beta}^{k)} +  i   \bigl(\chi \bar{\chi} \lambda \bigr)_{bc] \dot \beta}^{k)} + \bigl. \frac{i }{4} \bigl(\lambda \lambda \bar \lambda \bigr)_{bc] \dot \beta}^{k)}  + 3 i \bar \rho_{bc] \dot \beta}^{k)} \biggr)  \\
  + \frac{1}{6}  \varepsilon^{i(j} \bigl(H \lambda \bigr)_{abc \alpha}^{k)} + \frac{1}{2} \bigl(H \lambda \bigr)_{abc \alpha}^{ijk} -\frac{i}{16} \bigl(\gamma^d\bigr)_{\alpha}^{\;\;\;\dot\beta}  \bigl( \lambda \lambda \bar \lambda \bigr)_{abcd \dot \beta}^{ijk}\ . 
\end{multline}

\end{appendix}

\end{document}